\documentclass[fleqn,usedcolumn]{mnras}
\usepackage[british]{babel}             
\usepackage{newtxtext}                  
\usepackage[slantedGreek]{newtxmath}    
\let\la=\lesssim 
\usepackage[T1]{fontenc}                
\usepackage{graphicx}                   

\title[Dust temperature and time-dependent effects in photodissociation regions]{Dust temperature and time-dependent effects in the chemistry of photodissociation regions}

\author[G. Esplugues.]{G. Esplugues$^{1,2}$\thanks{E-mail: gisela@mpe.mpg.de}, S. Cazaux$^{3,4}$, P. Caselli$^{1}$, S. Hocuk$^{5,1}$, M. Spaans$^{2}$\\
$^{1}$Max Planck Institute for Extraterrestrial Physics, Giessenbachstrasse 1, D-85748 Garching, Germany\\
$^{2}$Kapteyn Astronomical Institute, University of Groningen, P.O. Box 800, NL 9700 AV Groningen, The Netherlands\\
$^{3}$Faculty of Aerospace Engineering, Delft University of Technology, Kluyverweg 1, 2629 HS, Delft, The Netherlands\\
$^{4}$University of Leiden, P.O. Box 9513, NL, 2300 RA, Leiden, The Netherlands\\
$^{5}$CentERdata, Tilburg University, P.O. Box 90153, 5000 LE, Tilburg, The Netherlands\\
}

\begin{document}

\pagerange{\pageref{firstpage}--\pageref{lastpage}} \pubyear{2018}

\maketitle

\label{firstpage}

\begin{abstract}

\noindent When studying chemistry of PDRs, time dependence becomes important as visual extinction increases, since certain chemical timescales are comparable to the cloud lifetime. Dust temperature is also a key factor, since it significantly influences gas temperature and mobility on dust grains, determining the chemistry occurring on grain surfaces. We present a study of the dust temperature impact and time effects on the chemistry of different PDRs, using an updated version of the Meijerink PDR code and combining it with the time-dependent code Nahoon. We find the largest temperature effects in the inner regions of high $G$$_{\mathrm{0}}$ PDRs, where high dust temperatures favour the formation of simple oxygen-bearing molecules (especially that of O$_2$), while the formation of complex organic molecules is much more efficient at low dust temperatures. We also find that time-dependent effects strongly depend on the PDR type, since long timescales promote the destruction of oxygen-bearing molecules in the inner parts of low $G$$_{\mathrm{0}}$ PDRs, while favouring their formation and that of carbon-bearing molecules in high $G$$_{\mathrm{0}}$ PDRs. From the chemical evolution, we also conclude that, in dense PDRs, CO$_2$ is a late-forming ice compared to water ice, and confirm a layered ice structure on dust grains, with H$_2$O in lower layers than CO$_2$. Regarding steady state, the PDR edge reaches chemical equilibrium at early times ($\lesssim$10$^5$ yr). This time is even shorter ($<$10$^4$ yr) for high $G$$_{\mathrm{0}}$ PDRs. By contrast, inner regions reach equilibrium much later, especially low $G$$_{\mathrm{0}}$ PDRs, where steady state is reached at $\sim$10$^6$-10$^7$ yr.

\end{abstract}


\section{Introduction}

Photodissociation regions (PDRs) are characterised by their exposure to strong far-ultraviolet (FUV) radiation fields (6$<$h$\nu$$<$13.6 eV), which results in the heating of gas up to relatively high temperatures. These regions are important from a chemical point of view, since they play a key role in the formation of new species along the molecular cloud, as the UV radiation penetrates into the region.

Photodissociation regions can be found in different environments of the Milky Way, such as in massive star-forming regions (e.g. Tauber et al. 1994; Hora et al. 2004), close to cooler stars which emit enough FUV radiation to form lower-density and lower-excitation PDRs (e.g. Wyrsowski et al. 2000; K\"ohler et al. 2014), on the surface of protoplanetary disks (e.g. van Dishoeck et al. 2006; Bergin et al. 2007), at the edge of molecular clouds (e.g. Spezzano et al. 2016), and also near evolved stars which emit strong FUV radiation (Meixner et al. 2001). This is also the case of the PDRs detected in planetary nebulae (PNe) through the emission of atomic fine structure lines, e.g., the ground state line of [CI] in NGC\ 6720 and in NGC\ 7293 (Bachiller et al. 1994, Young et al. 1997). Photodissociation regions are also responsible for most of the non-stellar infrared emission from galaxies (e.g. Fuente et al. 2008; Bayet et al. 2009).

The large variety of environments where PDRs are found determines their physical conditions. In particular, PDRs can be diffuse, with gas density $n$$\sim$10-10$^2$ cm$^{-3}$, or dense, with $n$$>$10$^4$ cm$^{-3}$, while the incident FUV flux may range from the interstellar radiation field (ISRF) to 10$^6$ times the ISRF in the surroundings of an O star. Photodissociation regions are characterised by a layered structure, as a result of the interaction of the radiation with the gas and dust. Typically, they contain an outer layer (the edge of the cloud where visual extinction is $A$$_{\mathrm{V}}$$<$1 mag) of partially ionized gas, where hydrogen is atomic and carbon is predominantly in the  form of C$^+$. The transition to molecular hydrogen occurs in a region where carbon is still ionized, while the neutral carbon layer and the transition to CO occur where hydrogen is already fully molecular (e.g. Tielens $\&$ Hollenbach et al. 1985; Joblin et al. 2018).

Ultraviolet photons dominating the energy balance of PDRs do not only influence significantly their chemical structure, but also the time evolution of the interstellar medium (ISM) conditions regulating the star formation processes. At low visual extinctions, physical and chemical processes dominated by interactions with photons are fast compared to dynamical processes. However, at large visual extinctions, certain chemical timescales are comparable to cloud lifetimes and time dependence becomes a key factor in the study of PDRs. 

There are several authors (e.g. Bertoldi \& Draine 1996; Morata et al. 2008; Hollenbach et al. 2009; Kirsanova et al. 2009; Motoyama et al. 2015; Le Gal et al. 2017) who include time dependence in their PDR codes, however they do consider a simpler treatment of surface chemistry than in the present study. In Esplugues et al. (2016), we showed not only the effects of varying the density and the intensity of the radiation field on the chemical evolution of different PDRs, but also the importance of considering surface chemistry when studying the chemical structure of molecular clouds exposed to different UV radiation fields. We derived that some parameters (such as the type of grain substrate and the probability of desorption) can alter the chemistry occurring on grain surfaces, leading to significant differences in the abundances of gas-phase species. Esplugues et al. (2016) also showed that many of these differences become even larger as the visual extinction increases, making evident the need of considering time dependence.    

In this paper, we focus on time dependence and its effects on the chemical evolution of different PDRs, as well as on the role of dust temperature ($T$$_{\mathrm{dust}}$) in the PDR chemistry. 
We carry out this study using an updated version of the {\it{Meijerink}} PDR code (presented in Sect. \ref{Numerical_mode}) with new solid species and surface chemical reactions, as well as with a new way to calculate the chemical desorption\footnote{Chemical desorption process occurs when there is excess energy after the two-body reaction on dust grains. In order to desorb, the newly formed molecule has to convert a fraction of this excess formation energy into kinetic energy and, in particular, into motion perpendicular to the substrate (Minissale \& Dulieu 2014, Minissale et al. 2016).} probabilities for two-body reactions. In Sect. \ref{Dust_temperat}, we present the temperature study considering two different expressions for $T$$_{\mathrm{dust}}$. In Sect. \ref{Time-depend.}, we combine our steady-state code with the time-dependent code Nahoon to anlayse the chemical evolution as a function of time and visual extinction. Section \ref{Results} contains the discussion of results, and a comparison with observations. In addition, we provide results for the time at which steady state is reached in each PDR type. A summary of the main conclusions is presented in Sect. \ref{conclusions}.

\section{The steady-state PDR code}
\label{Numerical_mode}

\subsection{Gas chemistry}
\label{Gas_chemistry}

The updated {\it{Meijerink}} PDR code consists of 7503 chemical gas-phase reactions from the Kinetic Database for Astrochemistry (KIDA)\footnote{http://kida.obs.u-bordeaux1.fr}. They include bimolecular reactions, charge-exchange reactions, radiative associations, associative detachment, dissociative recombination, neutralisation reactions, ion-neutral reactions, ionisation or dissociation of neutral species by UV photons, and ionisation or dissociation of species by direct collision with cosmic-ray particles or by secondary UV photons following H$_2$ excitation.

The heating mechanisms considered in the thermal balance of the code are photoelectric effect on grains, carbon ionisation heating, H$_2$ photodissociation heating by UV photons, H$_2$ collisional de-excitation heating, gas-grain collisional heating, gas-grain viscous heating, and cosmic-ray heating. As cooling mechanisms, we consider fine-structure line cooling (being [CII] at 158 $\mu$m and [OI] at 63 $\mu$m and at 146 $\mu$m the most prominent cooling lines), metastable-line cooling (including lines of C, C$^+$, Si, Si$^+$, O, O$^+$, S, S$^+$, Fe, and Fe$^+$), recombination cooling, and molecular cooling by H$_2$, CO, and H$_2$O (see Meijerink \& Spaans 2005 and Esplugues et al. 2016 for more details).

\subsection{Dust chemistry}
\label{Dust_chemistry_changes}

\begin{table}
\caption{Solid species in our PDR code.}             
\centering 
\begin{tabular}{l l l l l l l}     
\hline 
H                   & \vline & H$_2$      &   \vline & HCO       &  \vline & C \\
H$_{\mathrm{c}}$    & \vline & HO$_2$     &   \vline & H$_2$CO   &  \vline & CH \\
O                   & \vline & H$_2$O     &   \vline & CH$_3$O   &  \vline & CH$_2$\\
O$_2$               & \vline & H$_2$O$_2$ &   \vline & CH$_3$OH  &  \vline & CH$_3$\\
O$_3$                  & \vline & CO         &   \vline & N         &  \vline & CH$_4$ \\
OH               & \vline & CO$_2$     &   \vline & N$_2$     &  \vline & S \\
\hline 
\label{table:solid_species}               
\end{tabular}

\medskip
H$_{\mathrm{c}}$ refers to the strong interaction between hydrogen and the grain surface (chemisorption), where the forces involved are similar to valence forces (see Cazaux $\&$ Tielens 2002 for more details).  
\end{table}

In a precedent study, we updated the Meijerink PDR core with the chemistry occurring on grain surfaces and added 18 solid species. In the present study, we have included 6 additional solid species: S, C, CH, CH$_2$, CH$_3$, and CH$_4$ (see all the solid species considered in Table \ref{table:solid_species}). We have also updated the surface chemical network implemented in the {\it{Meijerink}} code taking recent laboratory experiments (e.g. Dulieu et al. 2013; Minissale et al. 2015, 2016) into account. The surface processes considered in the code are adsorption, thermal desorption, chemical desorption, two-body reactions, photo processes, and cosmic-ray processes. All these processes are described in detail in Esplugues et al. (2016). The other main change introduced in this new version of the code is the way to calculate the chemical desorption probabilities for two-body reactions in order to take more scenarios for the formation of chemical products into account. In particular, in the previous version of the Meijerink code, given the surface chemical reaction JA + JB $\rightarrow$ JC + JD (where J$i$ means solid $i$), we considered only two possibilities based on an empirical physical model adjusted on experimental data:

\begin{equation}
\mathrm{R1)}\ \quad \mathrm{JA + JB \rightarrow JC + JD} 
\label{equation:1}
\end{equation}

\noindent and    

\begin{equation}
\mathrm{R2)}\ \quad \mathrm{JA + JB \rightarrow C + D}. 
\label{equation:2}
\end{equation}

\noindent In this new version of the Meijerink code, however, we propose one way to extend it by considering chemical desorption per product, which implies four possibilities:

\begin{equation}
\mathrm{R1)}\ \quad \mathrm{JA + JB \rightarrow JC + JD}, 
\label{equation:3}
\end{equation}

\begin{equation}
\mathrm{R2)}\ \quad \mathrm{JA + JB \rightarrow C + D},  
\label{equation:4}
\end{equation}

\begin{equation}
\mathrm{R3)}\ \quad \mathrm{JA + JB \rightarrow JC + D}, 
\label{equation:5}
\end{equation}
    
\noindent and

\begin{equation}
\mathrm{R4)}\ \quad \mathrm{JA + JB \rightarrow C + JD},   
\label{equation:6}
\end{equation}

\noindent where the chemical desorption coefficients CD$_{\mathrm{JC}}$ and CD$_{\mathrm{JD}}$ of the species JC and JD, respectively, are independent and calculated using:

\begin{equation}
\mathrm{CD}=  exp\left(\displaystyle{ \frac{-E_{\mathrm{binding}}}{\epsilon \Delta H_{\mathrm{R}}/N}}\right).  
\end{equation}

\noindent The factor $E$$_{\mathrm{binding}}$ is the binding energy of the desorbed product ($E$$_{\mathrm{binding}}$(JC) for the case of CD$_{\mathrm{JC}}$ and $E$$_{\mathrm{binding}}$(JD) for the case of CD$_{\mathrm{JD}}$, using values shown in Table \ref{table:binding_energies}), $\epsilon$$\Delta$H$_{\mathrm{R}}$/$N$ represents the total chemical energy available for the kinetic energy perpendicular to the grain surface, $\Delta$H$_{\mathrm{R}}$ being the reaction enthalpy, $N$=3$\times$$n$$_{\mathrm{atoms}}$ is the degree of freedom considering the atoms of the two newly formed molecules, and $\epsilon$ the fraction of kinetic energy retained by the product of mass $m$ colliding with the surface, which has an effective mass $M$ (see Minissale et al. 2016, Cazaux et al. 2016 for more details):

\begin{equation}
\epsilon=  \frac{(M-m)^2}{(M+m)^2}.   
\end{equation}

\noindent The desorption probabilities for the four chemical reactions are:

\begin{equation}
  \mathrm{\delta_{R1} = 100 - max(CD_{JC},CD_{JD})}, 
\label{equation:3_1}
\end{equation}

\begin{equation}
  \mathrm{\delta_{R2} = min(CD_{JC},CD_{JD})}, 
\label{equation:3_2}
\end{equation}

\begin{equation}
  \mathrm{\delta_{R3} = CD_{JD} - min(CD_{JC},CD_{JD})}, 
\label{equation:3_3}
\end{equation}

\noindent and

\begin{equation}
\mathrm{\delta_{R4} = CD_{JC} - min(CD_{JC},CD_{JD})}, 
\label{equation:4_1}
\end{equation}

\noindent where  $\delta$$_{\mathrm{R1}}$+$\delta$$_{\mathrm{R2}}$+$\delta$$_{\mathrm{R3}}$+$\delta$$_{\mathrm{R4}}$=100$\%$.
In this case, unlike Esplugues et al. (2016), we calculate the desorption probabilities for each reaction using the binding energies of both products. This new approach considers therefore the fact that C and D are different products, with different energies and different degrees of freedom, and that, in exothermic reactions, the energy released is dissipated in a different manner for C and D. In addition, this formulation also reproduces the experimental results where only one product is observed, even if the considered reactions would have two products.
See Appendix \ref{Tables} for the list of chemical reactions occurring on grain surfaces that are included in the Meijerink PDR code.

\section{Dust temperature}
\label{Dust_temperat}

\begin{figure*}
\centering
\includegraphics[scale=0.395, angle=0]{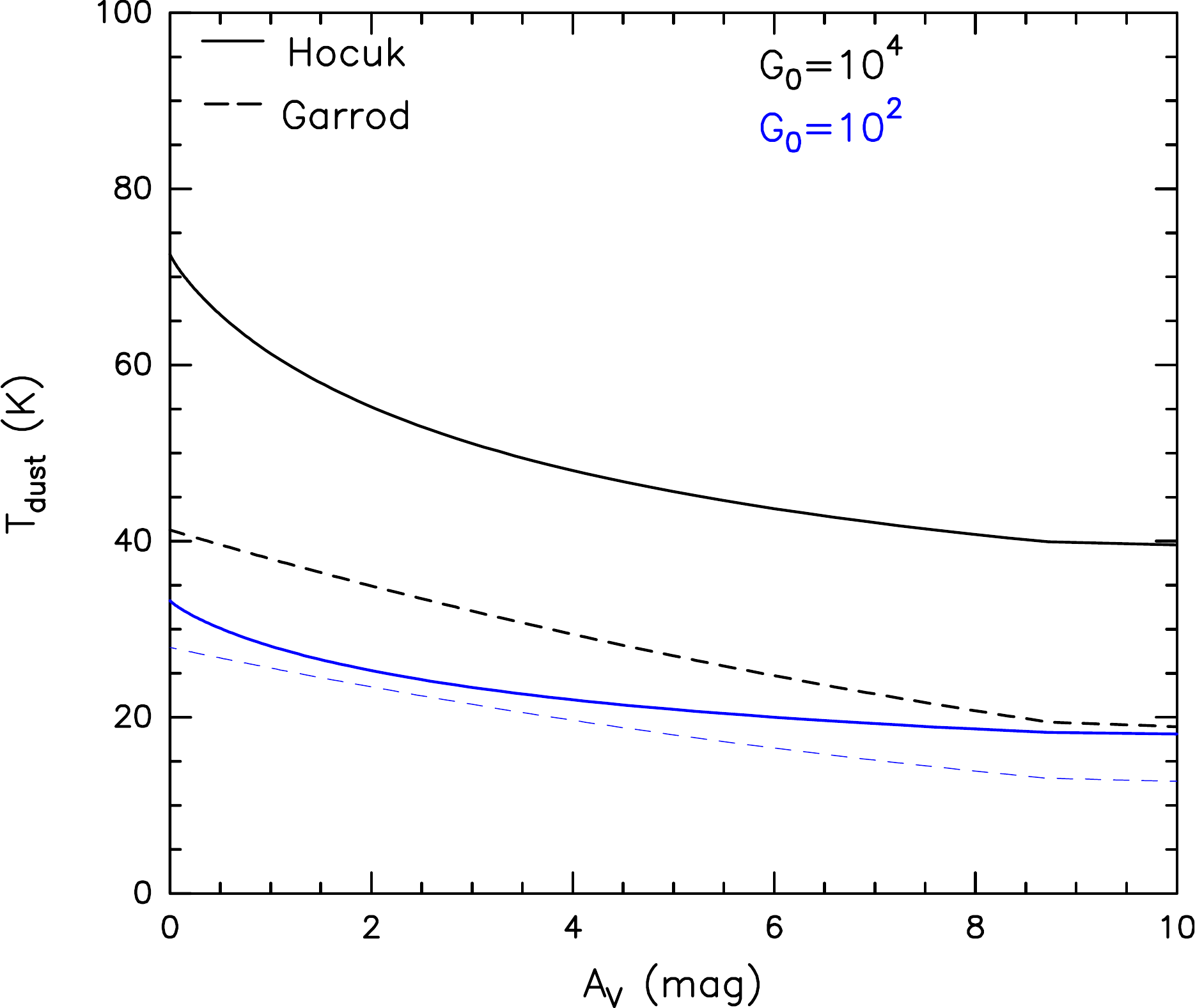}  
\hspace{1cm}
\includegraphics[scale=0.395, angle=0]{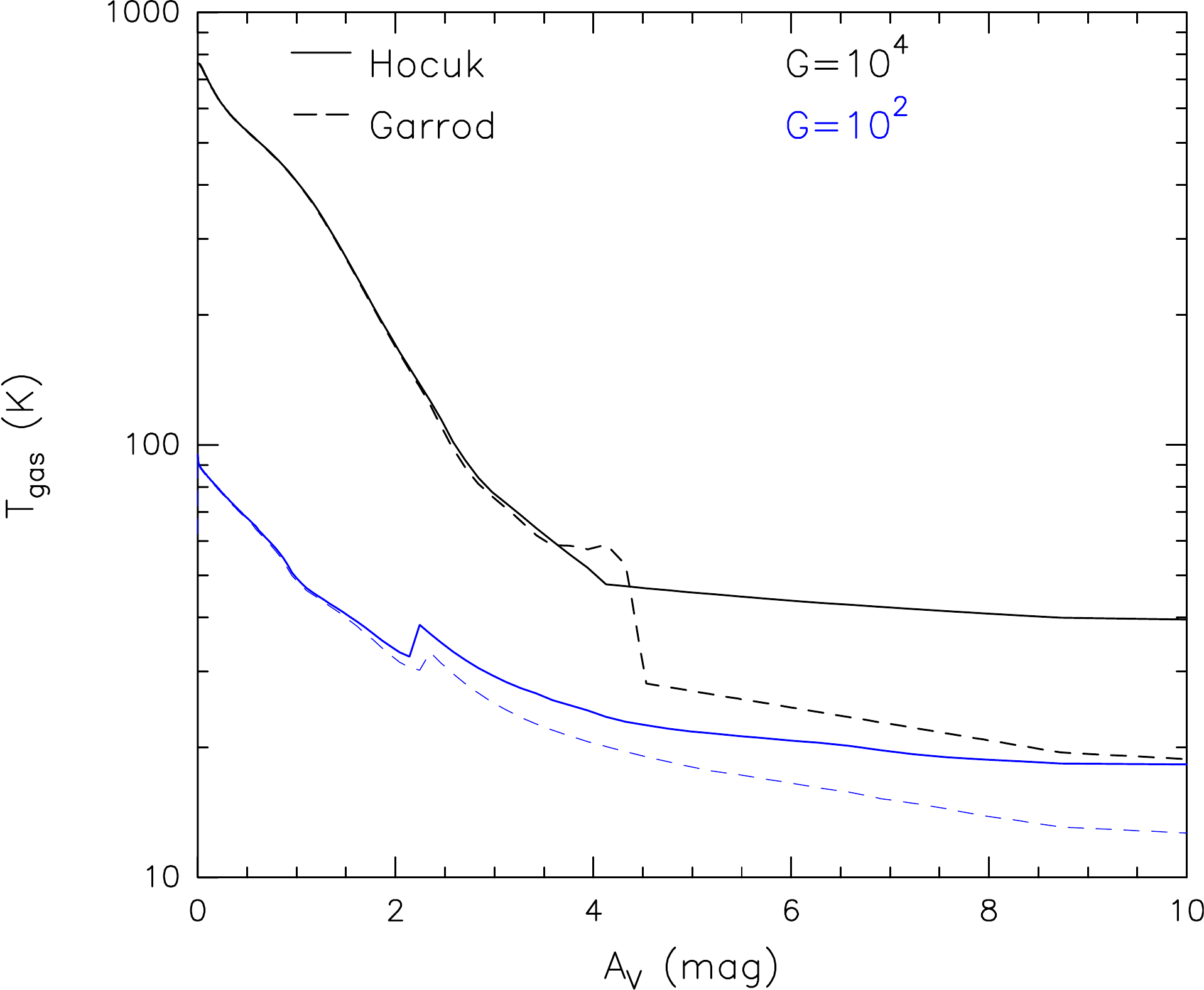}  
\hspace{0.7cm}\\
\caption{Dust (left) and gas (right) temperature for two PDR models with $G$$_0$=10$^2$ and $n$=10$^5$ cm$^{-3}$ (blue lines), and with $G$$_0$=10$^4$ and $n$=10$^5$ cm$^{-3}$ (black lines), considering analytical expressions from Garrod $\&$ Pauly (2011) (dashed line) and from Hocuk et al. (2017) (solid line).}
\label{figure:Tdust_comparison}
\end{figure*}

Interstellar dust is an ubiquitous component of the interstellar medium (ISM), whose mass is only about 0.7$\%$ of the gas (Fisher et al. 2014). In spite of this low value, dust grains have an important impact on the chemistry and thermodynamics of molecular clouds. In particular, the temperature of dust grains influences the gas temperature through heating and cooling processes along with chemical reaction rates. In addition, dust grain surfaces are also powerful interstellar catalysts since they are responsible for most of the production of the simplest (H$_2$) to the most complex (pre-biotic) molecules observed in the Universe.

Several analytical expressions for the dust temperature can be found in the literature, such as those from Hollenbach et al. (1991), Zucconi et al. (2001), and Garrod $\&$ Pauly (2011). These expressions are calculated in different ways. The solution by Hollenbach et al. (1991) assumes a one-sided slab geometry and combine the heating by ultraviolet (UV) photons, cosmic microwave background (CMB), and the re-processed infrared (IR). The derived temperature is a function of the intensity of the radiation field ($G$$_{\mathrm{0}}$) and of the visual extinction ($A$$_{\mathrm{V}}$), although the $A$$_{\mathrm{V}}$ dependence only takes into account the attenuation of UV photons. The expression provided by Zucconi et al. (2001) considers the contributions from the visual/near-infrared, mid-infrared, and far-infrared, and the dust temperature solution is given for the range 10$\lesssim$$A$$_{\mathrm{V}}$$\lesssim$400 mag. This expression is based on the observed dust temperature of L1544 at various $A$$_{\mathrm{V}}$ and it is only a function of the visual extinction. To obtain it, the authors solve the thermal balance without considering the UV field. They only include the visual and infrared part of the spectrum. The dust temperature expression provided by Garrod $\&$ Pauly (2011) was designed for low $A$$_{\mathrm{V}}$ regions and to be combined with that from Zucconi et al. (2001) for larger extinctions. This expression is only a function of $A$$_{\mathrm{V}}$.  

A recent analytical expression for the dust temperature ($T$$_{\mathrm{dust}}$) has been determined by Hocuk et al. (2017) from first principles for dust in thermal equilibrium by considering in detail the interstellar radiation field (ISRF), the attenuation of radiation, the dust opacities, and various grain material compositions (graphite, silicates SiO$_2$ and MgFeSiO, and carbonaceous silicate mixtures). This expression is:

\begin{eqnarray}
T_{\mathrm{dust}} = [11 + 5.7 \times tanh \left(\begin{array}{c}0.61 - \mathrm{log_{10}}(A_{\mathrm{V}}) \end{array} \right)] \times \left(\begin{array}{c}\chi^{1/5.9} \end{array} \right),
\end{eqnarray}\\

\noindent with $\chi$ the intensity of the radiation field in Draine units{\footnote{Draine field$\simeq$1.7$\times$Habing field (Habing 1968, Draine 1978).}}. The final solutions were compared with those obtained from the Monte Carlo radiative transfer code RADMC-3D{\footnote{http://www.ita.uni-heidelberg.de/$\sim$dullemond/software/radmc-3d}} and with observational results from several interstellar regions observed with Herschel. See Hocuk et al. (2017) for more details.

Depending on the size of dust grains, their temperature can present significant variations on short timescales (seconds to minutes) as derived by Cuppen et al. (2006) and Iqbal et al. (2014) using Monte Carlo simulations, and by Bron et al. (2014) using an analytical approach. In particular, the smallest grains (radii $a$$\lesssim$50 $\AA$) undergo very large temperature fluctuations (more than 30 K). This variations are equivalent to consider PDRs with radiation intensities of two different orders of magnitude, which significantly varies the chemistry (see Figure \ref{figure:Tdust_comparison} and Sections below). Therefore, in the case of very small grains, it is not realistic to consider an average temperature. However, larger dust grains (and especially those with a size $a$$\geq$200 $\AA$) can be approximated as having a steady temperature (Draine $\&$ Li 2001), since their temperature fluctuations are lower than 3 K (Cuppen et al. 2006). Nevertheless, it should be noted that recent studies (Pauly $\&$ Garrod 2016) show that the dust temperature choice is far from being trivial, since other factors, such as the mantle growth and its time evolution, can also vary the dust temperature. In particular, they find dust temperature variations of $\sim$11 K for grains with $a$$\lesssim$0,01 $\mu$m, while the temperature variation is only $\lesssim$5 K for grains with $a$$\gtrsim$0,1 $\mu$m. In any case, these results make also evident the fact that the larger the grain sizes, the lower the dust temperature variations. Considering this fact and in order to avoid large local dust temperature fluctuations in short timescales that could significantly alter the chemistry when studying the effects of other parameters (e.g. the effect of increasing the radiation field intensity), we have assumed a MRN grain size distribution (Mathis et al. 1977) in the Meijerink PDR code, with grain radius limited to 50 $\AA$$<$a$<$0.25 $\mu$m, for which is reasonable to consider an average dust grain temperature.

\begin{figure}
\centering
\includegraphics[scale=0.39, angle=0]{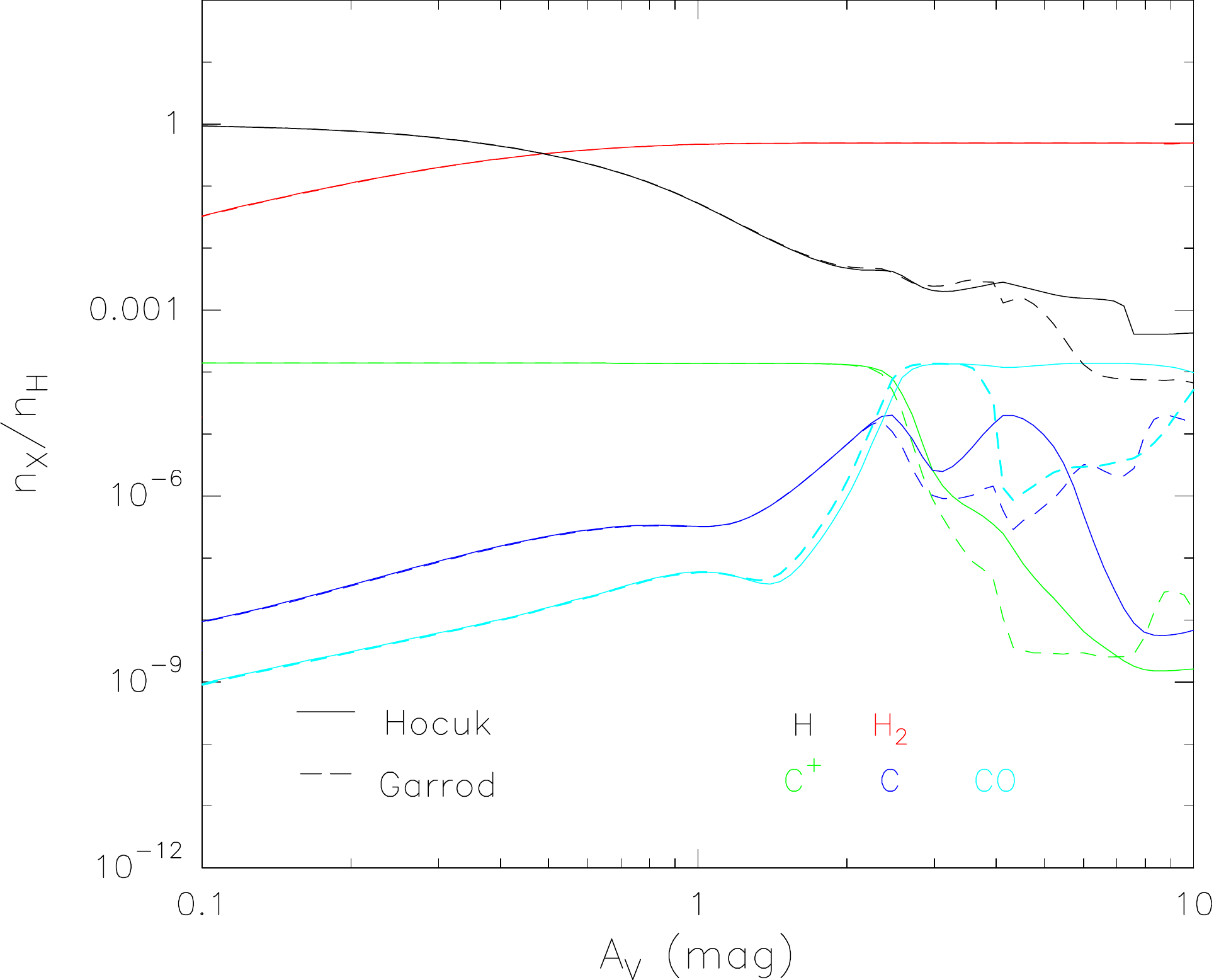}\\
\caption{Abundances of H, H$_2$, C$^+$, C, and CO obtained with the updated Meijerink PDR code using $T$$_{\mathrm{dust}}$ from Hocuk et al. (2017) (solid lines) and from Garrod $\&$ Pauly (2011) (dashed lines). Results are obtained considering $G$$_0$=10$^4$ and $n$=10$^5$ cm$^{-3}$.}
\label{figure:H-H2_comparison}
\end{figure}

In Esplugues et al. (2016), we calculated $T$$_{\mathrm{dust}}$ through the expression from Garrod $\&$ Pauly (2011)\footnote{Dust temperature expression derived from Garrod $\&$ Pauly (2011), but with an adaptation to include dependence with the intensity of the radiation field (Garrod private comm., see Esplugues et al. 2016 for more details).}. Here, we also consider in our analysis the $T$$_{\mathrm{dust}}$ expression from Hocuk et al. (2017). Figure \ref{figure:Tdust_comparison} shows the dust temperature values from these two expressions for two PDRs with different intensity of radiation field\footnote{We use $G$$_0$, the Habing field (Habing 1968), as the normalisation in which we express the incident FUV radiation field, where $G$$_0$=1 corresponds to a flux of 1.6$\times$10$^{-3}$ erg cm$^{-2}$ s$^{-1}$.} ($G$$_{\mathrm{0}}$=10$^2$ and $G$$_{\mathrm{0}}$=10$^4$) in the interval 0$\leq$$A$$_{\mathrm{V}}$$\leq$10 mag. For the PDR with the lowest $G$$_{\mathrm{0}}$ (blue), the differences for $T$$_{\mathrm{dust}}$ between both expressions are lower than 10 K. However, for the most extreme PDR (black), these differences are of up to 30 K, leading to significant differences in the chemistry of the considered regions at intermediate and large visual extinctions (see Figure \ref{figure:H-H2_comparison}). In Sect. \ref{Dust_effects}, we analyse in detail the impact of considering both dust temperature expressions on the chemistry of several molecule families.

\section{Time dependence}
\label{Time-depend.}

In a molecular cloud, as the visual extinction increases ($A$$_{\mathrm{V}}$$>$1 mag), certain chemical timescales become comparable to cloud lifetimes (10$^6$-10$^7$ yr) and steady-state chemistry does not apply. In these cases, time-dependent solutions to the chemistry are therefore needed. This is the case of photodissociation regions. At low visual extinctions ($A$$_{\mathrm{V}}$$\lesssim$1 mag), the energy balance is dominated by FUV photons and the chemical timescales are very short ($\lesssim$10$^5$ yr) compared to the molecular cloud lifetime. However, in the opaque interiors of the cloud ($A$$_{\mathrm{V}}$$>$6 mag), the chemistry is dominated by a low FUV flux and by long chemical timescales (e.g., the corresponding timescale to cosmic-ray desorption of CO ice is from 3$\times$10$^5$ to 3$\times$10$^9$ yr, depending on the assumptions regarding the CO desorption process, Hollenbach et al. 2009). At intermediate depths, UVs are attenuated by dust extinction, but photodesorption still prevents total freeze-out. 

To study the effects of time dependence on the chemistry of photodissociation regions, we have coupled the Meijerink PDR code with the time-dependent code Nahoon. In this way, the PDR code provides a fixed physical structure (density, temperature) and we perform post-processing computing by calculating the time-dependent chemistry of the medium with Nahoon. Grains are initially bare and the formation of ices takes place during the evolution of the interstellar gas cloud, starting from a diffuse, fully atomic stage to a molecular phase illuminated and warmed up by a nearby star. We follow the composition at any time with chemical network using rate equations that incorporates grain surface reactions on two different substrates (bare and icy grains). The chemistry evolves over a period of 10$^7$ yr. The Nahoon code has been modified to have the same chemical network and chemical processes as those included in the Meijerink PDR code. In particular, to the gas-phase chemistry network provided by KIDA, we added our grain surface chemistry network as detailed in Sect. \ref{Dust_chemistry_changes}. The grain surface processes taken into account are identical to those used in the PDR code: adsorption, thermal desorption, two-body reactions, chemical desorption, desorption by UV photons and cosmic rays, and dissociation by UV photons and cosmic-ray-induced UV photons. A more detailed description of Nahoon, which is publicly available on KIDA, can be found in Wakelam et al. (2012). In Sect. \ref{Time_dependence}, we analyse the time effects on the chemical evolution of different PDR types.

\section{Results and discussion}
\label{Results}

We show the results for several molecule families through three different PDR models: with density $n$=10$^5$ cm$^{-3}$ and $G$$_0$=10$^2$ (Model 1), with $n$=10$^5$ cm$^{-3}$ and $G$$_0$=10$^4$ (Model 2), and with $n$=10$^6$ cm$^{-3}$ and $G$$_0$=10$^4$ (Model 3), see Table \ref{table:parameters}. These models have been chosen to analyse how the dust temperature and time-dependent effects vary depending on the type of PDR.

\begin{table}[h!]
\caption{Adopted model parameters in our PDR code.}             
\centering   
\begin{tabular}{l l l l}     
\hline\hline       
Model & $G$$_{0}$    & $n$$_{\mathrm{H}}$ \\ 
      &              &  (cm$^{-3}$) \\
\hline                    
1  & 10$^2$       & 10$^{5}$   \\
2  & 10$^{4}$     & 10$^{5}$   \\
3  & 10$^4$       & 10$^{6}$    \\
\hline
\label{table:parameters}                  
\end{tabular}
\end{table}

\subsection{Dust temperature effects}
\label{Dust_effects}

\begin{figure*}
\centering
\includegraphics[scale=0.29, angle=0]{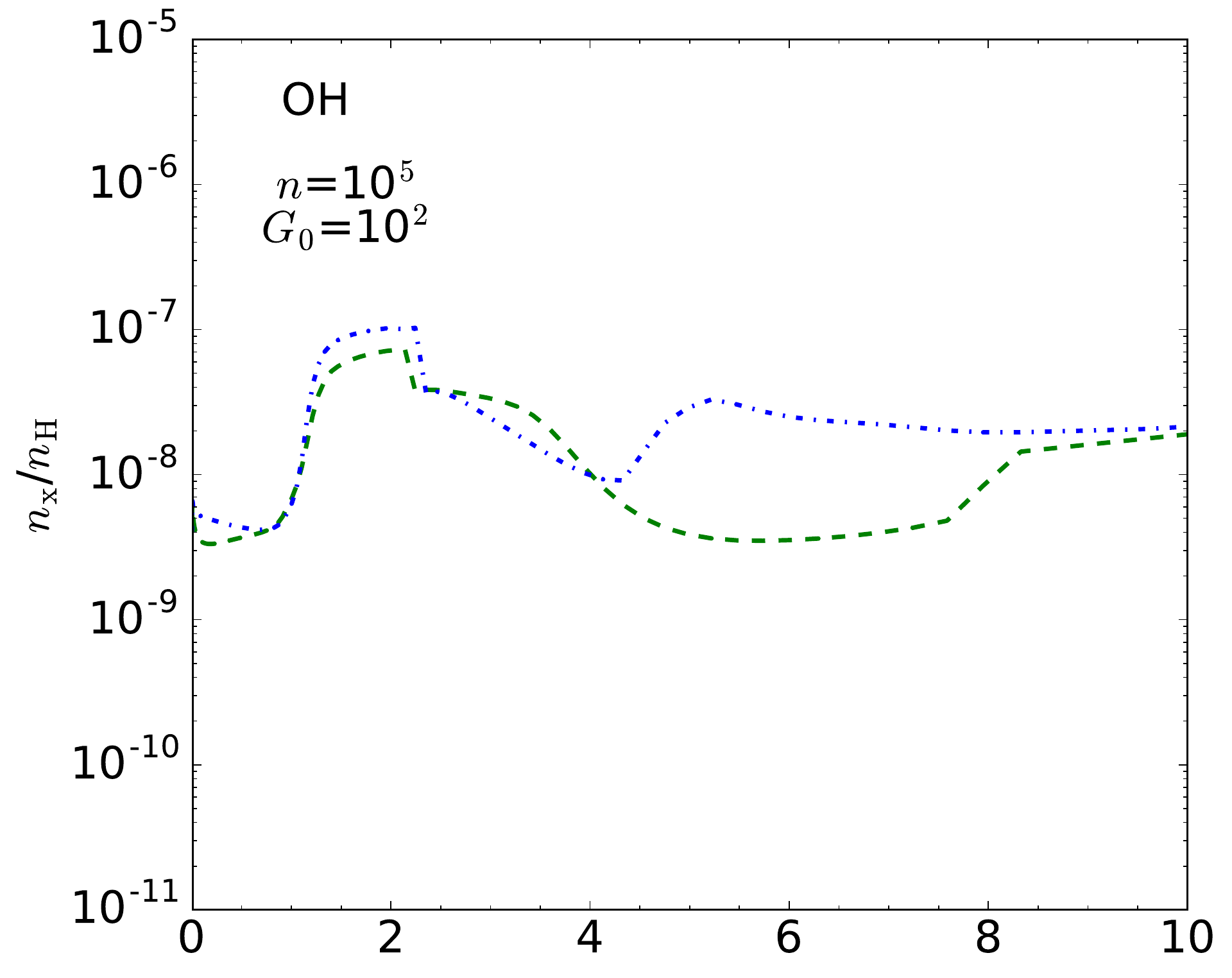}
\includegraphics[scale=0.29, angle=0]{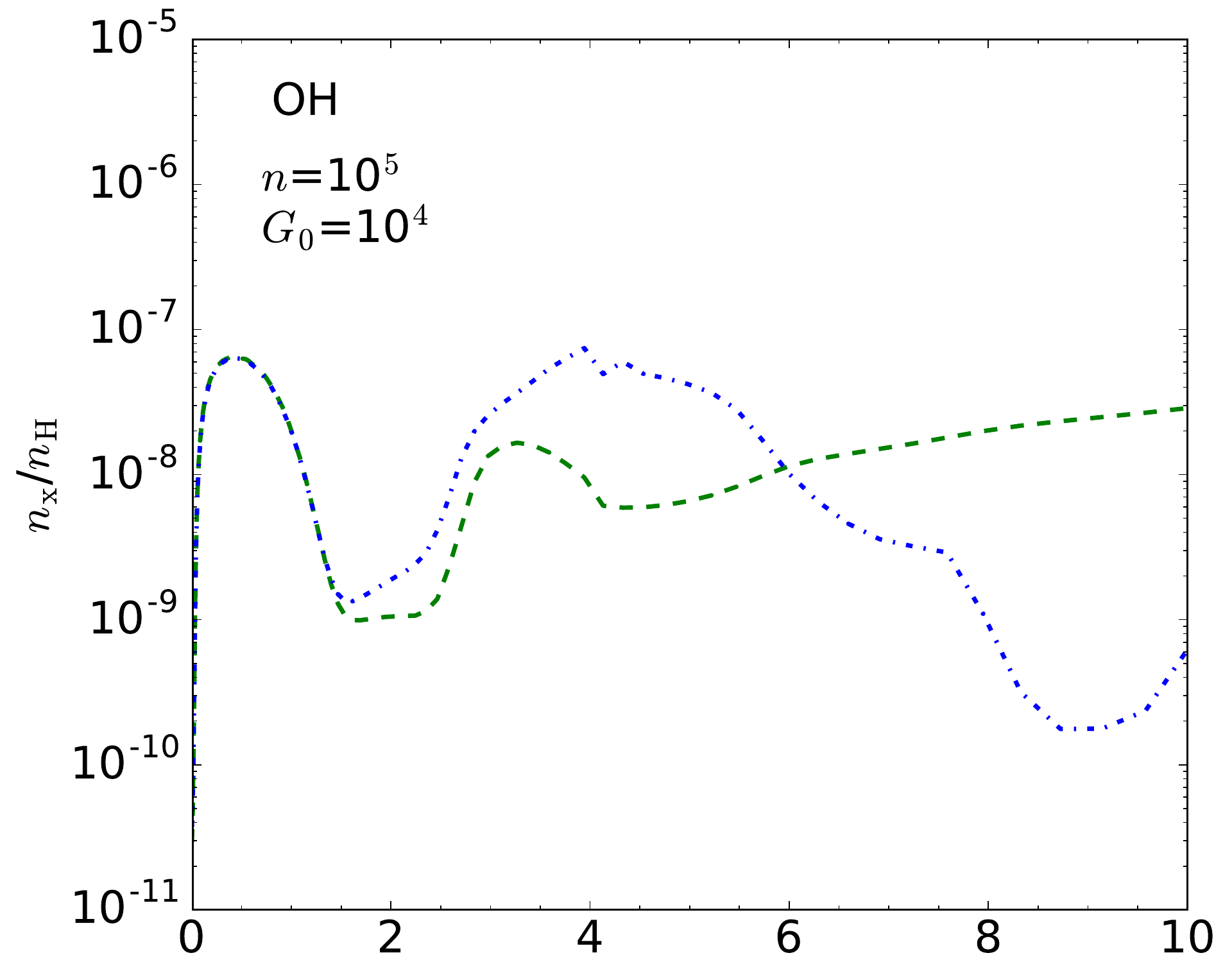}
\includegraphics[scale=0.29, angle=0]{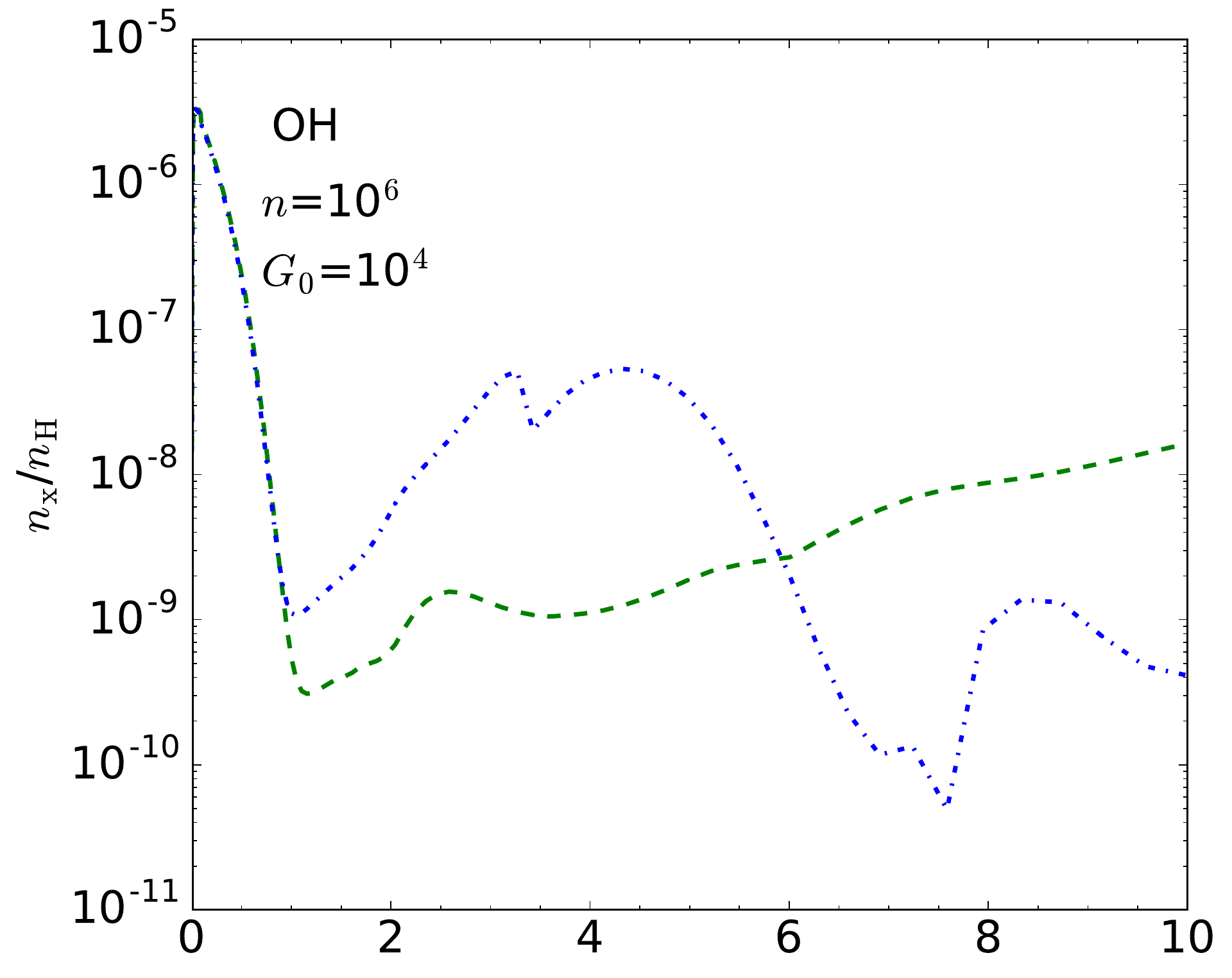} \\
\includegraphics[scale=0.29, angle=0]{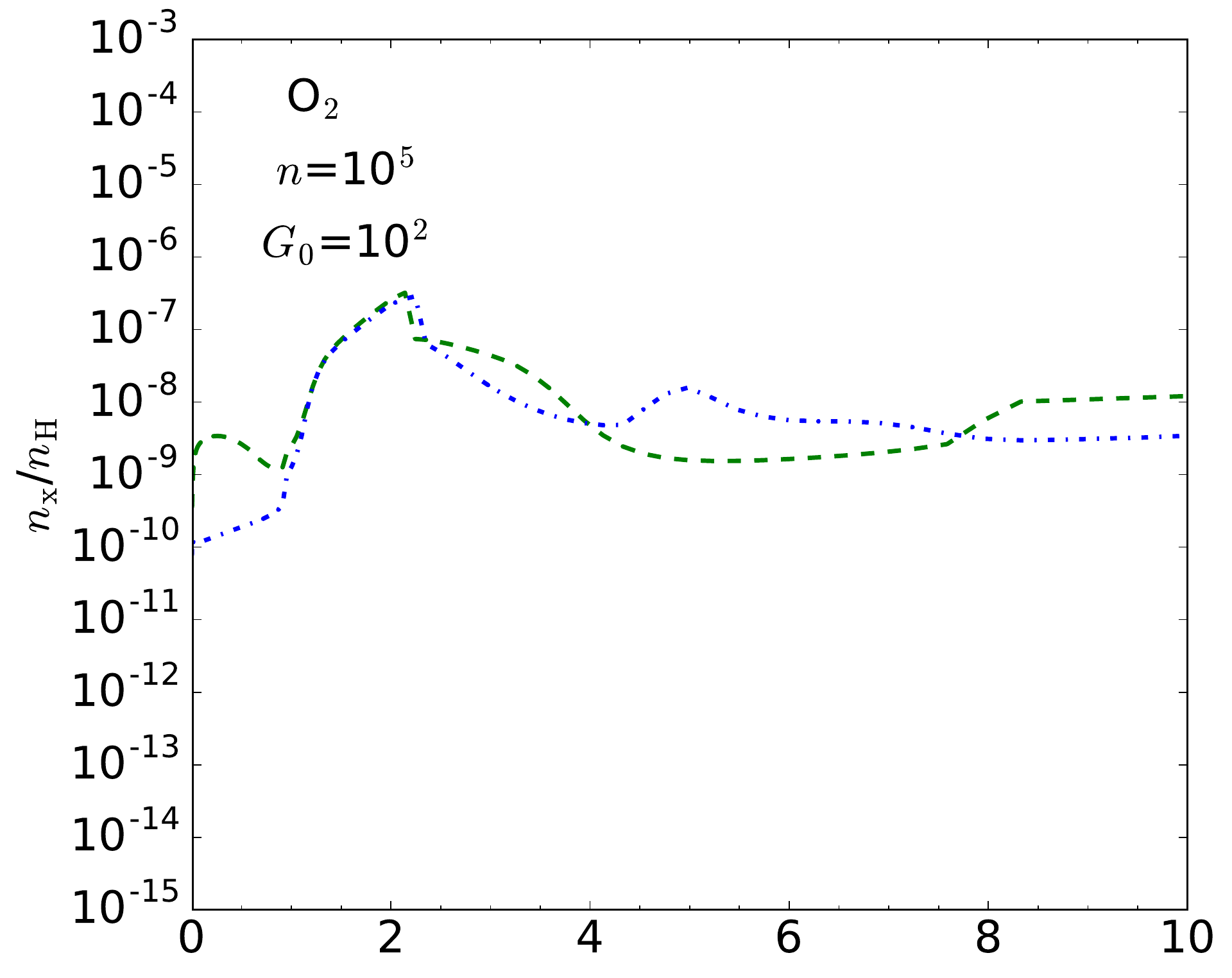}
\includegraphics[scale=0.29, angle=0]{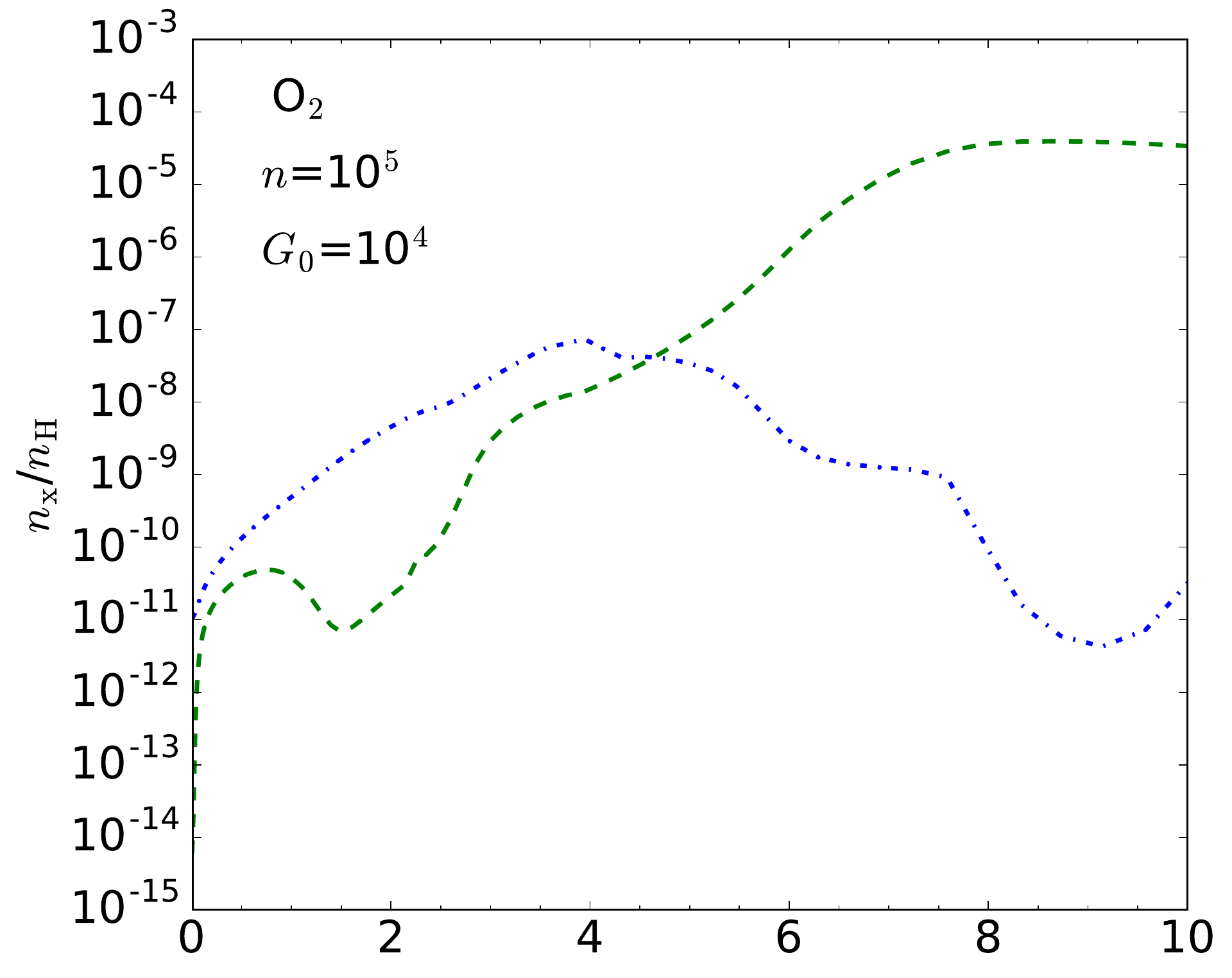}
\includegraphics[scale=0.29, angle=0]{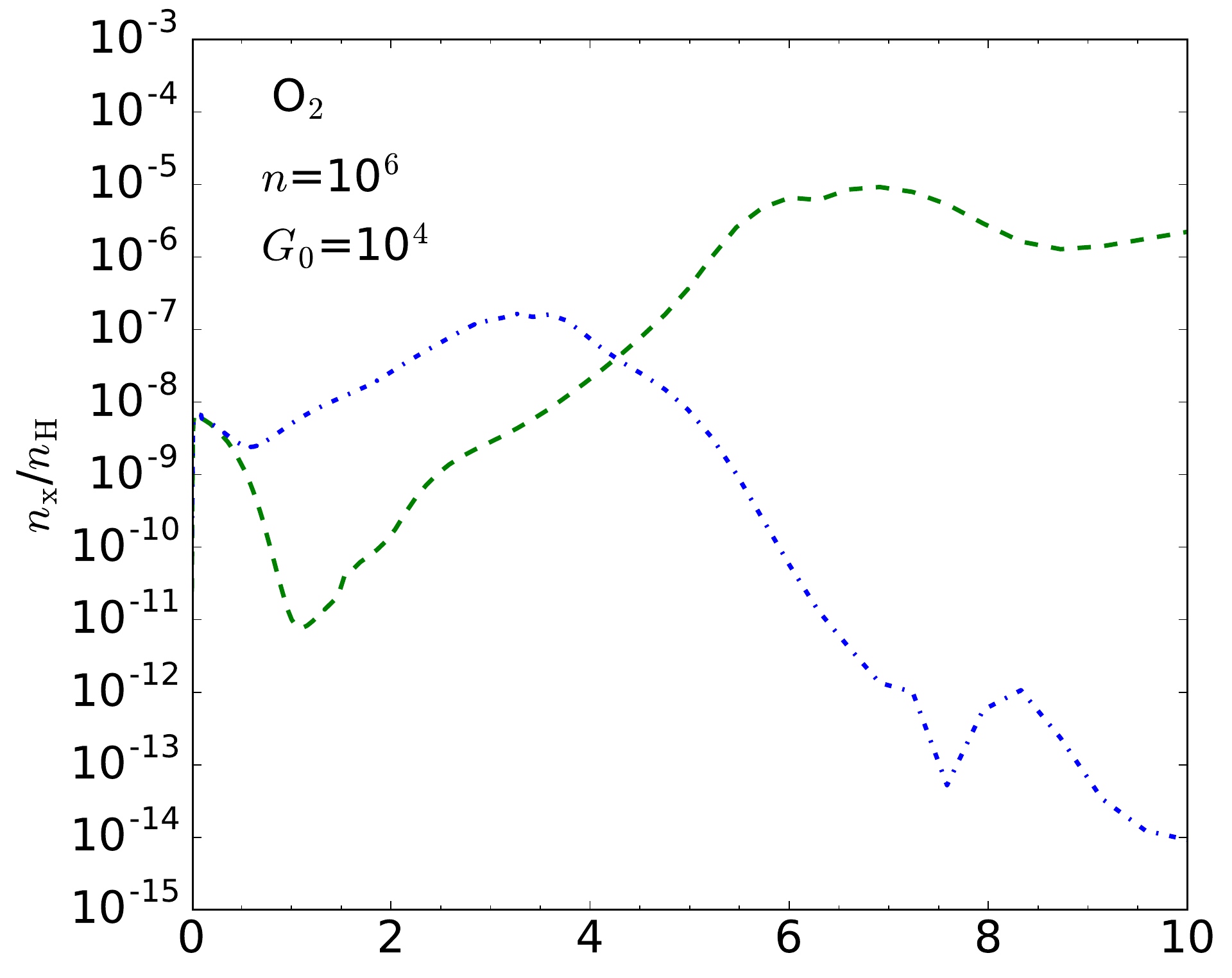} \\
\includegraphics[scale=0.29, angle=0]{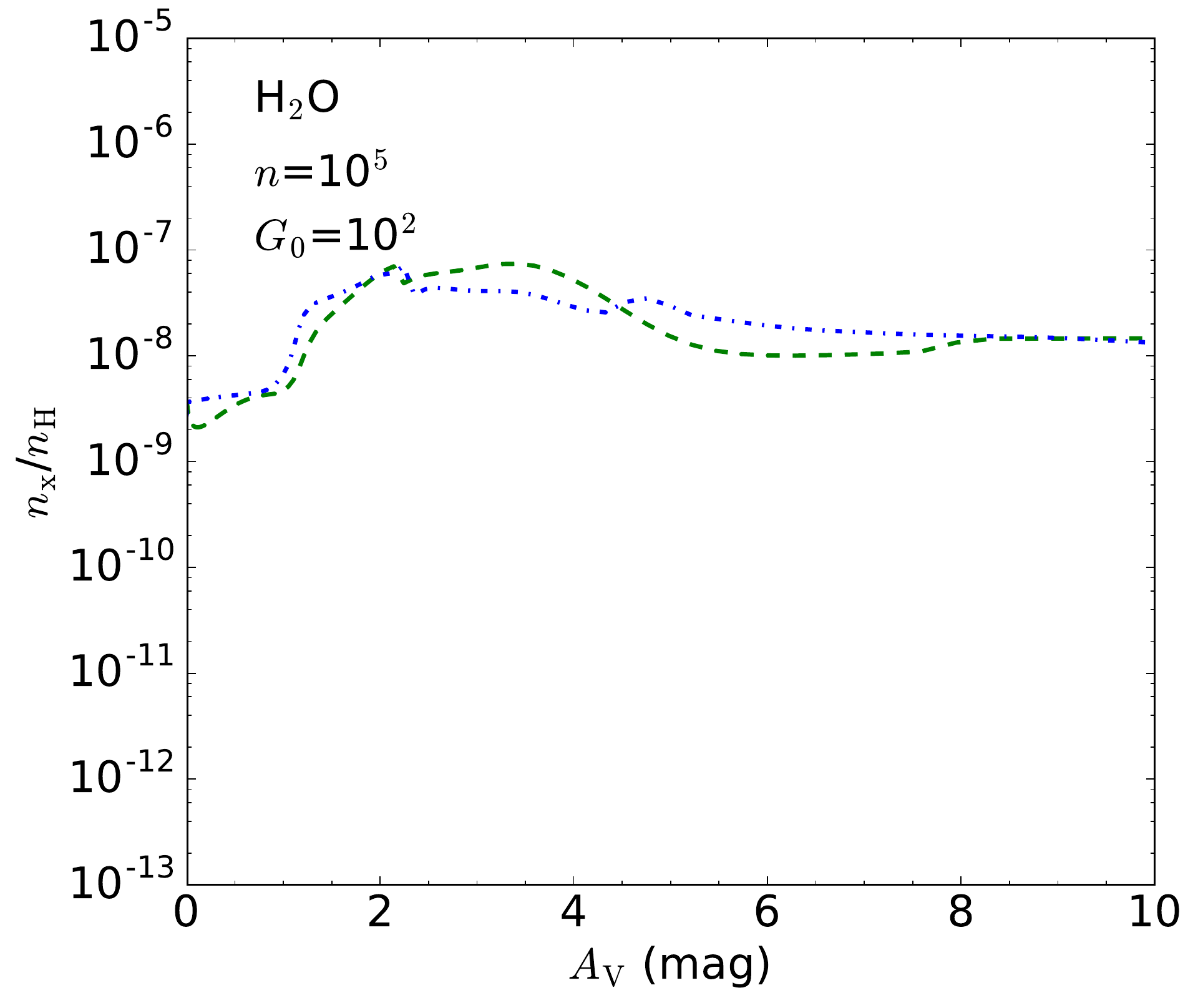}
\includegraphics[scale=0.29, angle=0]{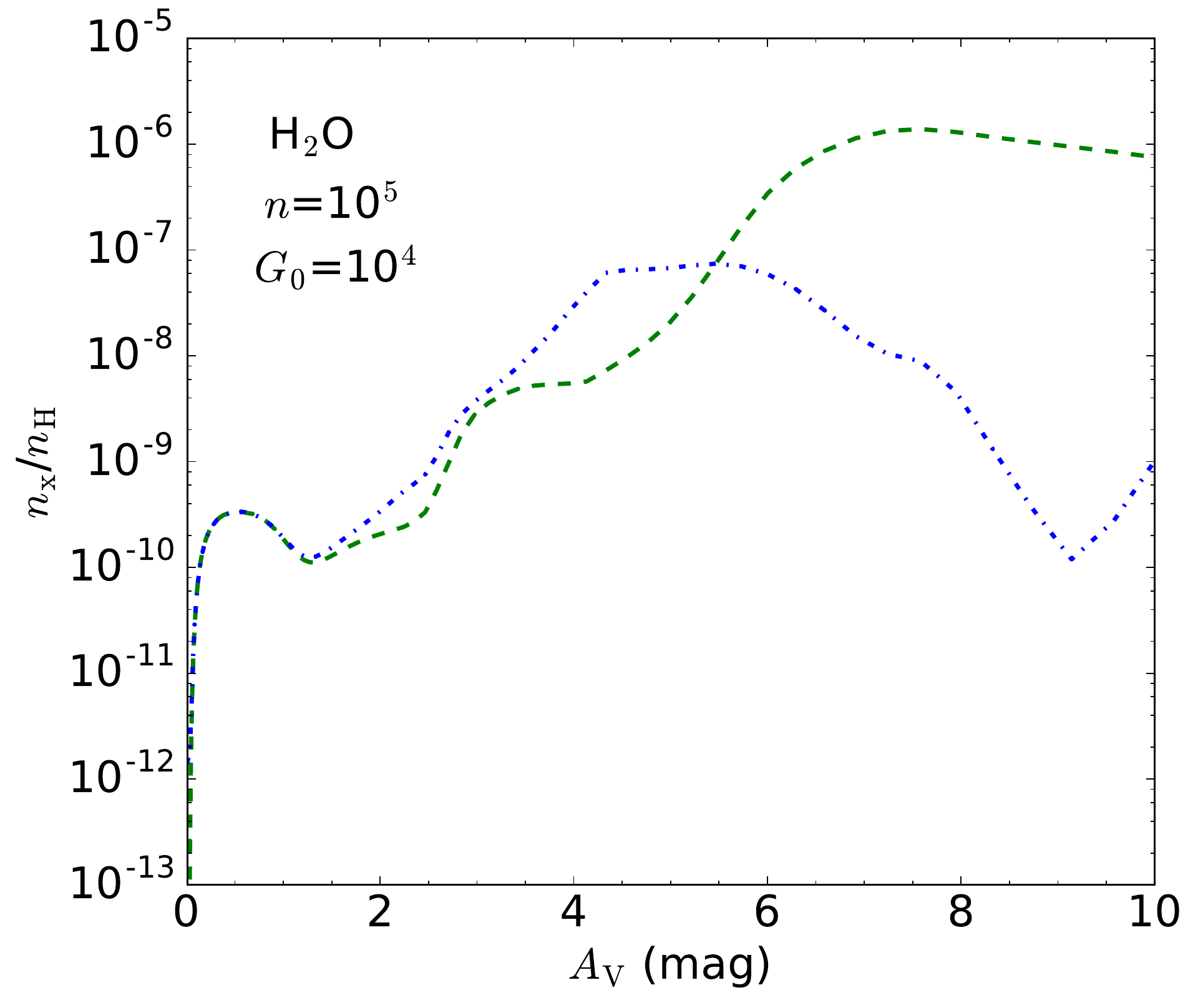}
\includegraphics[scale=0.29, angle=0]{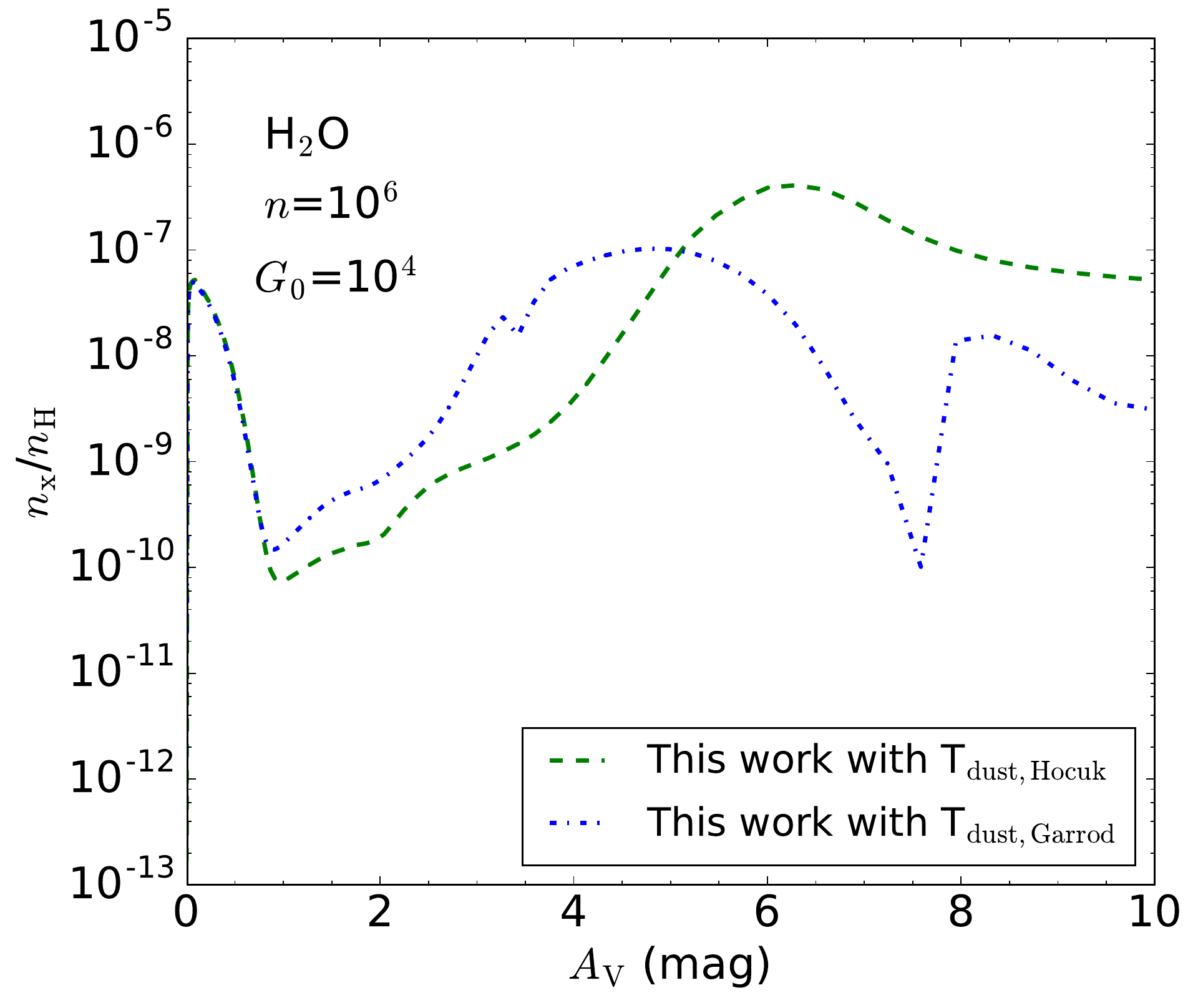}
\caption{Abundances of OH, O$_2$, and H$_2$O obtained with the updated Meijerink PDR code using $T$$_{\mathrm{dust}}$ from Hocuk et al. (2017) (green dashed lines) and from Garrod $\&$ Pauly (2011) (blue dotted lines). Results for Model 1 are shown on the left panels, for Model 2 in the middle panels, and for Model 3 on the right panels.}
\label{figure:O_bearing_comparison}
\end{figure*}

\begin{figure*}
\centering
\includegraphics[scale=0.282, angle=0]{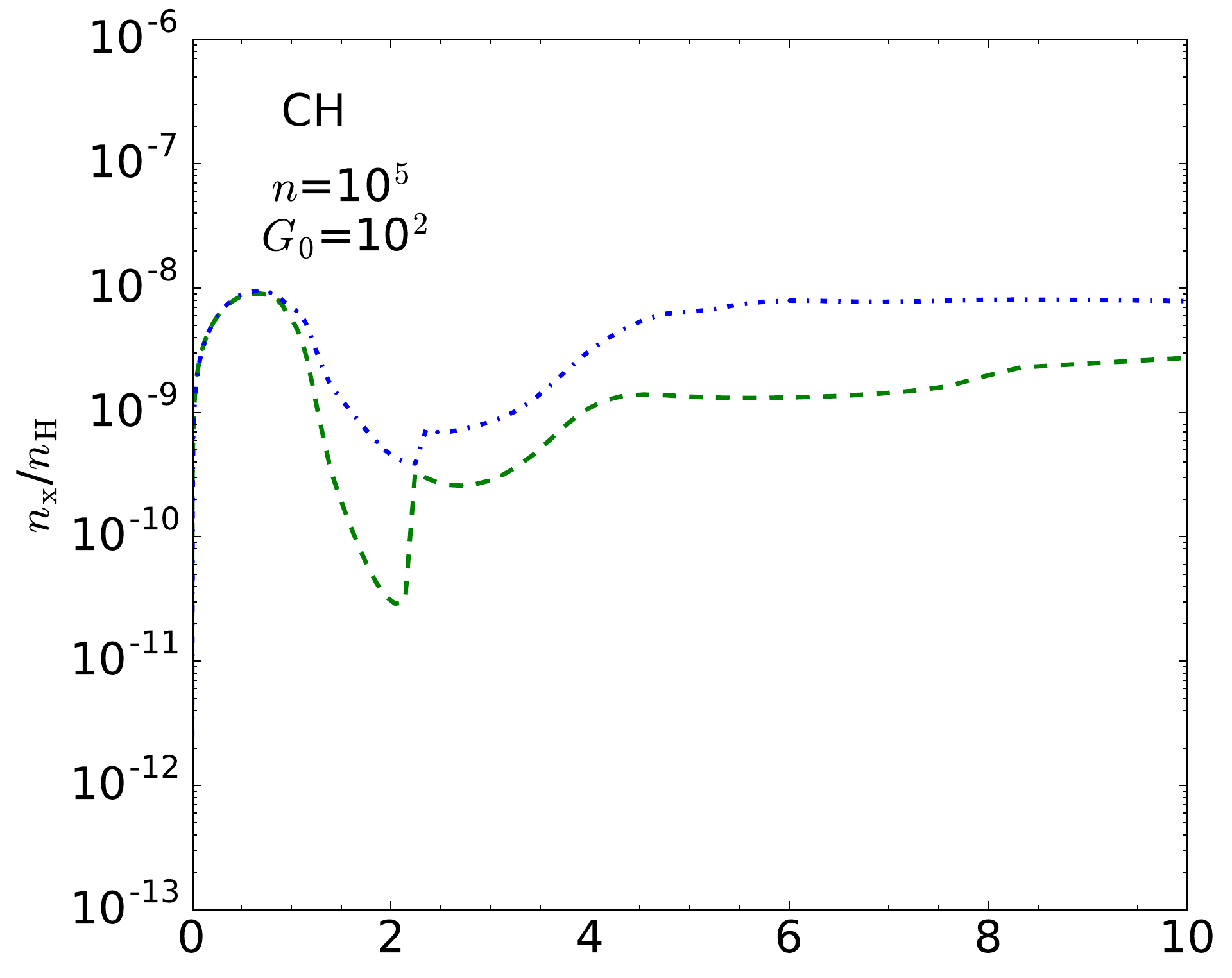}  
\includegraphics[scale=0.282, angle=0]{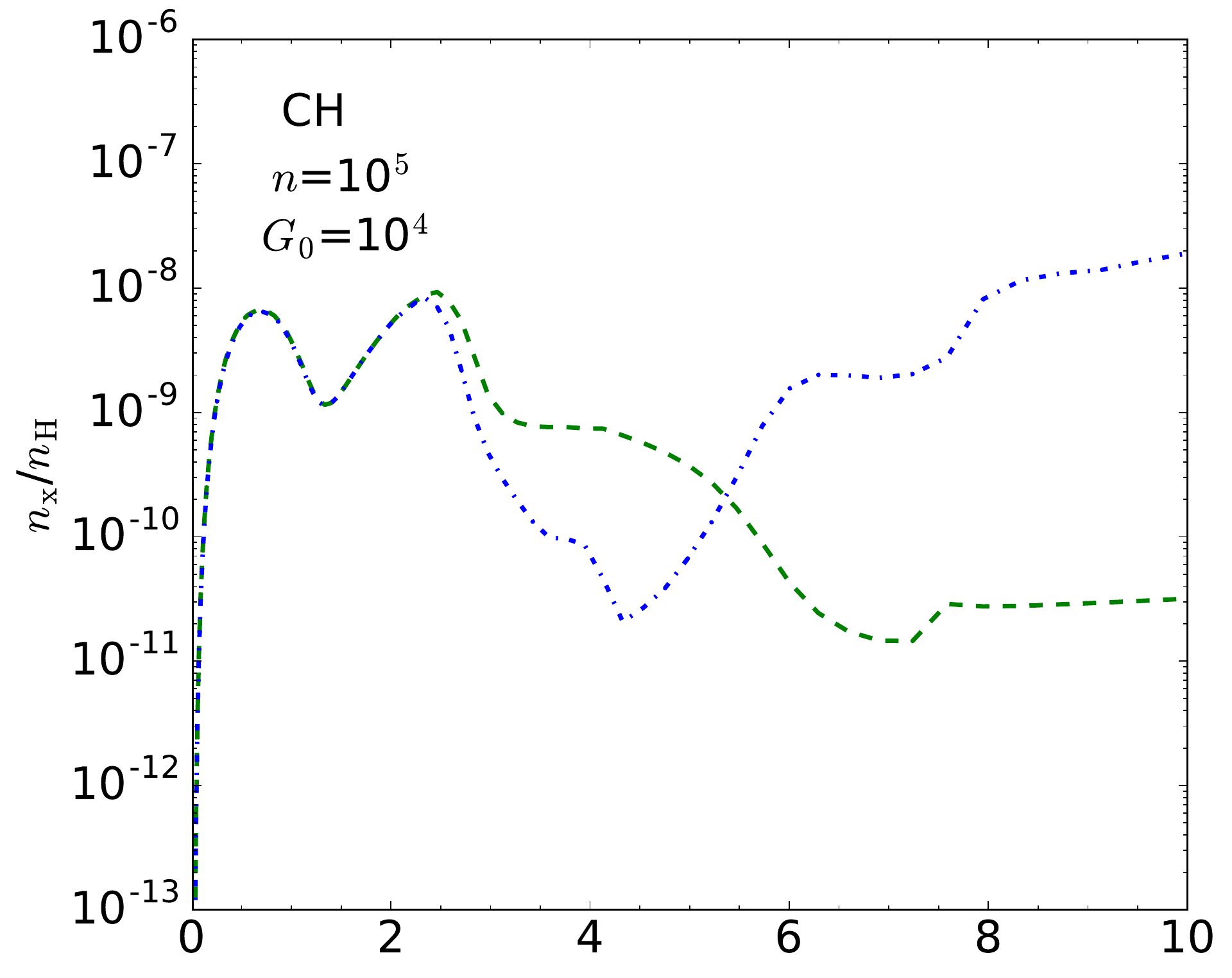}  
\includegraphics[scale=0.282, angle=0]{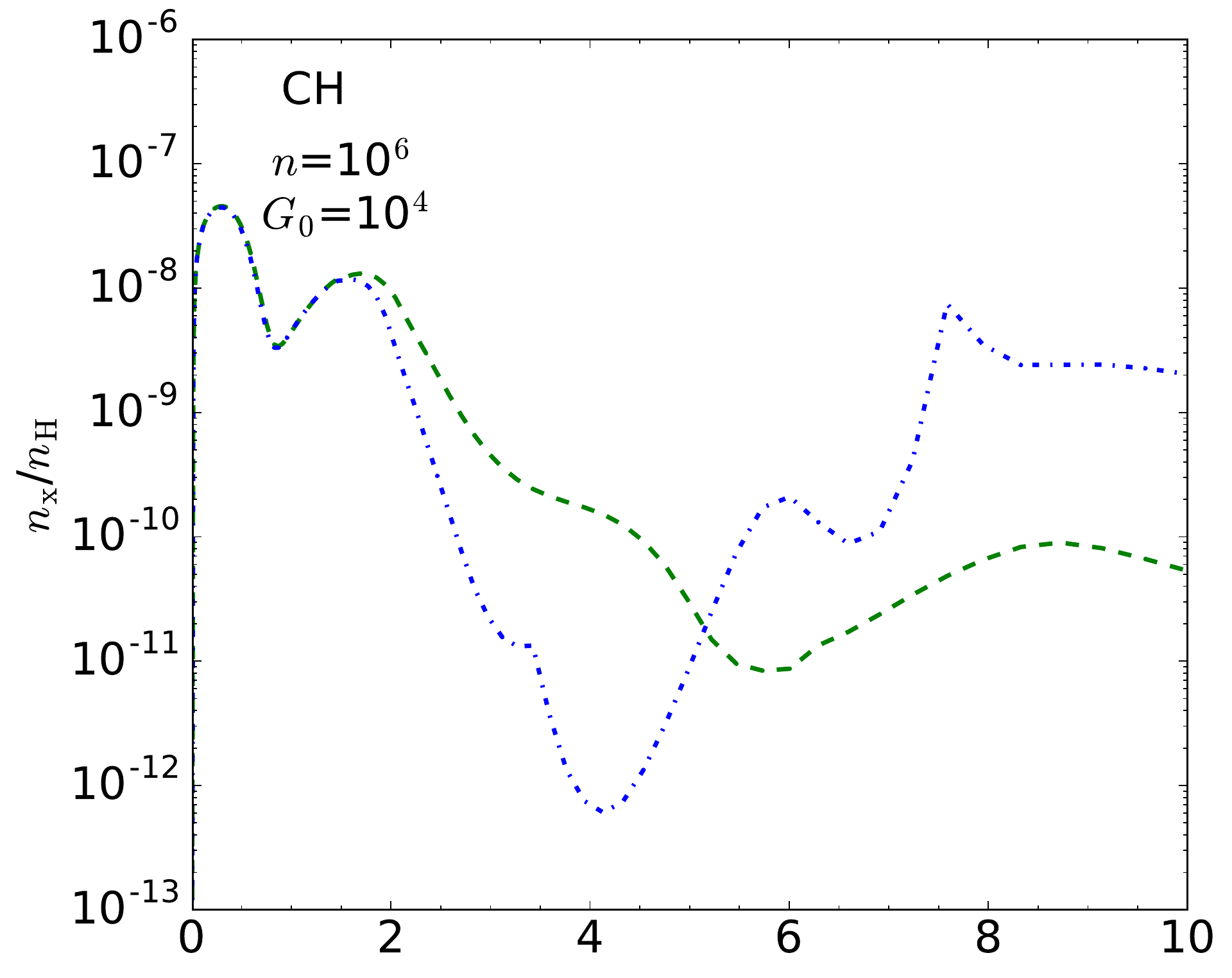} \\ 
\includegraphics[scale=0.282, angle=0]{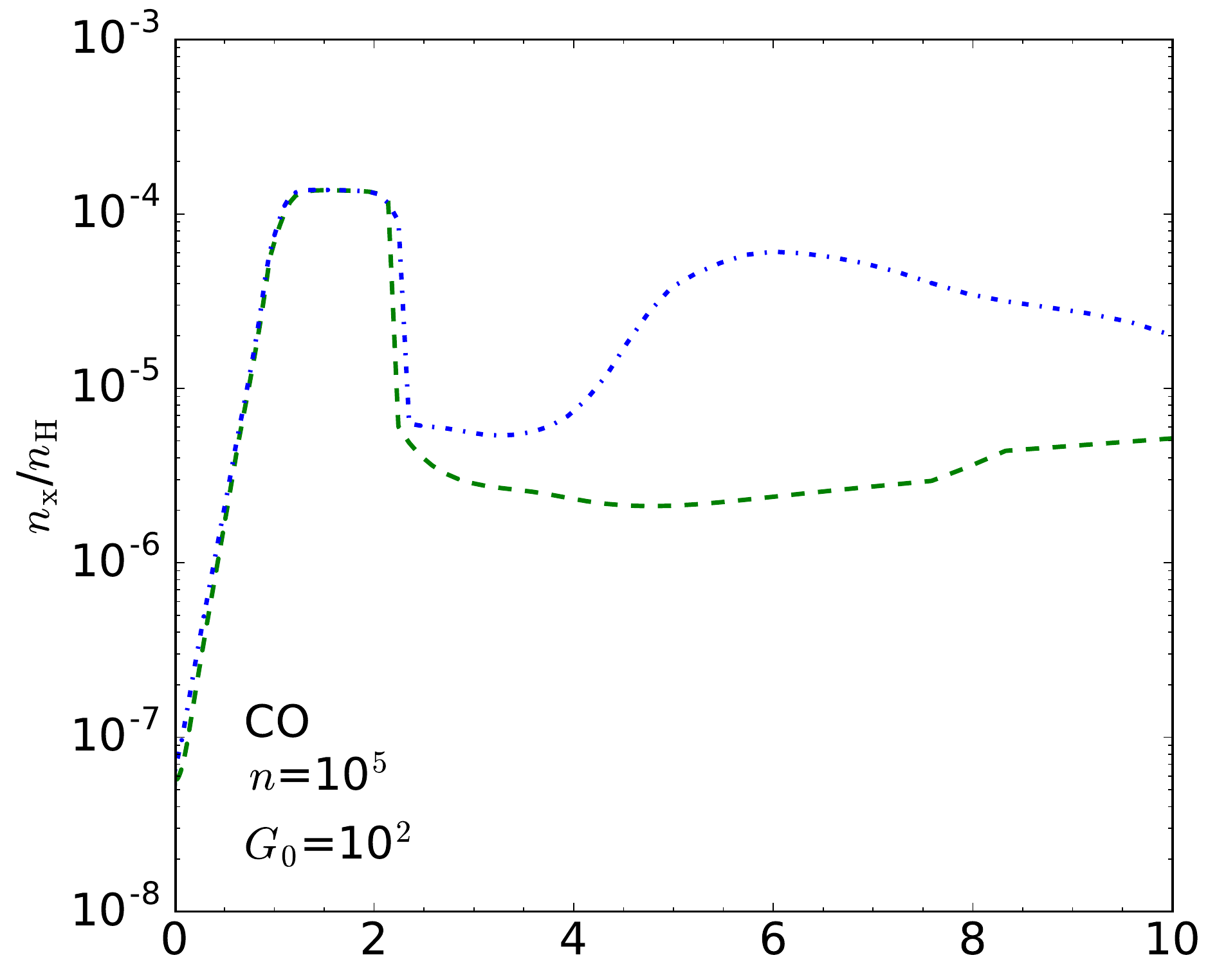}  
\includegraphics[scale=0.282, angle=0]{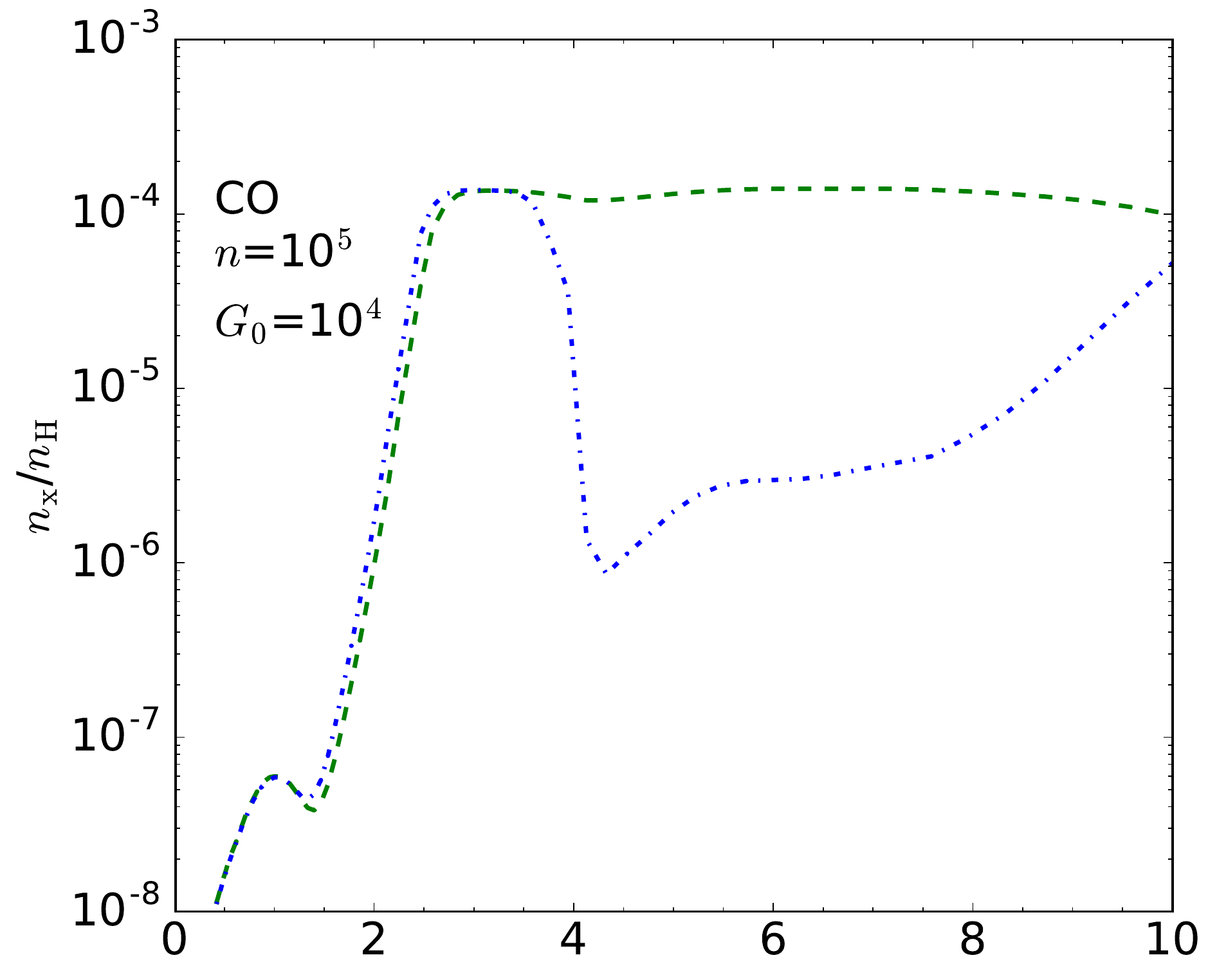}  
\includegraphics[scale=0.282, angle=0]{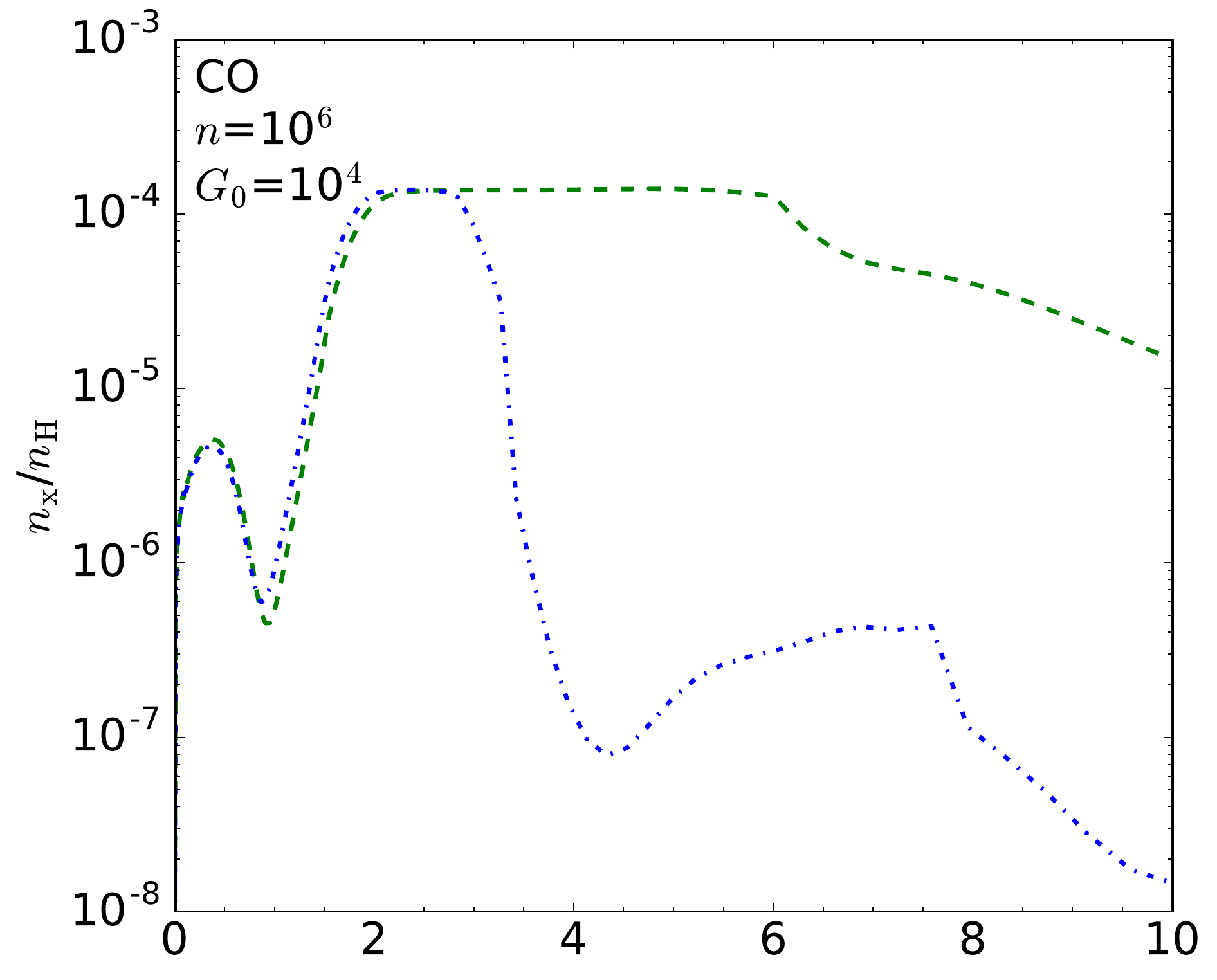}  \\
\includegraphics[scale=0.282, angle=0]{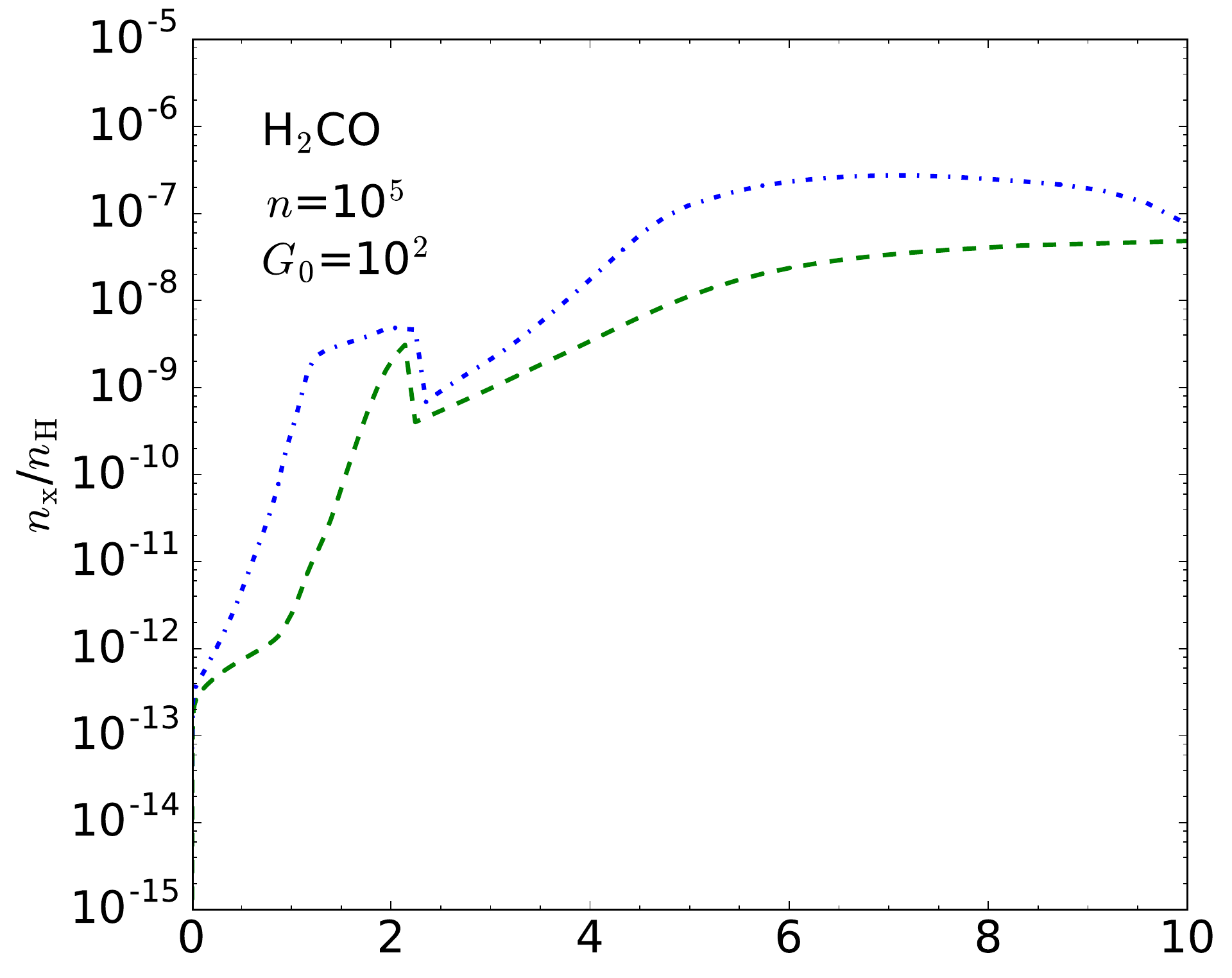}  
\includegraphics[scale=0.282, angle=0]{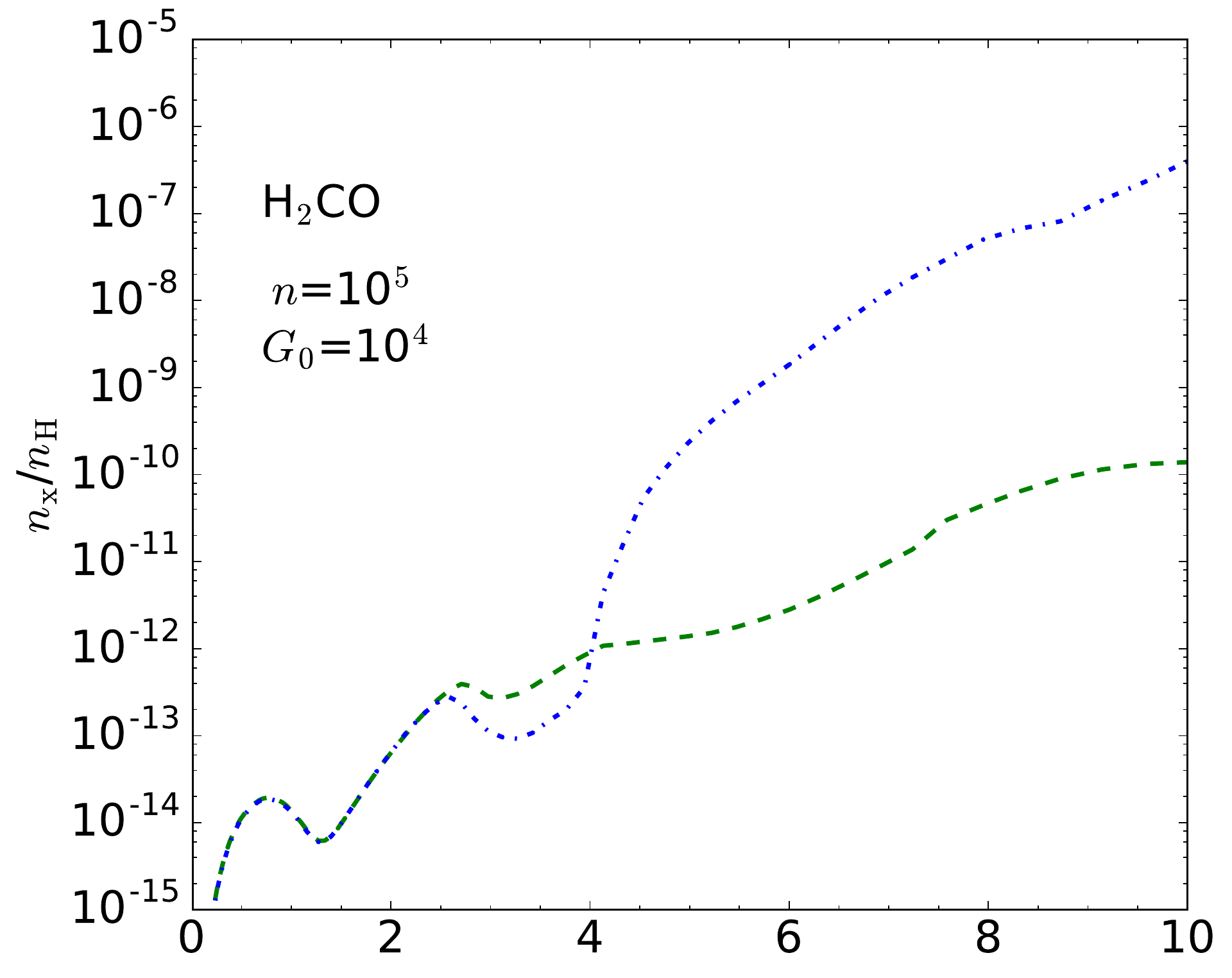} 
\includegraphics[scale=0.282, angle=0]{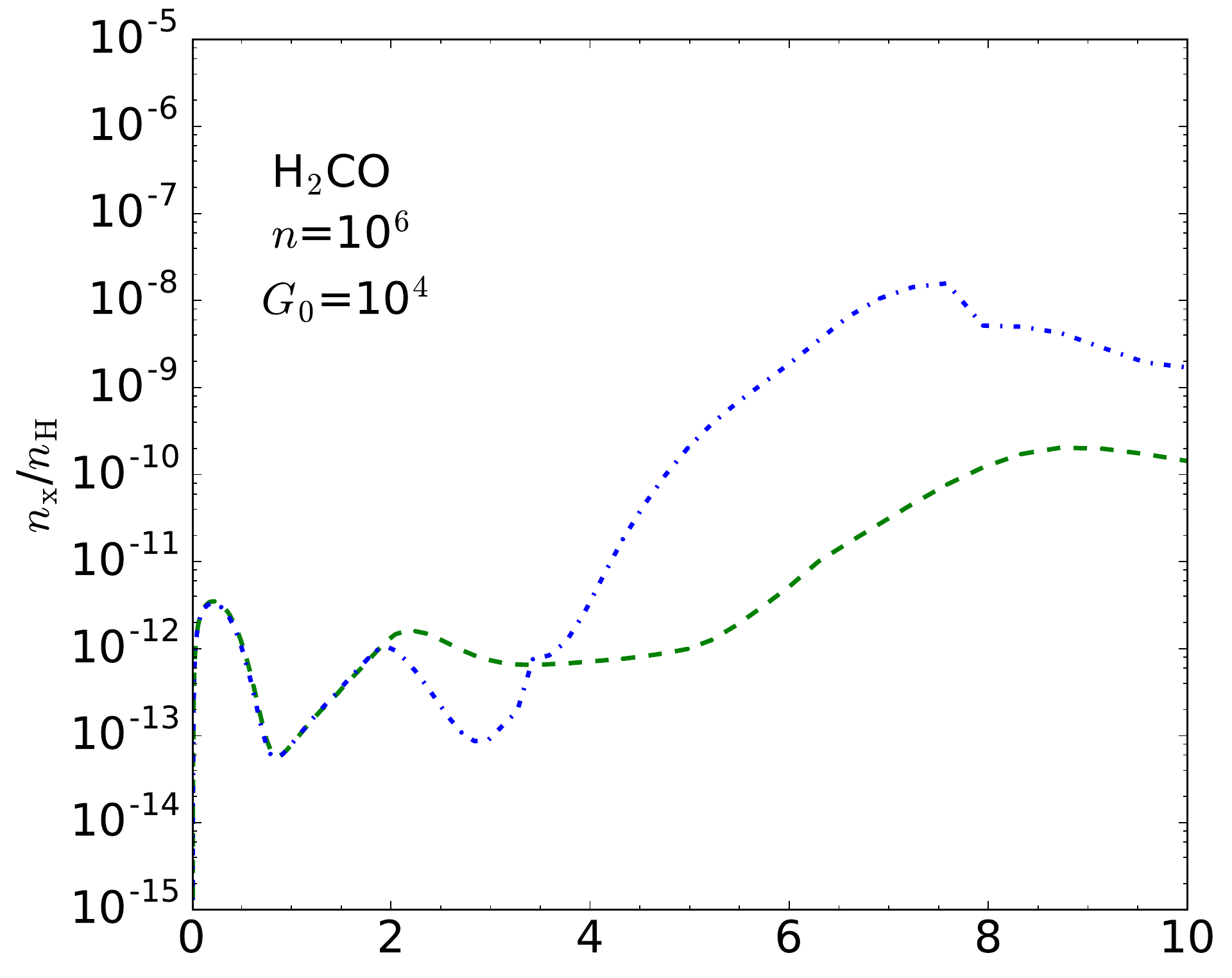}  \\
\includegraphics[scale=0.282, angle=0]{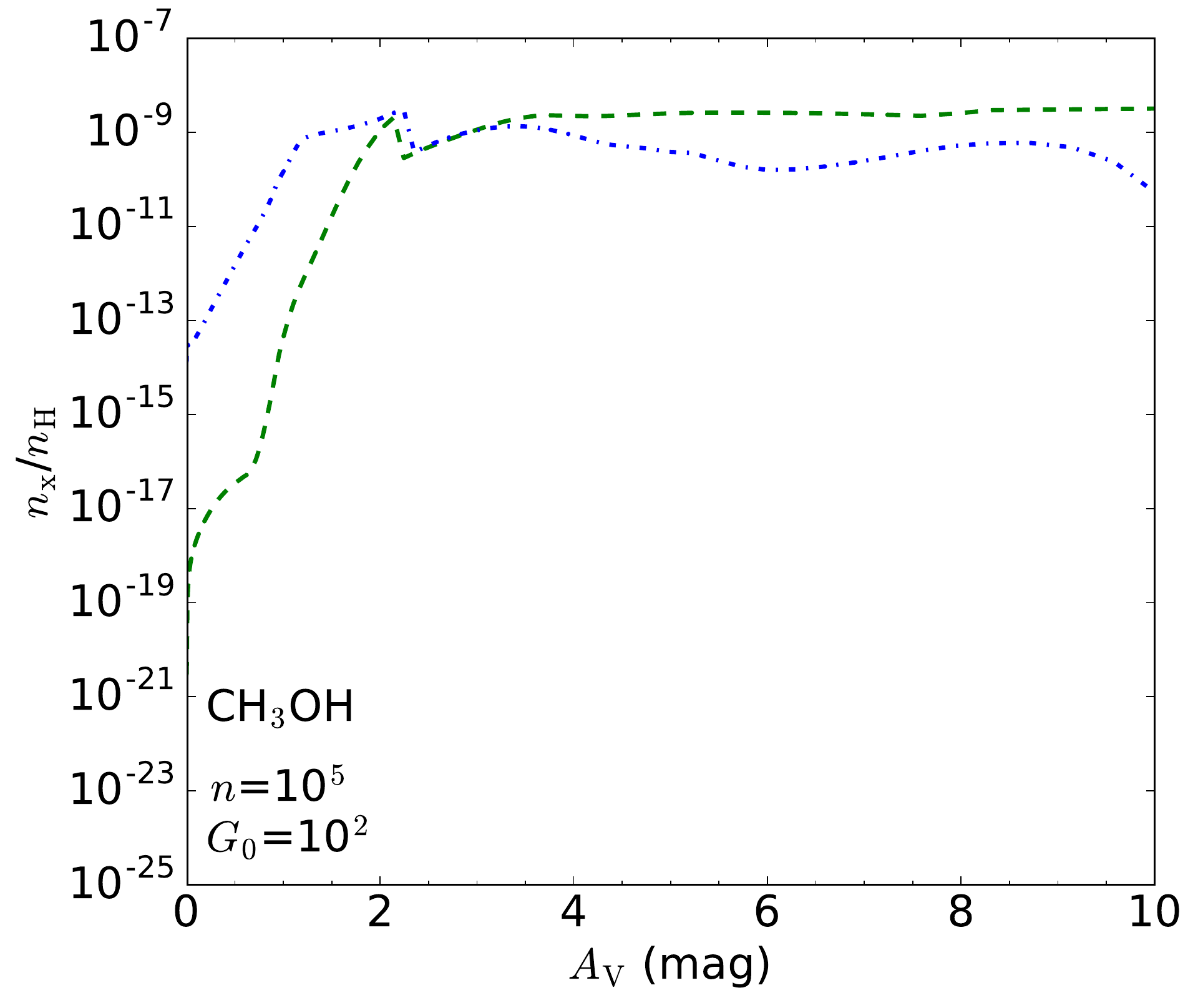}  
\includegraphics[scale=0.282, angle=0]{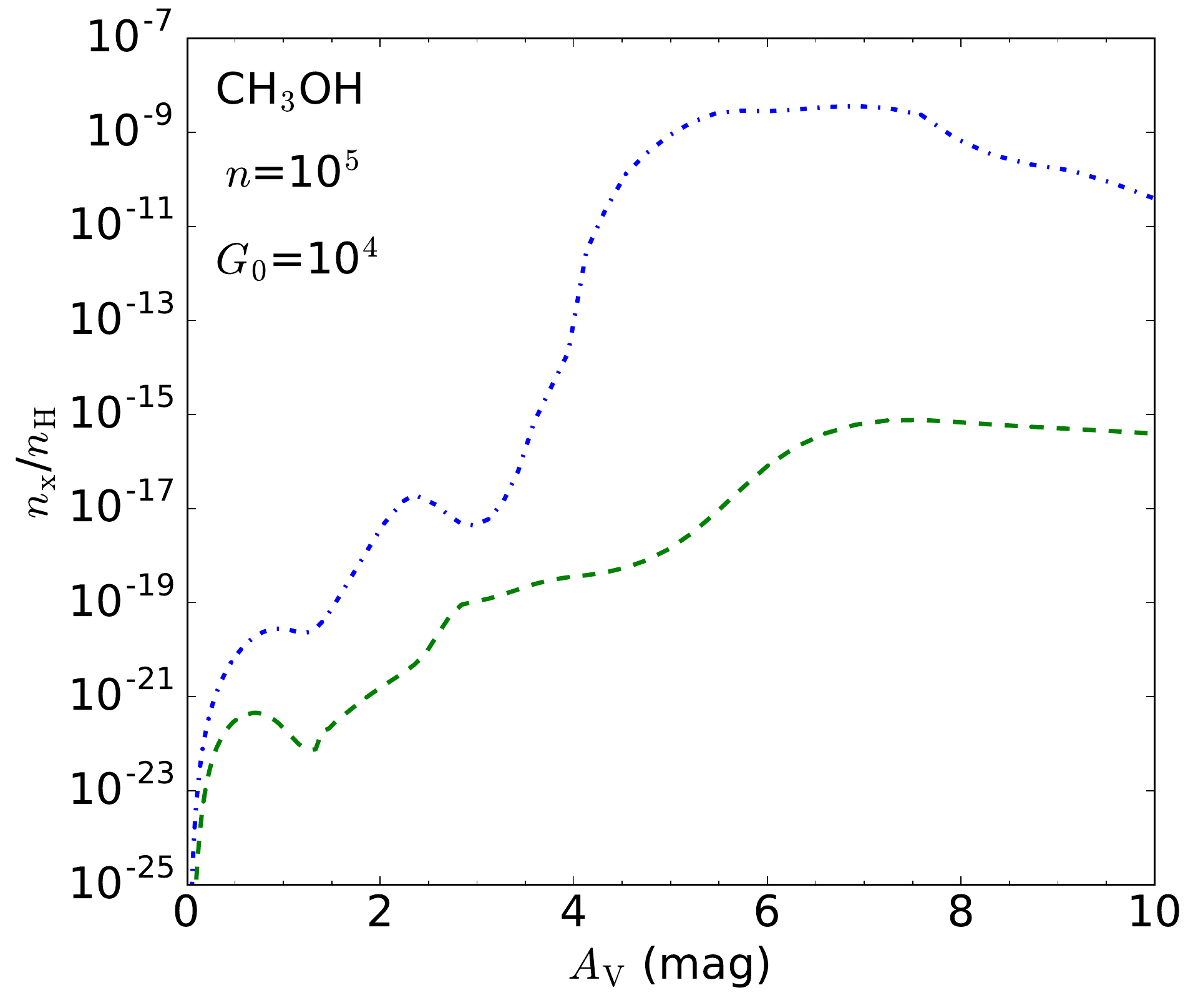}  
\includegraphics[scale=0.282, angle=0]{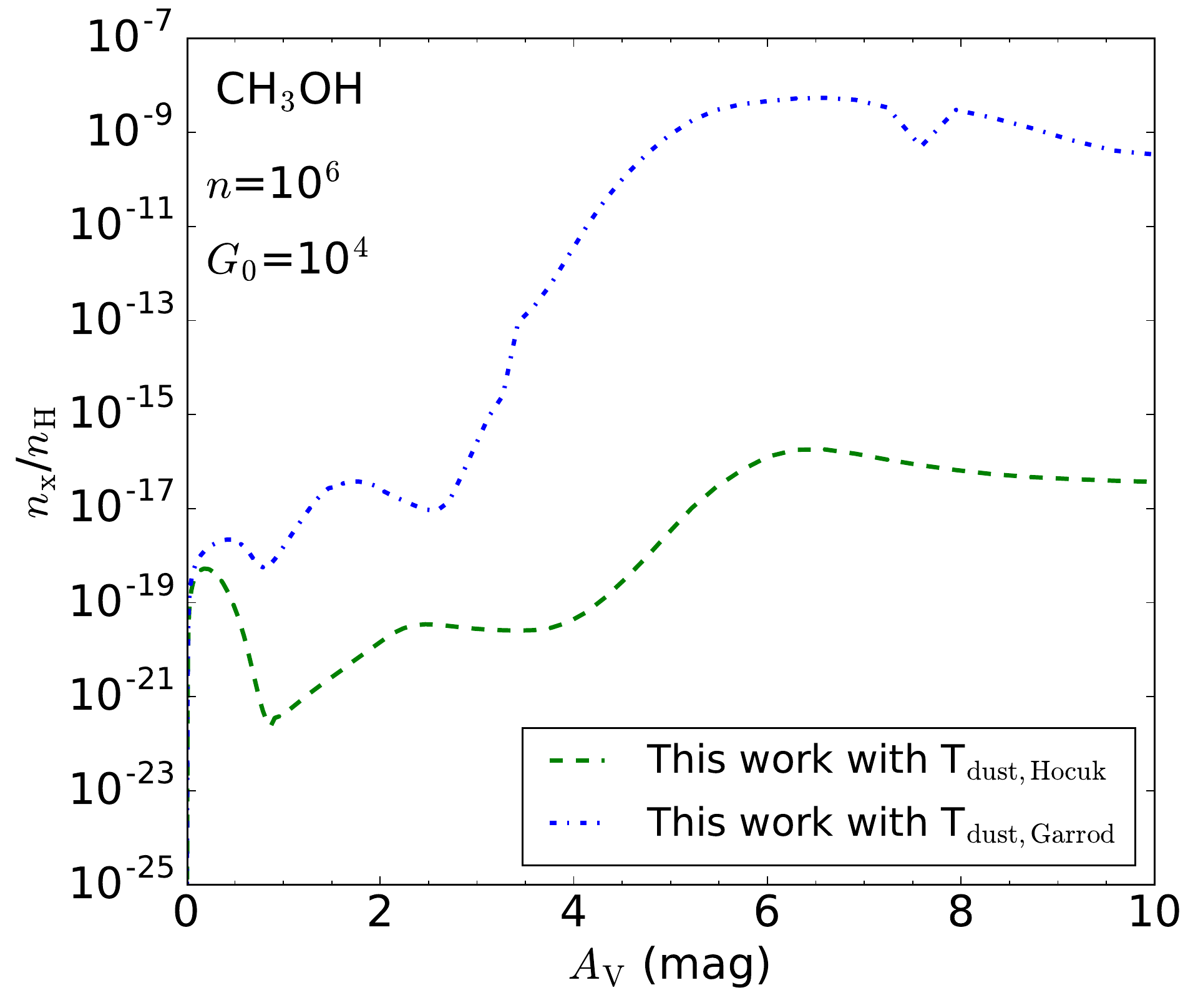}  
\caption{Abundances of CH, CO, H$_2$CO, and CH$_3$OH obtained with the updated Meijerink PDR code using $T$$_{\mathrm{dust}}$ from Hocuk et al. (2017) (green dashed lines) and from Garrod $\&$ Pauly (2011) (blue dotted lines). Results for Model 1 are shown on the left panels, for Model 2 in the middle panels, and for Model 3 on the right panels.}
\label{figure:C_bearing_comparison}
\end{figure*}

Figures \ref{figure:O_bearing_comparison}-\ref{figure:JH2O_comparison} show abundances for several species obtained with the most recent version of the Meijerink PDR code presented here, considering $T$$_{\mathrm{dust}}$ from Hocuk et al. (2017) (green dashed lines) and from Garrod $\&$ Pauly (2011) (blue dotted lines). We obtain that the chemical impact of considering different dust temperature significantly varies depending on the characteristics of the PDR, the visual extinction range, and the type of molecule. Below we analyse the dust temperature effects considering several molecule families.

\subsubsection{Simple oxygen-bearing molecules}
\label{simple oxygen molecules temperature}

Figure \ref{figure:O_bearing_comparison} shows the abundances of simple oxygen-bearing species (OH, O$_2$, and H$_2$O) considering the two distinct temperatures previously mentioned. The main abundance differences are found for high $G$$_{\mathrm{0}}$ PDRs, where the dust temperature varies up to $\sim$30 K depending on the considered $T$$_{\mathrm{dust}}$ expression as mentioned in Sect. \ref{Dust_temperat}. In particular, these abundance differences can be of up to 4 orders of magnitude in the inner regions of the cloud ($A$$_{\mathrm{V}}$$>$6 mag) for the case of water and of more than 6 orders of magnitude for molecular oxygen. In the edge of the cloud ($A$$_{\mathrm{V}}$$\leq$1 mag), however, the abundance differences are no more than one order of magnitude. 

For a high $G$$_{\mathrm{0}}$ PDR with density $n$=10$^5$ cm$^{-3}$ (middle panels), low dust temperatures (derived from Garrod$'$s expression) promote the formation of OH, O$_2$, and H$_2$O at 1$<$$A$$_{\mathrm{V}}$$\lesssim$5.5 mag, while for larger extinctions, high dust temperatures (obtained from Hocuk$'$s expression) lead to the highest oxygen-bearing molecule abundances with differences of up to 7 orders of magnitude between both expressions. This is the interesting case of molecular oxygen, an elusive molecule in the interstellar medium (Odin satellite only provided upper limits ($\leq$10$^{-7}$) for its abundances especially in cold dark clouds, e.g., Pagani et al. 2003) with only a few recent detections: the massive Orion star-forming region (with {\it{X}}(O$_2$)$\sim$10$^{-6}$, Goldsmith et al. 2011, Chen et al. 2014) and the low-mass dense core $\rho$ Oph A (with {\it{X}}(O$_2$)$\sim$5$\times$10$^{-8}$, Larsson et al. 2007; Liseau et al. 2012). Recently, this molecule has also been detected in surprisingly large quantities towards the Solar System comets 67P/Churyumov $-$ Gerasimenko (67P/C-G) by Bieler et al. (2015) with Rosetta, and in 1P/Halley by Rubin et al. (2015) with the Giotto mission. Their results confirm that O$_2$ is the fourth most abundant molecule in comets. In our PDR case, we find that, at $A$$_{\mathrm{V}}$$\gtrsim$4 mag, high dust temperatures allow to enhance the surface diffusion of O atoms that recombine to form solid O$_2$, which is then released into the gas-phase through thermal desorption. This is in agreement with Taquet et al. (2016). We also find chemical desorption from the reaction of two solid oxygens as an important way to form O$_2$ gas, especially at 4$\lesssim$$A$$_{\mathrm{V}}$$\lesssim$6 mag (see Fig. \ref{figure:O2_gas_formation}, left panel, in the Appendix \ref{Figures}). We highlight the need of carrying out an O$_2$ search in PDRs to make quantitative comparison with our predictions.

For the particular case of water in a high $G$$_{\mathrm{0}}$ PDR (middle bottom panel, Fig. \ref{figure:O_bearing_comparison}), it presents a low abundance variation for $A$$_{\mathrm{V}}$$<$1 mag when the dust temperature varies by $\sim$30 K, highlighting a gas-phase chemical formation route for this molecule via ion-chemistry at the edge of the cloud. For intermediate extinctions (1$<$$A$$_{\mathrm{V}}$$\lesssim$3 mag), the H$_2$O abundance variations are very small ($<$1 order of magnitude) between both $T$$_{\mathrm{dust}}$ expressions as also found for the OH abundances, while O$_2$ presents differences of about 2 orders of magnitude. This shows that at intermediate visual extinctions, OH is a more relevant reactant than O$_2$ to form water, and that the main H$_2$O formation route is through successive hydrogenation of atomic oxygen in agreement with Dulieu et al. (2010). In particular, we find this chemical reaction efficient for $A$$_{\mathrm{V}}$$<$5 mag (see Fig. \ref{figure:O2_gas_formation}, right panel, in the Appendix \ref{Figures}). For larger extinctions, the warmer the dust grains, the higher the water abundances with differences of up to four orders of magnitud between both $T$$_{\mathrm{dust}}$, being photo and cosmic ray desorption the most efficient reactions forming gaseous water at $A$$_{\mathrm{V}}$$\gtrsim$5 mag (Fig. \ref{figure:O2_gas_formation}, right panel, Appendix \ref{Figures}).

If the density of the PDR increases by one order of magnitude (right panels, Fig. \ref{figure:O_bearing_comparison}), the main effect with respect to the low density case is found at the edge of the cloud (at $A$$_{\mathrm{V}}$$\lesssim$0.5 mag) where the abundances of the three molecules (OH, O$_2$, and H$_2$O) increase by $\sim$2 orders of magnitude for both dust temperature expressions. In the case of a low $G$$_{\mathrm{0}}$ PDR (left panels, Fig. \ref{figure:O_bearing_comparison}), the temperature differences between both $T$$_{\mathrm{dust}}$ expressions are $\lesssim$10 K. These small differences lead to variations in the abundances of OH, O$_2$, and H$_2$O of no more than one order of magnitude for 0$\leq$$A$$_{\mathrm{V}}$$\leq$10 mag.

From these results, we therefore conclude that the largest impact in the chemistry of simple oxygen-bearing molecules is found in high $G$$_{\mathrm{0}}$ PDRs, which present the largest dust temperature differences between the two approaches for $T$$_{\mathrm{dust}}$. In these PDRs, low dust temperatures promote the formation of OH, O$_2$, and H$_2$O at intermediate visual extinctions ($A$$_{\mathrm{V}}$$\lesssim$5 mag), while high values of $T$$_{\mathrm{dust}}$ promote their formation at larger $A$$_{\mathrm{V}}$.

\subsubsection{Carbon-bearing molecules}
\label{carbon_dust-temperature}

Figure \ref{figure:C_bearing_comparison} shows the abundances of carbon-bearing molecules (CH, CO, H$_2$CO, and CH$_3$OH) considering two different dust temperatures (from Hocuk$'$s and Garrod$'$s expressions). For the simplest species (CH and CO shown in the two top panels), we distinguish two regimes for any PDR type: the low visual extinction regime ($A$$_{\mathrm{V}}$$\lesssim$2 mag), where the variation of dust temperature does not have a significant impact on the abundances of these molecules since they mainly form in the gas phase, and the high visual extinction range ($A$$_{\mathrm{V}}$$>$2 mag), where their abundances can vary by up to three orders of magnitude.  

For a high $G$$_{\mathrm{0}}$ PDR (middle panels), Hocuk$'$s expression produces the highest $T$$_{\mathrm{dust}}$ values, which lead to a low CO depletion on grain surfaces and, therefore, to large CO gas-phase abundances (up to two orders of magnitude larger than those obtained using Garrod$'$s expression). The large CO gas-phase abundance at $A$$_{\mathrm{V}}$$\gtrsim$4 mag obtained with Hocuk$'$s expression implies low abundances of solid CO and, therefore, a restriction in the formation of more complex molecules on the grain surfaces through CO ice, such as H$_2$CO and CH$_3$OH, as we observe in Fig. \ref{figure:C_bearing_comparison} (two bottom panels). In particular, we obtain that the abundance of H$_2$CO at $A$$_{\mathrm{V}}$$\gtrsim$4 mag is lower for Hocuk$'$s expression than for Garrod$'$s expression by up to $\sim$3 orders of magnitude. This difference is even larger (up to 6 orders of magnitude) in the case of the complex molecule CH$_3$OH. 

The increase of the PDR density favours the formation of all the carbon-bearing molecules at $A$$_{\mathrm{V}}$$\lesssim$1 mag as shown in Fig. \ref{figure:C_bearing_comparison} (right panels). In particular, we find that the abundances of CH, CO, H$_2$CO, and CH$_3$OH increase by about two orders of magnitude in the edge of the cloud without finding significant differences between both $T$$_{\mathrm{dust}}$ expressions. At intermediate and large extinctions ($A$$_{\mathrm{V}}$$\gtrsim$3 mag), we observe that the density increase mainly affects the abundances obtained with the lowest $T$$_{\mathrm{dust}}$ values (blue dotted curves), with CO being the most affected molecule. In particular, the increase of density by one order of magnitude leads to a CO abundances decrease of about three orders of magnitude due to a more efficient depletion. This promotes the formation of complex molecules. In fact, the abundances of CH$_3$OH are slightly larger at $A$$_{\mathrm{V}}$$>$7 mag in the PDR with density 10$^6$ cm$^{-3}$ (right bottom panel) than in the PDR with $n$=10$^5$ cm$^{-3}$ (middle bottom panel).

\begin{figure*}
\centering
\includegraphics[scale=0.285, angle=0]{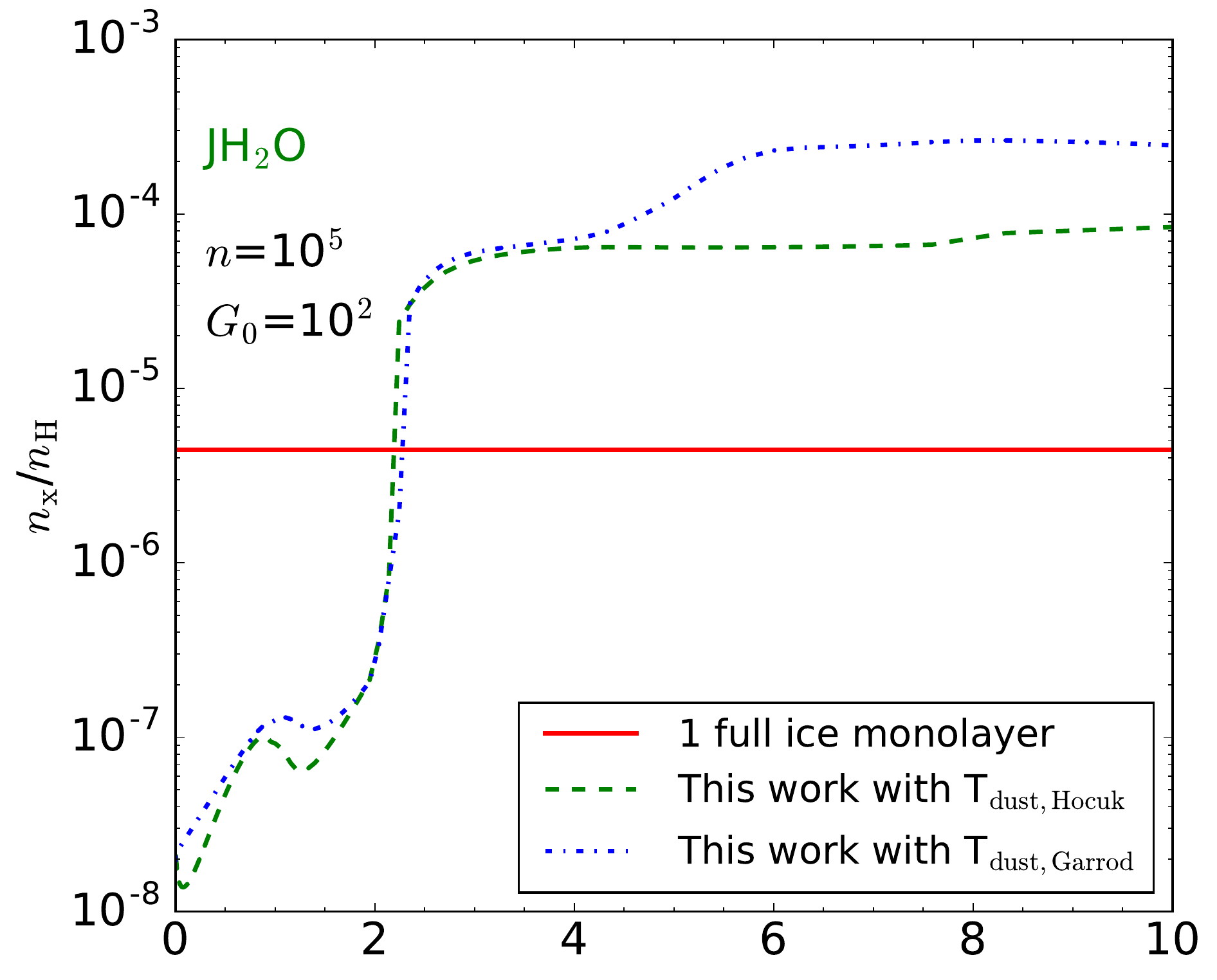}  \hspace{0.0cm}
\includegraphics[scale=0.285, angle=0]{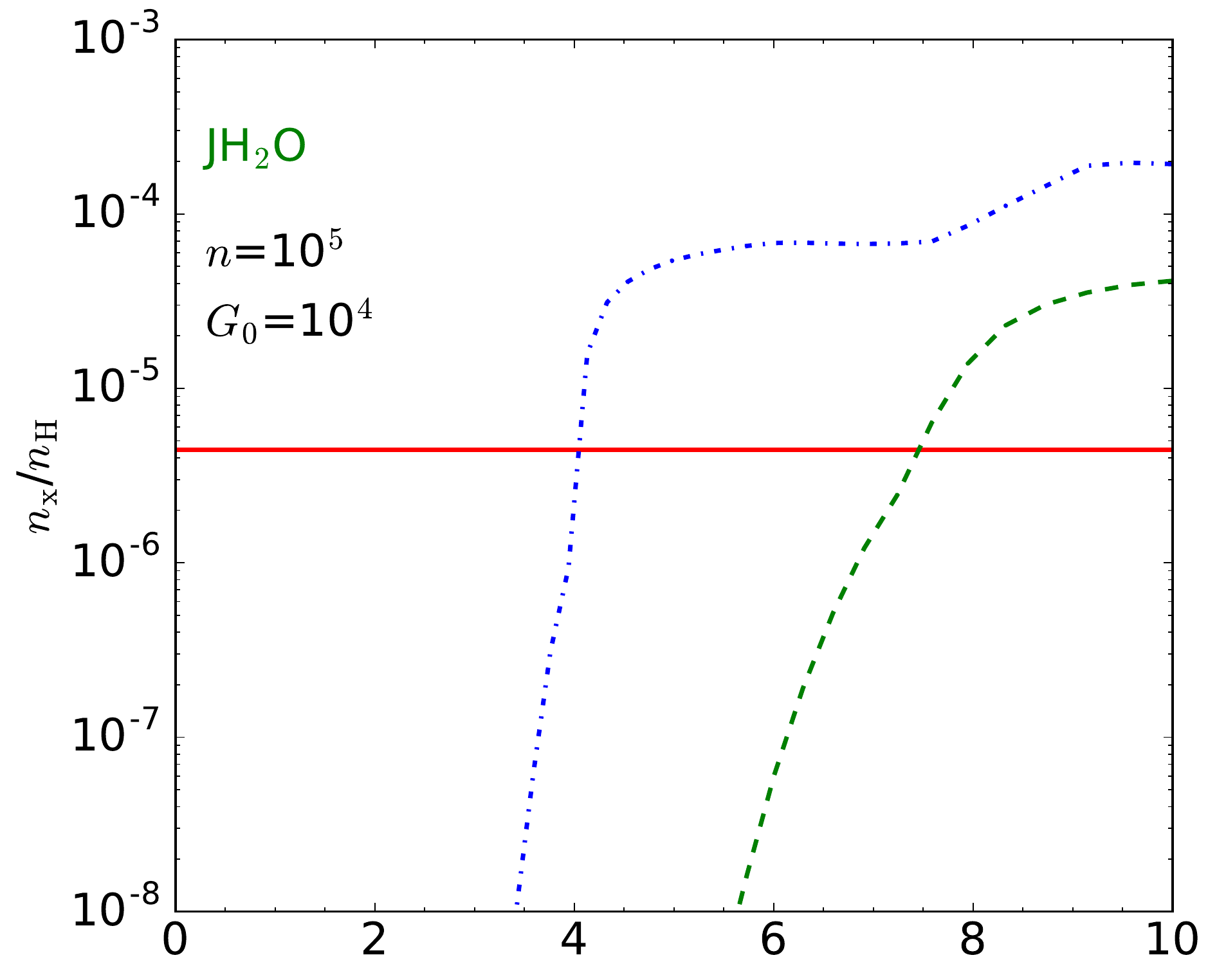}  \hspace{0.0cm}
\includegraphics[scale=0.285, angle=0]{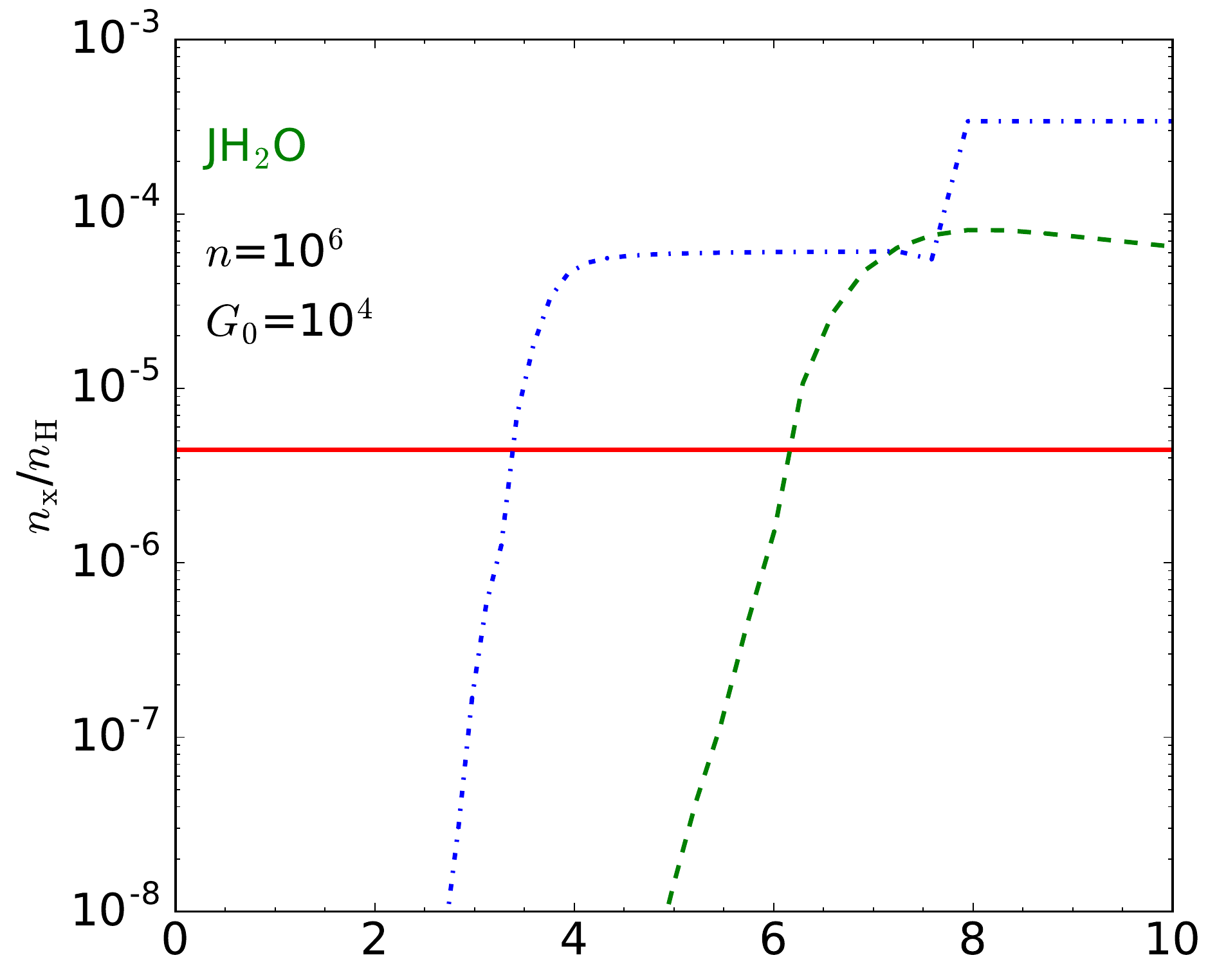}  \hspace{0.0cm}
\includegraphics[scale=0.285, angle=0]{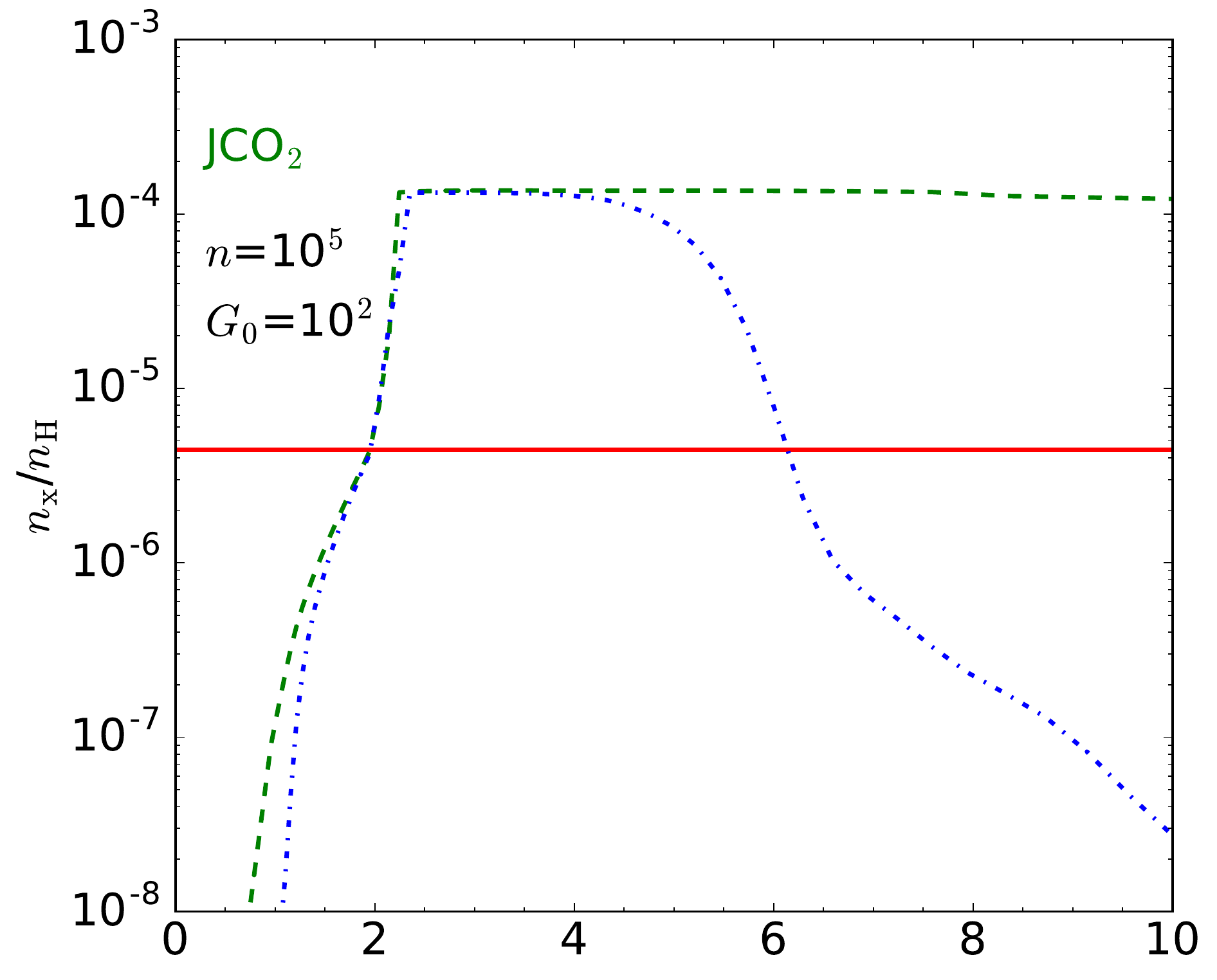}  \hspace{0.0cm}
\includegraphics[scale=0.285, angle=0]{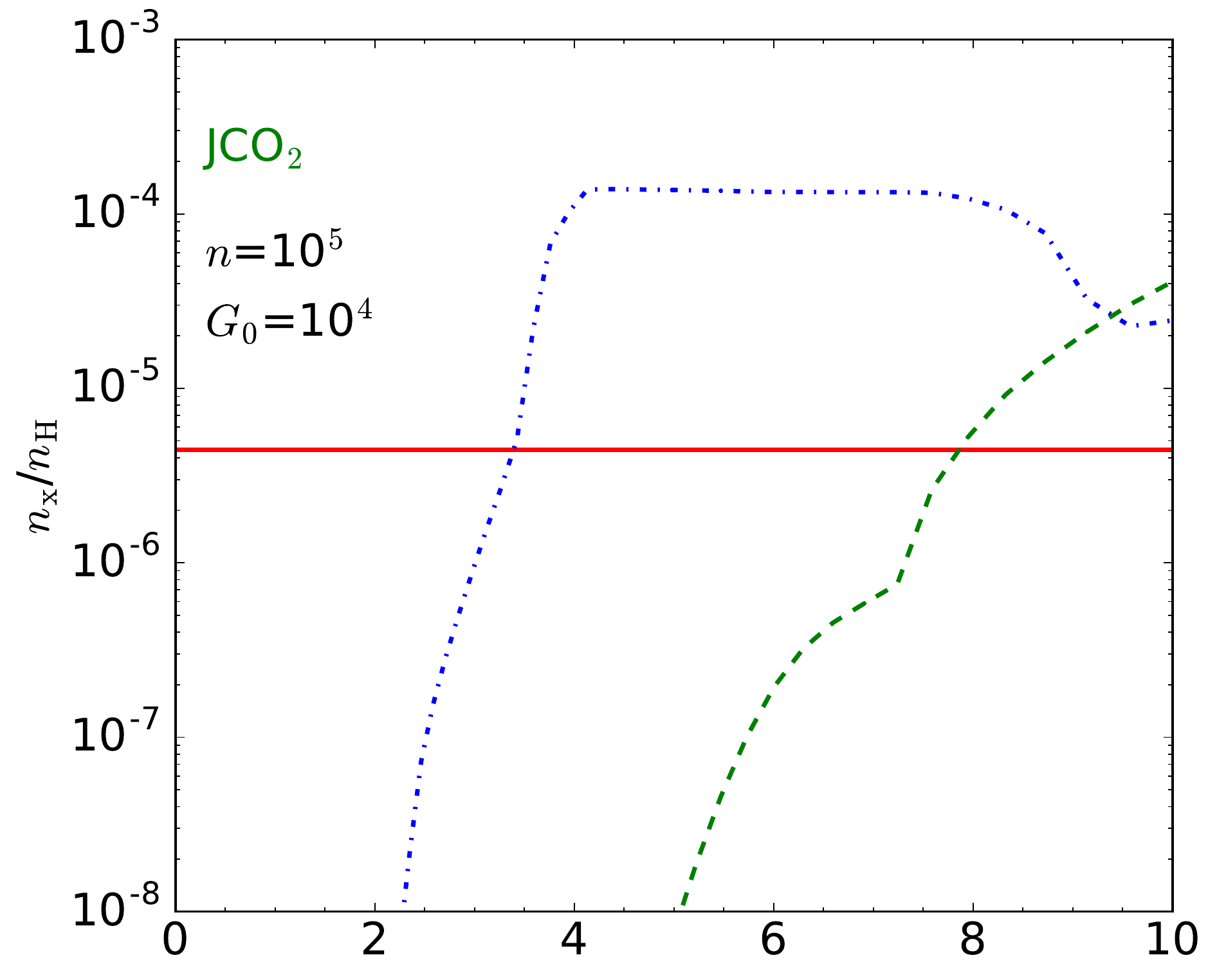}  \hspace{0.0cm}
\includegraphics[scale=0.285, angle=0]{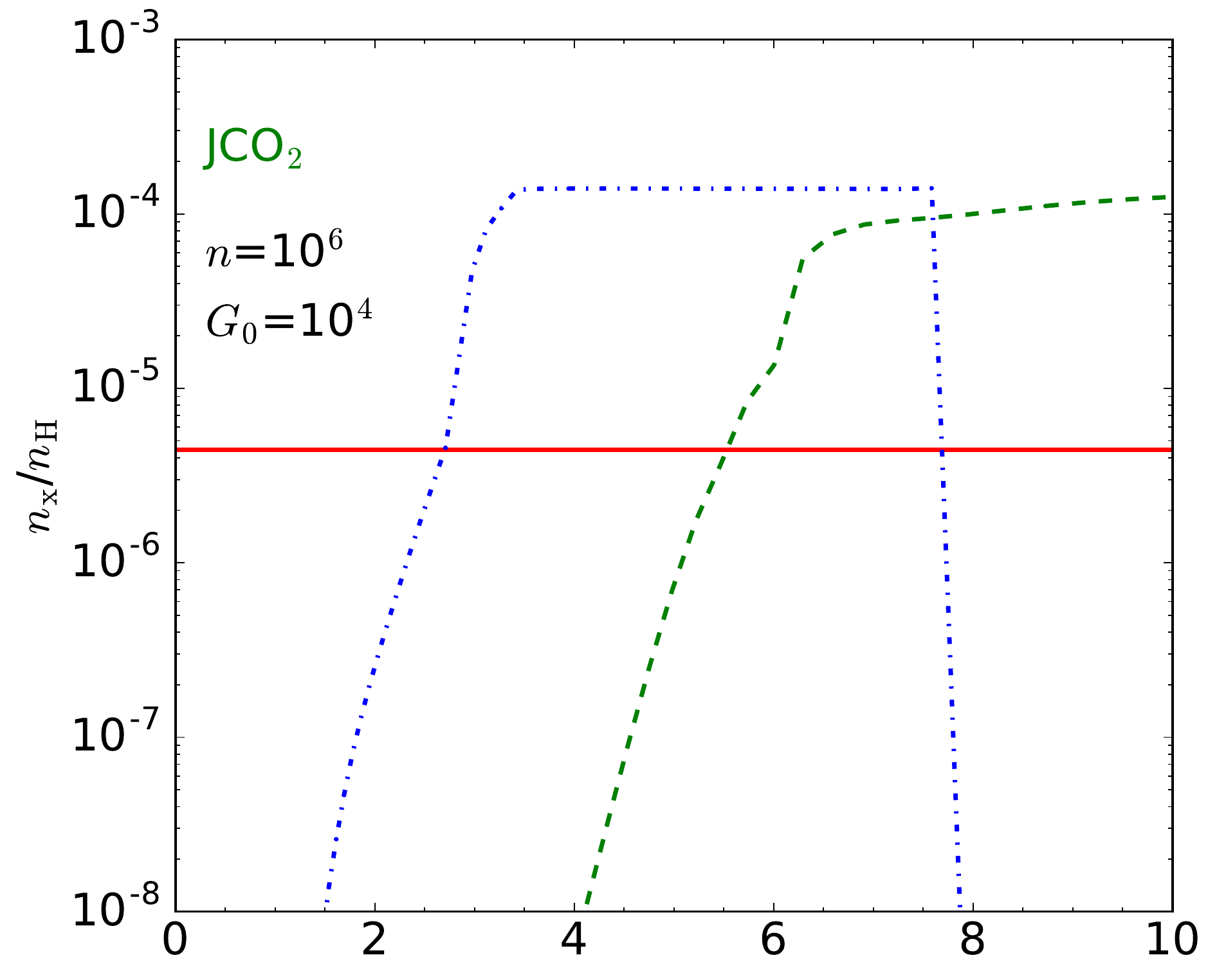}  \hspace{0.0cm}
\includegraphics[scale=0.285, angle=0]{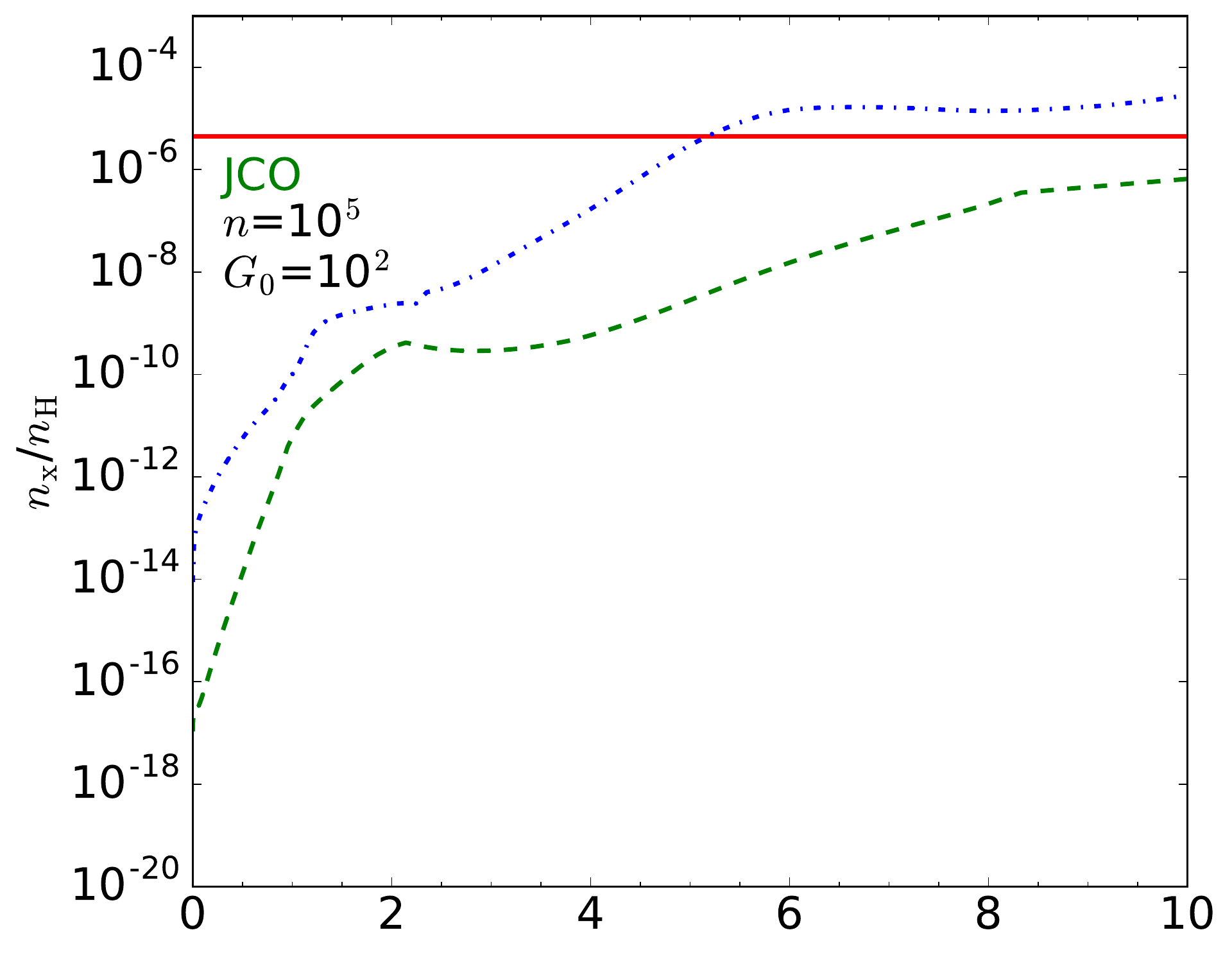}  \hspace{0.0cm}
\includegraphics[scale=0.285, angle=0]{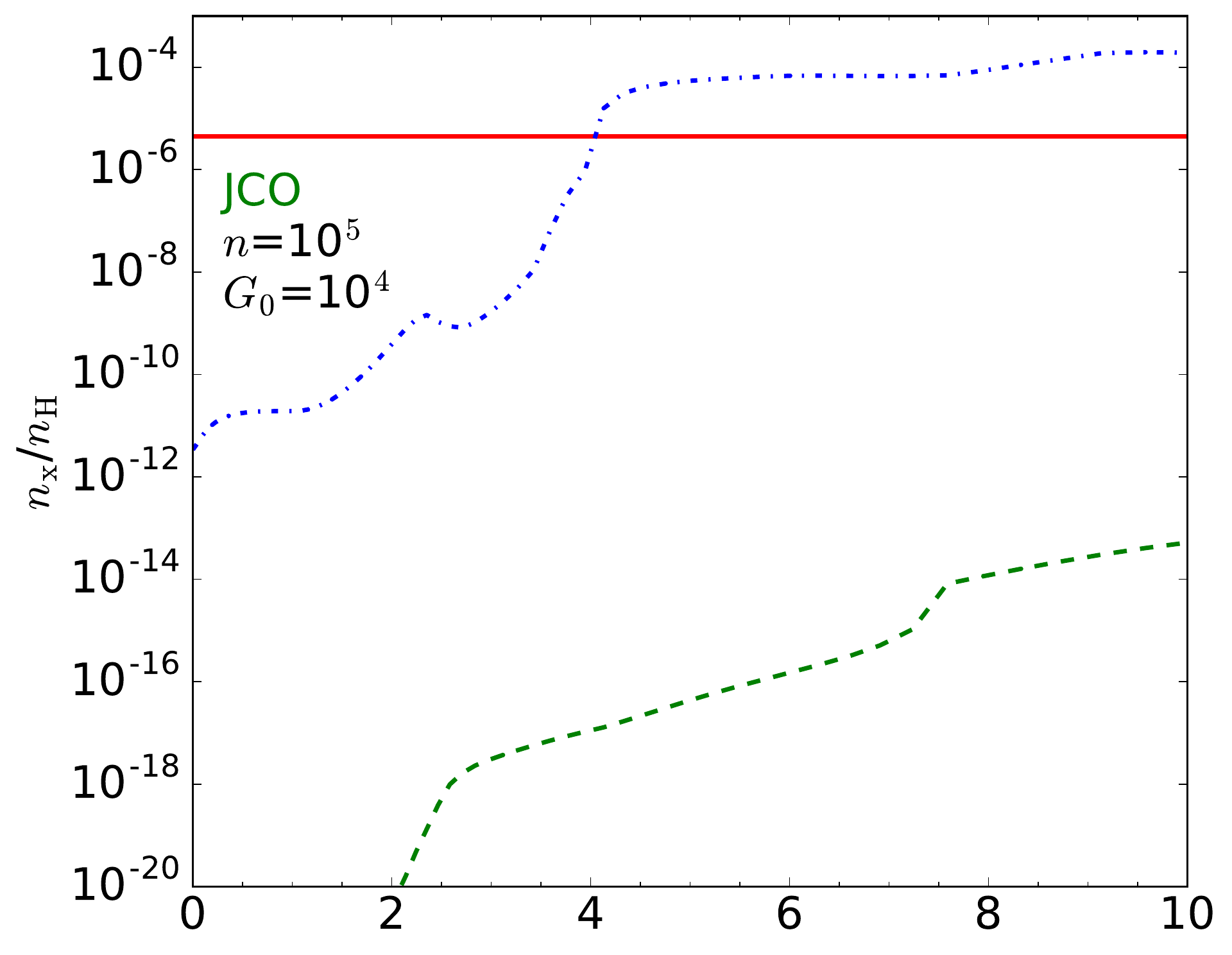}  \hspace{0.0cm}
\includegraphics[scale=0.285, angle=0]{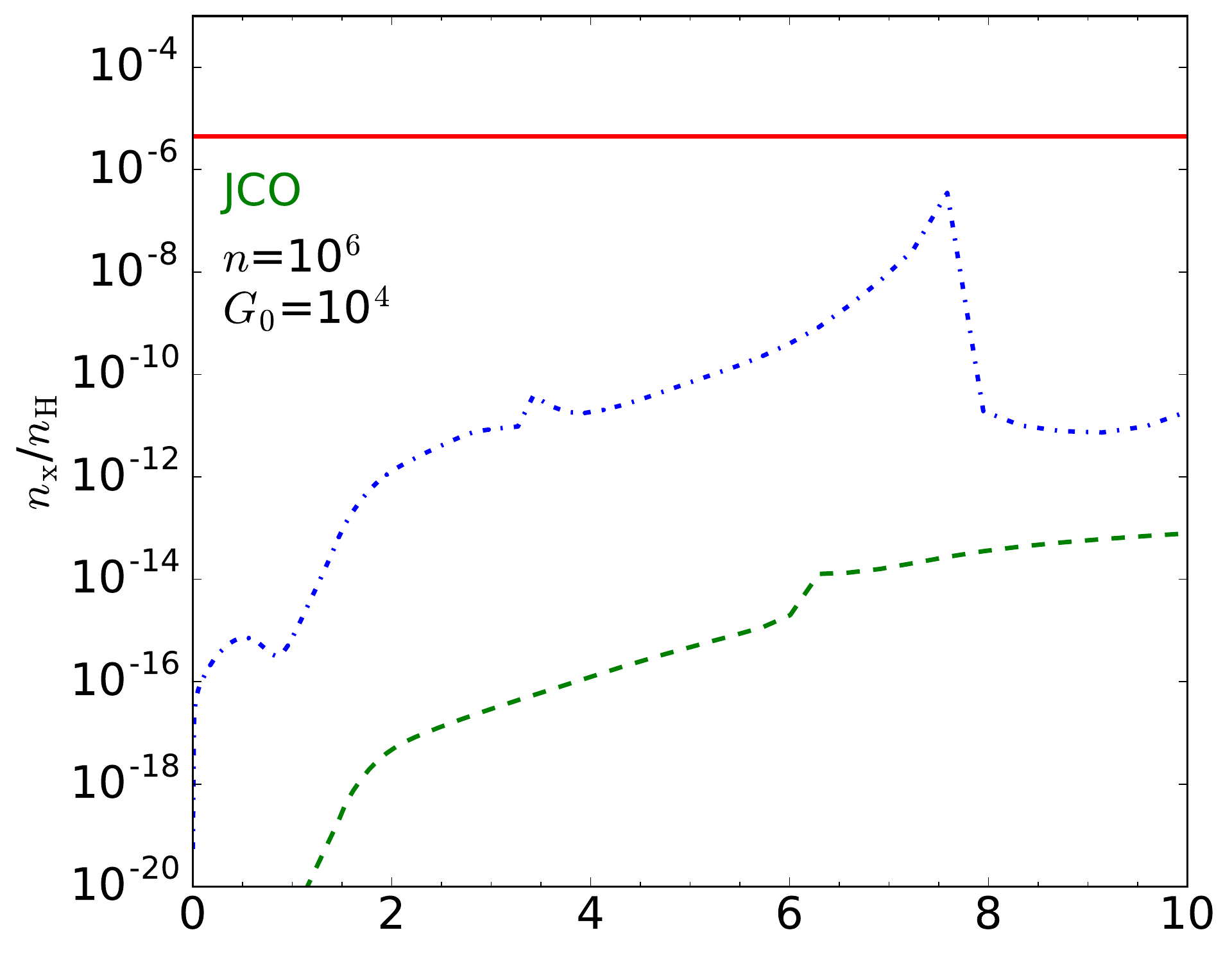}  \hspace{0.0cm}
\includegraphics[scale=0.285, angle=0]{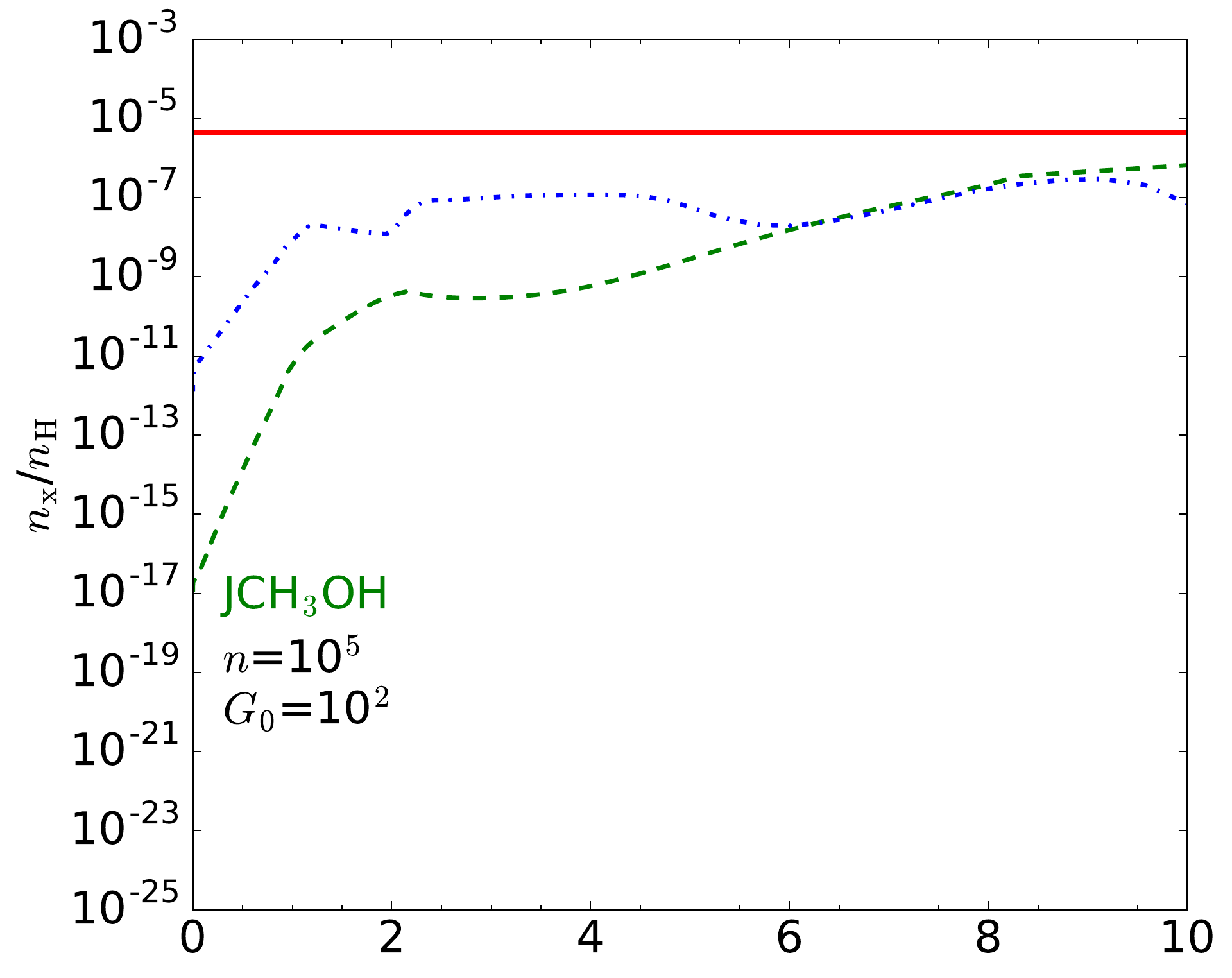}  \hspace{0.0cm}
\includegraphics[scale=0.285, angle=0]{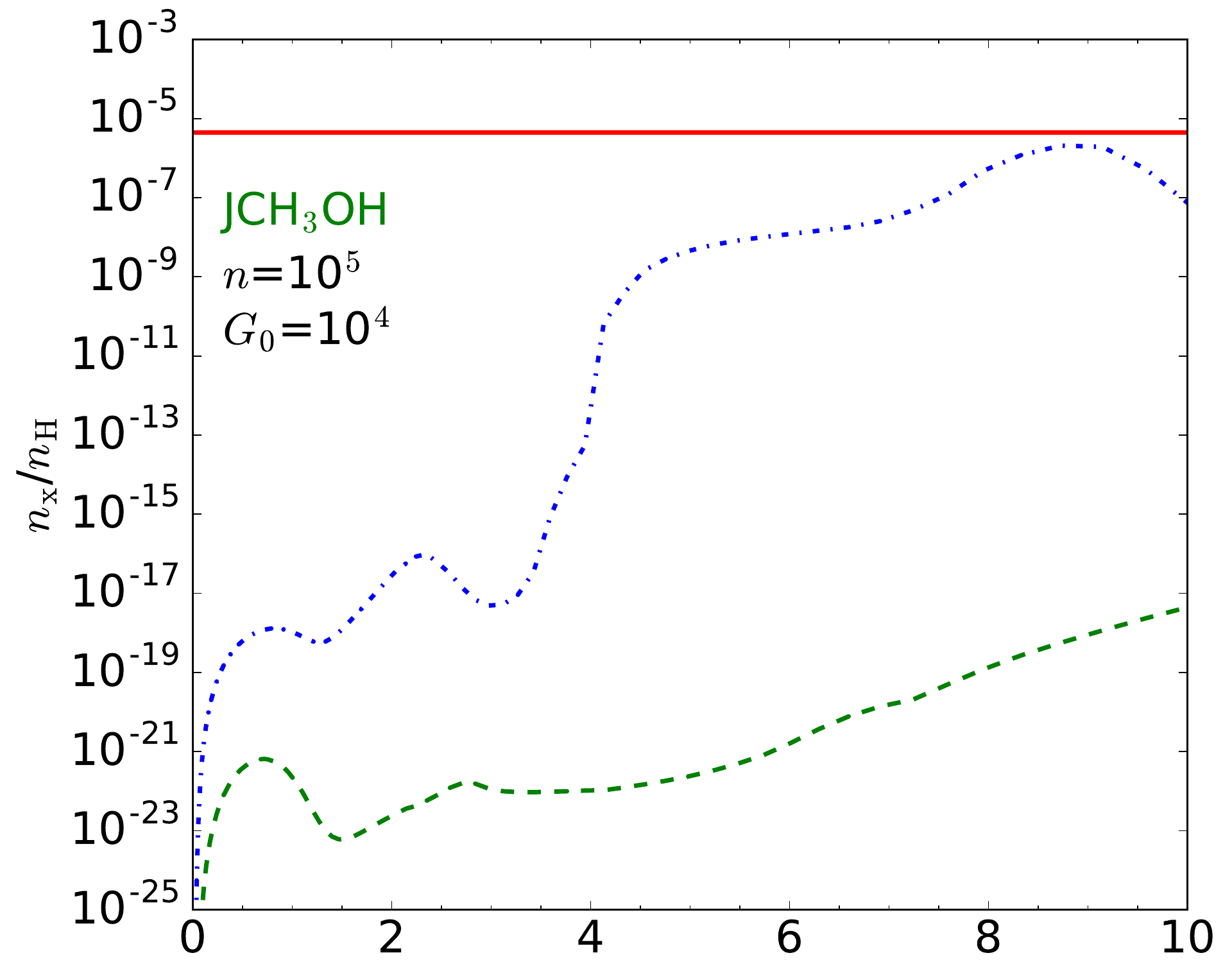}  \hspace{0.0cm}
\includegraphics[scale=0.285, angle=0]{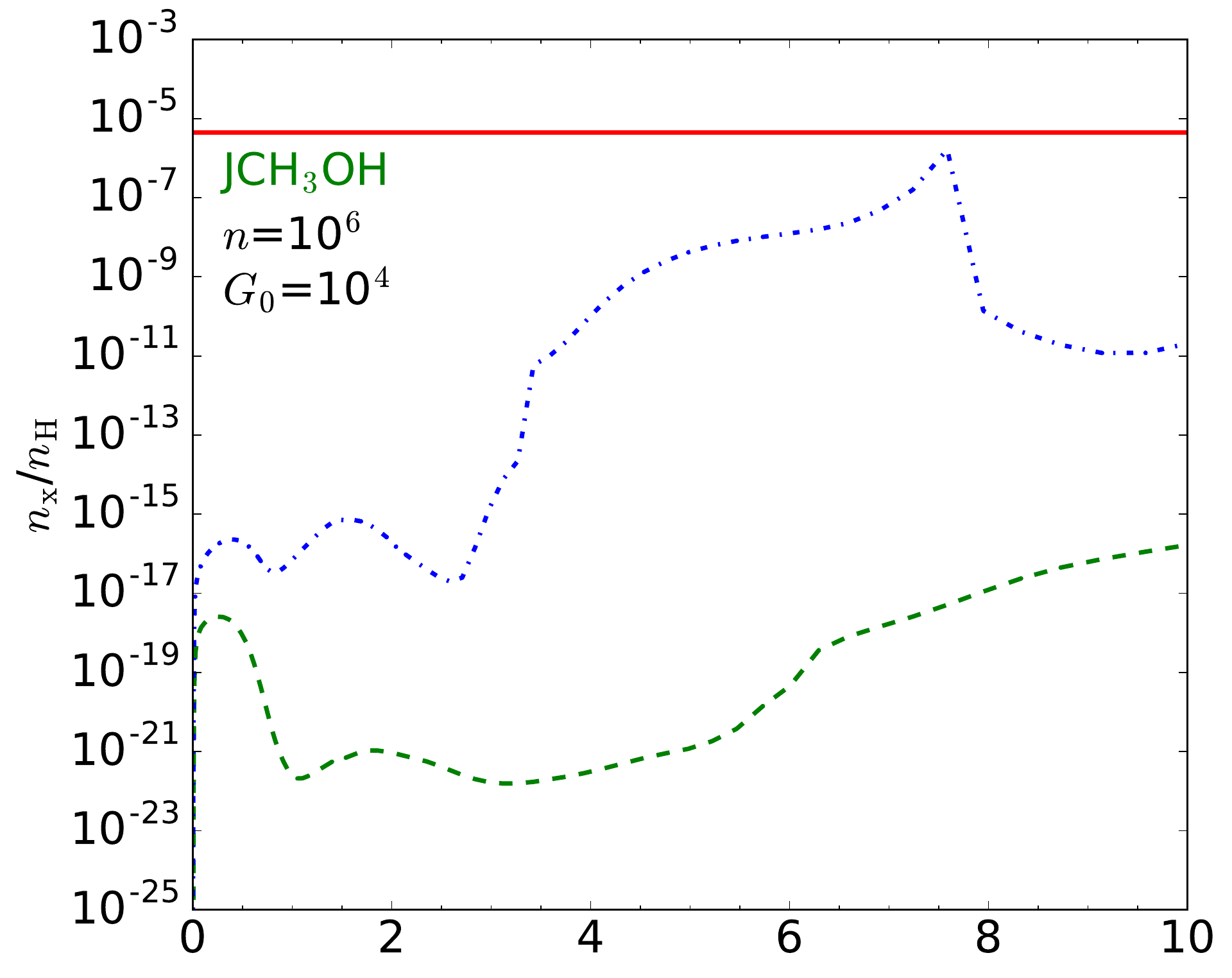}  \hspace{0.0cm}
\includegraphics[scale=0.285, angle=0]{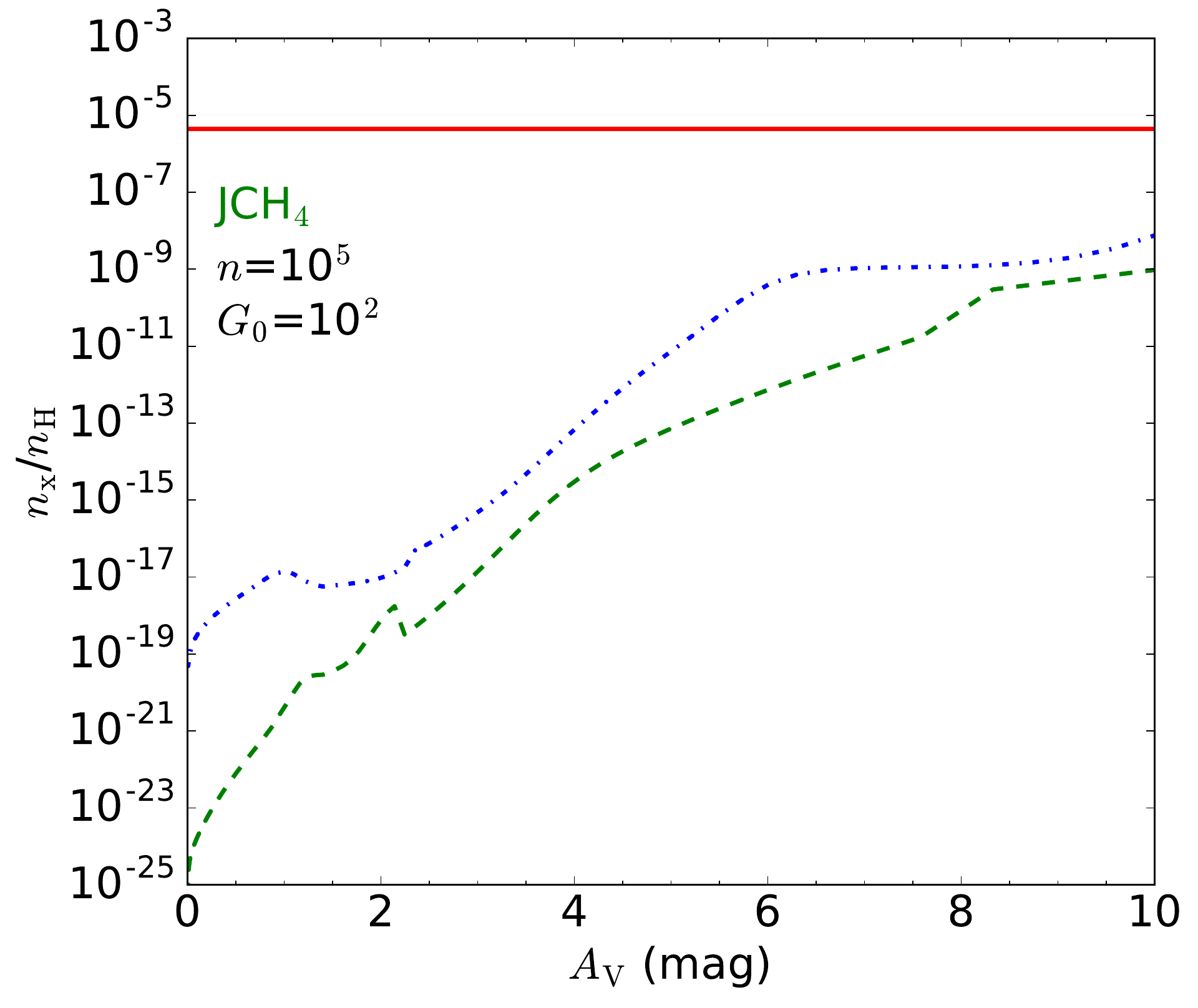}  \hspace{0.0cm}
\includegraphics[scale=0.285, angle=0]{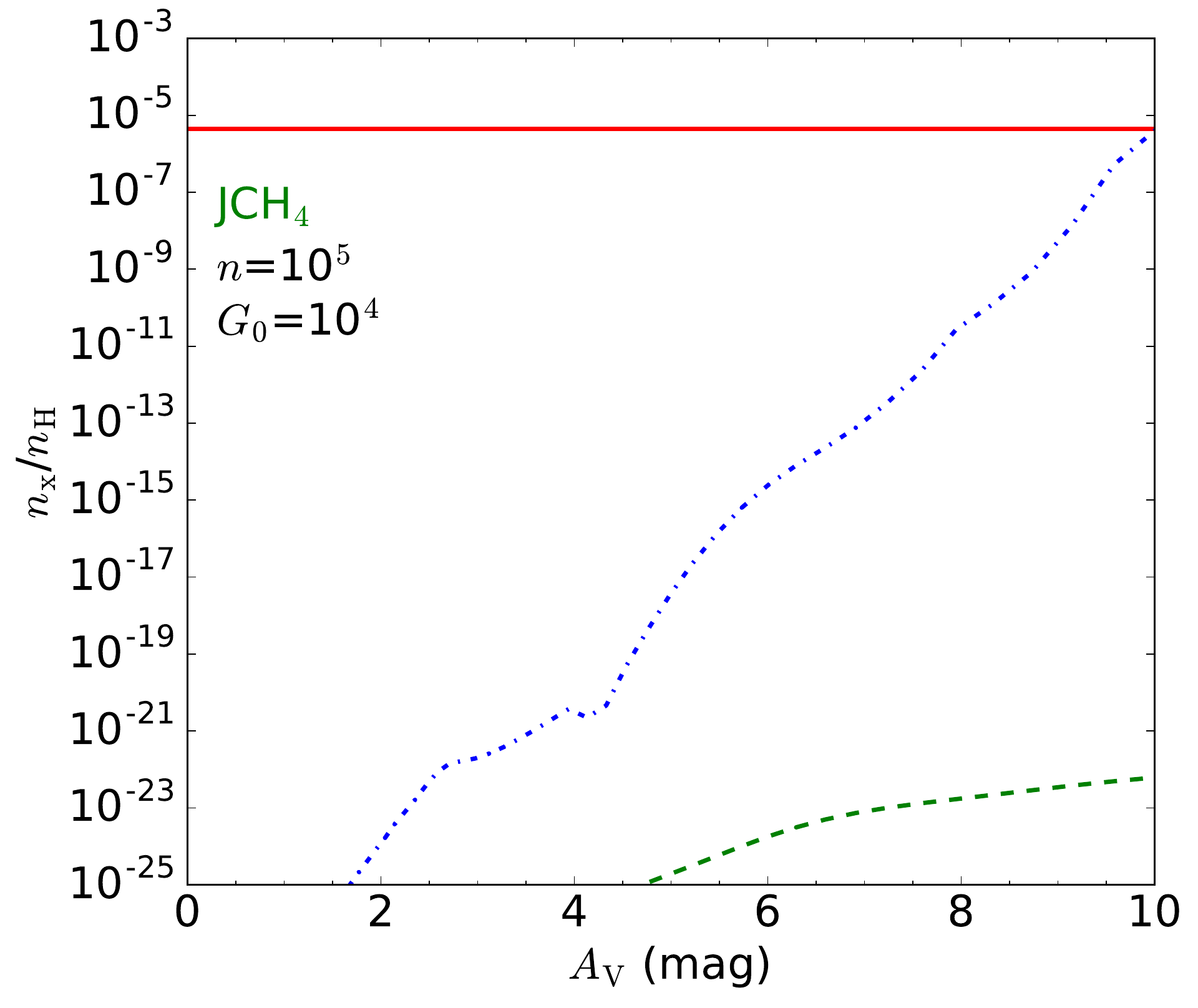}  \hspace{0.0cm}
\includegraphics[scale=0.285, angle=0]{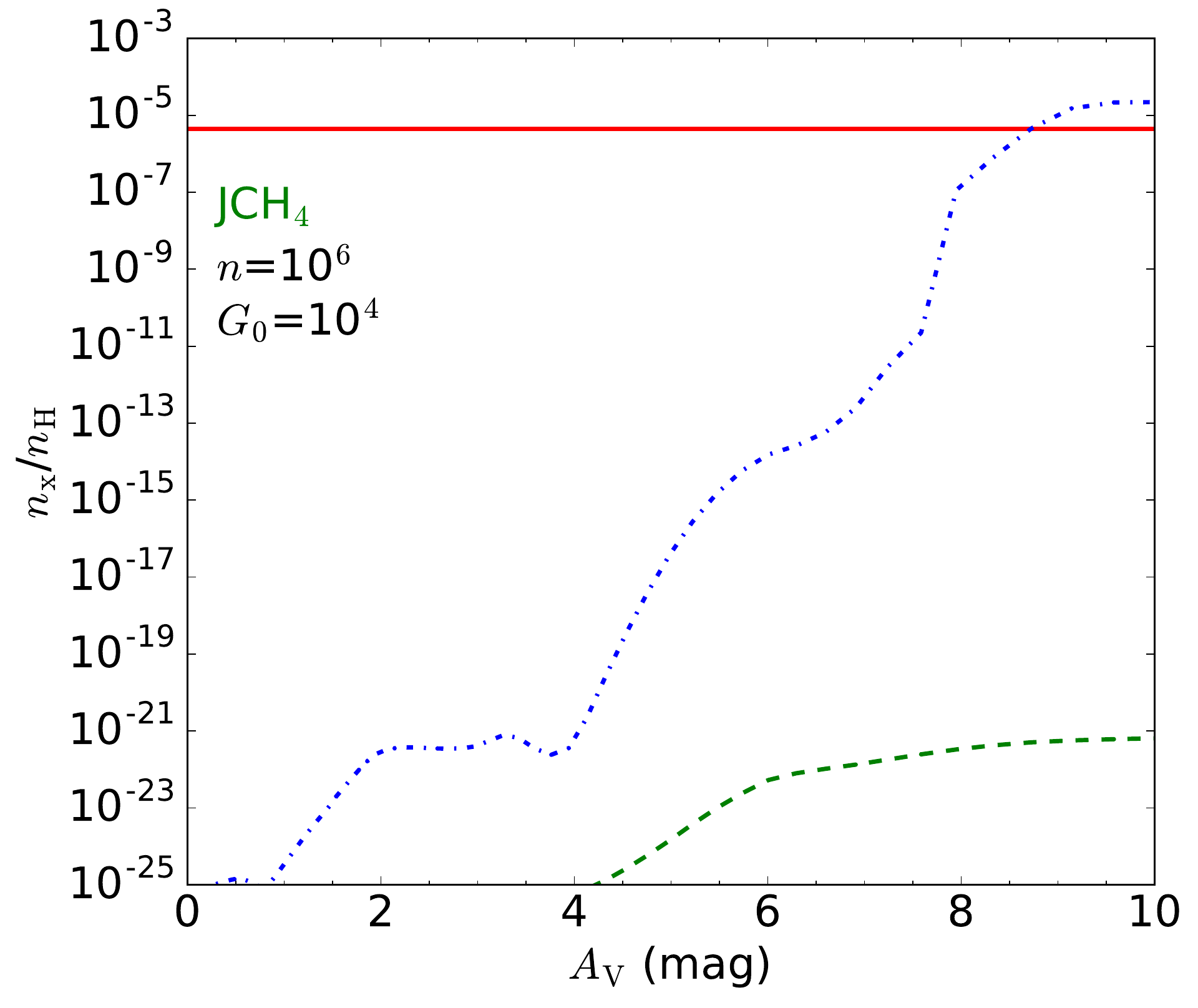}  \hspace{0.0cm}\\
\caption{Abundances of JH$_2$O, JCO$_2$, JCO, JCH$_3$OH, and JCH$_4$ obtained with the updated Meijerink PDR code using $T$$_{\mathrm{dust}}$ from Hocuk et al. (2017) (green dashed lines) and from Garrod $\&$ Pauly (2011) (blue dotted lines). Results for Model 1 are on the left panels, for Model 2 in the middle panels, and for Model 3 on the right panels. J$i$ means solid $i$. The red solid line represents the number of possible adsorption sites on grain surfaces per cm$^{2}$.}
\label{figure:JH2O_comparison}
\end{figure*}

For a low $G$$_{\mathrm{0}}$ PDR (left panels of Fig. \ref{figure:C_bearing_comparison}), in the edge of the cloud ($A$$_{\mathrm{V}}$$\lesssim$1 mag), we only find significant differences for H$_2$CO and CH$_3$OH when changing $T$$_{\mathrm{dust}}$, since these molecules are mainly formed on dust grains, which makes them very sensitive to variations of dust temperature. For these two molecules, the lower the dust temperatures, the higher the abundances, since H atoms can reside on dust grains longer if temperatures are low. This is in disagreement with Le Gal et al. (2017), who suggested that the warming up of grain surfaces speeds up chemical surface processes forming complex organic molecules (COMs), explaining thus the high abundance of some COMs observed in the PDR of the Horsehead ($G$$_{\mathrm{0}}$$\sim$10$^2$ and $n$$\sim$10$^5$ cm$^{-3}$, Habart et al. 2005, Guzm\'an et al. 2013) with respect to the core of the cloud (Gratier et al. 2013). It must be noted that while their conclusions were deduced considering a grain warm up from $\sim$10 K to $\sim$25 K, the $T$$_{\mathrm{dust}}$ difference in our comparison is $\sim$5 K for $A$$_{\mathrm{V}}$$\lesssim$1 mag. Nevertheless, we obtain the same trend for CH$_3$OH in the high $G$$_{\mathrm{0}}$ case (Fig. \ref{figure:C_bearing_comparison}, middle and right bottom panels), where the $T$$_{\mathrm{dust}}$ difference for $A$$_{\mathrm{V}}$$\lesssim$1 mag is about $\sim$25 K.
Figure \ref{figure:density_effects_low_G0} (see Appendix \ref{Figures}) shows abundances of H$_2$CO and CH$_3$OH at $A$$_{\mathrm{V}}$$\leq$1 mag when the density is increased by one order of magnitude ($n$$\sim$10$^6$ cm$^{-3}$), while the radiation intensity remains as $G$$_{\mathrm{0}}$$\sim$10$^2$. When density increases, we obtain an abundance increase for both molecules between one and three orders of magnitude. From these results, we propose an alternative stage where the presence of higher abundances of some COMs in the PDR than in the core of the Horsehead is the result of the presence of clumps with very high densities (of at least $n$=10$^6$ cm$^{-3}$) and low dust temperature values ($T$$_{\mathrm{dust}}$$<$25 K) in the edge of the cloud. See also Sect. \ref{COMS_time} for a more detailed explanation of the density role in the significant enrichment of some COMs in the PDR with respect to the cloud core.

From all these results, we derive that low dust temperatures significantly promote the formation of COMs in the inner regions of high $G$$_{\mathrm{0}}$ PDRs, as well as in the edge of clouds with low $G$$_{\mathrm{0}}$ PDRs.

\subsubsection{Solid molecules}

Figure \ref{figure:JH2O_comparison} shows the abundances of solid H$_2$O, CO$_2$, CO, CH$_3$OH, and CH$_4$ obtained using different analytical expression for $T$$_{\mathrm{dust}}$ (green dashed lines for Hocuk's expression and blue dotted lines for Garrod$'$s expression). 

For a low $G$$_{\mathrm{0}}$ PDR (left panels), the visual extinctions at which the first full ice monolayers of H$_2$O and CO$_2$ are formed barely changes with the $T$$_{\mathrm{dust}}$ considered due to the small difference between both expression ($\lesssim$10 K). For both molecules, this formation occurs at $A$$_{\mathrm{V}}$$\sim$2-3 mag, as in diffuse molecular clouds (Boogert et al. 2015). For the case of CO, only the lowest $T$$_{\mathrm{dust}}$ leads to the formation of CO ice at intermediate and large extinctions ($A$$_{\mathrm{V}}$$\geq$5 mag), while other more complex molecules, such as CH$_3$OH and CH$_4$, present abundances lower than 10$^{-6}$ and do not form ice at $A$$_{\mathrm{V}}$$\leq$10 mag for any of the two $T$$_{\mathrm{dust}}$ considered. This is in disagreement with Hollenbach et al. (2009), who obtained similar maxima for the CO and CH$_4$ ice abundances in a PDR with $G$$_{\mathrm{0}}$=100 and $n$=10$^4$ cm$^{-3}$, suggesting an overproduction of methane, since \"Oberg et al. (2008) and Boogert et al. (2015) observed solid CH$_4$/H$_2$O abundances of $\sim$0.05 and 0.01 in low- and high-mass young stellar objects, respectively.  

In a PDR with the same density, but a $G$$_{\mathrm{0}}$ two orders of magnitude higher (middle panels), the difference between both dust temperature expressions is $\sim$30 K (see Fig. \ref{figure:Tdust_comparison}), which is high enough to make molecular depletion onto dust grains less efficient in the warmest case. This leads to the formation of H$_2$O and CO$_2$ ices at larger extinctions (between $\sim$3 and 5 mag) for both $T$$_{\mathrm{dust}}$ expressions. For the case of solid CO, CH$_3$OH, and CH$_4$, the increase of the radiation intensity from $G$$_{\mathrm{0}}$=10$^2$ to $G$$_{\mathrm{0}}$=10$^4$ produces a significant drop in their abundances of at least 5 orders of magnitude in the highest $T$$_{\mathrm{dust}}$ case (green lines), highlighting the need of cool grains to form ices of carbon monoxide, methanol and methane.  

When density increases (right panels), the visual extinction at which H$_2$O and CO$_2$ ices are formed slightly decreases for both dust temperature expressions. This is due to the increase in the rate at which atoms and molecules hit dust grains, which is linearly dependent on the gas number density. Regarding minor ice mantle components, the density increase in a very high $G$$_{\mathrm{0}}$ PDR allows the formation of methane ice only at large extinctions ($A$$_{\mathrm{V}}$$>$8 mag) when the dust grain temperature remains low ($\lesssim$10 K). No formation of methanol ice is found in any of the considered PDR types, although a low $T$$_{\mathrm{dust}}$ significantly promotes its formation. 

We therefore conclude that low dust temperatures promote the formation of solid H$_2$O, CO, CH$_3$OH, and CH$_4$ in all type of PDRs, while warm grains promote the formation of solid CO$_2$ at any $A$$_{\mathrm{V}}$ for low $G$$_{\mathrm{0}}$, and only at very large extinctions ($A$$_{\mathrm{V}}$$>$8 mag) for high $G$$_{\mathrm{0}}$ PDRs.

\subsubsection{Comparison with observations: dust temperature}

\begin{figure}
\centering
\includegraphics[scale=0.4, angle=0]{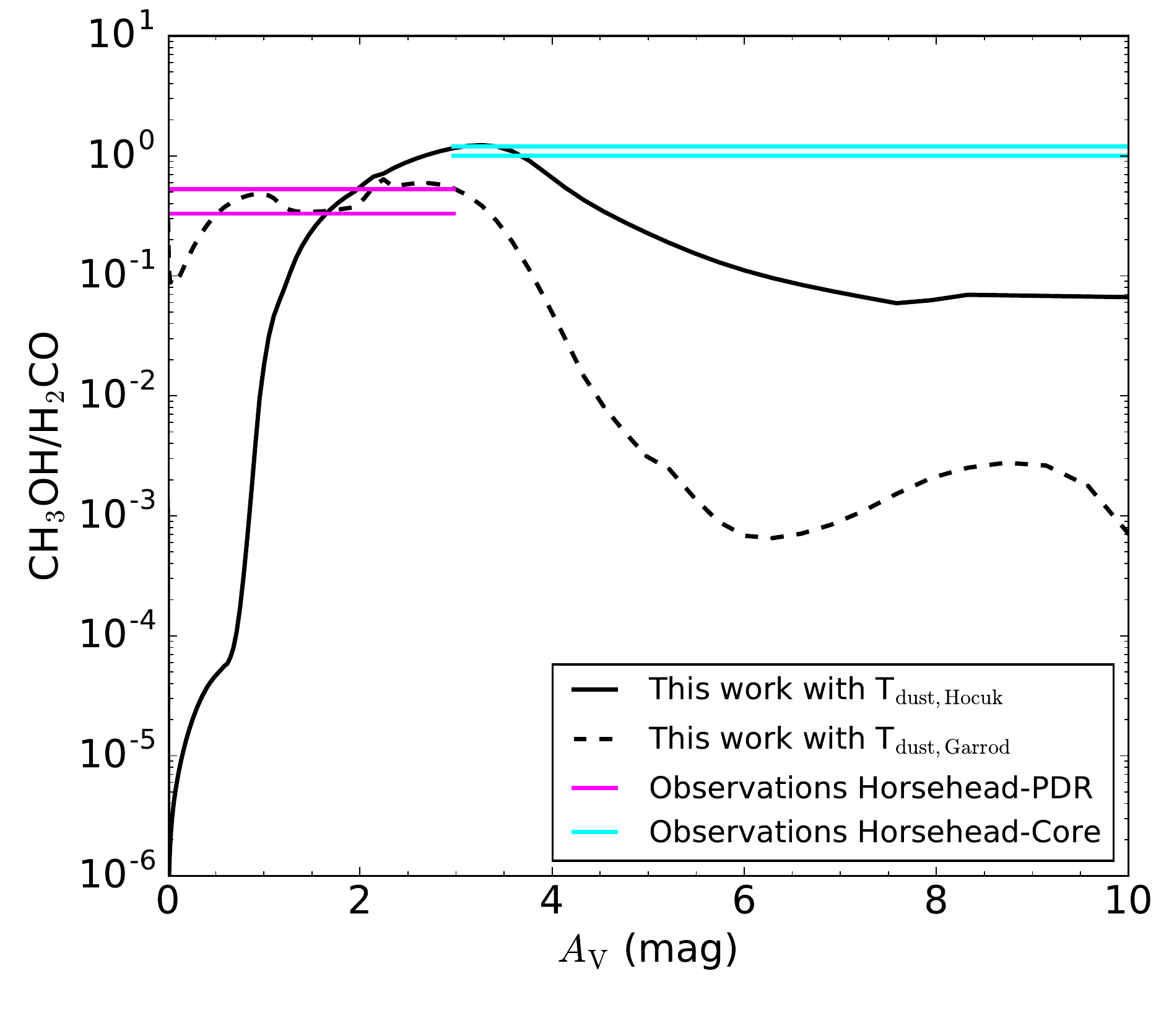}  
\hspace{0.7cm}\\
\caption{CH$_3$OH/H$_2$CO ratio obtained for a PDR with $G$$_{\mathrm{0}}$=10$^2$ and $n$=10$^5$ cm$^{-3}$ considering two different $T$$_{\mathrm{dust}}$ expression: from Garrod $\&$ Pauly (2011) (black dashed line) and from Hocuk et al. (2017) (black solid line). Observations (Guzm\'an et al. 2011, 2013) of the PDR and the core of the Horsehead are also shown with magenta and cyan lines, respectively, considering their uncertainties through a double line.}
\label{figure:comparison_different_Tdust_observations}
\end{figure}

\begin{figure*}
\centering
\includegraphics[scale=0.42, angle=0]{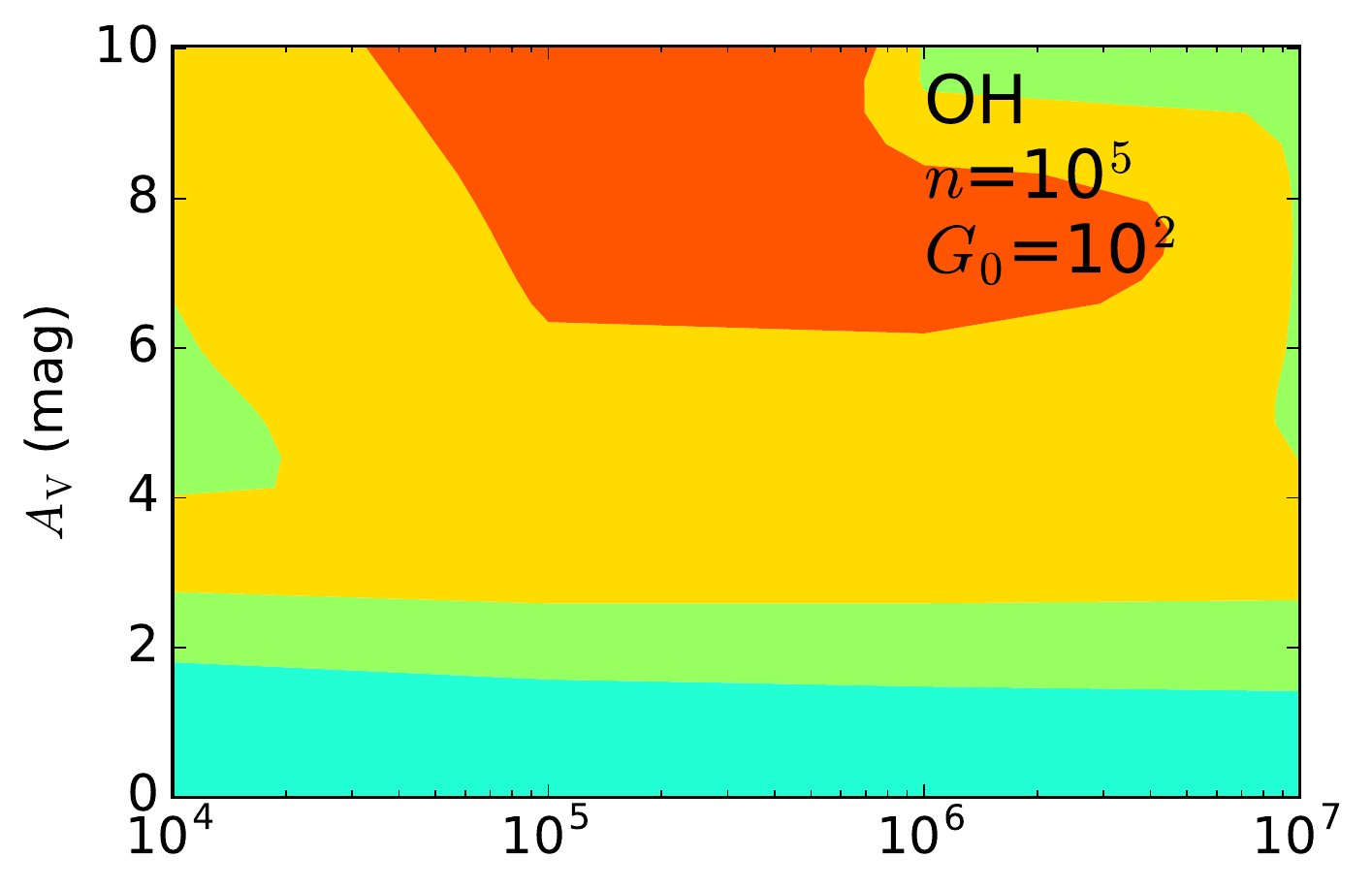}  \hspace{0.00cm}
\includegraphics[scale=0.42, angle=0]{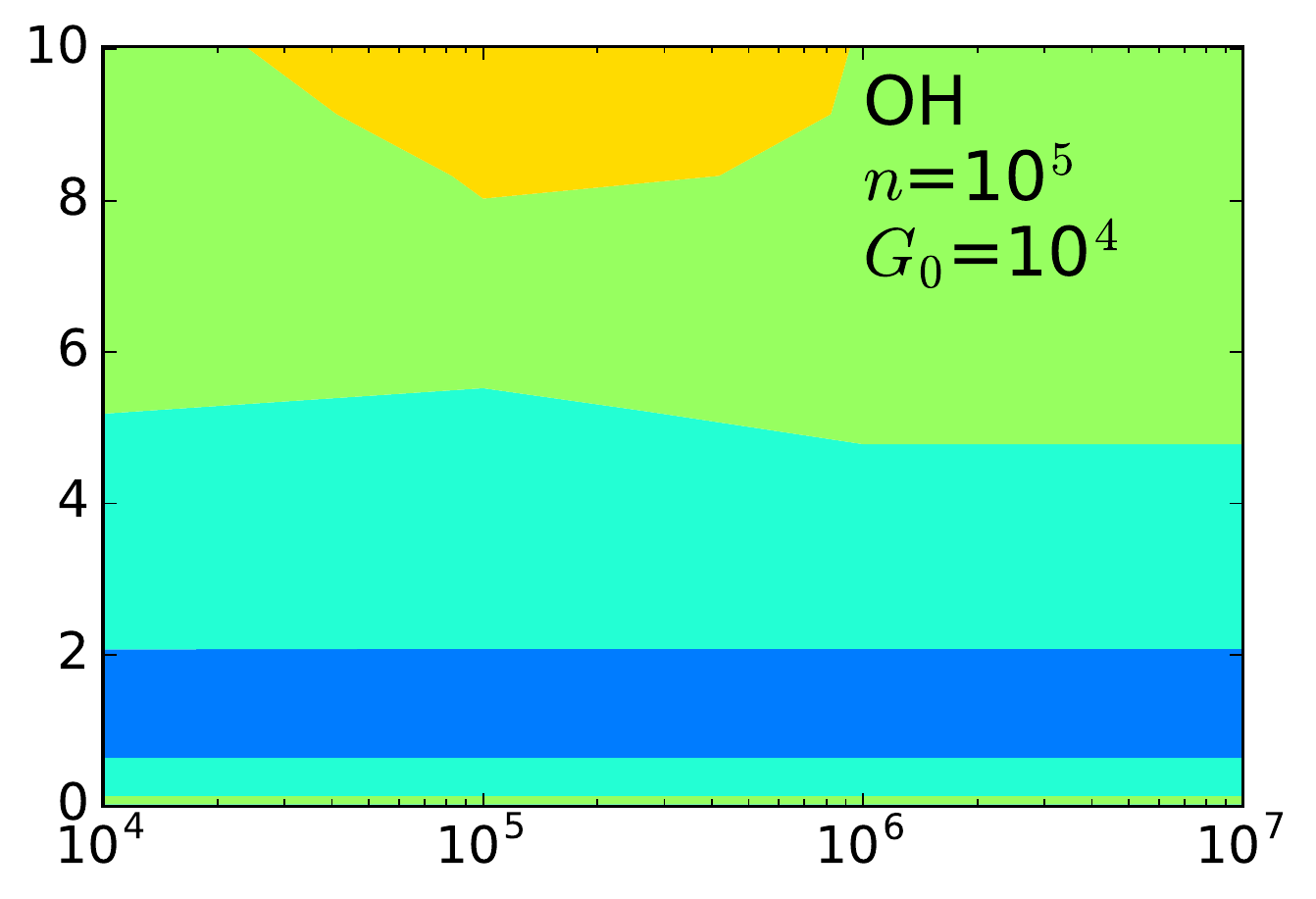}  \hspace{0.00cm}
\includegraphics[scale=0.42, angle=0]{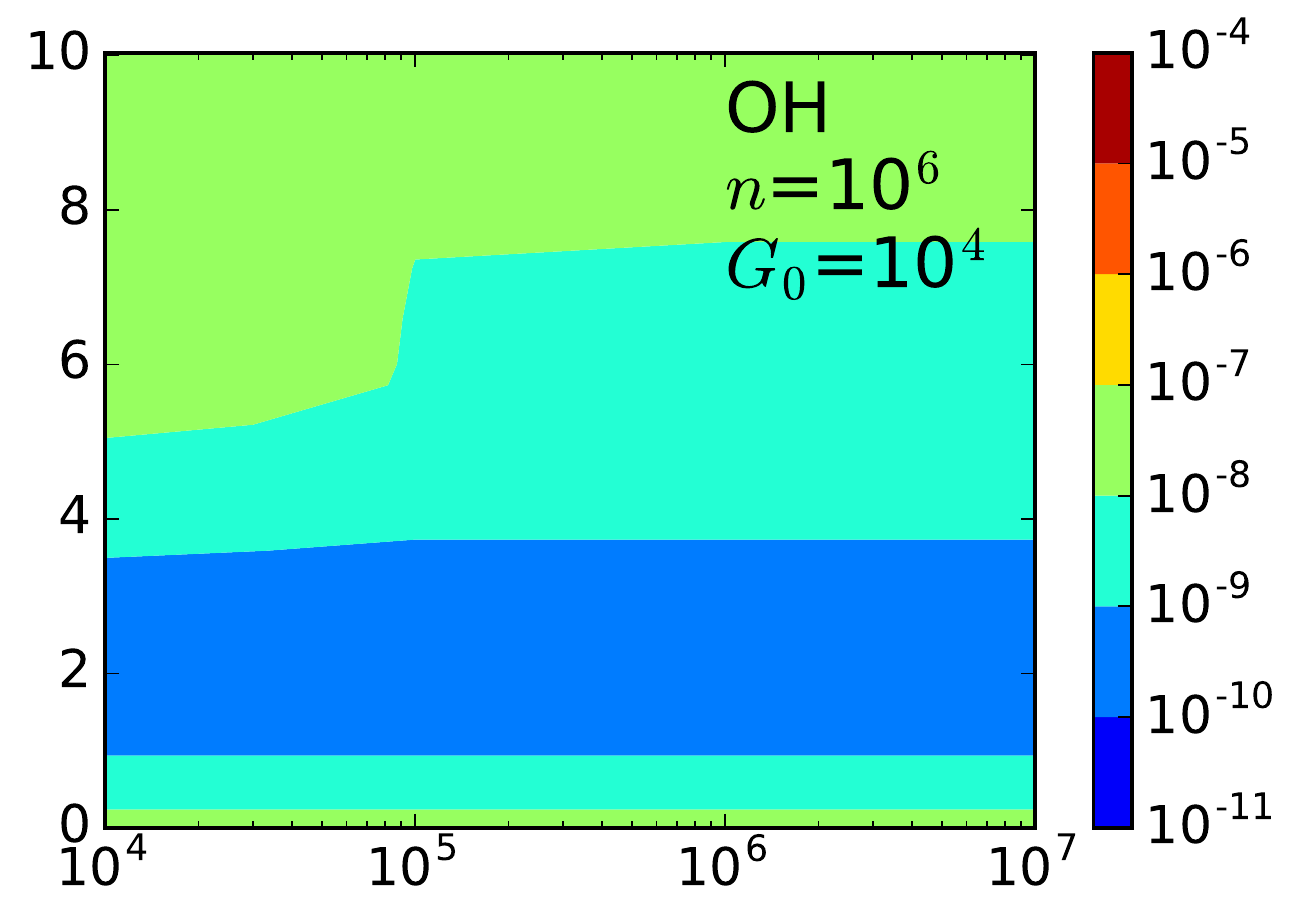}  \hspace{0.00cm}
\includegraphics[scale=0.42, angle=0]{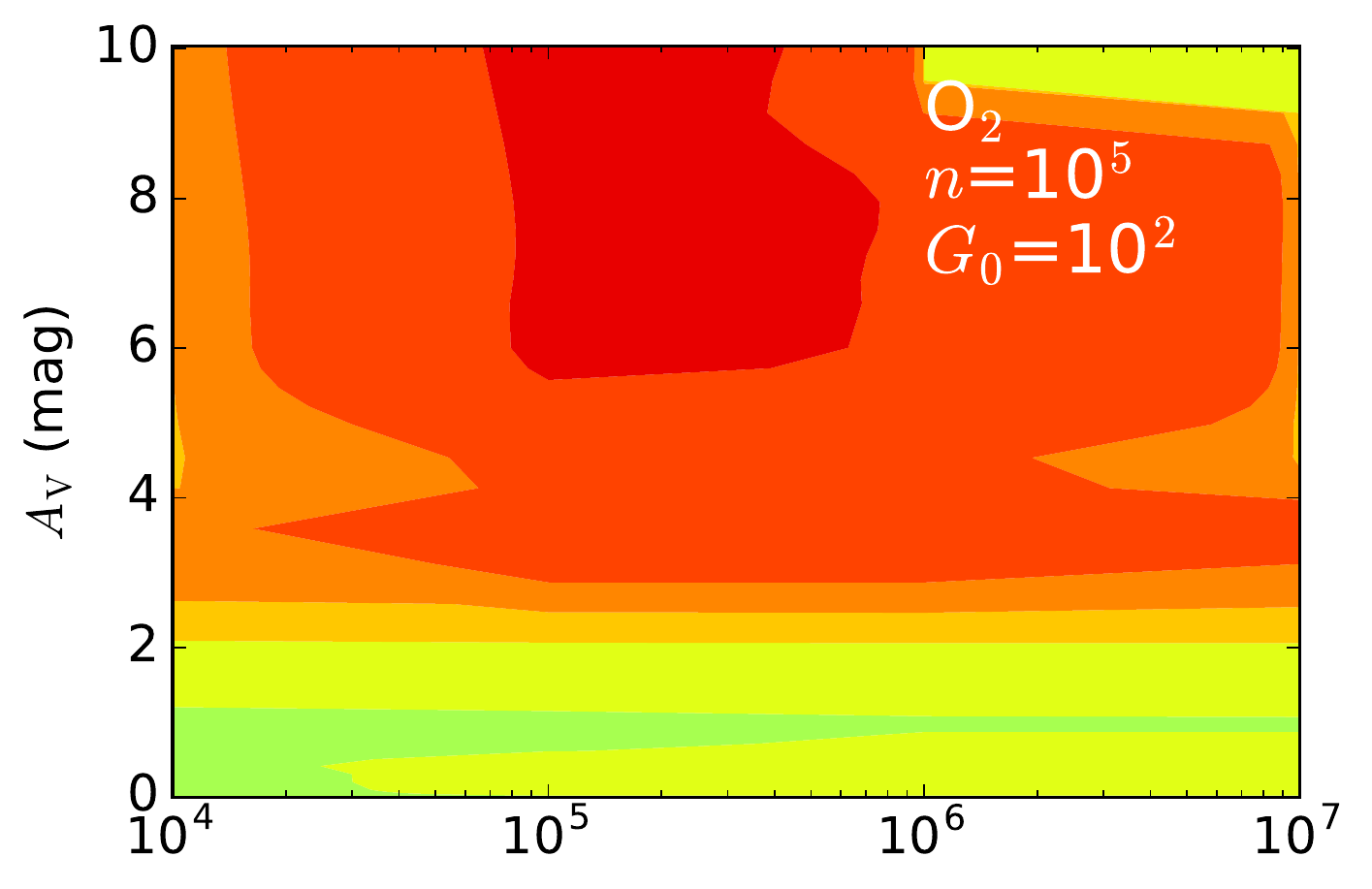}  \hspace{0.00cm}
\includegraphics[scale=0.42, angle=0]{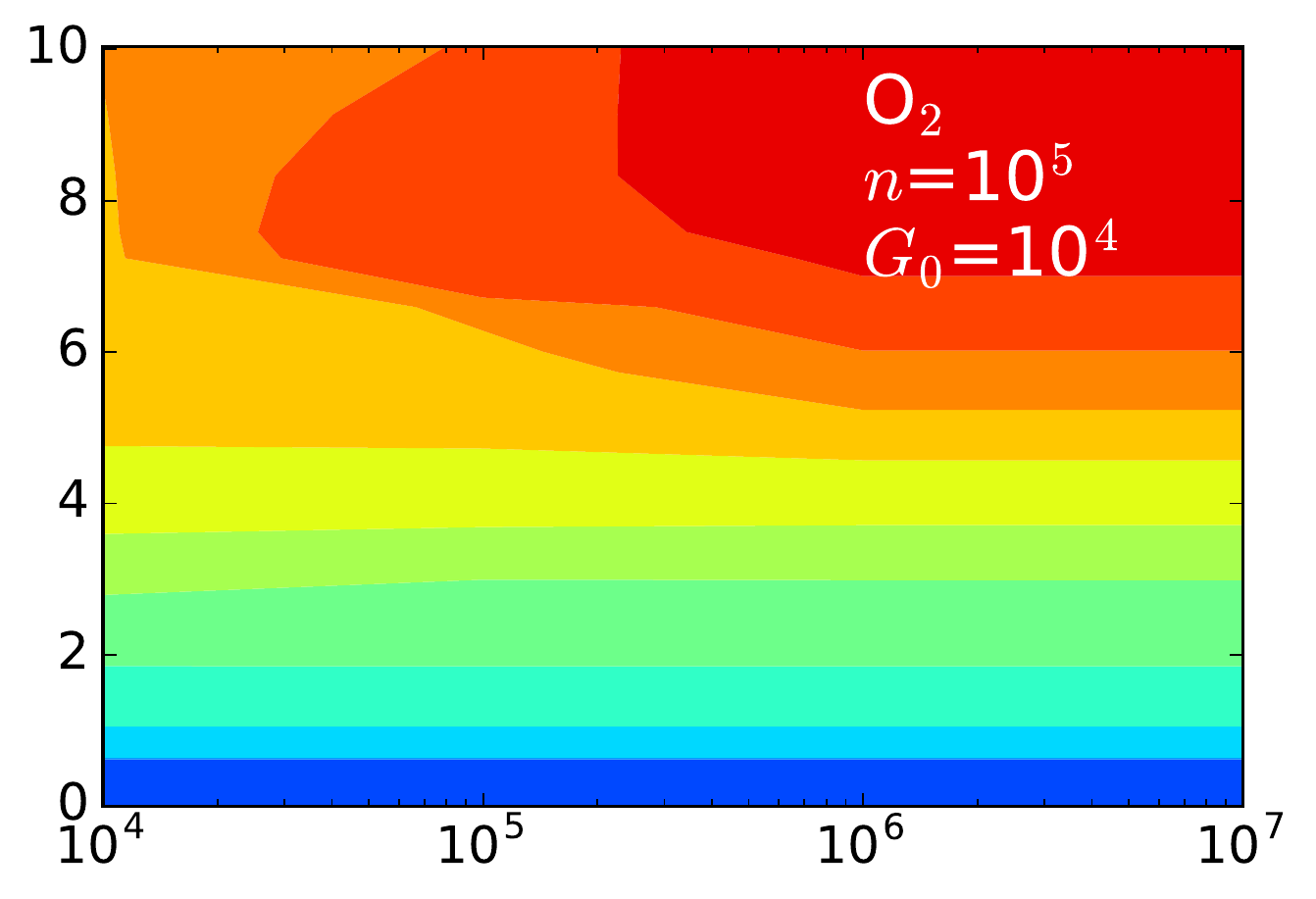}  \hspace{0.00cm}
\includegraphics[scale=0.42, angle=0]{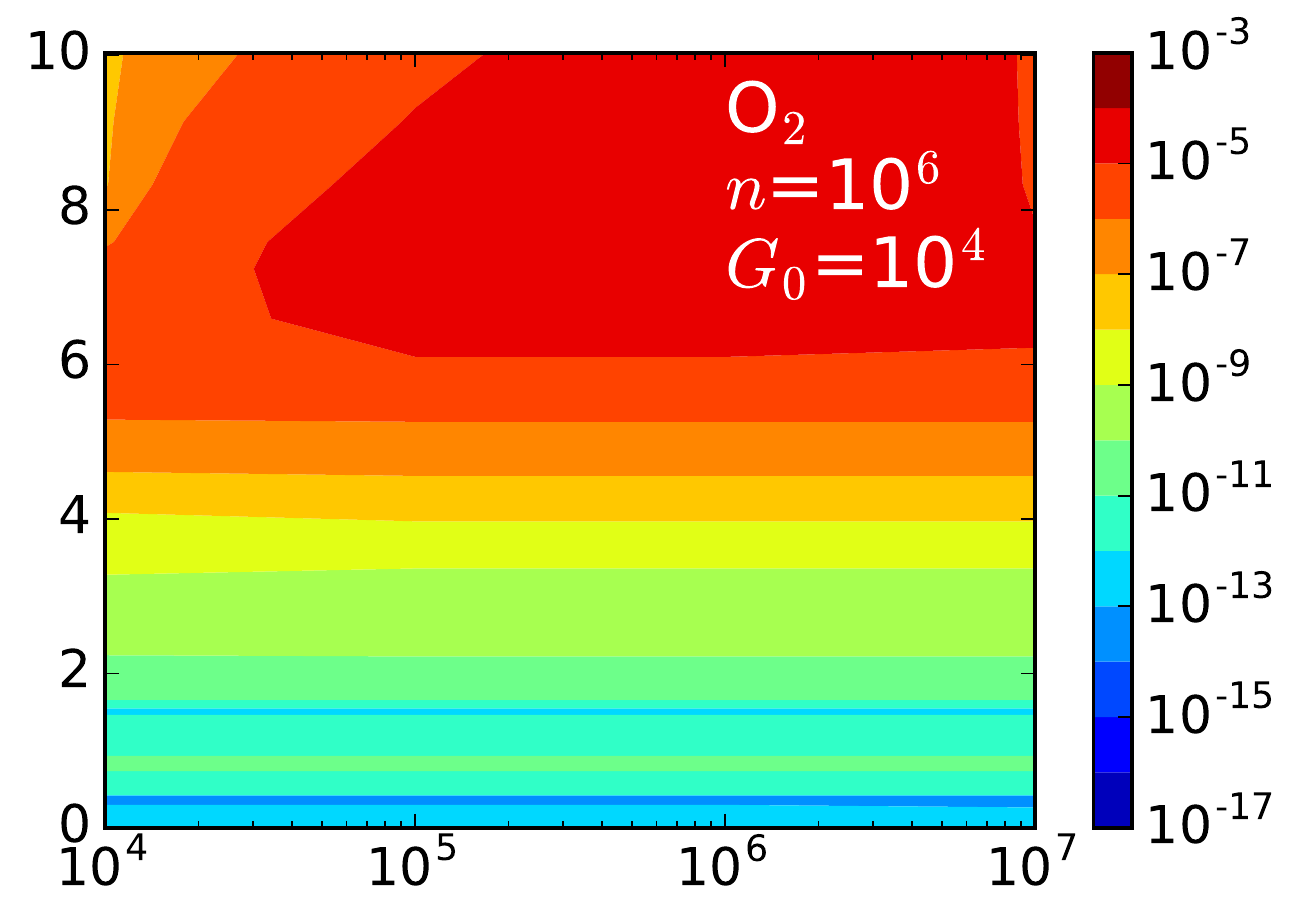}  \hspace{0.00cm}
\includegraphics[scale=0.42, angle=0]{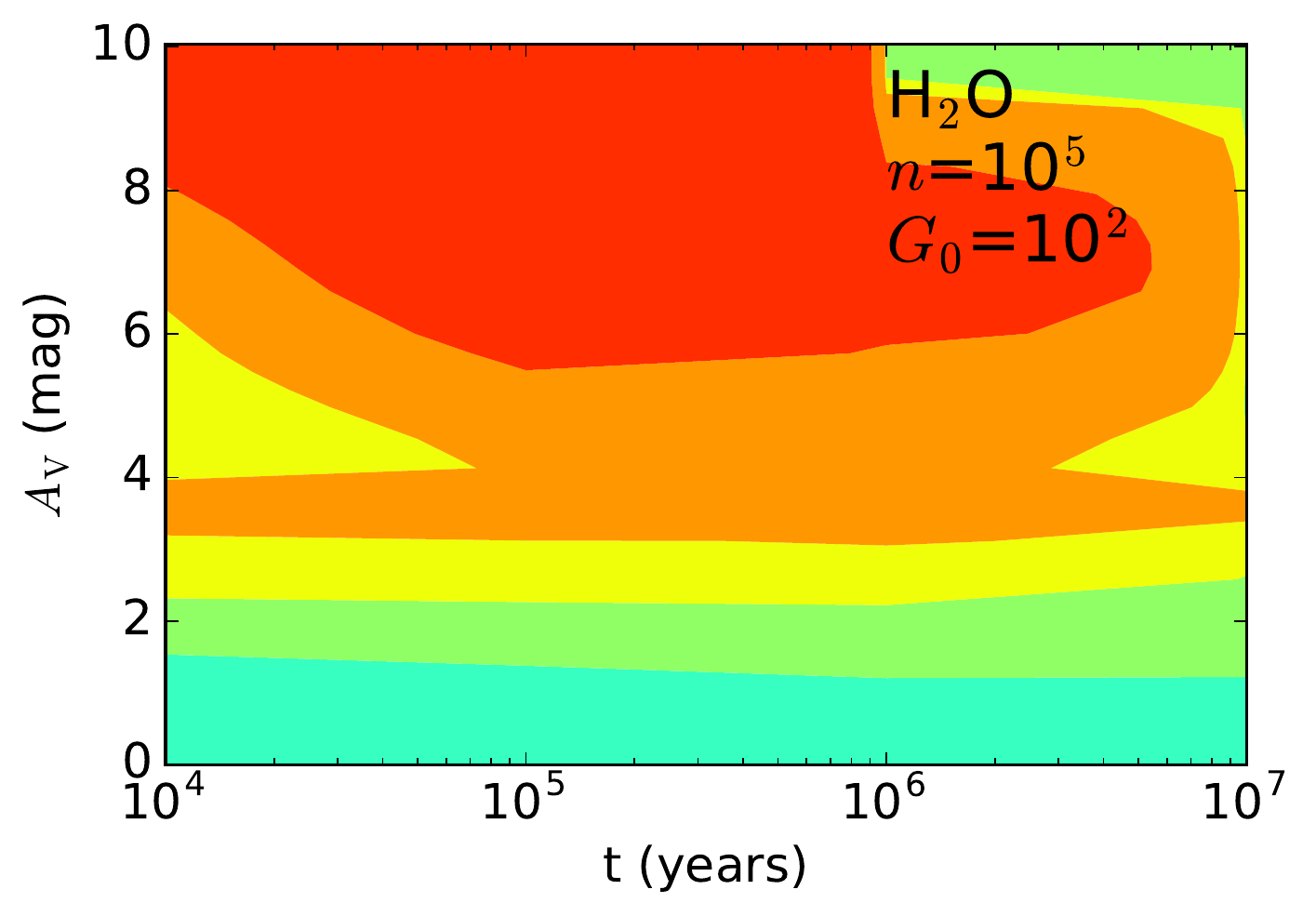}  \hspace{0.0cm}
\includegraphics[scale=0.42, angle=0]{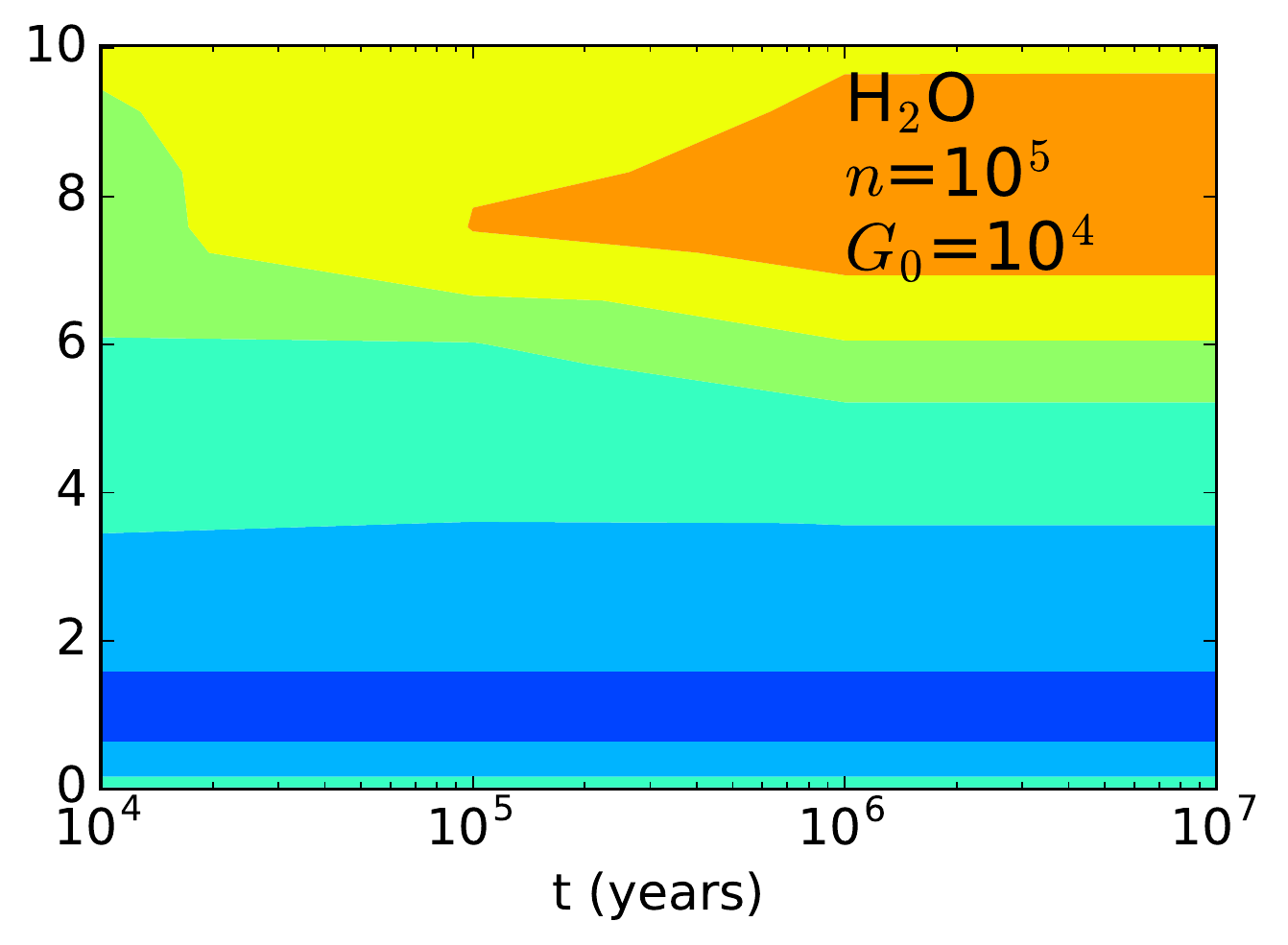}  \hspace{0.0cm}
\includegraphics[scale=0.42, angle=0]{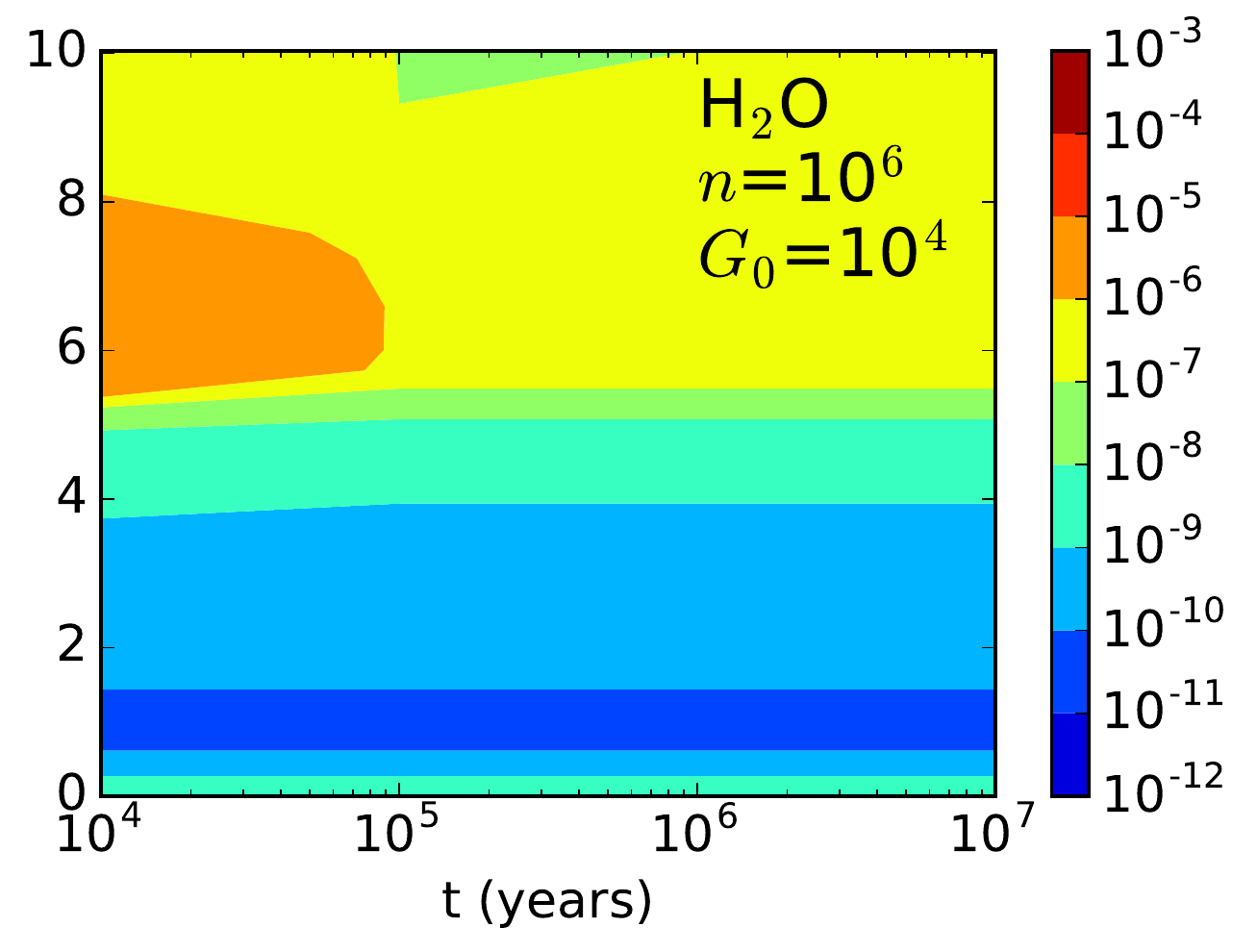}  \hspace{0.0cm}\\
\caption{Contour maps with the abundances of OH, O$_2$, and H$_2$O with respect to H nuclei for Models 1 (left panel), 2 (middle panel), and 3 (right panel) as a function of time (x-axis) and visual extinction (y-axis).}
\label{figure:contours1}
\end{figure*}

After the analysis of the chemical impact produced by the variation of the dust temperature by using the expressions from Garrod $\&$ Pauly (2011) and Hocuk et al. (2017) in the PDR code, we compare our predictions with observations to find the temperature of the best agreement.

For the case of a low $G$$_{\mathrm{0}}$ PDR, we have compared our results obtained using both $T$$_{\mathrm{dust}}$ expressions with observations of two molecules, CH$_3$OH and H$_2$CO, in the Horsehead ($n$=10$^5$ cm$^{-3}$ and $G$$_{\mathrm{0}}$$\sim$10$^2$, Habart et al. 2005). In particular, we have compared with the ratio of these two molecules, since the estimation of their abundances with respect to H$_2$ presents large uncertainties due to the strong dependence of the H$_2$ density on the dust temperature considered\footnote{Leurini et al. (2010) found a variation in the density of H$_2$ larger than a factor of 2 when the difference considered in the dust temperature is 20 K.}. Figure \ref{figure:comparison_different_Tdust_observations} shows this comparison considering observations of the H$_2$CO and CH$_3$OH in the PDR (the IR peak at $A$$_{\mathrm{V}}$$\sim$1 mag) and the core ($A$$_{\mathrm{V}}$$\sim$8 mag, Pety et al. 2012). The kinetic temperatures assumed to infer the observational results were $T$$_{\mathrm{kin}}$=40-65 K and 20 K, for the PDR and the core respectively (Guzm\'an et al. 2011, 2013), which are consistent with the PDR model temperatures for both regions (Fig. \ref{figure:Tdust_comparison}, right panel). The results show that the CH$_3$OH/H$_2$CO ratio in the PDR region is reproduced by either expressions, however none of them reproduces the observations in the core. Nevertheless, the difference between observations and model is about one order of magnitude using Hocuk$'$s expression, and about three orders of magnitude using Garrod$'$s expression at $A$$_{\mathrm{V}}$=8 mag.   

For the case of a high $G$$_{\mathrm{0}}$ PDR, observations of the densest parts of the Orion Bar ($n$=10$^5$-10$^6$ cm$^{-3}$ and $G$$_{\mathrm{0}}$$\sim$10$^4$, Marconi et al. 1998, Leurini et al. 2010) carried out with the Herschel space telescope reveal a dust temperature gradient from $\sim$70 K to $\sim$48 K for the largest grains at different positions in the Bar (Arab et al. 2012). Millar $\&$ Williams (1993) also show through far-infrared (FIR) observations that the temperatures of dust grains with size$\sim$3000$\AA$ in the Bar region are about 75 K. Comparing these results with those shown in Fig. \ref{figure:Tdust_comparison} (left panel), we clearly see that the $T$$_{\mathrm{dust}}$ expression from Hocuk et al. (2017) provides dust temperature values in full agreement with the observations of the Orion Bar. In the following, we consider the dust temperature expression from Hocuk et al. (2017).

\subsection{Time-dependent effects}
\label{Time_dependence}

\begin{figure*}
\centering
\includegraphics[scale=0.42, angle=0]{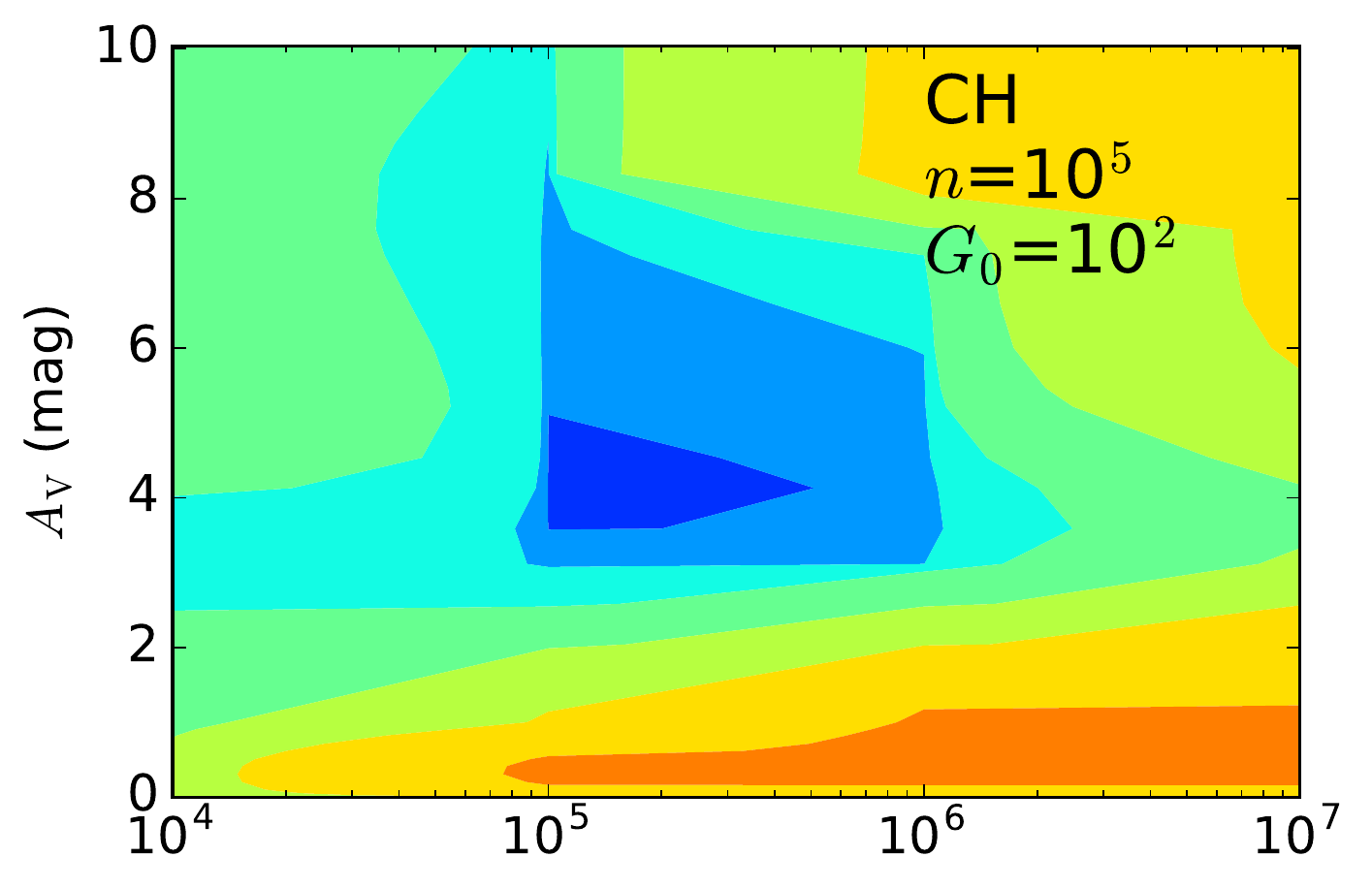}  \hspace{0.0cm}
\includegraphics[scale=0.42, angle=0]{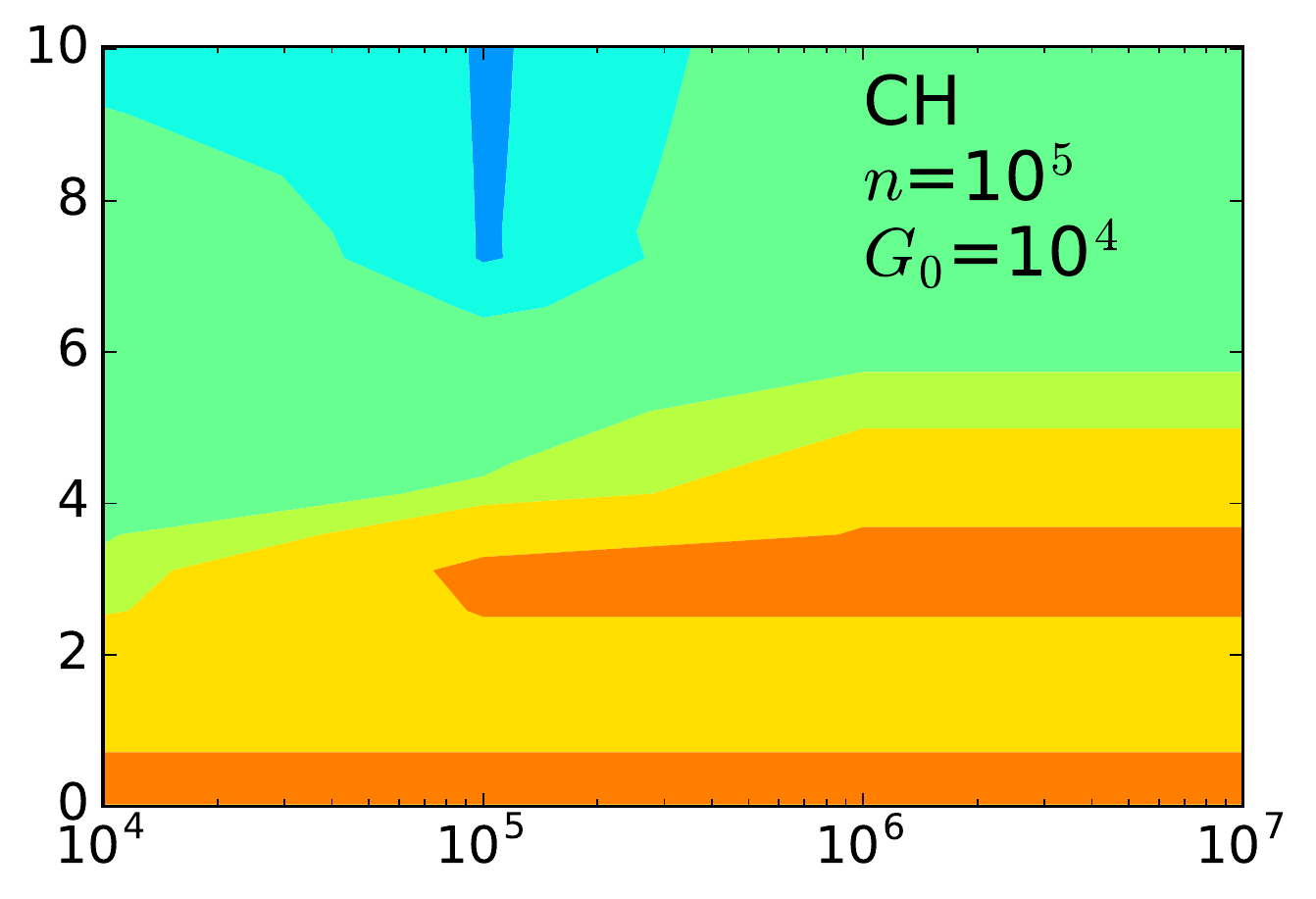}  \hspace{0.0cm}
\includegraphics[scale=0.42, angle=0]{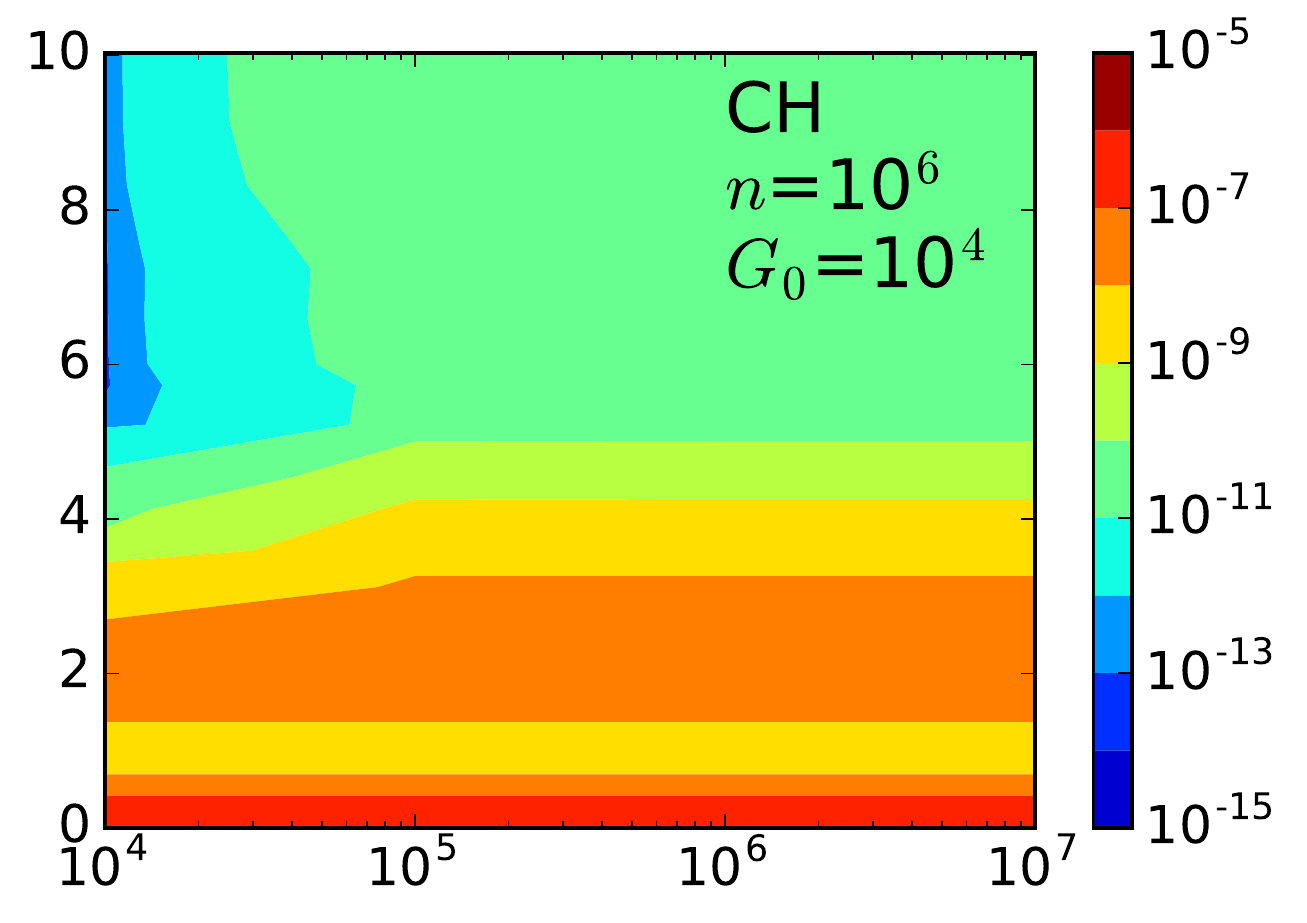}  \hspace{0.0cm}
\includegraphics[scale=0.42, angle=0]{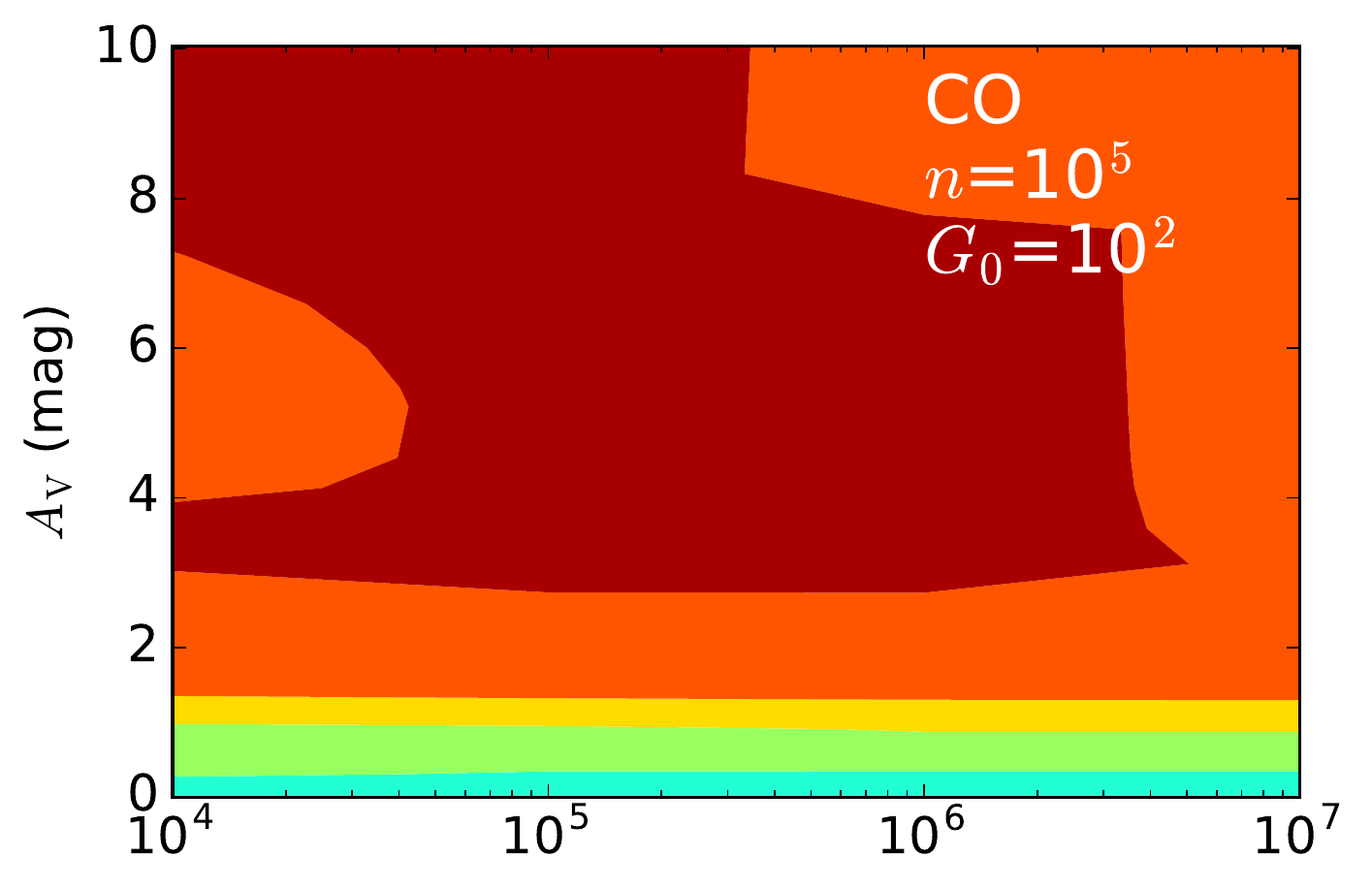}  \hspace{0.00cm}
\includegraphics[scale=0.42, angle=0]{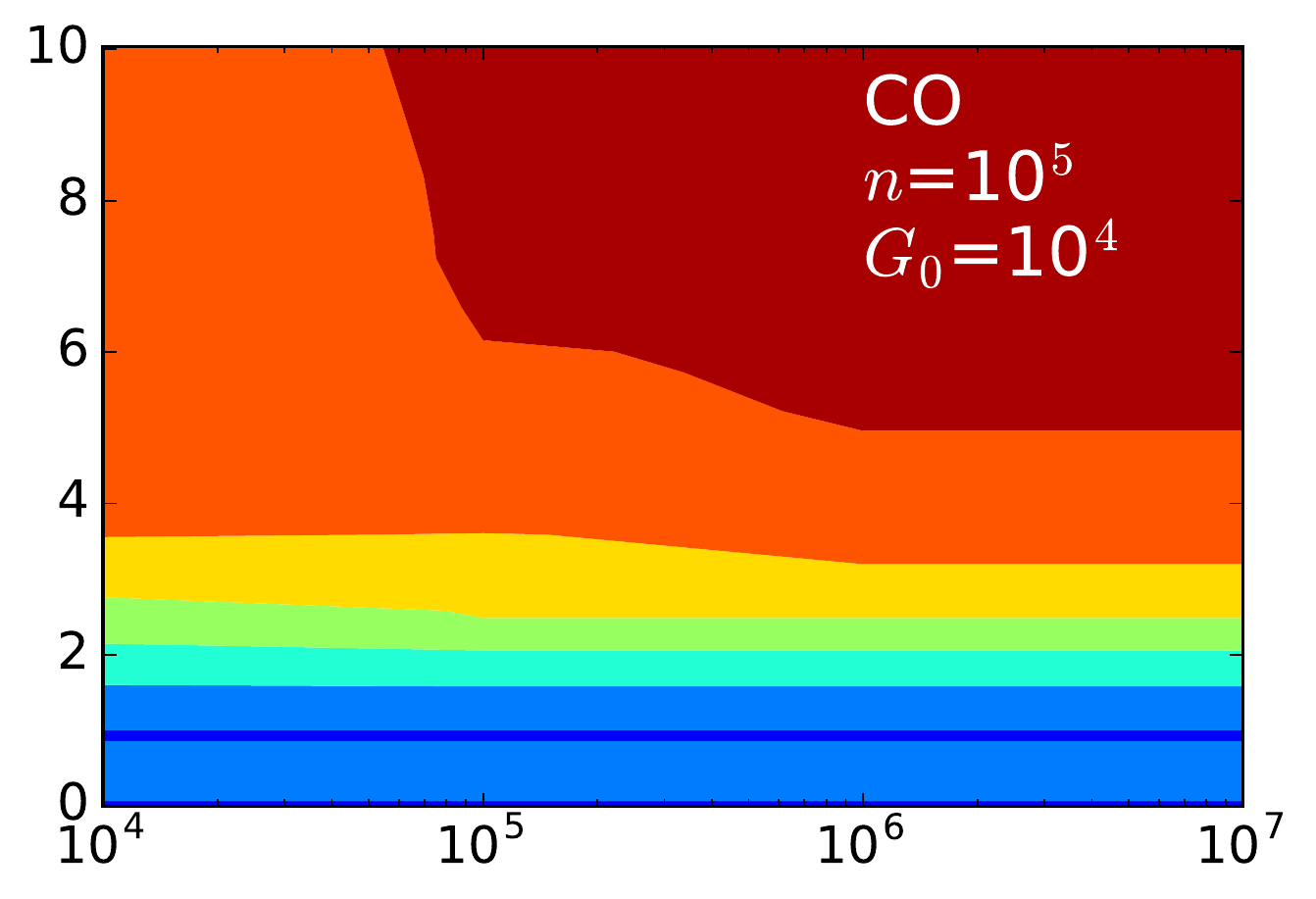}  \hspace{0.00cm}
\includegraphics[scale=0.42, angle=0]{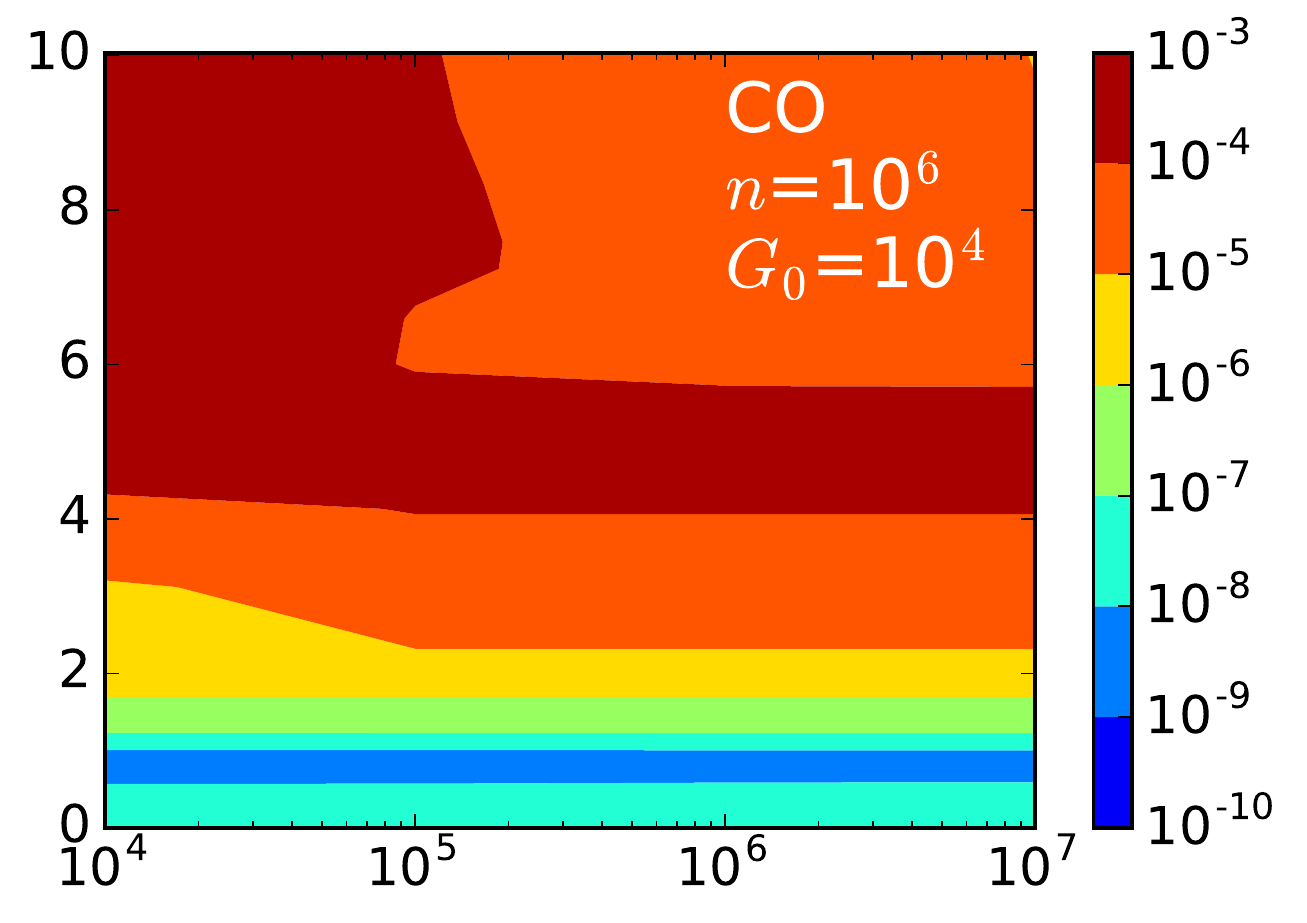}  \hspace{0.00cm}
\includegraphics[scale=0.42, angle=0]{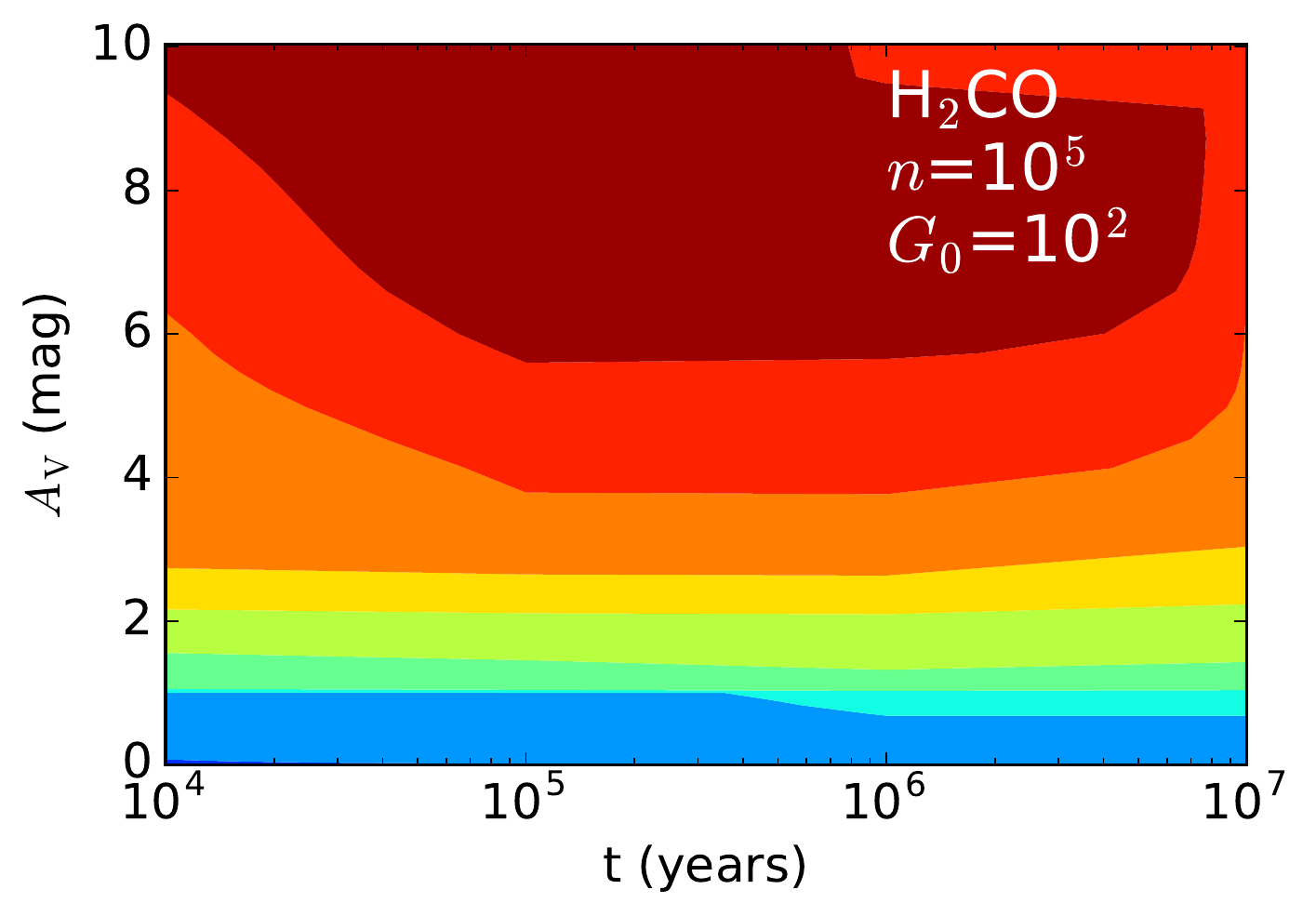}  \hspace{0.00cm}
\includegraphics[scale=0.42, angle=0]{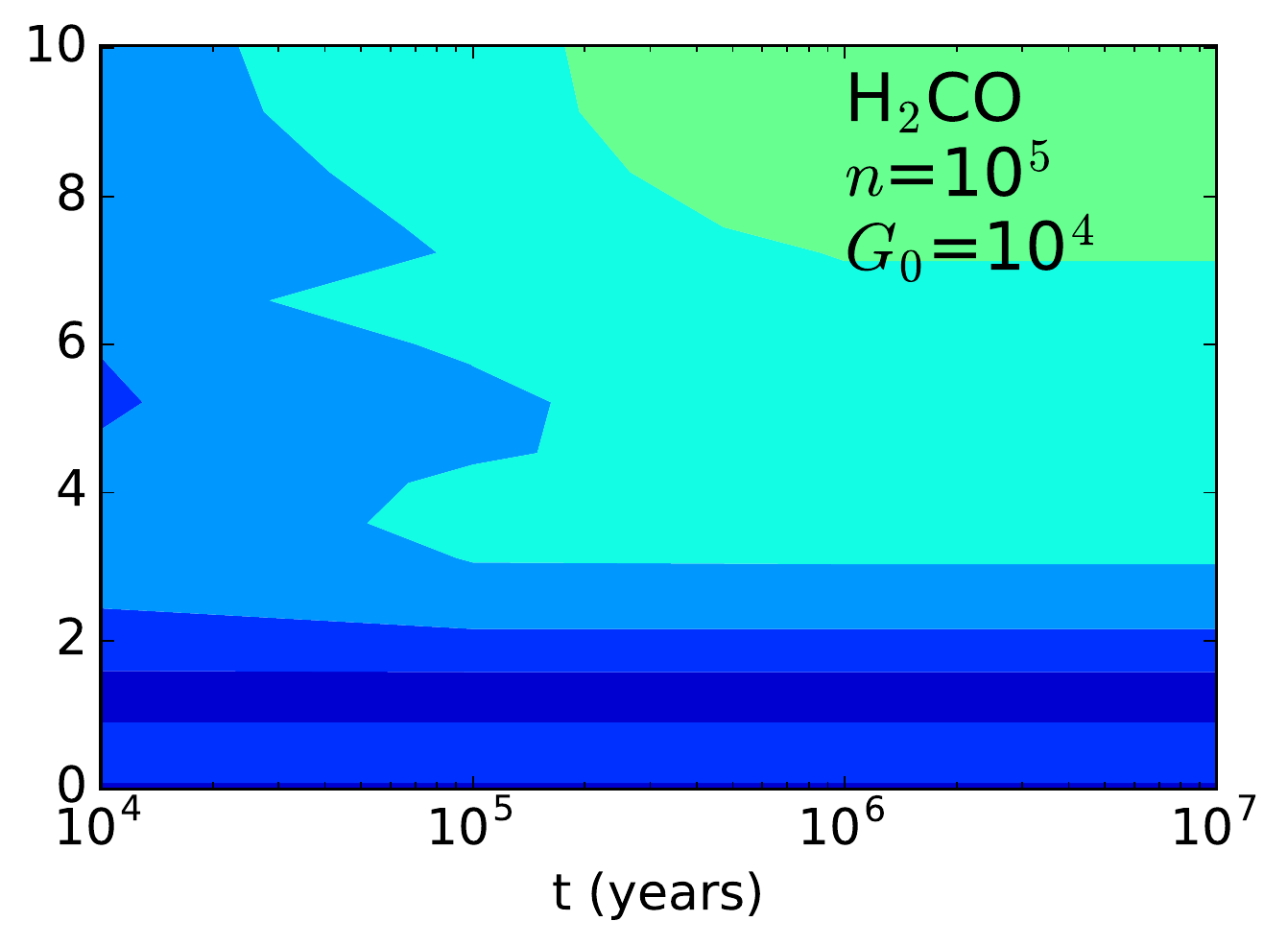}  \hspace{0.00cm}
\includegraphics[scale=0.42, angle=0]{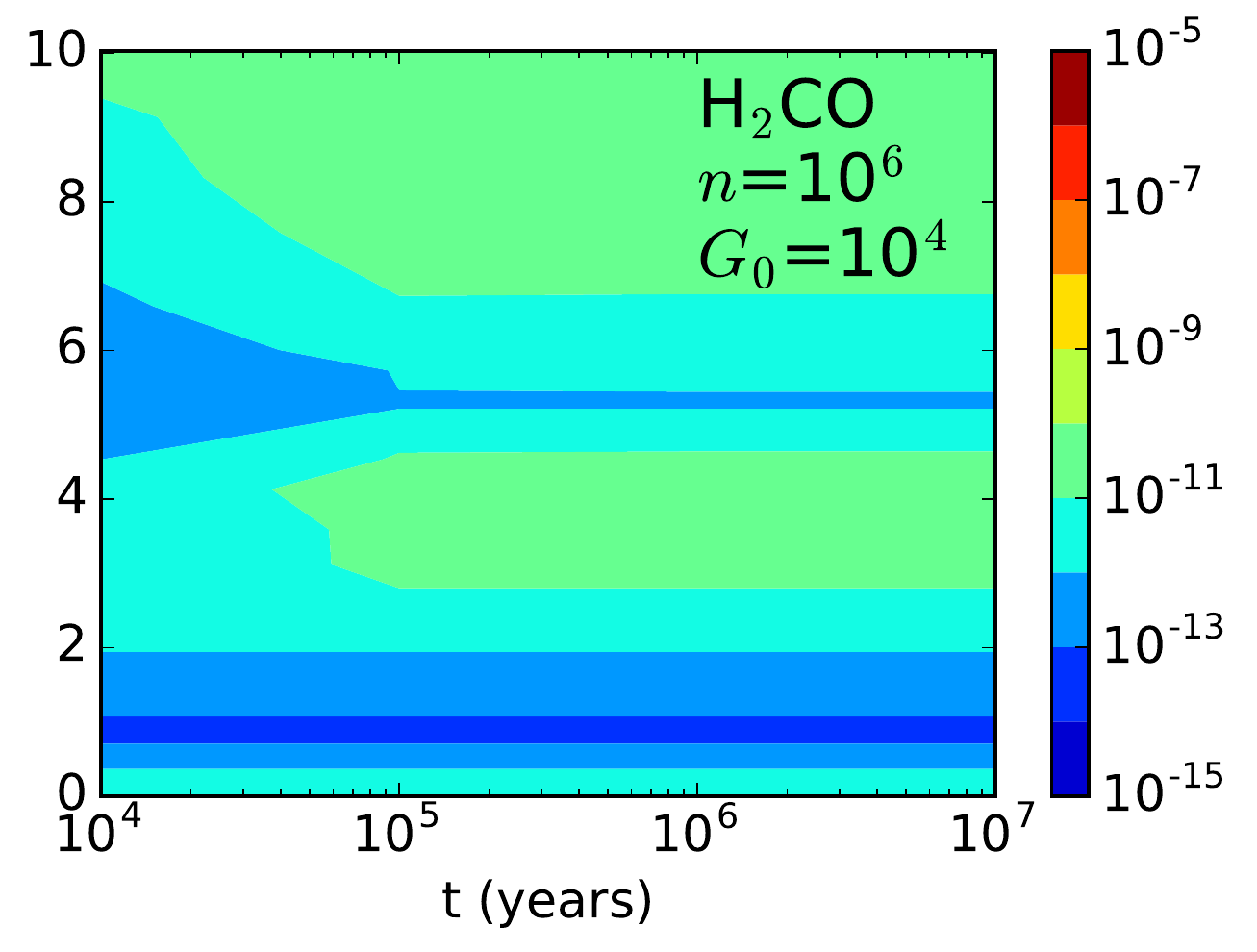}  \hspace{0.00cm}\\
\caption{Contour maps with the abundances of CH, CO, H$_2$CO, CH$_3$OH, and CH$_3$OH with respect to H nuclei for Models 1 (left panel), 2 (middle panel), and 3 (right panel) as a function of time (x-axis) and visual extinction (y-axis).}
\label{figure:contours2}
\end{figure*}

\begin{figure*}
\centering
\includegraphics[scale=0.42, angle=0]{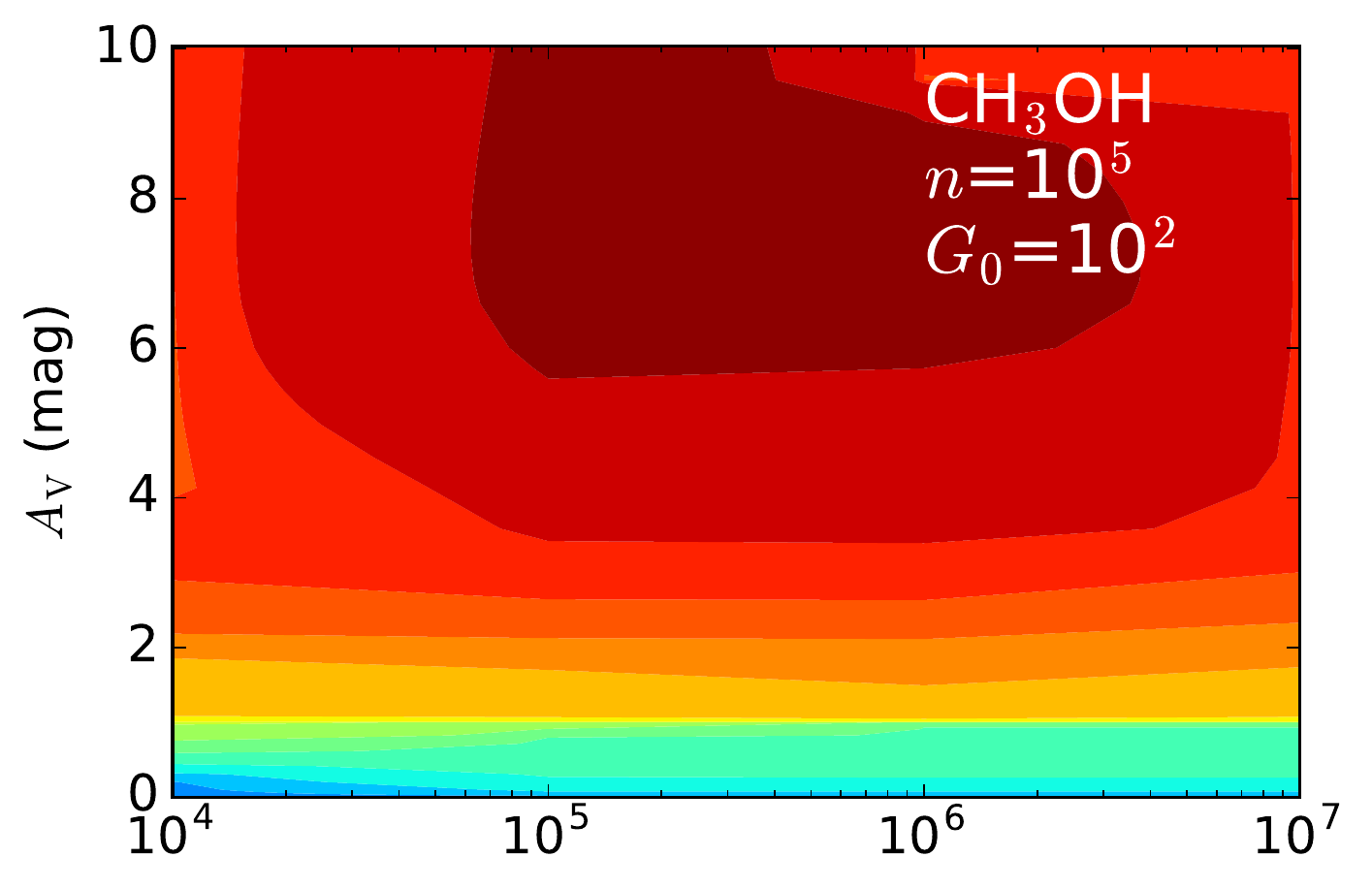}  \hspace{0.0cm}
\includegraphics[scale=0.42, angle=0]{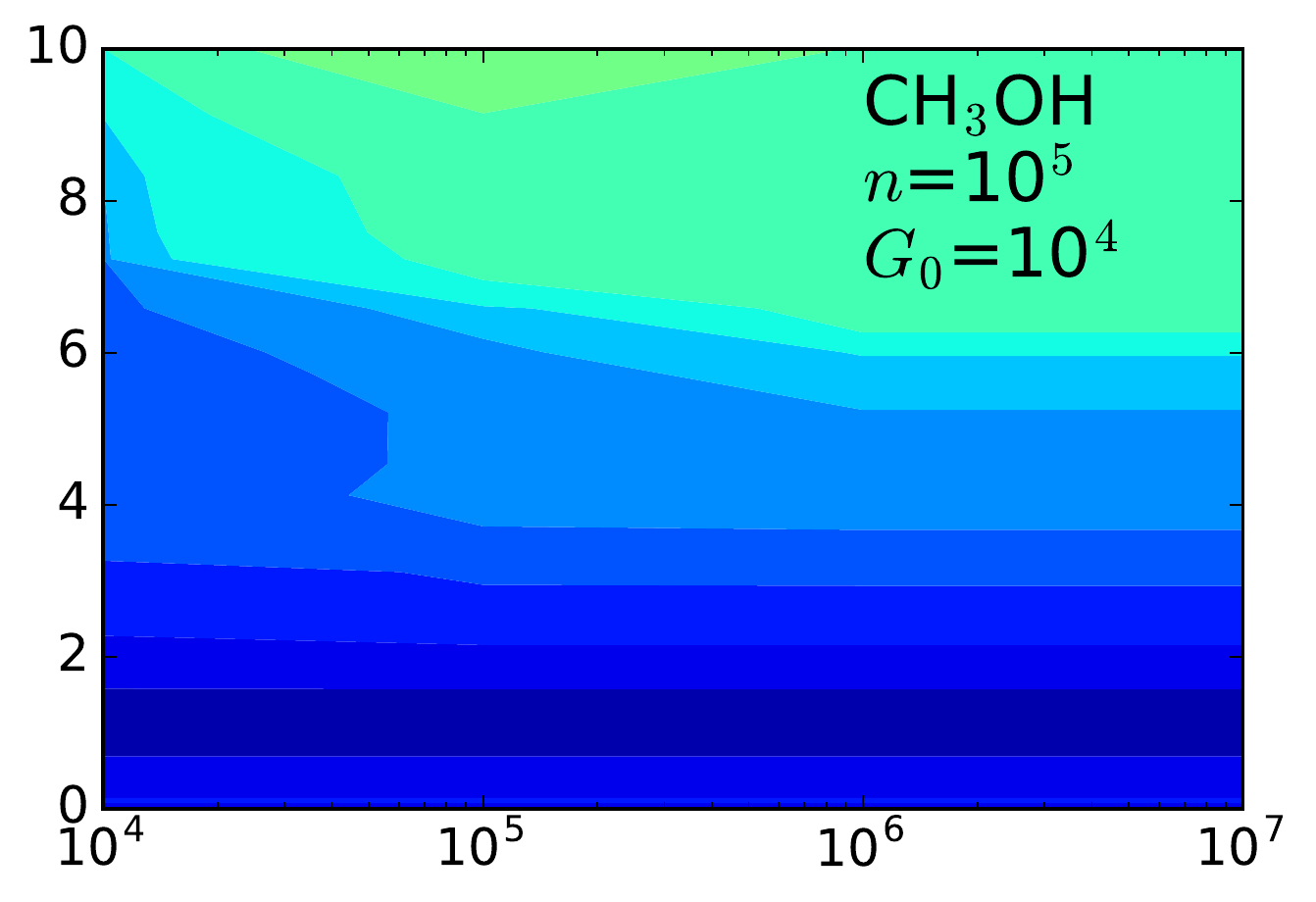}  \hspace{0.0cm}
\includegraphics[scale=0.42, angle=0]{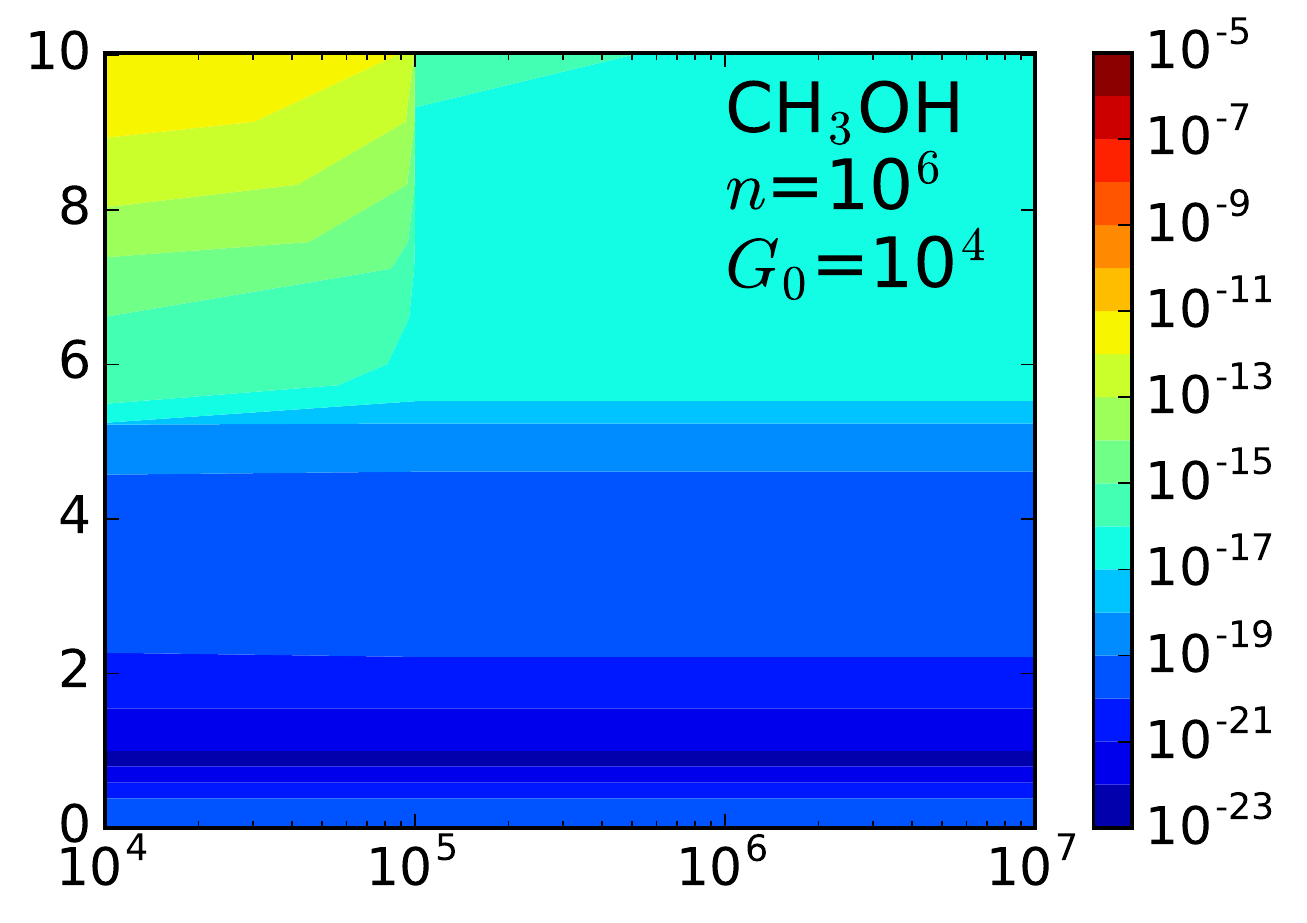}  \hspace{0.0cm}
\includegraphics[scale=0.42, angle=0]{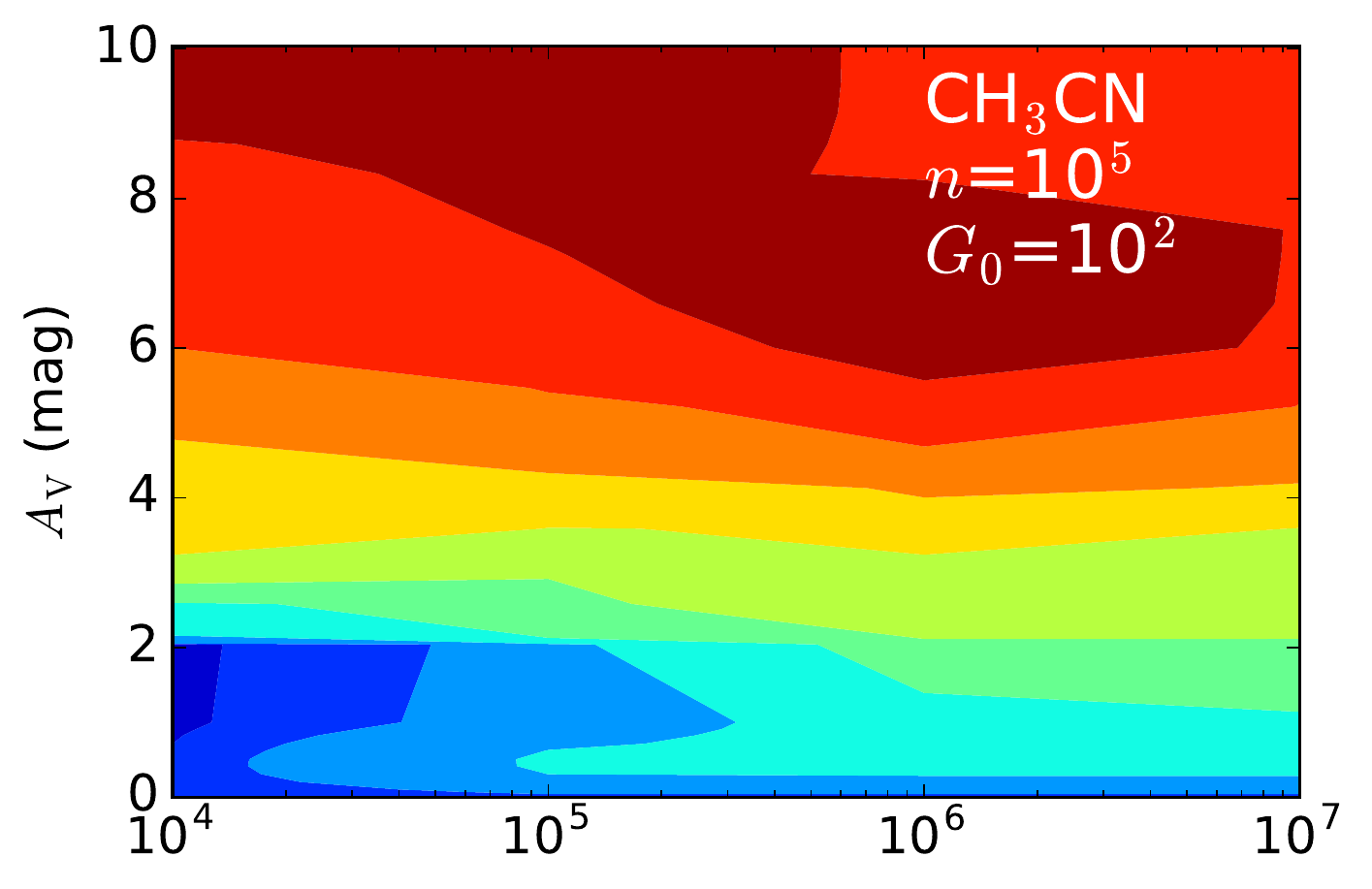}  \hspace{0.0cm}
\includegraphics[scale=0.42, angle=0]{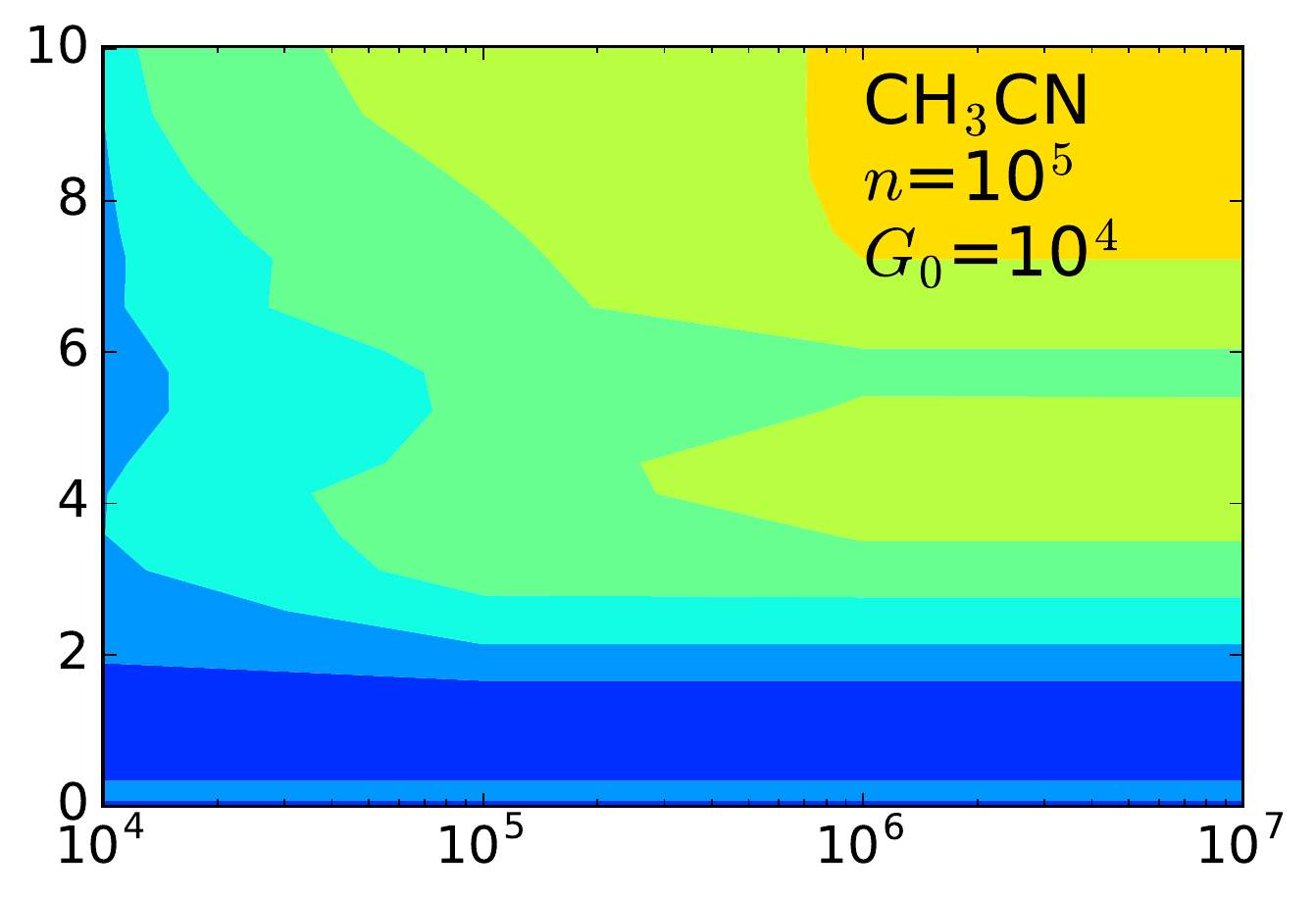}  \hspace{0.0cm}
\includegraphics[scale=0.42, angle=0]{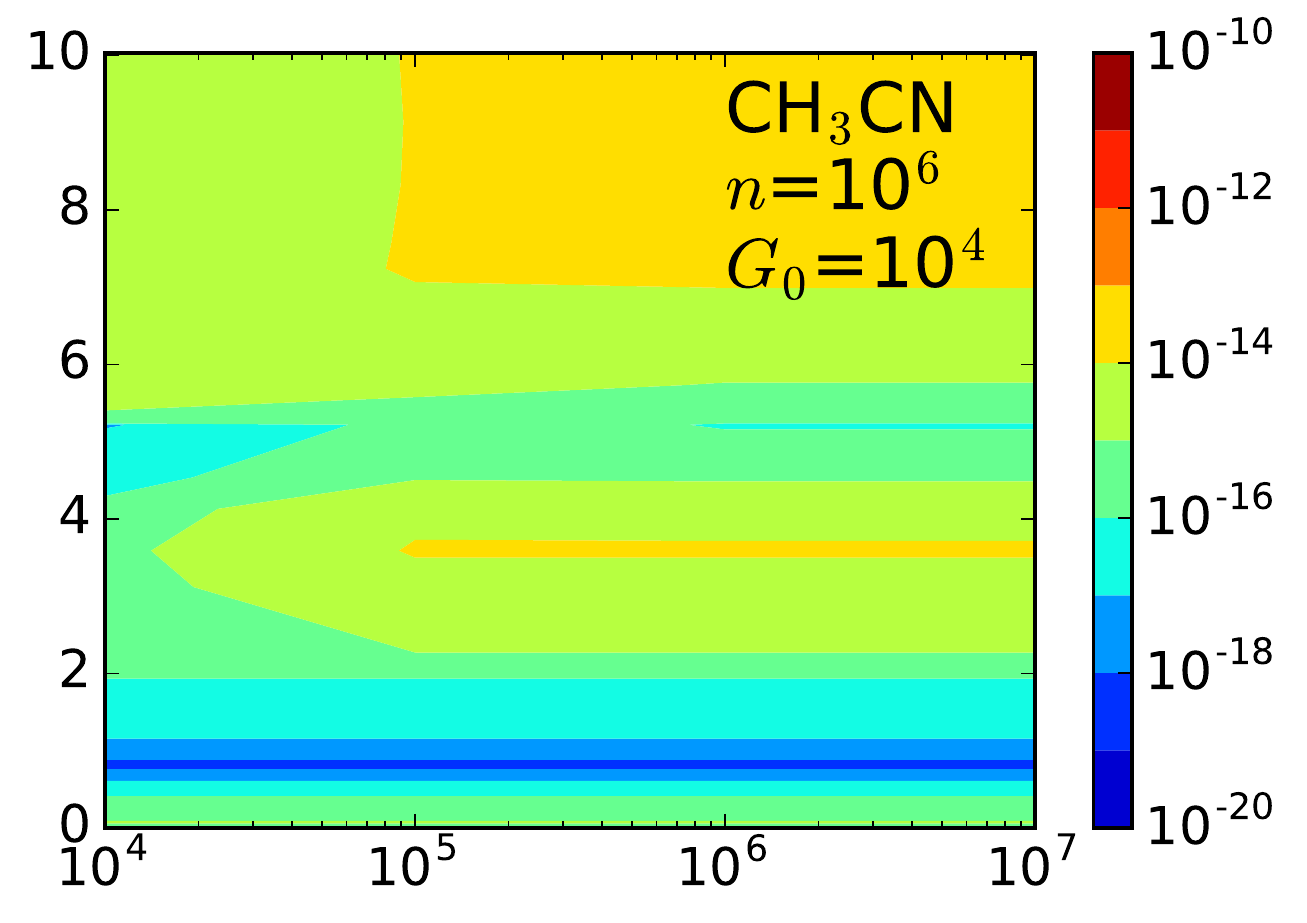}  \hspace{0.0cm}
\includegraphics[scale=0.42, angle=0]{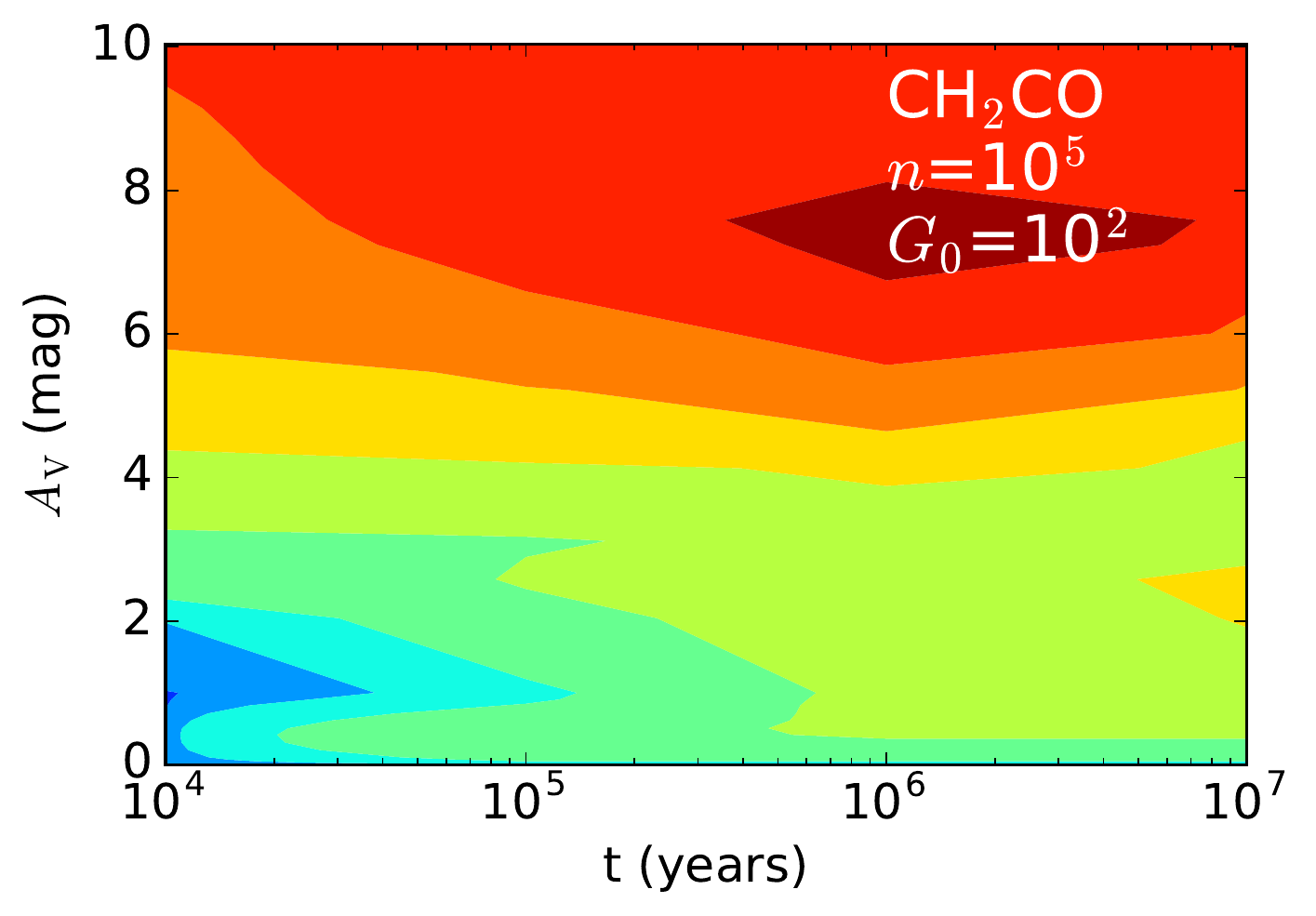}  \hspace{0.0cm}
\includegraphics[scale=0.42, angle=0]{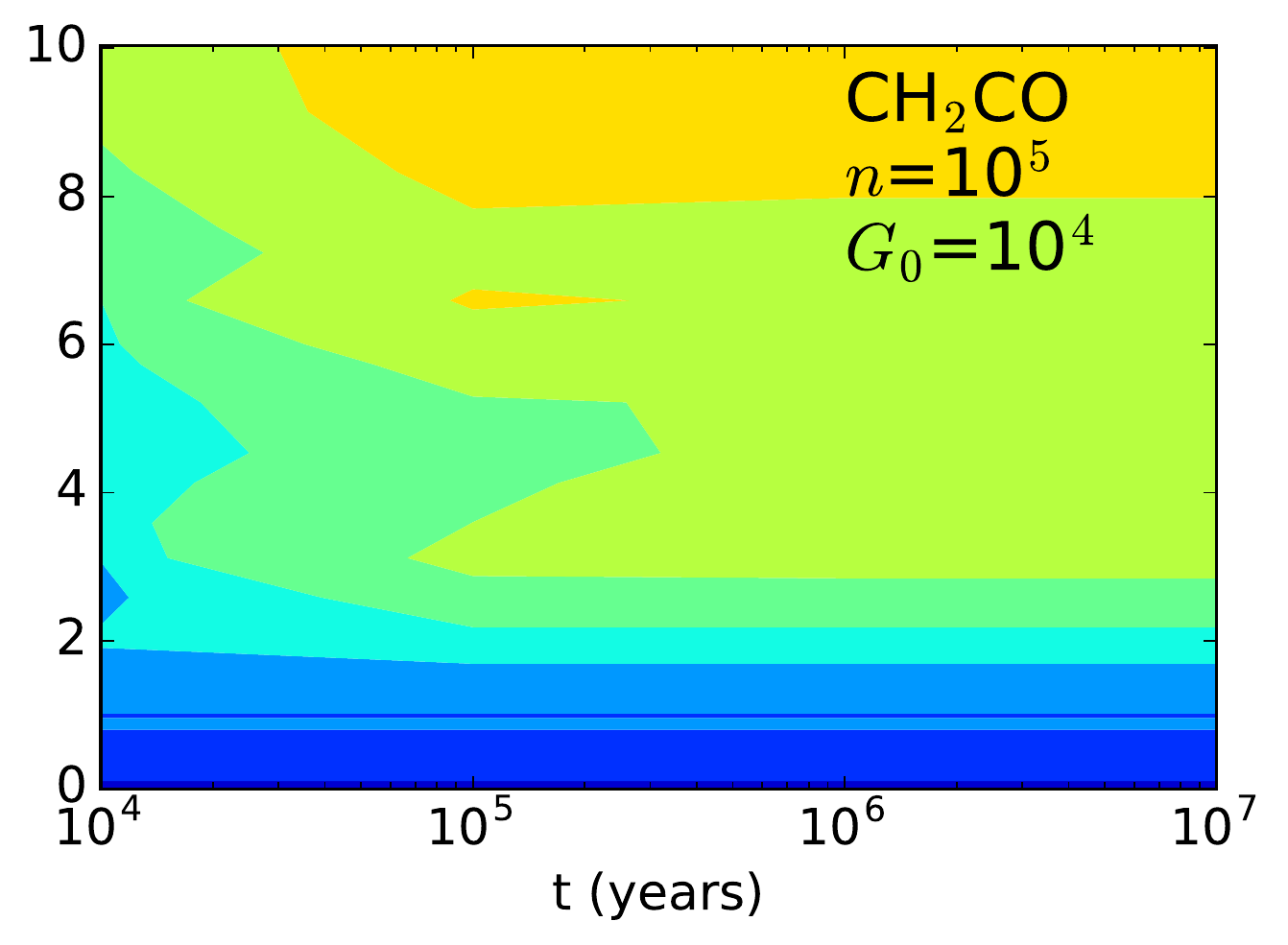}  \hspace{0.0cm}
\includegraphics[scale=0.42, angle=0]{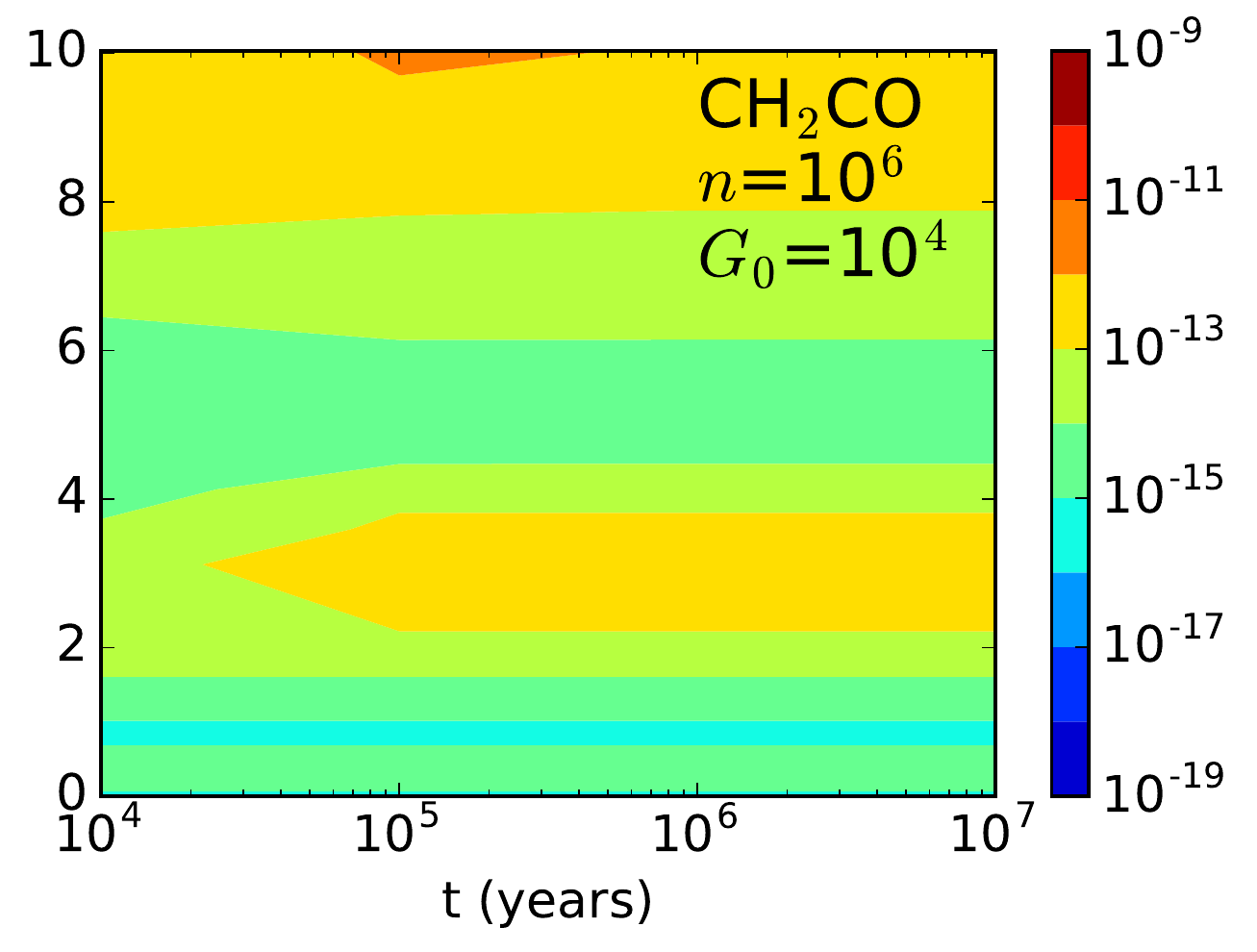}  \hspace{0.0cm}\\
\caption{Contour maps with the abundances of CH$_3$OH, CH$_3$CN, and CH$_2$CO with respect to H nuclei for Models 1 (left panel), 2 (middle panel), and 3 (right panel) as a function of time (x-axis) and visual extinction (y-axis).}
\label{figure:contours3}
\end{figure*}

\begin{figure*}
\centering
\includegraphics[scale=0.42, angle=0]{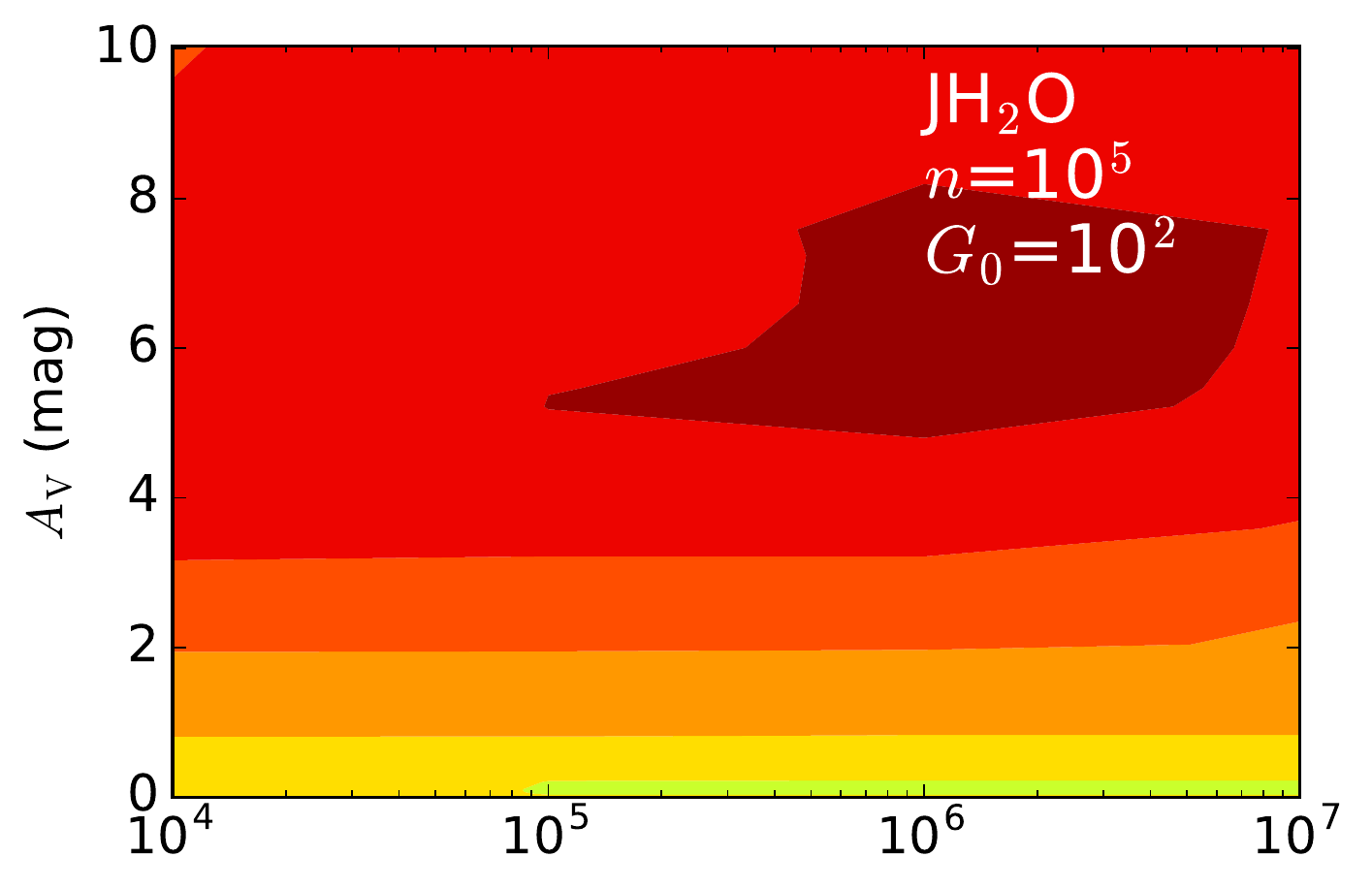}  \hspace{0.00cm}
\includegraphics[scale=0.42, angle=0]{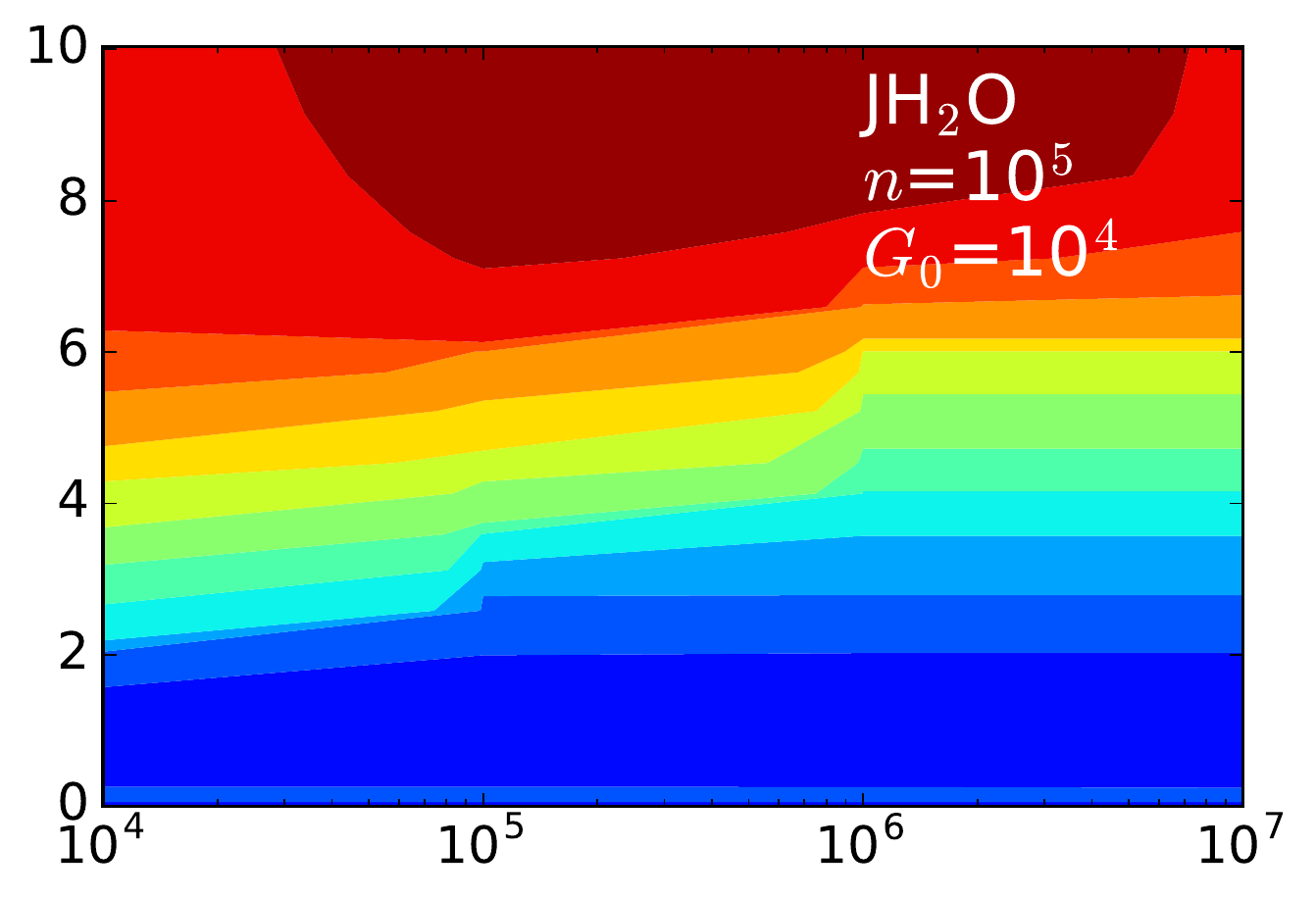}  \hspace{0.00cm}
\includegraphics[scale=0.42, angle=0]{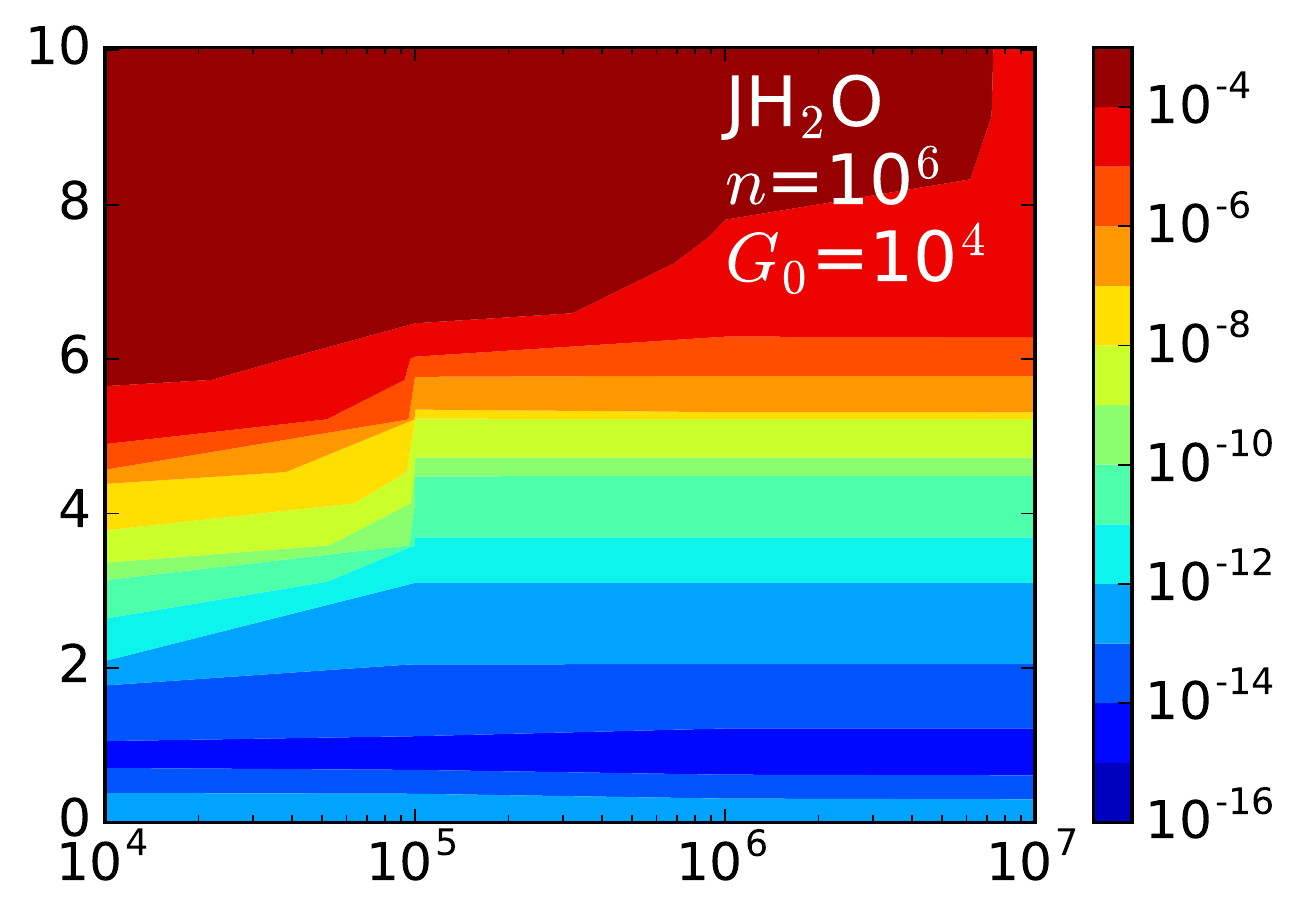}  \hspace{0.00cm}
\includegraphics[scale=0.42, angle=0]{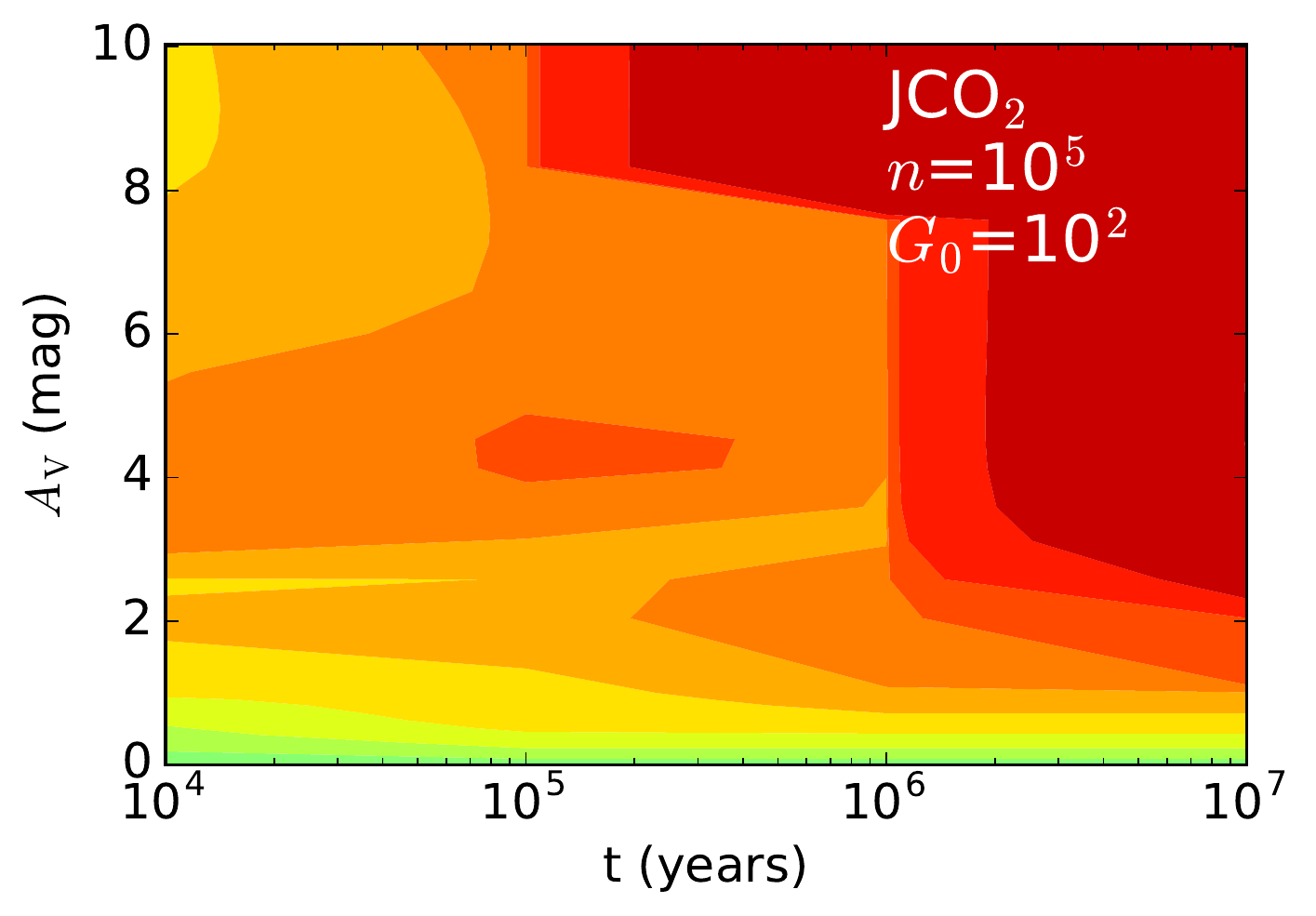}  \hspace{0.00cm}
\includegraphics[scale=0.42, angle=0]{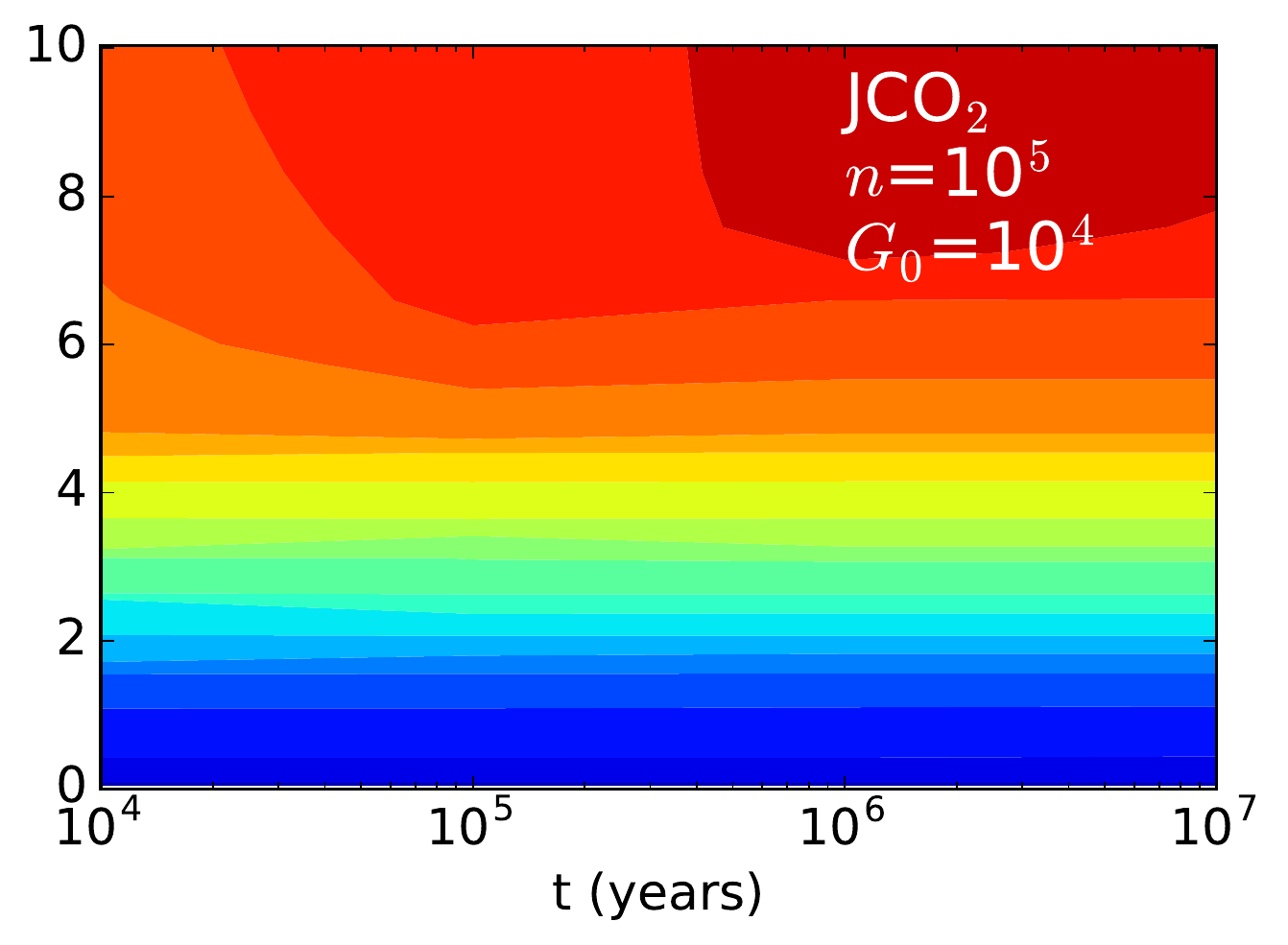}  \hspace{0.00cm}
\includegraphics[scale=0.42, angle=0]{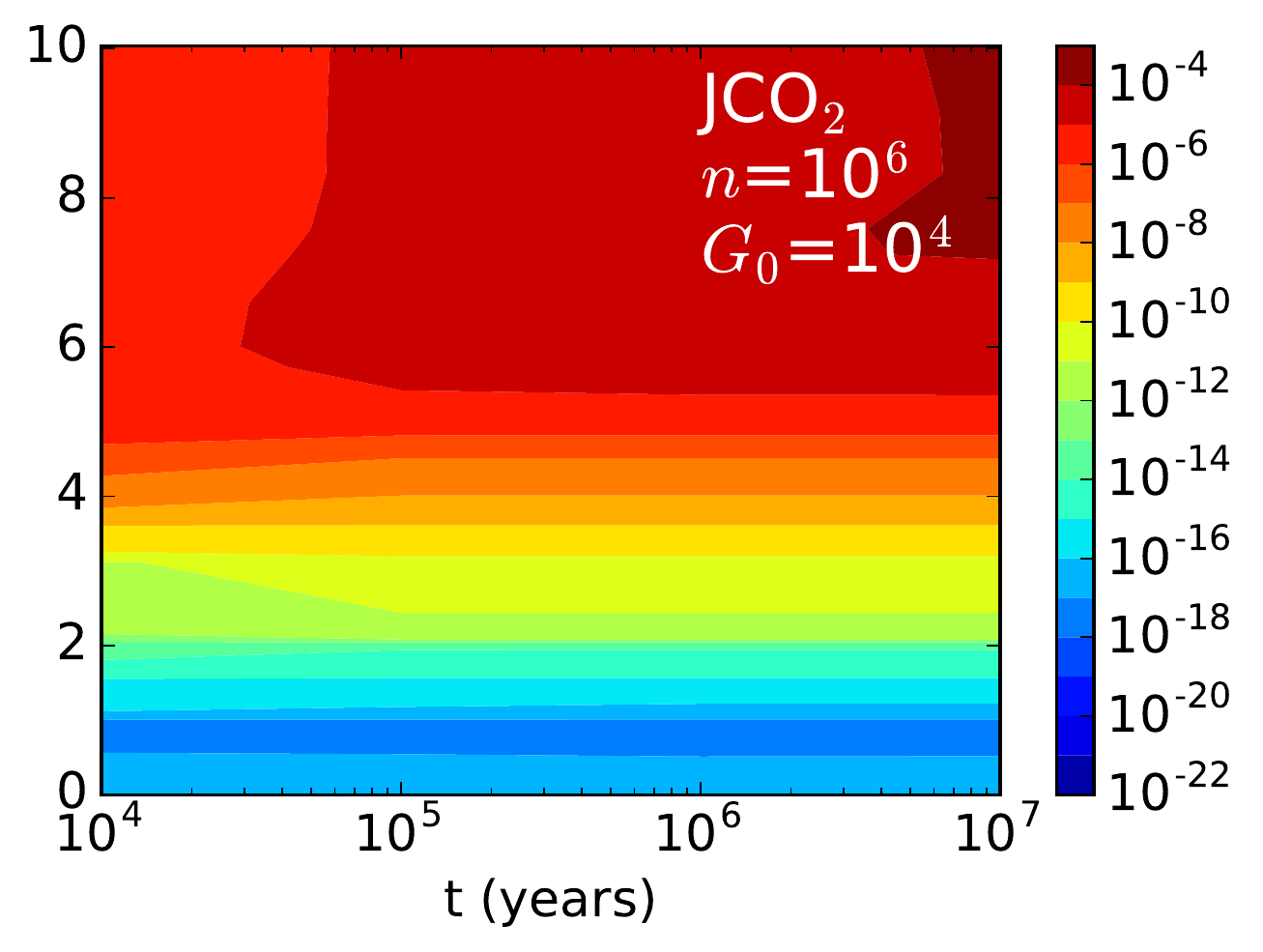}  \hspace{0.00cm}
\includegraphics[scale=0.42, angle=0]{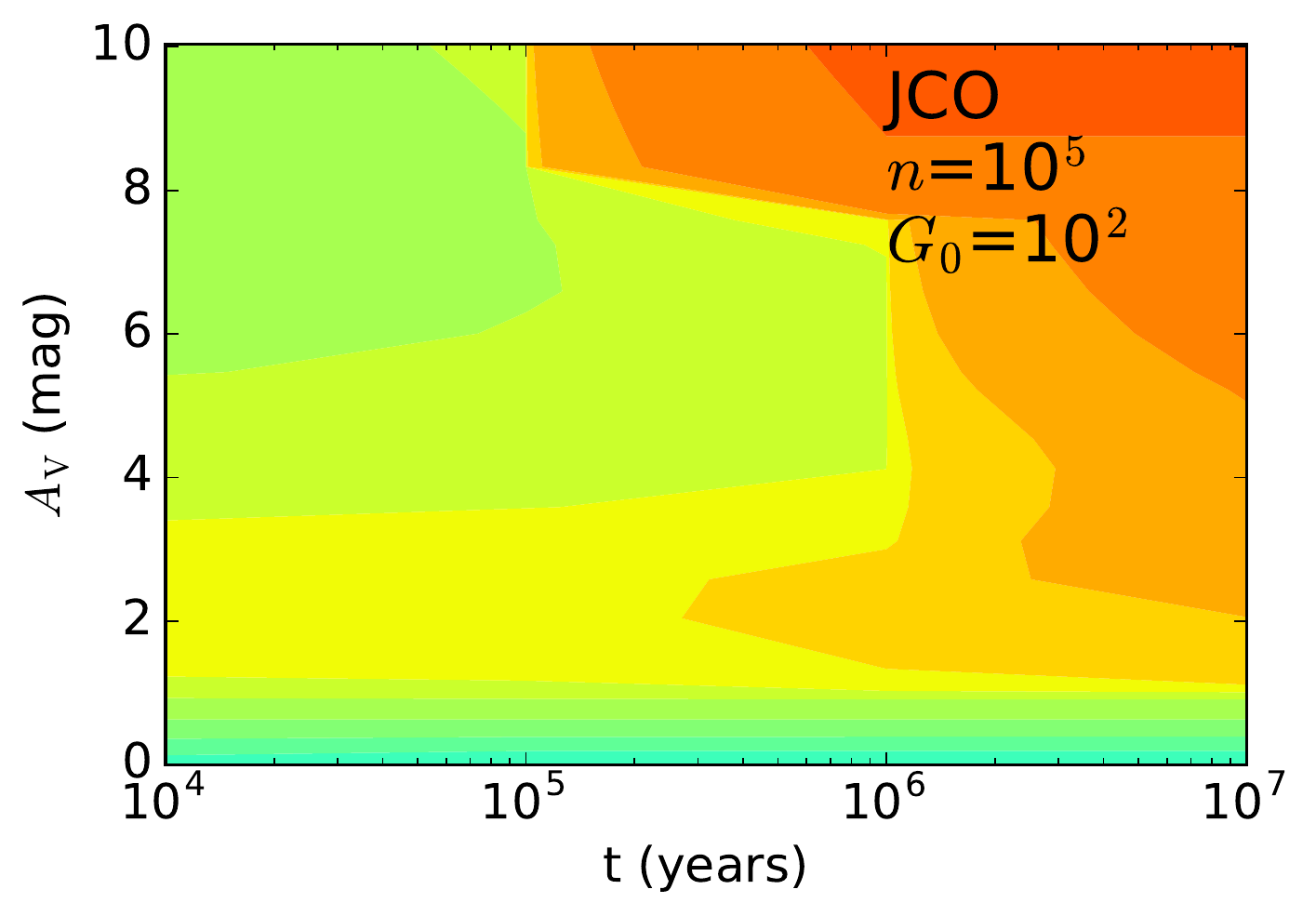}  \hspace{0.00cm}
\includegraphics[scale=0.42, angle=0]{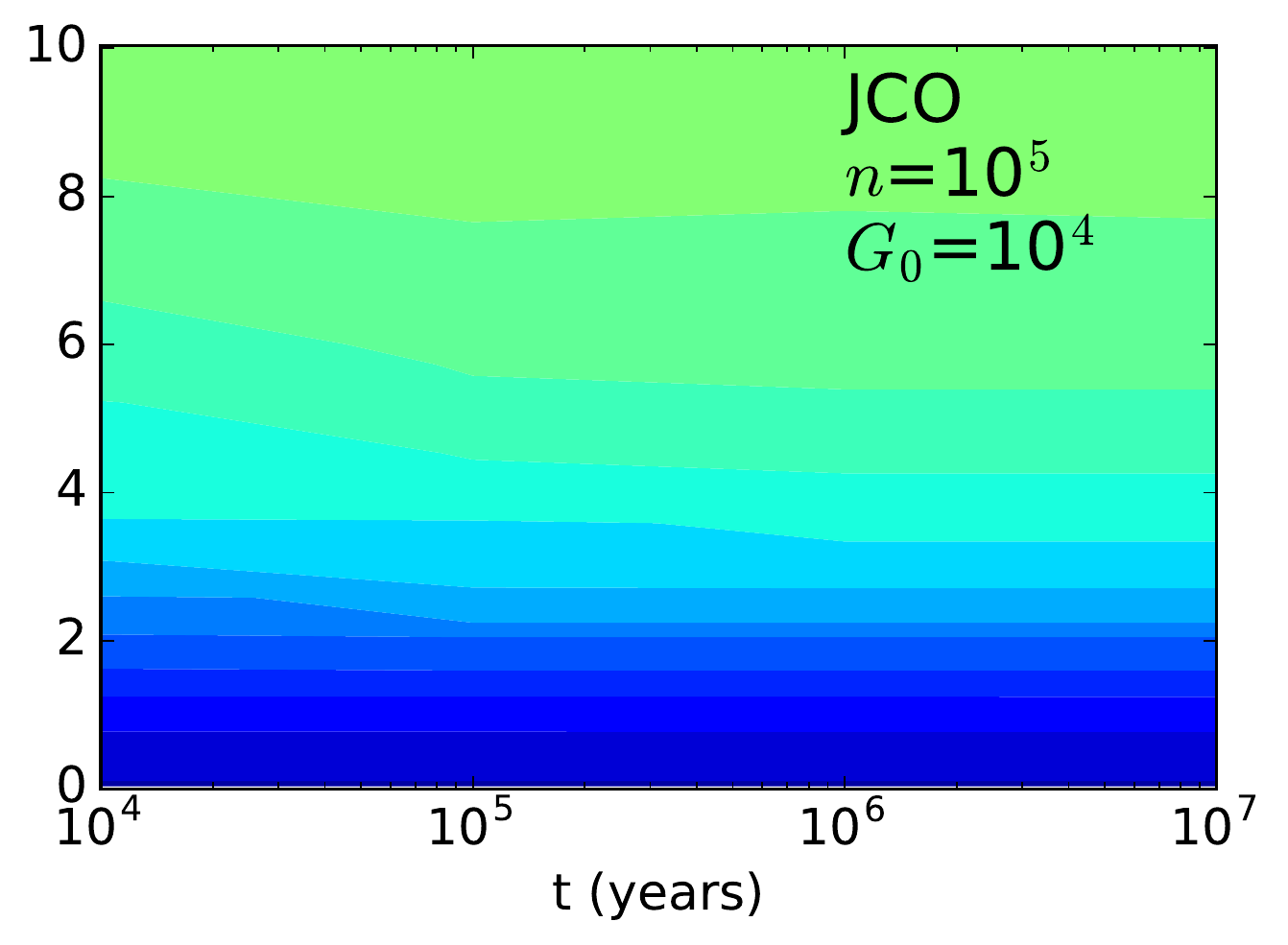}  \hspace{0.00cm}
\includegraphics[scale=0.42, angle=0]{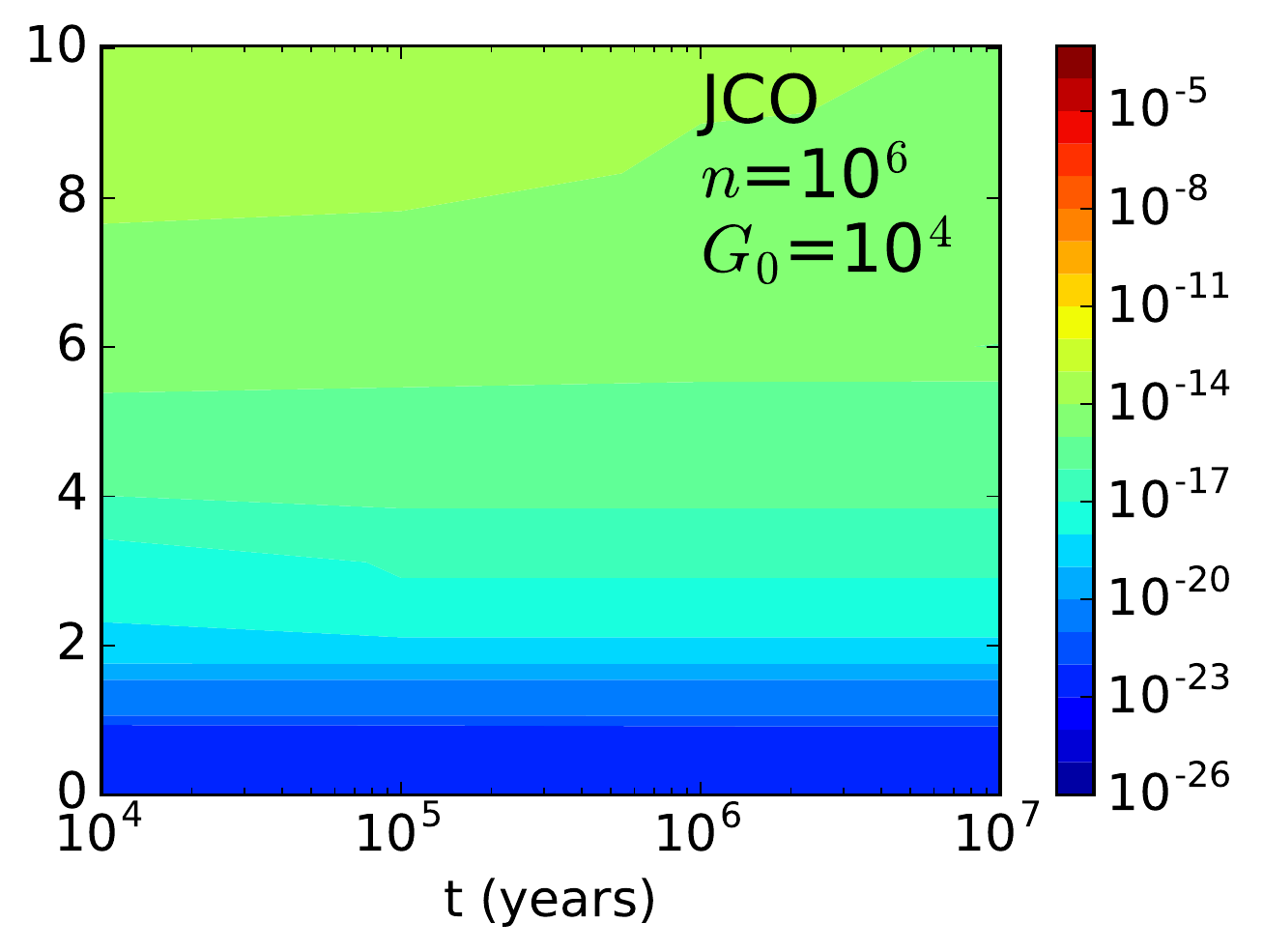}  \hspace{0.00cm}
\includegraphics[scale=0.42, angle=0]{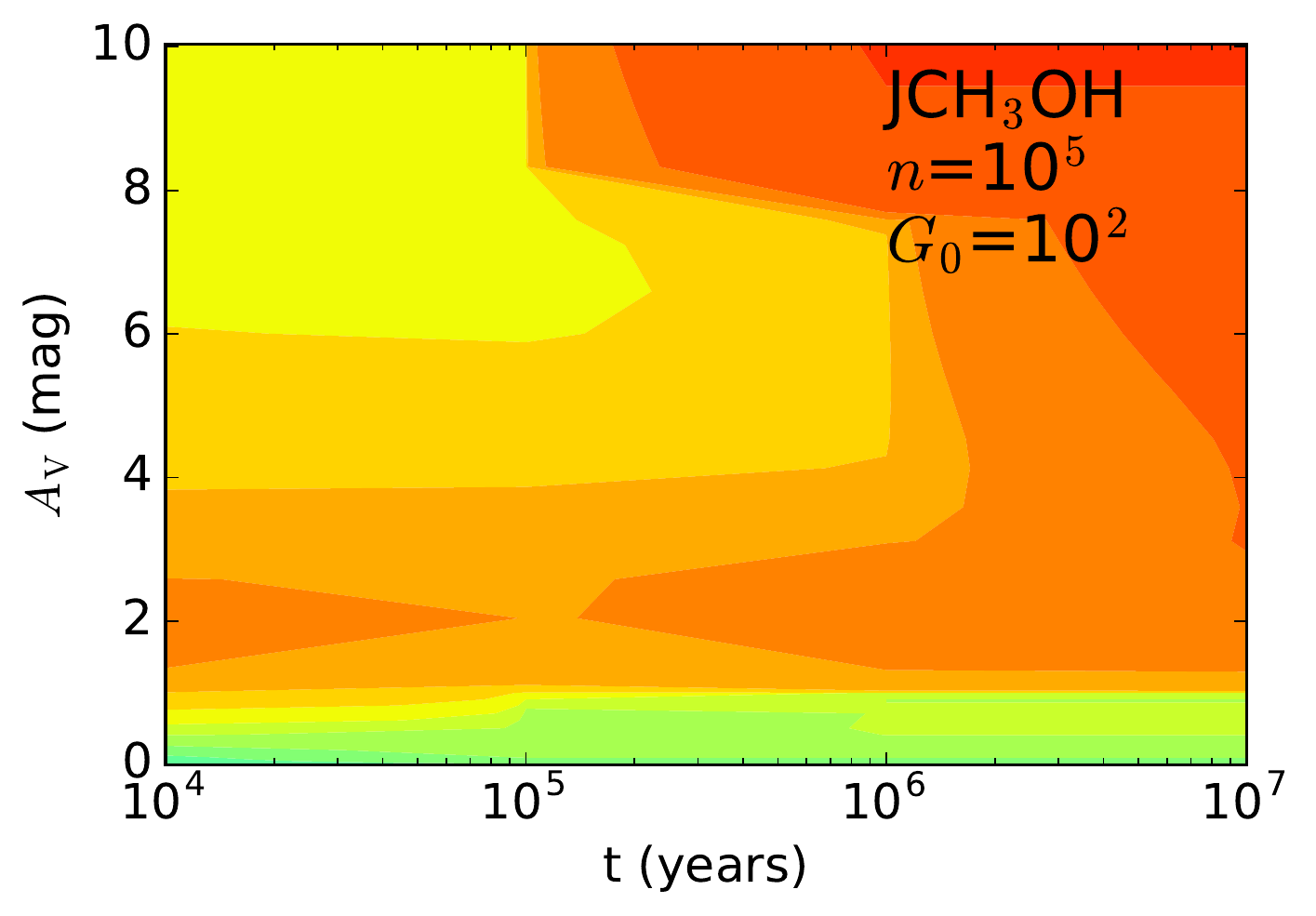}  \hspace{0.00cm}
\includegraphics[scale=0.42, angle=0]{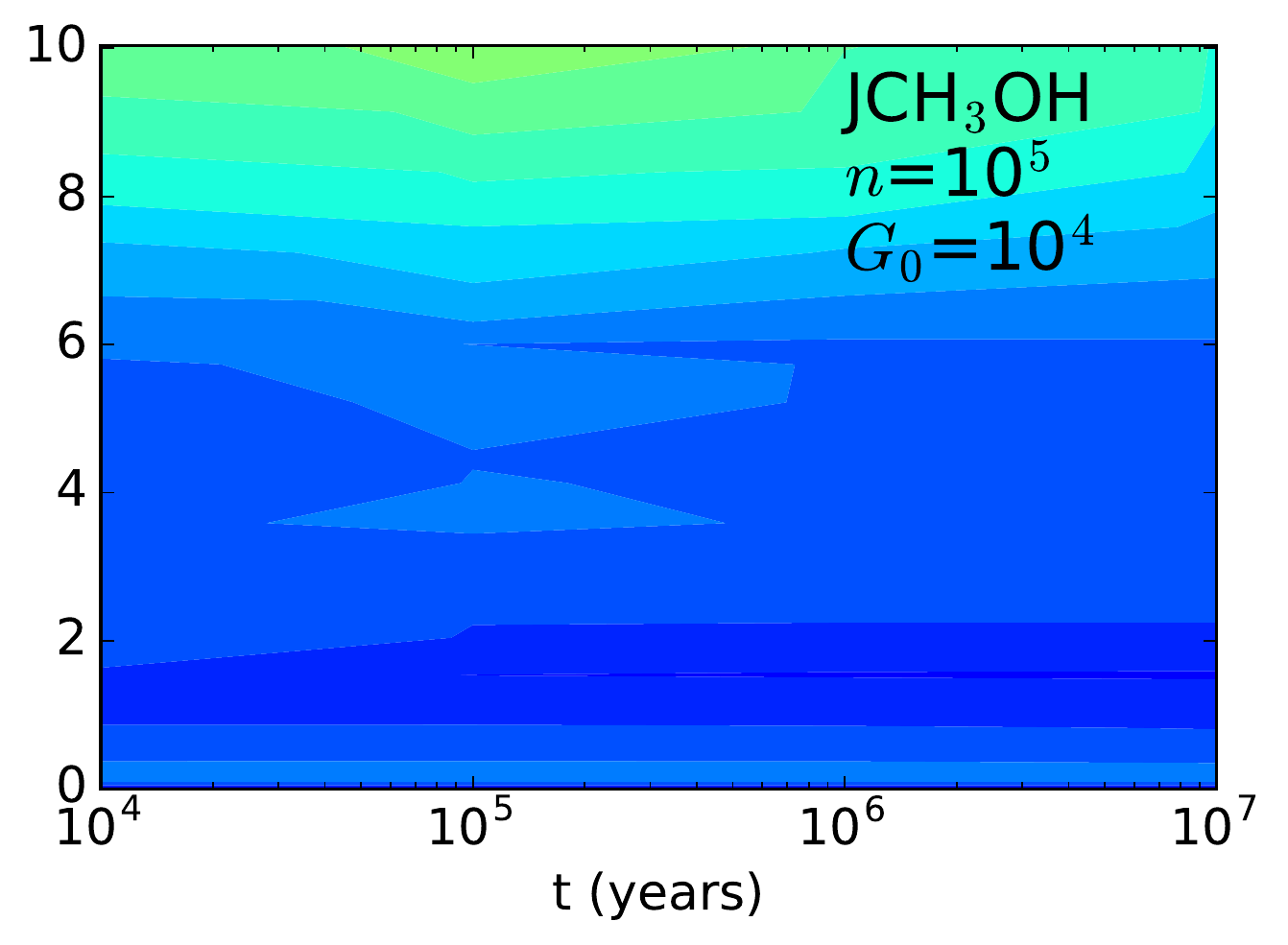}  \hspace{0.00cm}
\includegraphics[scale=0.42, angle=0]{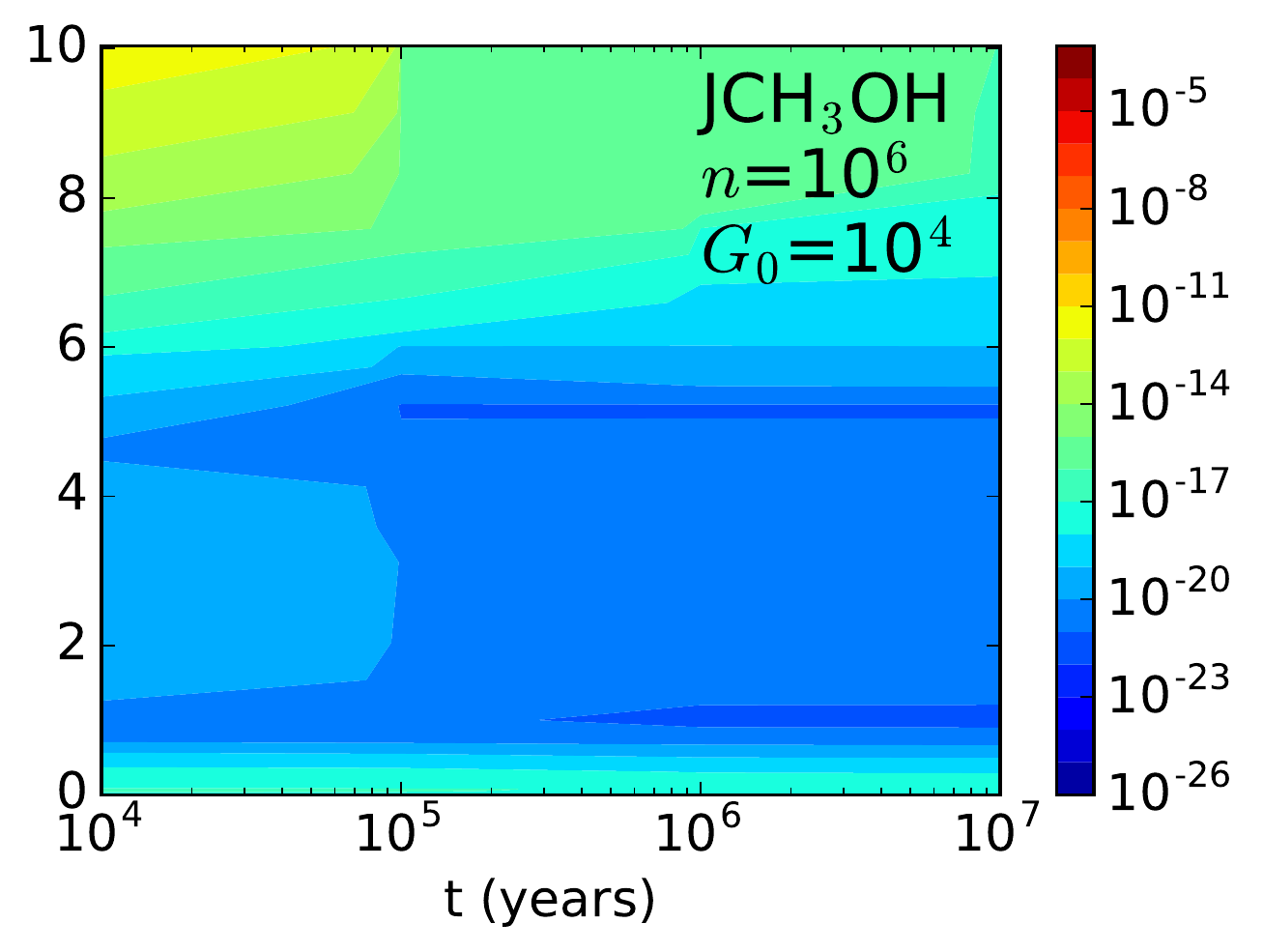}  \hspace{0.00cm}\\
\caption{Contour maps with the abundances of solid H$_2$O, CO$_2$, CO, and CH$_3$OH with respect to H nuclei for Models 1 (left panel), 2 (middle panel), and 3 (right panel) as a function of time (x-axis) and visual extinction (y-axis).}
\label{figure:contours4}
\end{figure*}

Figures \ref{figure:contours1}-\ref{figure:contours4} show the abundances of several families of molecules as a function of time (10$^4$$\leq$$t$$\leq$10$^7$ yr) and visual extinction (0$\leq$$A$$_{\mathrm{V}}$$\leq$10 mag) for three different type of PDRs (Models 1, 2, and 3 defined in Sect. \ref{Results}).

\subsubsection{Simple oxygen-bearing molecules}

Figure \ref{figure:contours1} shows the time evolution of the chemical abundances of OH (top), O$_2$ (middle), and H$_2$O (bottom). For any type of PDR, we obtain that the abundances of these three molecules at the edge of the cloud are $\lesssim$10$^{-8}$ for any evolutionary time. However, as the visual extinction increases, their abundances increase with a difference of up to 10 orders of magnitude between the edge ($A$$_{\mathrm{V}}$$\leq$1 mag) and the inner ($A$$_{\mathrm{V}}$$>$6 mag) part of the cloud depending on the type of PDR. 

For a low $G$$_{\mathrm{0}}$ PDR (left panel), although these three molecules present their highest abundances at $A$$_{\mathrm{V}}$$\gtrsim$6 mag, there are significant time differences between them. In the case of water, the abundance peak ($\sim$10$^{-5}$) is reached at an early evolutionary stage ($t$$\sim$10$^4$ yr), indicating that the reactions forming gas-phase water are fast. This abundance peak only presents variations lower than one order of magnitude for $A$$_{\mathrm{V}}$$\gtrsim$6 mag until $t$$\sim$10$^6$ yr, while for longer times water gas starts being significantly destroyed to form water ice. This abundance decrease is also found for O$_2$ and OH, which, after depletion, represents an important reactant to form CO$_2$ ice through the surface reaction JOH+JCO $\rightarrow$ JCO$_2$+JH. This is in agreement with Hollenbach et al. (2009) who also found that most of the gas-phase oxygen goes to H$_2$O ice and CO$_2$ ice at $t$$\sim$10$^7$ yr for $A$$_{\mathrm{V}}$$>$8 mag. Long timescales promote, therefore, the destruction of simple oxygen-bearing molecules at intermediate and large extinctions.   

When the intensity of the radiation field increases by two orders of magnitude (middle panels, Fig. \ref{figure:contours1}), the time for which the maximum abundances of H$_2$O are reached also increases by about one order of magnitude ($t$$\sim$10$^5$ yr). Similar behaviour is found for the other two species (OH and O$_2$), which indicates that, in high $G$$_{\mathrm{0}}$ PDR, long timescales promote the formation of simple oxygen-bearing molecules only in the inner regions of the cloud. In this case, the maximum abundances found for H$_2$O and OH are about one order of magnitude lower than those found in the low $G$$_{\mathrm{0}}$ PDR case. Molecular oxygen, however, reaches the same maximum abundance ($\sim$10$^{-5}$) independently on the intensity of the radiation, time being the main difference at which this value is reached. This shows that water formation is more linked to successive hydrogenations of atomic oxygen than to reactions involving molecular oxygen, as also found in Sect. \ref{simple oxygen molecules temperature}.

The main effect of increasing the density by one order of magnitude in a high $G$$_{\mathrm{0}}$ PDR is found for H$_2$O (right bottom panel, Fig. \ref{figure:contours1}). For this molecule, this increase allows water to reach its maximum abundance at an early evolutionary stage ($t$$<$10$^4$ yr), while in the lower density PDR (middle bottom panel) its maximum abundance is reached at $t$$\gtrsim$10$^5$ yr. At low visual extinctions ($A$$_{\mathrm{V}}$$<$5 mag), the abundances of these three simple oxygen-bearing molecules remain very low ($\lesssim$10$^{-8}$) at any evolutionary stage.     

From these results, we deduce that at intermediate and large $A$$_{\mathrm{V}}$, long timescales promote the formation of simple oxygen-bearing molecules in high $G$$_{\mathrm{0}}$ PDRs, while favour their destruction if $G$$_{\mathrm{0}}$ is low. In the edge of the cloud, no significant time effects are found on the evolution of these species.  

\subsubsection{Simple carbon-bearing molecules}

Figure \ref{figure:contours2} shows the chemical evolution of CH (top), CO (middle), and H$_2$CO (bottom) as a function of time and visual extinction. We obtain that CH is mainly formed at the edge of the cloud for either a low and a high $G$$_{\mathrm{0}}$ PDR (top left and top middle panels), while in the core its presence is much less significant. For a high intensity radiation field (top middle panel), abundances of CH barely change with time for $A$$_{\mathrm{V}}$$\lesssim$2 mag. In this case, CH mainly forms through the very endothermic reaction between H$_2$ and the ion C$^+$, which is efficiently formed with high radiation intensity. When $G$$_{\mathrm{0}}$ decreases (top left panel), the abundance of C$^+$ also decreases. In this case, we find that CH abundances progressively increase with time.

Unlike CH, we do not find significant time dependence in the abundances of CO at the edge of the three considered PDRs. For this molecule, time only becomes important at $A$$_{\mathrm{V}}$$\gtrsim$3 mag and presents different effects depending on the type of PDR. For a low $G$$_{\mathrm{0}}$ PDR (left middle panel, Fig. \ref{figure:contours2}), long timescales promote its destruction since it freezes out, while in a high $G$$_{\mathrm{0}}$ PDR (middle panel) the opposite behaviour is found. We also obtain that formaldehyde presents a similar time dependence (see left and middle bottom panels). In the case of CO, the abundance variation due to time effects is only $\lesssim$1 order of magnitude for any $A$$_{\mathrm{V}}$, suggesting that this species is very stable with time. In other words, for each visual extinction, it is formed almost as much CO as is destroyed for any type of PDR. Similar results are also found for H$_2$CO, whose abundances change by no more than two orders of magnitude over 10$^7$ yr, especially at low and intermediate $A$$_{\mathrm{V}}$.          

When the density of the PDR increases by one order of magnitude (right panels), we can distinguish two regimes. For a low extinction regime ($A$$_{\mathrm{V}}$$\lesssim$1 mag), the main effect found is the increase of the abundances of the three species (CH, CO, and H$_2$CO) by up to two orders of magnitude, since, close to the edge, destruction of molecules is dominated by photodissociation whose rates varies as $n$, while formation rates vary as $n$$^2$, resulting in abundances that increase with density. In a higher extinction regime, the main effect is found on the abundances of CO at $A$$_{\mathrm{V}}$$>$5 mag. In particular, while CO reaches its maximum abundance at $t$$\gtrsim$10$^5$ yr in the lowest density PDR (middle panel), this value is reached at $t$$<$10$^5$ yr in the high density PDR. This is due to a variation in the efficiency of the chemical reactions forming CO. In the low density case, CO mainly forms through dissociative recombination of HCO$^+$, which is formed through the reaction between C$^+$ and H$_2$O (H$_2$O being more efficiently formed at large extinctions and long timescales as shown in Fig. \ref{figure:contours1}, middle bottom panel). However, as the density increases, the formation of HCO$^+$ through the ionisation of chemically desorbed HCO becomes more efficient.

\subsubsection{Complex organic molecules}
\label{COMS_time} 
 
Figure \ref{figure:contours3} shows the chemical evolution of CH$_3$OH, CH$_3$CN, and CH$_2$CO as a function of time and visual extinction.  

The CH$_3$OH formation starts being efficient at $A$$_{\mathrm{V}}$$>$2 mag, since it mainly forms on grain surfaces through chemical desorption upon the surface reaction between solid H with solid H$_3$CO. In a low $G$$_{\mathrm{0}}$ PDR, the maximum abundances of methanol are obtained at 10$^5$$\lesssim$$t$$\lesssim$5$\times$10$^6$ yr, when enough time has passed to form solid H$_3$CO, its precursor. In the case of considering a more intense radiation field (two orders of magnitude higher, top middle panel), we obtain that the abundances of methanol sharply decrease at any visual extinction and evolutionary stage by several (up to ten) orders of magnitude. Only the increase of density (top right panel, Fig. \ref{figure:contours3}) promotes the formation of methanol deep in the cloud at $t$$<$10$^5$ yr.

Results for molecule CH$_3$CN are shown in the middle panels of Fig. \ref{figure:contours3}. In the gas phase, one of the principal precursors of CH$_3$CN is HCN, which mainly forms through an exchange chemical reaction whose activation energy barrier is 100 K (MacKay 1999). This barrier is slightly higher than the gas temperature of the low $G$$_{\mathrm{0}}$ PDR at low extinctions (see right panel of Fig. \ref{figure:Tdust_comparison}), which explains the low ($\lesssim$10$^{-16}$) abundances of methyl cyanide at the edge of this type of PDRs (left middle panel in Fig. \ref{figure:contours3}). As $A$$_{\mathrm{V}}$ increases, the abundance of CH$_3$CN increases by up to 10 orders of magnitude in the core, indicating that this molecule is significantly enhanced by reactions occurring on grain surfaces. Regarding time effects, Fig. \ref{figure:contours3} (left panel) shows that CH$_3$CN presents differences in its abundances no larger than one order of magnitude over time at $A$$_{\mathrm{V}}$$\gtrsim$3 mag. The opposite effect is, however, found when $G$$_{\mathrm{0}}$ increases by two orders of magnitude (middle panel). In this case, time effects become more important for the evolution of CH$_3$CN as the visual extinction increases. In particular, its abundances increase by up to three orders of magnitude from $t$=10$^4$ yr to $t$=10$^6$ yr, showing that CH$_3$CN is a late-forming molecule in high $G$$_{\mathrm{0}}$ PDRs. 

Particularly interesting is the effect of increasing the density of the PDR (right middle panel, Fig. \ref{figure:contours3}) by one order of magnitude for CH$_3$CN at $A$$_{\mathrm{V}}$$\lesssim$0.5 mag. This produces an increase of the CH$_3$CN abundances of at least two orders of magnitude with respect to the low density case (middle panel), becoming even higher than some values found at intermediate extinctions ($A$$_{\mathrm{V}}$$\sim$5 mag) for an early ($t$$<$10$^5$ yr) evolutionary stage. We find that this effect, which has also been observationally detected by Gratier et al. (2013), is mainly due to a more efficient formation of HCN (precursor of CH$_3$CN in the gas phase) in the edge of the cloud produced by the density increase. In particular, at $A$$_{\mathrm{V}}$$\lesssim$0.5 mag, we obtain an increase of HCN of about two orders of magnitude when the density increases by one order of magnitude (comparison between middle and right panels of Fig. \ref{figure:contours5} in the Appendix \ref{Figures}). In this case of a very high density PDR, CH$_3$CN mainly forms through radiative association (CH$_3$$^+$ + HCN, with CH$_3$$^+$ being quite abundant at low $A$$_{\mathrm{V}}$ with respect to large extinctions due to the high UV radiation), followed by dissociative recombination.

Ketene has long been identified in the interstellar medium and different gas-phase pathways (with ethylene ions as precursors) have been proposed for its formation (Millar et al. 1991). However, its detection in the cold prestellar cores L1544 (Spezzano et al. 2017) and L1689B at temperatures of $\sim$10 K suggests a formation with grain surface chemistry (through methane-carbon monoxide ices) and subsequent non-thermal desorption via induced UV photons and cosmic ray impacts (Bacmann et al. 2012, Maity et al. 2014). In a low $G$$_{\mathrm{0}}$ PDR (left bottom panel, Fig. \ref{figure:contours3}), we find the maximum CH$_2$CO abundances at late times (5$\times$10$^5$$\lesssim$$t$$\lesssim$5$\times$10$^6$ yr) in the inner regions ($A$$_{\mathrm{V}}$$>$6 mag) of the PDR. 

As $G$$_{\mathrm{0}}$ increases (middle bottom panel), the abundance of CH$_2$CO sharply decreases for all visual extinctions and for any evolutionary stage due to the increase of radiation, which prevents the formation of ketene ice precursors. In this case, the formation of ketene at the edge of the PDR becomes inefficient and time effects are only important at intermediate and large extinctions. In general, we observe that abundances of CH$_2$CO change by no more than one order of magnitude at $t$$>$10$^5$ yr for each visual extinction, independently on the type of PDR. We therefore deduce that visual extinction is a more important factor than time for the formation of ketene.

 \subsubsection{Solid molecules}

The two top rows of Fig. \ref{figure:contours4} show the chemical evolution of solid water (JH$_2$O) and solid carbon dioxide (JCO$_2$) for Models 1, 2, and 3, as a function of time.  

For a low $G$$_{\mathrm{0}}$ PDR (left top panel), the abundances of solid water increase with visual extinction. In particular, the maximun abundance ($\sim$10$^{-4}$) of JH$_2$O with respect to hydrogen is reached at $t$$\gtrsim$10$^5$ yr for $A$$_{\mathrm{V}}$$\sim$5-8 mag and remains roughly constant over more than one million years. When $G$$_{\mathrm{0}}$ increases, radiation effects prevent solid water formation at low $A$$_{\mathrm{V}}$. In this case, the maximum solid water abundance is roughly the same as in the low $G$$_{\mathrm{0}}$ PDR, but found at larger extinctions ($A$$_{\mathrm{V}}$$>$7 mag). We also find that this abundance peak is reached at an earlier evolutionary stage ($\sim$5$\times$10$^4$ yr) than in the low $G$$_{\mathrm{0}}$ case. This time becomes significantly lower as the density of the PDR increases (right top panel). We therefore conclude that high $G$$_{\mathrm{0}}$ values promote the formation of solid water at earlier evolutionary stages and larger extinctions than low $G$$_{\mathrm{0}}$ values.

Figure \ref{figure:solid_H2O} (Appendix \ref{Figures}) shows the threshold to form one full monolayer of water ice (see Esplugues et al. 2016 for more details about the calculation of this limit), together with the fractional water abundances over time. The first water ice monolayer is formed at a very early stage (10$^4$ yr) for any type of PDR, although the visual extinction varies between 3-4 mag depending on the considered $G$$_{\mathrm{0}}$ value (the higher $G$$_0$, the larger $A$$_{\mathrm{V}}$ due to the increase of UV radiation, which prevents ice formation). For longer timescales, we also find formation of full monolayers of water ice for any PDR, but at larger extinctions than in the early stage.

Results for solid CO$_2$ are also shown in Fig. \ref{figure:contours4} (second row). The abundances of this species for a low $G$$_0$ PDR (left panel) at $A$$_{\mathrm{V}}$$<$1 mag are very low ($\lesssim$10$^{-10}$) and independent on the evolutionary stage. By contrast, for larger extinctions, abundances of JCO$_2$ present a strong time-dependence with variations of up to 4 orders of magnitude between the early and the evolved stages. The maximum abundance ($\sim$10$^{-5}$) of solid CO$_2$ is first reached very deep in the cloud ($A$$_{\mathrm{V}}$$>$8 mag) at 10$^5$$<$$t$$<$10$^{6}$ yr, while for a more evolved cloud (10$^6$$\lesssim$$t$$\lesssim$10$^{7}$ yr), this abundance peak is reached at much lower extinctions ($A$$_{\mathrm{V}}$$\lesssim$3 mag). In the late stage, we find that the formation of JCO$_2$ mainly occurs through the reaction between solid CO and solid O at low $A$$_{\mathrm{V}}$, but, as the visual extinction increases, we obtain that solid OH also becomes an important precursor of JCO$_2$. For this type of low $G$$_{\mathrm{0}}$ PDR, the maximum number of full CO$_2$ ice monolayers is reached at a late stage ($t$=10$^6$ yr), according to results from Fig. \ref{figure:solid_CO2} (Appendix \ref{Figures}). 

For a higher $G$$_{\mathrm{0}}$ PDR (Fig. \ref{figure:contours4}, second row, middle panel), we obtain that the abundances of solid CO$_2$ are only significant at $A$$_{\mathrm{V}}$$\gtrsim$5 mag and that they increase as the cloud evolves. We also find that the density increase promotes the formation of JCO$_2$ at earlier stages (right bottom panel) than the low density case, as also found for JH$_2$O, allowing to reach JCO$_2$ abundances $\gtrsim$10$^{-5}$ at $t$$\sim$5$\times$10$^4$ yr at intermediate extinctions.

Comparing the results for both molecules (solid H$_2$O and solid CO$_2$), we deduce that carbon dioxide is a more time-dependent species than water in low $G$$_{\mathrm{0}}$ PDRs. We also derive that CO$_2$ is a late-forming ice with respect to water ice in dense PDRs, since the formation of the first water ice monolayer occurs at $t$$\sim$10$^4$ yr, while for CO$_2$ occurs at $t$$\sim$10$^5$-10$^6$ yr for $A$$_{\mathrm{V}}$$\leq$10 mag. We here highlight that ices form in layers, with water ice as first layers, and CO$_2$ ice on the top in PDRs.      

Other ices, such as CO and CH$_3$OH (Fig. \ref{figure:contours4}, the two bottom rows), present significant abundances only in the low $G$$_{\mathrm{0}}$ case at large visual extinctions ($A$$_{\mathrm{V}}$$>$6 mag) for $t$$\gtrsim$10$^6$ yr. Nevertheless, their abundance peaks are $\sim$10$^{-6}$-10$^{-8}$, i.e. up to four orders of magnitude lower than the maximum water ice abundance. They are, therefore, minor and late-forming ice constituents in PDRs.

 \subsubsection{Steady state}

A chemical system reaches equilibrium when the rate at which each molecule is formed is equal to the rate at which it is destroyed, keeping its abundance constant over time. As previously stated, at low visual extinctions ($A$$_{\mathrm{V}}$$\lesssim$1 mag) in a molecular cloud, the energy balance is dominated by FUV photons and the chemical timescales are very short compared to the molecular cloud lifetime (10$^6$-10$^7$ yr). However, as the visual extinction increases, certain chemical timescales become comparable to cloud lifetimes and steady-state chemistry does not apply. Time at which steady state is reached can also be affected by several mechanisms, such as turbulent motions (which can mix external regions exposed to the UV field with the inner regions of the cloud), star formation and the violent phenomena associated to its early stages. Here we only analyse when steady state is reached depending on the PDR type.

In Figures \ref{figure:contours1}-\ref{figure:contours4}, we have shown the chemical evolution of different molecule families over time (10$^4$-10$^7$ yr) for low and high $G$$_{\mathrm{0}}$ PDRs, and also varying the density. For the low visual extinction case ($A$$_{\mathrm{V}}$$\lesssim$1 mag), we observe that steady state is reached at early times ($t$$\lesssim$10$^5$ yr) by all the considered molecules in the different PDRs, and, in particular, at $t$$<$10$^4$ yr in high $G$$_{\mathrm{0}}$ PDRs. Only a few complex molecules (CH$_3$CN and CH$_2$CO) present a slower chemical evolution in the $G$$_{\mathrm{0}}$=100 case with equilibrium times 10$^5$$\lesssim$$t$$<$10$^6$ yr.

For larger extinctions, however, chemical equilibrium is reached at very different times, which strongly depends on the PDR characteristics. In a high $G$$_{\mathrm{0}}$ PDR with density $n$=10$^5$ cm$^{-3}$ (middle panels of Figs. \ref{figure:contours1}-\ref{figure:contours4}), steady state is reached at $t$$\lesssim$10$^6$ yr for all molecules in the range 0$\leq$$A$$_{\mathrm{V}}$$\leq$10 mag. If the density is increased by one order of magnitude (right panels), the chemical equilibrium is reached even at shorter times ($t$$<$5$\times$10$^5$ yr) for most of the species. By contrast, in a cloud associated to a PDR with low intensity radiation field (left panels, Figs. \ref{figure:contours1}-\ref{figure:contours3}), chemical equilibrium is reached at very long timescales (10$^6$$\lesssim$$t$$\lesssim$10$^7$ yr) for $A$$_{\mathrm{V}}$$>$2 mag, i.e. at a time comparable to the cloud lifetime. This large time difference to reach equilibrium is mainly due to the temperature variation between both (low and high $G$$_{\mathrm{0}}$) PDRs; in the low $G$$_{\mathrm{0}}$ case, the temperature is significantly lower than in the high $G$$_{\mathrm{0}}$ case (by $\sim$50 K and $\sim$25 K for the gas temperature at 3 mag and 8 mag, respectively, and by $\sim$25 K and $\sim$20 K for the dust temperature at those $A$$_{\mathrm{V}}$, Fig. \ref{figure:Tdust_comparison}), producing a decrease in the reaction rate, and some chemical barriers cannot be overcome.

\subsubsection{Comparison with observations: abundances}

\begin{table}
\caption{Observational abundances with respect to total hydrogen nuclei in the Horsehead and the Orion Bar.}             
\centering 
\begin{tabular}{l l l l}     
\hline\hline       
Species         & \multicolumn{2}{c}{Horsehead} & {Orion Bar} \\ 
  & PDR & Core  \\
\hline 
   H$_2$CO   &  (2.9$\pm$0.4)$\times$10$^{-10}$ & (2.0$\pm$0.3)$\times$10$^{-10}$  & (1.8$\pm$0.9)$\times$10$^{-9}$ \\
   CH$_3$OH  &  (1.2$\pm$0.2)$\times$10$^{-10}$ & (2.3$\pm$0.3)$\times$10$^{-10}$ & (1.5$\pm$0.9)$\times$10$^{-9}$  \\
   CO        &  (5$\pm$3)$\times$10$^{-5}$ &   -  &  (1.5$\pm$0.6)$\times$10$^{-4}$ \\
   CH        &   -    &   -   &  (6.0$\pm$0.9)$\times$10$^{-8}$ \\
   H$_2$O    &   -    &   -   &  (9$\pm$3)$\times$10$^{-10}$ \\
\hline 
\label{table:observations}              
\end{tabular}

\medskip
Data for the Horsehead are obtained from Pety et al. (2005) and Guzm\'an et al. (2013, 2014). Data for the Orion Bar are obtained from Nagy et al. (2017) and Cuadrado et al. (2017). Abundances for CO, CH, and H$_2$O have been obtained considering $N$(H)=3$\times$10$^{21}$ cm$^{-2}$ (van der Werf et al. 2013).  
\end{table}

In this section, we compare our model abundances with observations (Table \ref{table:observations}) of CO, CH, H$_2$O, H$_2$CO, and CH$_3$OH in the Horsehead ($G$$_{\mathrm{0}}$$\sim$100 and $n$$\sim$10$^5$ cm$^{-3}$, Habart et al. 2005, Guzm\'an et al. 2013) and the Orion Bar ($G$$_{\mathrm{0}}$$\sim$10$^4$ and $n$$\sim$10$^4$-10$^6$ cm$^{-3}$, Marconi et al. 1998, Leurini et al. 2010). According to these physical conditions, our Model 1 would correspond to the Horsehead PDR and Model 3 to the Orion Bar.

For the case of simple molecules in the Horsehead PDR, Pety et al. (2005) derived a C$^{18}$O abundance of 1.9$\times$10$^{-7}$ in the IR peak ($A$$_{\mathrm{V}}$$\sim$1 mag). Tercero et al. (2010) and Esplugues et al. (2013) obtained a $^{16}$O/$^{18}$O ratio $\sim$250, which is lower than the Solar isotopic abundance ($\sim$500, Anders \& Grevesse 1989). Considering a $^{16}$O/$^{18}$O ratio of 250, we derive an abundance for CO in the Horsehead PDR of 5$\times$10$^{-5}$. We reproduce this value in the Model 1 at a visual extinction of  1$\lesssim$$A$$_{\mathrm{V}}$$<$3.5 mag for any evolutionary time (Fig. \ref{figure:contours2}, left middle panel), in agreement with results from Pety et al. (2005). In the Orion Bar, Nagy et al. (2017) observed CO, CH, and H$_2$O with abundances of 1.5$\times$10$^{-4}$, 6$\times$10$^{-8}$, and 9$\times$10$^{-10}$, respectively. We reproduce these values with the Model 3 ($n$=10$^6$ cm$^{-3}$ and $G$$_0$=10$^4$) at $A$$_{\mathrm{V}}$$\lesssim$2.5 mag for any evolutionary stage as well, since the abundances of theses species barely change over time in this PDR model. The $T$$_{\mathrm{kin}}$ considered in Nagy et al. (2017) to obtain the observational CO abundance is $\sim$137 K, in agreement with the gas temperature considered in our model (Fig. \ref{figure:Tdust_comparison}, right panel, solid black line) for the range ($A$$_{\mathrm{V}}$$\lesssim$2.5 mag) where the observations are reproduced.

The H$_2$CO abundance (2.9$\times$10$^{-10}$) in the Horsehead PDR was observed at the IR peak at $A$$_{\mathrm{V}}$$\sim$1 mag (Guzm\'an et al. 2011, Pety et al. 2012, Guzm\'an et al. 2013). We reproduce this value for any evolution time at $A$$_{\mathrm{V}}$$\sim$1.5 mag (see Fig. \ref{figure:contours2}, left bottom panel), which represents an extinction of about 1 mag lower than in Esplugues et al. (2016). The H$_2$CO observational abundance was derived using a non-local excitation and radiative transfer model considering $n$$_{\mathrm{H}}$=5$\times$10$^4$-10$^5$ cm$^{-3}$, assuming a kinetic temperature $T$$_{\mathrm{kin}}$=40-65 K, which are consistent with our PDR model parameters (Fig. \ref{figure:Tdust_comparison}, right panel).

The observed H$_2$CO abundance in the core ($A$$_{\mathrm{V}}$$\sim$8 mag, Pety et al. 2012, Guzm\'an et al. 2013) of the Horsehead is $\sim$2$\times$10$^{-10}$(obtained considering $T$$_{\mathrm{kin}}$=20 K), however we predict abundances at least three orders of magnitude higher than this value at 3$<$$A$$_{\mathrm{V}}$$\leq$10 mag. The dust temperature considered in our PDR model (slightly lower than 20 K, Fig. \ref{figure:Tdust_comparison} left panel) could be a main factor producing this overestimation, since the typical dust temperature considered in this extinction range of the Horsehead is $T$$_{\mathrm{dust}}$$\sim$20-30 K (Goicoechea et al. 2009, Guzm\'an et al. 2013), and the lower the dust temperature, the larger the H$_2$CO abundances at intermediate and large extinctions as found in Sect. \ref{carbon_dust-temperature}. In particular, we find that a $T$$_{\mathrm{dust}}$ difference of only $\sim$4 K (difference found at $A$$_{\mathrm{V}}$=8 mag from Fig. \ref{figure:Tdust_comparison}) leads to an H$_2$CO abundance difference of one order of magnitude (Fig. \ref{figure:C_bearing_comparison}).

For the Orion Bar case, Leurini et al. (2006, 2010) observationally deduced that H$_2$CO traces the warm interclump close to the strong FUV-field in the Orion Bar. We obtain a difference between the observed and the modelled H$_2$CO abundance of less than two orders of magnitude at $A$$_{\mathrm{V}}$$\sim$2.5-4.5 mag and $t$$\geq$5$\times$10$^4$ yr, for a high (10$^6$ cm$^{-3}$) density PDR model (see Fig. \ref{figure:contours2}, right bottom panel). This difference is much lower than that obtained in Esplugues et al. (2016) and similar to that obtained by Cuadrado et al. (2017), who also found that H$_2$CO survives in the extended gas directly exposed to the strong FUV flux. Observational H$_2$CO in the Orion Bar was derived using a non-LTE LVG model with $T$$_{\mathrm{kin}}$=150-250 K, $T$$_{\mathrm{dust}}$$\geq$60 K, and $n$(H$_2$)=10$^6$ cm$^{-3}$  (Cuadrado et al. 2017), consistent with our density, dust and gas model temperatures.

The observed abundance of methanol in the PDR of the Horsehead (at $A$$_{\mathrm{V}}$$\sim$1 mag) is 1.2$\times$10$^{-10}$ (Guzm\'an et al. 2013). We reproduce this value at $A$$_{\mathrm{V}}$$\sim$1.5-2.5 mag independently on the stage of evolution, since our models show that CH$_3$OH is formed as fast as is destroyed for this visual extinction range over time (see Fig. \ref{figure:contours3}, left top panel). 
In the case of the cloud core of the Horsehead, however, we obtain an overestimation of the CH$_3$OH abundance with respect to the observed value (2.3$\times$10$^{-10}$ at $A$$_{\mathrm{V}}$$\sim$8 mag) of at least two orders of magnitude.

In the case of the Orion Bar, Leurini et al. (2006, 2010) deduced that CH$_3$OH traces the denser and cooler clumps observed in its inner region. The observed CH$_3$OH abundance in the Orion Bar is 1.5$\times$10$^{-9}$ (Table \ref{table:observations}). As previous studies (e.g. Cuadrado et al. 2017), we also underestimate this value by several orders of magnitude (see Fig. \ref{figure:contours3}, right top panel). At present, no model seems to reproduce the inferred abundances of CH$_3$OH nor H$_2$CO towards the Orion Bar. Nevertheless, it is interesting to highlight the effect of increasing by one order of magnitude the density of a high $G$$_{\mathrm{0}}$ PDR model. It leads, in an early stage ($t$$<$10$^5$ yr), to a sharp increase of the CH$_3$OH abundance of about 6 orders of magnitude between $A$$_{\mathrm{V}}$=4 mag and $A$$_{\mathrm{V}}$=9 mag, suggesting that the observations of methanol in the Orion Bar would correspond to the presence of a very dense clump ($n$$\geq$10$^7$ cm$^{-3}$) formed at an early stage ($t$$\lesssim$10$^5$ yr). On the other hand, turbulent diffusion could also be an important process, directly affecting the abundances of methanol and leading to these differences between model results and observations, since this mechanism significantly increases the abundances of its precursors (CO) in the inner parts of the cloud (Bell et al. 2010).

\section{Summary and conclusions}
\label{conclusions}

We have presented for the first time a study about the effects of dust temperature and time on the chemistry of several types of PDRs using an updated version of the Meijerink PDR code. 

Considering two distinct dust temperature expressions (from Garrod $\&$ Pauly 2011, and Hocuk et al. 2017), which differ from each other up to $\sim$30 K depending on the PDR characteristics, we have found the most significant chemical impact in high $G$$_{\mathrm{0}}$ PDRs. In this case, the formation of complex molecules, such as methanol, is clearly more efficient (up to 8 orders of magnitude) for the lowest dust temperature in both the PDR and the core, while the formation of simple oxygen-bearing molecules (such as O$_2$) is much more efficient for the highest $T$$_{\mathrm{dust}}$ values in the core. Large temperature effects have also been found in the visual extinction threshold at which ice formation occurs, with variations of up to $\sim$5 mag in the formation of water and CO$_2$ ices. 

In this paper, we have also combined the steady-state Meijerink PDR code with the Nahoon time-dependent cloud code to realistically study the chemical evolution of several molecule families in the inner regions of different PDRs over 10$^7$ yr. The model results show that time dependence mainly affects the chemical evolution at intermediate and large visual extinctions, where long timescales promote the destruction of oxygen-bearing molecules in low $G$$_{\mathrm{0}}$ PDRs, and favour their formation and that of carbon-bearing molecules in high $G$$_{\mathrm{0}}$ PDRs. Regarding solid species, we have found that CO$_2$ is a strongly time-dependent molecule, especially in low $G$$_{\mathrm{0}}$ PDRs, as well as a late-forming ice compared to water, since the formation of its first ice monolayer occurs at $t$$\geq$10$^5$ yr for $A$$_{\mathrm{V}}$$\leq$10 mag. Formation of the first water ice monolayer occurs, however, at $t$$\sim$10$^4$ yr for the same visual extinction range. This confirms the layered ice structure on dust grains (with H$_2$O in lower layers than CO$_2$) previously deduced by Cuppen et al. (2009) using CO, H$_2$CO, and CH$_3$OH ices. 

Time evolution results also show that, at the edge of the cloud ($A$$_{\mathrm{V}}$$\leq$1 mag), chemical steady state is reached at short times ($t$$\lesssim$10$^5$ yr) in any PDR type, being this time even shorter ($t$$<$10$^4$ yr) for high $G$$_{\mathrm{0}}$ PDRs. By contrast, for larger visual extinctions, steady state strongly depends on the PDR characteristics. In high $G$$_{\mathrm{0}}$ PDRs with densities $n$$\geq$10$^5$cm$^{-3}$, chemistry reaches equilibrium at $t$$<$10$^6$ yr, while in low $G$$_{\mathrm{0}}$ PDRs the time at which steady state is reached is $t$$>$10$^6$ yr.

The consideration of different types of PDRs in this study has also allowed us to shed light on the observed abundance enhancement of some COMs (e.g. CH$_3$CN) in PDR regions with respect to the cloud core (Gratier et al. 2013). From our theoretical results, we conclude that this enrichment is mainly due to the effect of increased density (which favours the formation of COM precursors, e.g. HCN in the case of CH$_3$CN) rather than to a direct UV radiation effect or the warming up of grain surfaces as previously suggested (Guzm\'an et al. 2014, Le Gal et al. 2017). 

The results presented here, together with those from Esplugues et al. (2016), show how sensitive the chemistry of PDRs is to the variation of their physical and chemical properties (density and intensity of radiation field), to the dust properties (dust temperature, type of grain substrate, efficiency of desorption), and to the evolutionary stage at each visual extinction. Our results also demonstrate how strongly coupled all these parameters are and the need of detailed observations of the physical and chemical structure of PDRs in order to put constraints on the chemical processes. This, together with the further exploration of the impact of other mechanisms associated to the PDR dynamics (e.g. turbulent diffusion or advection), as well as the implementation of more complex dust temperature treatments (e.g. mantle growth effects), will allow us to better reproduce observations and to gain a deeper understanding of the chemical evolution of photodissociation regions.

\section*{Acknowledgments}

The authors would like to thank the anonymous referee for valuable suggestions and comments. This work is supported by the European Research Council (ERC; project PALs 320620) and by the Netherlands Organisation for Scientific Research (NWO). S.C. is supported by the Netherlands Organization for Scientific Research (NWO; VIDI project 639.042.017) and by the European Research Council (ERC; project PALs 320620). P.C. and M.S. acknowledge the financial support of the European Research Council (ERC; project PALs 320620).

{}

\begin{appendix}
\section{Tables}
\label{Tables}

\noindent This Appendix lists (Tables \ref{table:adsorption_reactions}-\ref{table:cosmic_ray_reactions}) all the surface reactions considered in the code. The new reactions included in this new version of the Meijerink PDR code are in italic. Binding energies are also shown (Table \ref{table:binding_energies}) for each species depending on the type of the grain substrate (bare or icy surface). 

\clearpage

\begin{table}
\caption{Adsorption reactions.}             
\centering 
\begin{tabular}{l l l}     
\hline\hline       
Reaction$^{(a)}$ &   & Reaction          \\ 
\hline 
H $\rightarrow$   JH         & \vline & CO$_2$ $\rightarrow$ JCO$_2$       \\ 
H $\rightarrow$  JH$_c$      &\vline &  HCO $\rightarrow$ JHCO         \\
H$_2$ $\rightarrow$  JH$_2$  &\vline &  H$_2$CO $\rightarrow$ JH$_2$CO           \\
O $\rightarrow$   JO         &\vline &  CH$_3$O $\rightarrow$ JCH$_3$O        \\
O$_2$ $\rightarrow$  JO$_2$  &\vline &  CH$_3$OH $\rightarrow$ JCH$_3$OH            \\
O$_3$ $\rightarrow$  JO$_3$  &\vline &  N $\rightarrow$  JN            \\
OH $\rightarrow$  JOH        &\vline &  N$_2$ $\rightarrow$  JN$_2$          \\
H$_2$O $\rightarrow$ JH$_2$O &\vline &  {\it{S $\rightarrow$  JS}}         \\
HO$_2$ $\rightarrow$ JHO$_2$ &  \vline  & {\it{CH $\rightarrow$ JCH}}  \\ 

H$_2$O$_2$ $\rightarrow$ JH$_2$O$_2$ & \vline & {\it{CH$_2$ $\rightarrow$ JCH$_2$}}      \\ 
{\it{C $\rightarrow$  JC}}          &\vline & {\it{CH$_3$ $\rightarrow$ JCH$_3$}}         \\
CO $\rightarrow$  JCO        &\vline & {\it{CH$_4$ $\rightarrow$ JCH$_4$}}             \\
\hline 
\label{table:adsorption_reactions}               
\end{tabular}

\medskip
$^{(a)}$ The expression J$i$ means solid $i$. The new reactions included in this version of the PDR code are in italic. 
\end{table}

\begin{table}
\caption{Desorption reactions.}             
\centering 
\begin{tabular}{l l l}     
\hline\hline       
Reaction$^{(a)}$ & & Reactions           \\ 
\hline 
JH          $\rightarrow$  H       &  \vline  & JCO$_2$     $\rightarrow$  CO$_2$   \\
JH$_c$          $\rightarrow$  H$_c$       &  \vline  & JHCO        $\rightarrow$  HCO \\
JH$_2$      $\rightarrow$  H$_2$     &  \vline  &  JH$_2$CO    $\rightarrow$  H$_2$CO   \\
JO          $\rightarrow$  O   &  \vline  &  JCH$_3$O    $\rightarrow$  CH$_3$O  \\
JO$_2$      $\rightarrow$  O$_2$        &  \vline  & JCH$_3$OH    $\rightarrow$  CH$_3$OH  \\
JO$_3$      $\rightarrow$  O$_3$  &  \vline  & JN      $\rightarrow$  N \\
JOH         $\rightarrow$  OH   &  \vline  & JN$_2$      $\rightarrow$  N$_2$\\
JH$_2$O     $\rightarrow$  H$_2$O       &  \vline  &  {\it{JS      $\rightarrow$  S}} \\ 
JHO$_2$     $\rightarrow$  OH + O &  \vline &  {\it{JCH    $\rightarrow$ CH}}\\
JH$_2$O$_2$ $\rightarrow$  H$_2$O$_2$  &  \vline &  {\it{JCH$_2$    $\rightarrow$  CH$_2$}}  \\
{\it{JC $\rightarrow$  C}}              &  \vline &  {\it{JCH$_3$    $\rightarrow$  CH$_3$}} \\
JCO         $\rightarrow$  CO    &  \vline &  {\it{JCH$_4$    $\rightarrow$  CH$_4$}}   \\\hline 
\label{table:desorption_reactions}               
\end{tabular} 

\medskip
$^{(a)}$ The expression J$i$ means solid $i$. The new reactions included in this version of the PDR code are in italic. 
\end{table}

\begin{table}
\caption{Binding energies for the bare grain ($E$$_{\mathrm{b}}$) and water ice ($E$$_{\mathrm{i}}$) substrates.}             
\centering 
\begin{tabular}{l l l l}     
\hline\hline       
Species & $E$$_{\mathrm{b}}$ (K) & $E$$_{\mathrm{i}}$ (K) & References          \\ 
\hline 
   H            & 500   & 650     &  (1), (2)      \\
   H$_c$        & 10000 & 10000   &  (3)               \\
   H$_2$        & 300   & 500     &  (4), (5)               \\
   O            & 1500  & 1420    &  (1), (6)                \\
   O$_2$        & 1250  & 1160    &  (7)        \\
   O$_3$        & 2200  & 2200    &  (8)            \\
   OH           & 4600  & 4600    &  (9)                 \\
   H$_2$O       & 4800  & 5700    &  (10), (11)           \\
   HO$_2$       & 4000  & 4000    &  (9)           \\
   H$_2$O$_2$   & 6000  & 6000    &  (9)          \\ 
   CO           & 1200  & 1300    &  (12), (7) \\
   CO$_2$       & 3000  & 2670    &  (7), (13)               \\
   HCO          & 1600  & 1600    &  (14)       \\
   H$_2$CO      & 3700  & 3250    &  (15)\\
   CH$_3$O      & 3700  & 3700    &  (16) \\
   CH$_3$OH     & 3700  & 3700    &  (16)\\
   N            & 720   & 720     &  (17)                  \\  
   N$_2$        & 790   & 1140    &  (18)             \\
   S            & 1100  & 1100    &  (19)             \\
   C            & 800   & 800     &  (20) \\
   CH           & 870   & 870     &  (21)            \\
   CH$_2$       & 945   & 945     &  (21)            \\
   CH$_3$       & 1017  & 1017    &  (21)             \\
   CH$_4$       & 1090  & 1090    &  (22)             \\
\\\hline 
\label{table:binding_energies}               
\end{tabular}

\medskip
(1) Bergeron et al. (2008); (2) Al-Halabi \& van Dishoeck (2007); (3) Cazaux \& Tielens (2004); (4) Pirronello et al. (1997); (5) Amiaud et al. (2006); (6) Minissale (2014); (7) Noble et al. (2012a); (8) Borget et al. (2001); (9) Dulieu et al. (2013); (10) Sandford \& Allamandola (1988); (11) Speedy et al. (1996); (12) Collings et al. (2003); (13) Karssemeijer et al. (2014); (14) Garrod \& Herbst (2006); (15) Noble et al. (2012b); (16) Collings et al. (2004); (17) Minissale et al. (2016); (18) Fuchs et al. (2006); (19) Aikawa (1997); (20) Tielens \& Allamandola (1987); (21) Taquet et al. (2014); (22) Herrero et al. (2010).
\end{table}

\begin{table*}
\caption{Reactions on grain surfaces.}
\label{table:two-body-reactions}
\begin{tabular}{lrlr}
\hline
Reaction & $\delta$$_{\mathrm{bare}}$$^{(a)}$ & $\delta$$_{\mathrm{ice}}$$^{(a)}$ & $\epsilon$$^{(b)}$ \\
\hline
JH + JH           $\rightarrow$  H$_2$            & 9.630E-1 & 9.640E-2 & 0\\
JH + JH           $\rightarrow$  JH$_2$         & 3.700E-2 & 9.036E-1 & 0\\
JH + JO           $\rightarrow$  OH               & 3.875E-1 & 3.880E-2 & 0\\
JH + JO           $\rightarrow$  JOH            & 6.125E-1 & 9.612E-1 & 0\\
JH + JOH          $\rightarrow$  H$_2$O           & 2.677E-1 & 2.680E-2 & 0\\
JH + JOH          $\rightarrow$  JH$_2$O        & 7.323E-1 & 9.732E-1 & 0 \\
{\it{JH + JHO$_2$      $\rightarrow$  H$_2$O$_2$}}       & 4.600E-3 & 5.000E-4 & 0\\
{\it{JH + JHO$_2$      $\rightarrow$  JH$_2$O$_2$}}    & 9.954E-1 & 9.995E-1 & 0\\
{\it{JH + JO$_2$       $\rightarrow$  HO$_2$}}           & 1.380E-2 & 1.400E-3 & 0\\
JH + JO$_2$       $\rightarrow$  JHO$_2$        & 9.862E-1 & 9.986E-1 & 0\\
JH + JCO          $\rightarrow$  HCO              & 6.700E-3 & 7.000E-4 & 2000\\
JH + JCO          $\rightarrow$  JHCO           & 9.933E-1 & 9.993E-1 & 2000\\
JH + JHCO         $\rightarrow$  H$_2$CO          & 6.610E-2 & 6.700E-3 & 0\\
JH + JHCO         $\rightarrow$  JH$_2$CO       & 9.339E-1 & 9.933E-1 & 0\\
{\it{JH + JH$_2$CO     $\rightarrow$  CH$_3$O}}          & 1.000E-4 & 1.000E-4 & 2000\\
JH + JH$_2$CO     $\rightarrow$  JCH$_3$O       & 9.999E-1 & 9.999E-1 & 2000\\
JH + JCH$_3$O     $\rightarrow$  CH$_3$OH         & 2.350E-2 & 2.400E-3 & 0 \\
JH + JCH$_3$O     $\rightarrow$  JCH$_3$OH      & 9.765E-1 & 9.976E-1 & 0\\
{\it{JH + JC           $\rightarrow$  CH}}               & 8.212E-1 & 8.220E-2 & 0\\
{\it{JH + JC           $\rightarrow$  JCH }}           & 1.788E-1 & 9.178E-1 & 0\\
{\it{JH + JCH          $\rightarrow$  CH$_2$}}           & 7.668E-1 & 7.670E-2 & 0\\
{\it{JH + JCH          $\rightarrow$  JCH$_2$}}        & 2.332E-1 & 9.233E-1 & 0\\
{\it{JH + JCH$_2$      $\rightarrow$  CH$_3$ }}          & 6.937E-1 & 6.940E-2 & 0\\
{\it{JH + JCH$_2$      $\rightarrow$  JCH$_3$}}        & 3.063E-1 & 9.306E-1 & 0\\
{\it{JH + JCH$_3$      $\rightarrow$  CH$_4$ }}          & 5.886E-1 & 5.890E-2 & 0\\
{\it{JH + JCH$_3$      $\rightarrow$  JCH$_4$}}        & 4.114E-1 & 9.411E-1 & 0\\
JO + JO           $\rightarrow$  O$_2$            & 6.884E-1 & 6.890E-2 & 0\\
JO + JO           $\rightarrow$  JO$_2$         & 3.116E-1 & 9.311E-1 & 0\\
{\it{JO + JC           $\rightarrow$  CO}}               & 8.659E-1 & 8.660E-2 & 0\\
{\it{JO + JC           $\rightarrow$  JCO}}            & 1.341E-1 & 9.134E-1 & 0\\
{\it{JO + JCO          $\rightarrow$  CO$_2$}}           & 1.403E-1 & 1.400E-2 & 650\\
{\it{JO + JCO          $\rightarrow$  JCO$_2$}}        & 8.597E-1 & 9.860E-1 & 650\\
{\it{JO + JO$_2$       $\rightarrow$  O$_3$}}            & 3.000E-4 & 1.000E-4 & 0\\
{\it{JO + JO$_2$       $\rightarrow$  JO$_3$}}         & 9.997E-1 & 9.999E-1 & 0\\
{\it{JO + JCH$_3$      $\rightarrow$  JCH$_3$O}}       & 1.000E+0 & 1.000E+0 & 0\\
{\it{JOH + JOH         $\rightarrow$  H$_2$O$_2$}}       & 2.000E-4 & 1.000E-4 & 0\\
JOH + JOH         $\rightarrow$  JH$_2$O$_2$    & 9.998E-1 & 9.999E-1 & 0\\
{\it{JOH + JCH$_2$     $\rightarrow$  JCH$_3$O}}       & 1.000E+0 & 1.000E+0 & 0\\
JN + JN           $\rightarrow$  N$_2$            & 8.977E-1 & 8.980E-2 & 0\\
JN + JN           $\rightarrow$  JN$_2$         & 1.023E-1 & 9.102E-1 & 0\\
\hline
JH + JO$_3$       $\rightarrow$  OH + O$_2$        & 8.020E-2 & 8.100E-3 & 480\\
{\it{JH + JO$_3$       $\rightarrow$  OH + JO$_2$}}       & 0.000E+0 & 0.000E+0 & 480\\
{\it{JH + JO$_3$       $\rightarrow$  JOH + O$_2$}}       & 2.346E-1 & 2.340E-2 & 480\\
JH + JO$_3$       $\rightarrow$  JOH + JO$_2$      & 6.852E-1 & 9.685E-1 & 480\\
{\it{JH + JH$_2$O       $\rightarrow$  OH + H$_2$ }}       & 0.000E+0 & 0.000E+0 & 9600\\
{\it{JH + JH$_2$O       $\rightarrow$  OH + JH$_2$}}       & 0.000E+0 & 0.000E+0 & 9600\\
{\it{JH + JH$_2$O       $\rightarrow$  JOH + H$_2$}}       & 0.000E+0 & 0.000E+0 & 9600\\
JH + JH$_2$O       $\rightarrow$  JOH + JH$_2$      & 1.000E+0 & 1.000E+0 & 9600\\
{\it{JH + JHO$_2$      $\rightarrow$  OH + OH}}            & 3.400E-3 & 4.000E-4 & 0\\
{\it{JH + JHO$_2$      $\rightarrow$  OH + JOH}}           & 0.000E+0 & 0.000E+0 & 0\\
{\it{JH + JHO$_2$      $\rightarrow$  JOH + OH}}           & 0.000E+0 & 0.000E+0 & 0\\
JH + JHO$_2$      $\rightarrow$  JOH + JOH          & 9.966E-1 & 9.996E-1 & 0\\
JH + JH$_2$O$_2$  $\rightarrow$  H$_2$O + OH        & 2.120E-2 & 2.100E-3 & 1000\\
{\it{JH + JH$_2$O$_2$  $\rightarrow$  JH$_2$O + OH }}      & 7.200E-3 & 8.000E-4 & 1000\\
{\it{JH + JH$_2$O$_2$  $\rightarrow$  H$_2$O + JOH }}      & 0.000E+0 & 0.000E+0 & 1000\\
JH + JH$_2$O$_2$  $\rightarrow$  JH$_2$O + JOH      & 9.716E-1 & 9.971E-1 & 1000\\
JH + JHCO         $\rightarrow$  CO + H$_2$         & 4.347E-1 & 4.360E-2 & 0\\
{\it{JH + JHCO         $\rightarrow$  CO + JH$_2$}}        & 0.000E+0 & 0.000E+0 & 0\\
{\it{JH + JHCO         $\rightarrow$  JCO + H$_2$}}        & 4.827E-1 & 4.820E-2 & 0\\
JH + JHCO         $\rightarrow$  JCO + JH$_2$       & 8.260E-2 & 9.082E-1 & 0\\
{\it{JH + JH$_2$CO      $\rightarrow$  HCO + H$_2$}}     & 2.000E-4 & 1.000E-4 & 2200\\
{\it{JH + JH$_2$CO      $\rightarrow$  HCO + JH$_2$}}     & 0.000E+0 & 0.000E+0 & 2200\\
{\it{JH + JH$_2$CO      $\rightarrow$  JHCO + H$_2$}}     & 5.050E-1 & 5.050E-2 & 2200\\
JH + JH$_2$CO      $\rightarrow$  JHCO + JH$_2$     & 4.948E-1 & 9.494E-1 & 2200\\
{\it{JH + JCH$_3$O      $\rightarrow$  H$_2$CO + H$_2$}}   & 1.260E-2 & 1.400E-3 & 150\\
{\it{JH + JCH$_3$O      $\rightarrow$  H$_2$CO + JH$_2$}}  & 0.000E+0 & 0.000E+0 & 150\\
{\it{JH + JCH$_3$O      $\rightarrow$  JH$_2$CO + H$_2$}}  & 8.596E-1 & 8.590E-2 & 150\\
\hline
\end{tabular}
\end{table*}

\begin{table*}
\contcaption{}
\label{table:two-body-reactions}
\begin{tabular}{lrlr}
\hline
Reaction & $\delta$$_{\mathrm{bare}}$$^{(a)}$ & $\delta$$_{\mathrm{ice}}$$^{(a)}$ & $\epsilon$$^{(b)}$  \\
\hline
JH + JCH$_3$O      $\rightarrow$  JH$_2$CO + JH$_2$ & 1.278E-1 & 9.127E-1 & 150\\
{\it{JH + JH$_4$CO      $\rightarrow$  CH$_3$O + H$_2$}}   & 0.000E+0 & 0.000E+0 & 3200\\
{\it{JH + JH$_4$CO      $\rightarrow$  CH$_3$O + JH$_2$}}  & 0.000E+0 & 0.000E+0 & 3200\\
{\it{JH + JH$_4$CO      $\rightarrow$  JCH$_3$O + H$_2$}}  & 0.000E+0 & 0.000E+0 & 3200\\
JH + JH$_4$CO      $\rightarrow$  JCH$_3$O + JH$_2$ & 1.000E+0 & 1.000E+0 & 3200\\
{\it{JH + JCO$_2$      $\rightarrow$  CO + OH}}            & 0.000E+0 & 0.000E+0 & 10000\\
{\it{JH + JCO$_2$      $\rightarrow$  CO + JOH}}           & 0.000E+0 & 0.000E+0 & 10000\\
{\it{JH + JCO$_2$      $\rightarrow$  JCO + OH}}           & 0.000E+0 & 0.000E+0 & 10000\\
JH + JCO$_2$      $\rightarrow$  JCO + JOH          & 1.000E+0 & 1.000E+0 & 10000\\
{\it{JH + JCH            $\rightarrow$  C + H$_2$}}        & 3.915E-1 & 3.915E-2 & 0\\
{\it{JH + JCH            $\rightarrow$  C + JH$_2$}}       & 0.000E+0 & 0.000E+0 & 0\\
{\it{JH + JCH            $\rightarrow$  JC + H$_2$}}       & 3.859E-1 & 3.865E-2 & 0\\
{\it{JH + JCH            $\rightarrow$  JC + JH$_2$}}      & 2.226E-1 & 9.222E-1 & 0\\
{\it{JH + JCH$_2$        $\rightarrow$  CH + H$_2$}}       & 8.000E-5 & 4.800E-5 & 0\\
{\it{JH + JCH$_2$        $\rightarrow$  CH + JH$_2$}}      & 0.000E+0 & 0.000E+0 & 0\\
{\it{JH + JCH$_2$        $\rightarrow$  JCH + H$_2$}}      & 4.452E-2 & 4.452E-3 & 0\\
{\it{JH + JCH$_2$        $\rightarrow$  JCH + JH$_2$}}     & 9.554E-1 & 9.955E-1 & 0\\
{\it{JH + JCH$_3$        $\rightarrow$  CH$_2$ + H$_2$}}   & 0.000E+0 & 0.000E+0 & 0\\
{\it{JH + JCH$_3$        $\rightarrow$  CH$_2$ + JH$_2$}}  & 0.000E+0 & 0.000E+0 & 0\\
{\it{JH + JCH$_3$        $\rightarrow$  JCH$_2$ + H$_2$}}  & 0.000E+0 & 0.000E+0 & 0\\
{\it{JH + JCH$_3$        $\rightarrow$  JCH$_2$ + JH$_2$}} & 1.000E+0 & 1.000E+0 & 0\\
{\it{JH + JCH$_4$        $\rightarrow$  CH$_3$ + H$_2$}}   & 0.000E+0 & 0.000E+0 & 0\\
{\it{JH + JCH$_4$        $\rightarrow$  CH$_3$ + JH$_2$}}  & 0.000E+0 & 0.000E+0 & 0\\
{\it{JH + JCH$_4$        $\rightarrow$  JCH$_3$ + H$_2$}}  & 0.000E+0 & 0.000E+0 & 0\\
{\it{JH + JCH$_4$        $\rightarrow$  JCH$_3$ + JH$_2$}} & 1.000E+0 & 1.000E+0 & 0\\
{\it{JO + JO$_3$        $\rightarrow$  O$_2$ + O$_2$}}     & 3.872E-1 & 3.880E-2 & 2500\\
{\it{JO + JO$_3$        $\rightarrow$  O$_2$ + JO$_2$}}    & 0.000E+0 & 0.000E+0 & 2500\\
{\it{JO + JO$_3$        $\rightarrow$  JO$_2$ + O$_2$ }}   & 0.000E+0 & 0.000E+0 & 2500\\
JO + JO$_3$        $\rightarrow$  JO$_2$ + JO$_2$   & 6.128E-1 & 9.612E-1 & 2500\\
{\it{JO + JOH           $\rightarrow$  O$_2$ + H}}         & 1.890E-2 & 2.000E-3 & 0\\
{\it{JO + JOH           $\rightarrow$  O$_2$ + JH}}        & 0.000E+0 & 0.000E+0 & 0\\
{\it{JO + JOH           $\rightarrow$  JO$_2$ + H}}        & 5.264E-1 & 5.260E-2 & 0\\
JO + JOH           $\rightarrow$  JO$_2$ + JH       & 4.547E-1 & 9.454E-1 & 0\\
{\it{JO + JHO$_2$       $\rightarrow$  O$_2$ + OH}}        & 2.190E-2 & 2.300E-3 & 0\\
{\it{JO + JHO$_2$       $\rightarrow$  O$_2$ + JOH }}      & 1.516E-1 & 1.510E-2 & 0\\
{\it{JO + JHO$_2$       $\rightarrow$  JO$_2$ + OH}}       & 0.000E+0 & 0.000E+0 & 0\\
{\it{JO + JHO$_2$       $\rightarrow$  JO$_2$ + JOH}}      & 8.265E-1 & 9.826E-1 & 0\\
{\it{JO + JHCO      $\rightarrow$  CO$_2$ + H}}            & 5.110E-2 & 5.200E-3 & 0\\
{\it{JO + JHCO      $\rightarrow$  CO$_2$ + JH}}           & 0.000E+0 & 0.000E+0 & 0\\
{\it{JO + JHCO      $\rightarrow$  JCO$_2$ + H}}           & 8.348E-1 & 8.340E-2 & 0\\
JO + JHCO      $\rightarrow$  JCO$_2$ + JH          & 1.141E-1 & 9.114E-1 & 0\\
{\it{JO + JH$_2$CO      $\rightarrow$  CO$_2$ + H$_2$}}    & 3.680E-2 & 3.700E-3 & 335\\
{\it{JO + JH$_2$CO      $\rightarrow$  CO$_2$ + JH$_2$}}   & 0.000E+0 & 0.000E+0 & 335\\
{\it{JO + JH$_2$CO      $\rightarrow$  JCO$_2$ + H$_2$}}   & 8.901E-1 & 8.900E-2 & 335\\
JO + JH$_2$CO      $\rightarrow$  JCO$_2$ + JH$_2$  & 7.310E-2 & 9.073E-1 & 335\\
{\it{JOH + JH$_2$       $\rightarrow$  H$_2$O + H}}        & 0.000E+0 & 0.000E+0 & 2100\\
{\it{JOH + JH$_2$       $\rightarrow$  H$_2$O + JH}}       & 0.000E+0 & 0.000E+0 & 2100\\
{\it{JOH + JH$_2$       $\rightarrow$  JH$_2$O + H}}       & 4.052E-1 & 4.060E-2 & 2100\\
JOH + JH$_2$       $\rightarrow$  JH$_2$O + JH      & 5.948E-1 & 9.594E-1 & 2100\\
{\it{JOH + JCO          $\rightarrow$  CO$_2$ + H}}        & 1.000E-4 & 6.000E-5 & 400\\
{\it{JOH + JCO          $\rightarrow$  CO$_2$ + JH}}       & 0.000E+0 & 0.000E+0 & 400\\
{\it{JOH + JCO          $\rightarrow$  JCO$_2$ + H}}       & 5.794E-1 & 5.794E-2 & 400\\
JOH + JCO          $\rightarrow$  JCO$_2$ + JH      & 4.205E-1 & 9.420E-1 & 400\\
{\it{JOH + JHCO          $\rightarrow$  CO$_2$ + H$_2$}}   & 2.577E-2 & 2.640E-3 & 0\\
{\it{JOH + JHCO          $\rightarrow$  CO$_2$ + JH$_2$}}  & 0.000E-0 & 0.000E-0 & 0\\
{\it{JOH + JHCO          $\rightarrow$  JCO$_2$ + H$_2$}}  & 8.936E-1 & 8.936E-2 & 0\\
JOH + JHCO          $\rightarrow$  JCO$_2$ + JH$_2$ & 8.063E-2 & 9.080E-1 & 0\\
JOH + JCH$_3$OH   $\rightarrow$  JCH$_3$O + JH$_2$O & 1.000E+0 & 1.000E+0 & 5000\\
JHO$_2$ + JH$_2$    $\rightarrow$  H$_2$O$_2$ + H   & 0.000E+0 & 0.000E+0 & 5000\\
JHO$_2$ + JH$_2$    $\rightarrow$  H$_2$O$_2$ + JH  & 0.000E+0 & 0.000E+0 & 5000\\
JHO$_2$ + JH$_2$    $\rightarrow$  JH$_2$O$_2$ + H  & 0.000E+0 & 0.000E+0 & 5000\\
JHO$_2$ + JH$_2$    $\rightarrow$  JH$_2$O$_2$ + JH & 1.000E+0 & 1.000E+0 & 5000\\
\hline
\end{tabular}

\medskip
The expression J$i$ means solid $i$. $^{(a)}$ The parameters $\delta$$_{\mathrm{bare}}$ and $\delta$$_{\mathrm{ice}}$ indicate the probabilities of desorption upon reaction for bare and icy substrates, respectively. 
$^{(b)}$ The parameter $\epsilon$ indicates the activation barrier for each reaction. The new reactions included in this version of the PDR code are in italic.

\end{table*}

\begin{table}
\caption{Photoreactions on dust grains.}             
\centering 
\begin{tabular}{l l l l}     
\hline\hline       
Reactions$^{(a)}$ & $\alpha$$_{{i}}$$^{(b)}$ (s$^{-1}$) & $\xi$$_{{i}}$$^{(b)}$         \\ 
\hline 
{\it{JCO  + Photon     $\rightarrow$ JC + JO}}         & 2.59$\times$10$^{-10}$ & 3.53 \\                
JH$_2$  + Photon     $\rightarrow$ JH + JH      & 8.00$\times$10$^{-10}$ & 2.20    \\
JO$_2$  + Photon     $\rightarrow$ JO + JO      & 7.90$\times$10$^{-10}$ & 2.13 \\
JOH     + Photon     $\rightarrow$ JH + JO      & 3.90$\times$10$^{-10}$ & 2.24  \\
JCO$_2$ + Photon     $\rightarrow$ JO + JCO     & 8.90$\times$10$^{-10}$ & 3.00   \\
JH$_2$O + Photon     $\rightarrow$ JH + JOH     & 8.00$\times$10$^{-10}$ & 2.20   \\
JHCO    + Photon     $\rightarrow$ JH + JCO     & 1.10$\times$10$^{-09}$ & 1.09  \\
JH$_2$CO+ Photon     $\rightarrow$ JH + JHCO    & 5.87$\times$10$^{-10}$ & 0.53   \\
JCH$_3$O+ Photon     $\rightarrow$ JH + JH$_2$CO & 5.87$\times$10$^{-10}$ & 0.53  \\
JCH$_3$OH+ Photon    $\rightarrow$ JH + JCH$_3$O & 5.87$\times$10$^{-10}$ & 0.53     \\
JN$_2$  + Photon     $\rightarrow$ JN + JN       & 2.30$\times$10$^{-10}$  & 3.88 \\
JHO$_2$ + Photon     $\rightarrow$ JO + JOH      & 3.28$\times$10$^{-10}$  & 1.63 \\
JHO$_2$ + Photon     $\rightarrow$ JO$_2$ + JH   & 3.28$\times$10$^{-10}$  & 1.63 \\
JH$_2$O$_2$ + Photon $\rightarrow$ JOH + JOH     & 8.30$\times$10$^{-10}$  & 1.80 \\
JO$_3$      + Photon $\rightarrow$ JO$_2$ + JO   & 3.30$\times$10$^{-10}$  & 1.40 \\
\hline 
{\it{JCH      + Photon $\rightarrow$ C + H}}       & 9.20$\times$10$^{-10}$  & 1.72 \\
{\it{JCH$_2$  + Photon $\rightarrow$ H + CH}}      & 5.80$\times$10$^{-10}$  & 2.02 \\
{\it{JCH$_3$  + Photon $\rightarrow$ H + CH$_2$}}  & 1.35$\times$10$^{-10}$  & 2.27 \\
{\it{JCH$_4$  + Photon $\rightarrow$ H$_2$ + CH$_2$}}  & 7.20$\times$10$^{-10}$  & 2.59 \\
\hline 
JCO     + Photon     $\rightarrow$ CO                & 3.67$\times$10$^{-10}$  & 2.54 \\
JH$_2$O + Photon     $\rightarrow$ H$_2$O            & 3.67$\times$10$^{-11}$  & 2.20 \\
{\it{JHCO  + Photon $\rightarrow$ HCO}}               & 3.67$\times$10$^{-11}$  & 0.53 \\
JH$_2$CO+ Photon     $\rightarrow$ H$_2$CO           & 3.67$\times$10$^{-11}$  & 0.53 \\
{\it{JCH$_3$O  + Photon $\rightarrow$ CH$_3$O}}       & 3.67$\times$10$^{-11}$  & 0.53 \\
{\it{JCH$_3$OH  + Photon $\rightarrow$ CH$_3$OH}}     & 5.00$\times$10$^{-13}$  & 0.53 \\
{\it{JCH$_3$OH  + Photon $\rightarrow$ CH$_3$O + H}}  & 5.00$\times$10$^{-13}$  & 0.53 \\
\\\hline 
\label{table:photo-reactions}              
\end{tabular}

\medskip
$^{(a)}$ The expression J$i$ means solid $i$. $^{(b)}$ Values for $\alpha$$_{{i}}$ and $\xi$$_{{i}}$ (dimensionless) are taken from KIDA.  The new reactions included in this version of the PDR code are in italic.
\end{table}

\begin{table}
\caption{Cosmic-ray reactions and photo processes induced by cosmic rays.}             
\centering 
\begin{tabular}{l l l }     
\hline\hline       
Reaction$^{(a)}$  &  $\kappa$$_{\mathrm{CR}}$$^{(b)}$ (s$^{-1}$)           \\ 
\hline 
JH$_2$  + CR       $\rightarrow$  JH + JH         &  5.00$\times$10$^{-17}$  \\
JO$_2$  + CRP      $\rightarrow$  JO  + JO        &  3.75$\times$10$^{-14}$   \\
JOH     + CRP      $\rightarrow$  JH + JO         &  2.55$\times$10$^{-14}$   \\
JCO$_2$ + CRP      $\rightarrow$  JO + JCO        &  8.55$\times$10$^{-14}$  \\
JH$_2$O + CRP      $\rightarrow$  JH + JOH        &  4.85$\times$10$^{-14}$   \\ 
JHCO    + CRP      $\rightarrow$  JH + JCO        &  2.11$\times$10$^{-14}$  \\
JH$_2$CO+ CRP      $\rightarrow$  JH + JHCO       &  2.11$\times$10$^{-14}$   \\
JCH$_3$O+ CRP      $\rightarrow$  JH + JH$_2$CO   &  2.11$\times$10$^{-14}$ \\
JCH$_3$OH+ CRP     $\rightarrow$  JH + JCH$_3$O   &  2.11$\times$10$^{-14}$ \\
JN$_2$  + CRP    $\rightarrow$  JN + JN     & 2.50$\times$10$^{-16}$  \\
JHO$_2$ + CRP    $\rightarrow$  JO + JOH    & 3.75$\times$10$^{-14}$ \\
JHO$_2$ + CRP    $\rightarrow$  JH + JO$_2$ & 3.75$\times$10$^{-14}$  \\
JH$_2$O$_2$ + CRP $\rightarrow$ JOH + JOH   & 7.50$\times$10$^{-14}$ \\
JO$_3$  + CRP    $\rightarrow$  JO$_2$ + JO & 3.75$\times$10$^{-14}$ \\

\hline
JCO     + CRP    $\rightarrow$  CO              & 1.08$\times$10$^{-14}$ \\
JH$_2$O + CRP    $\rightarrow$  H$_2$O          & 1.08$\times$10$^{-14}$\\
JH$_2$CO+ CRP    $\rightarrow$  H$_2$CO         & 1.08$\times$10$^{-14}$ \\
JCH$_3$OH+ CRP    $\rightarrow$  CH$_3$OH        & 1.08$\times$10$^{-14}$  \\
{\it{JCH+ CRP           $\rightarrow$  C + H }}         &  3.65$\times$10$^{-14}$ \\
{\it{JCH$_3$+ CRP       $\rightarrow$  H + CH$_2$}}     &  2.50$\times$10$^{-14}$ \\
{\it{JCH$_4$+ CRP       $\rightarrow$  H$_2$ + CH$_2$}} &  1.17$\times$10$^{-13}$ \\
\\\hline 
\label{table:cosmic_ray_reactions}              
\end{tabular}

\medskip
$^{(a)}$ The expression J$i$ means solid $i$. $^{(b)}$ Values for the cosmic ray rate coefficient, $\kappa$$_{\mathrm{CR}}$, are taken from KIDA. The new reactions included in this version of the PDR code are in italic.
\end{table}

\clearpage

\end{appendix}

\newpage

\begin{appendix}
\onecolumn
\section{Figures}
\label{Figures}

\clearpage

\begin{figure}
\centering
\includegraphics[scale=0.33, angle=0]{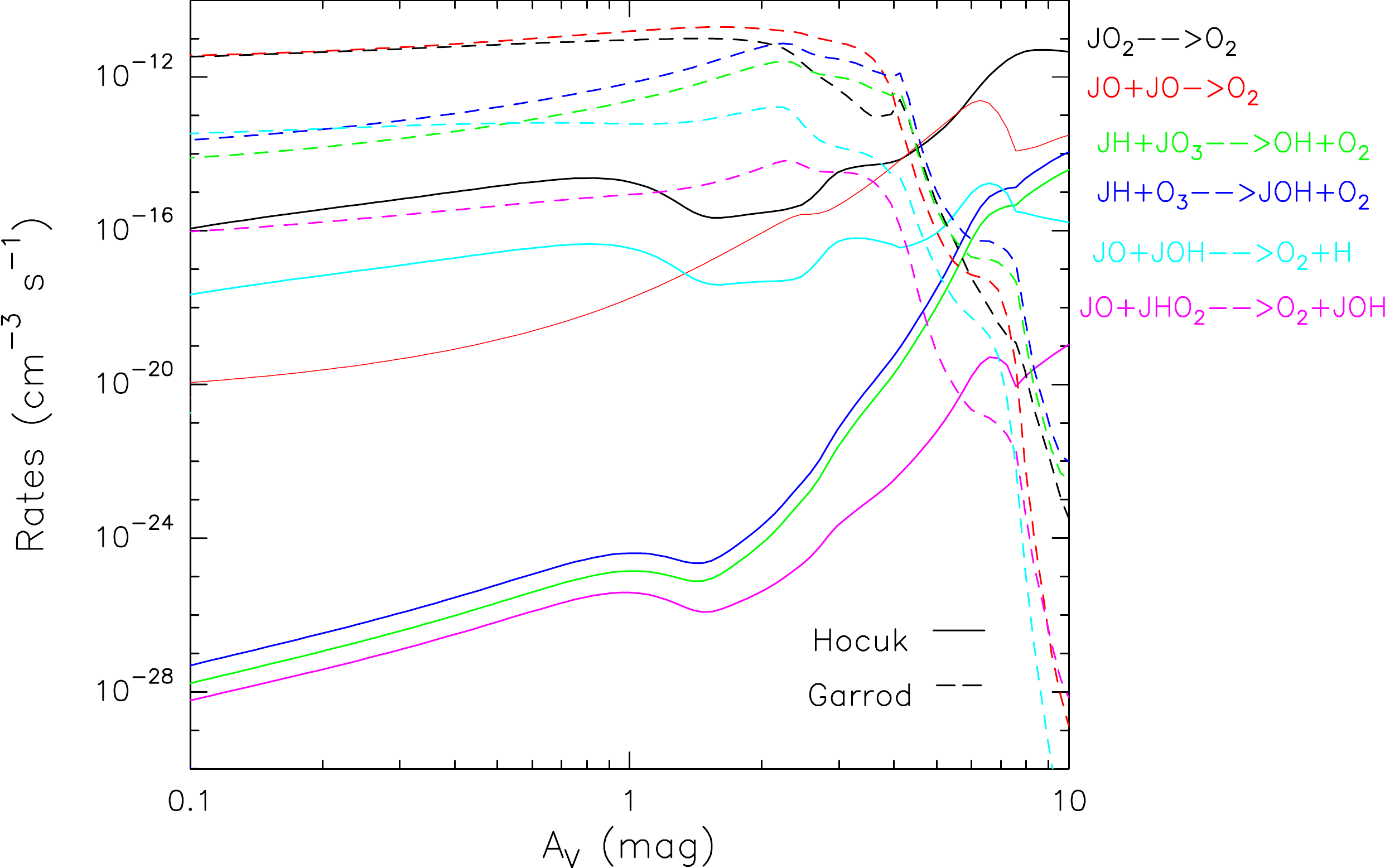}  \hspace{0.40cm}
\includegraphics[scale=0.33, angle=0]{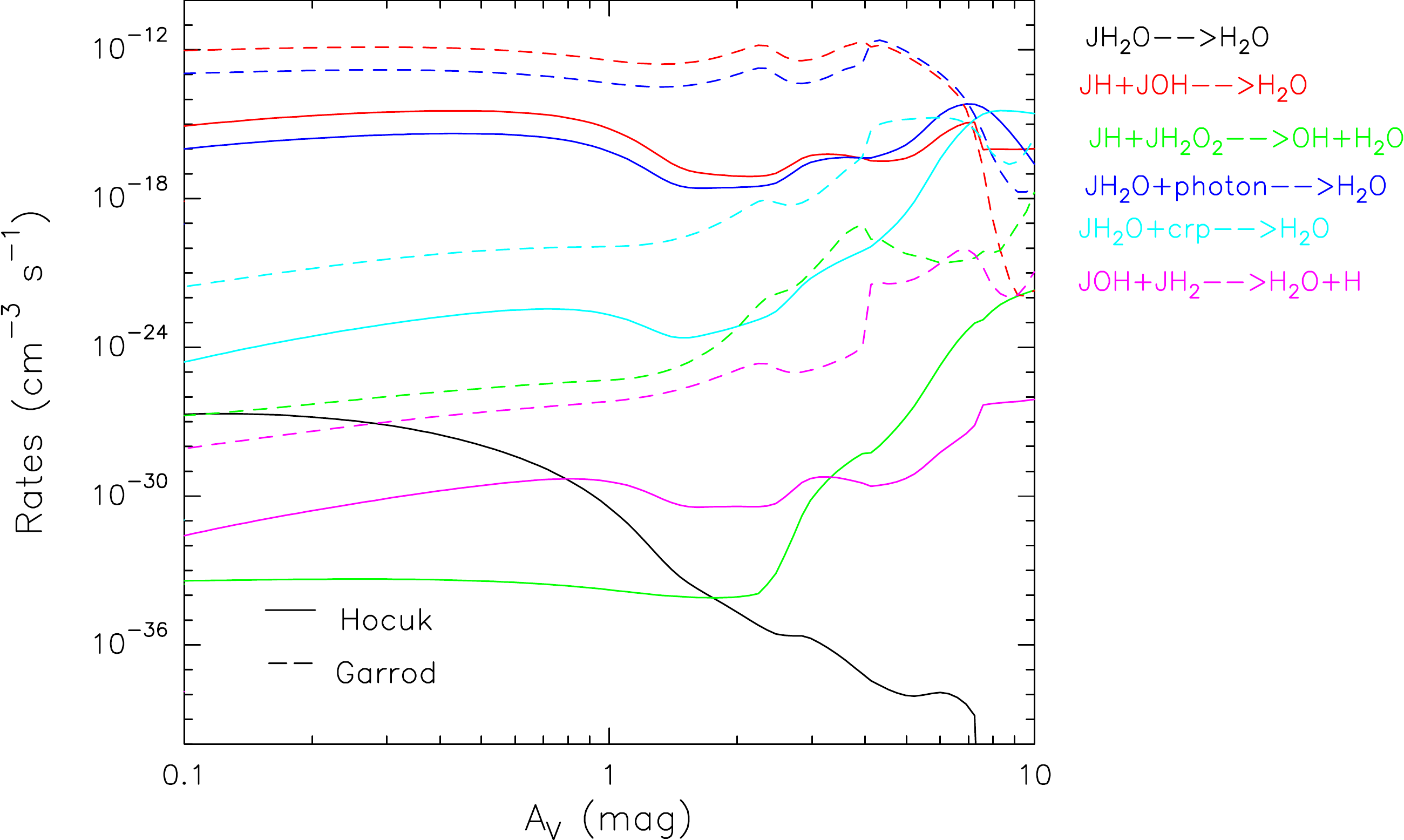}\\
\caption{Rates for surface reactions forming O$_2$ gas (left) and H$_2$O gas (right) for Model 2 ($G$$_{\mathrm{0}}$=10$^4$ and $n$=10$^5$ cm$^{-3}$) using $T$$_{\mathrm{dust}}$ from Hocuk et al. (2017) (solid lines) and from Garrod $\&$ Pauly (2011) (dashed lines). JX means solid X.}
\label{figure:O2_gas_formation}
\end{figure}

\begin{figure}
\centering
\includegraphics[scale=0.33, angle=0]{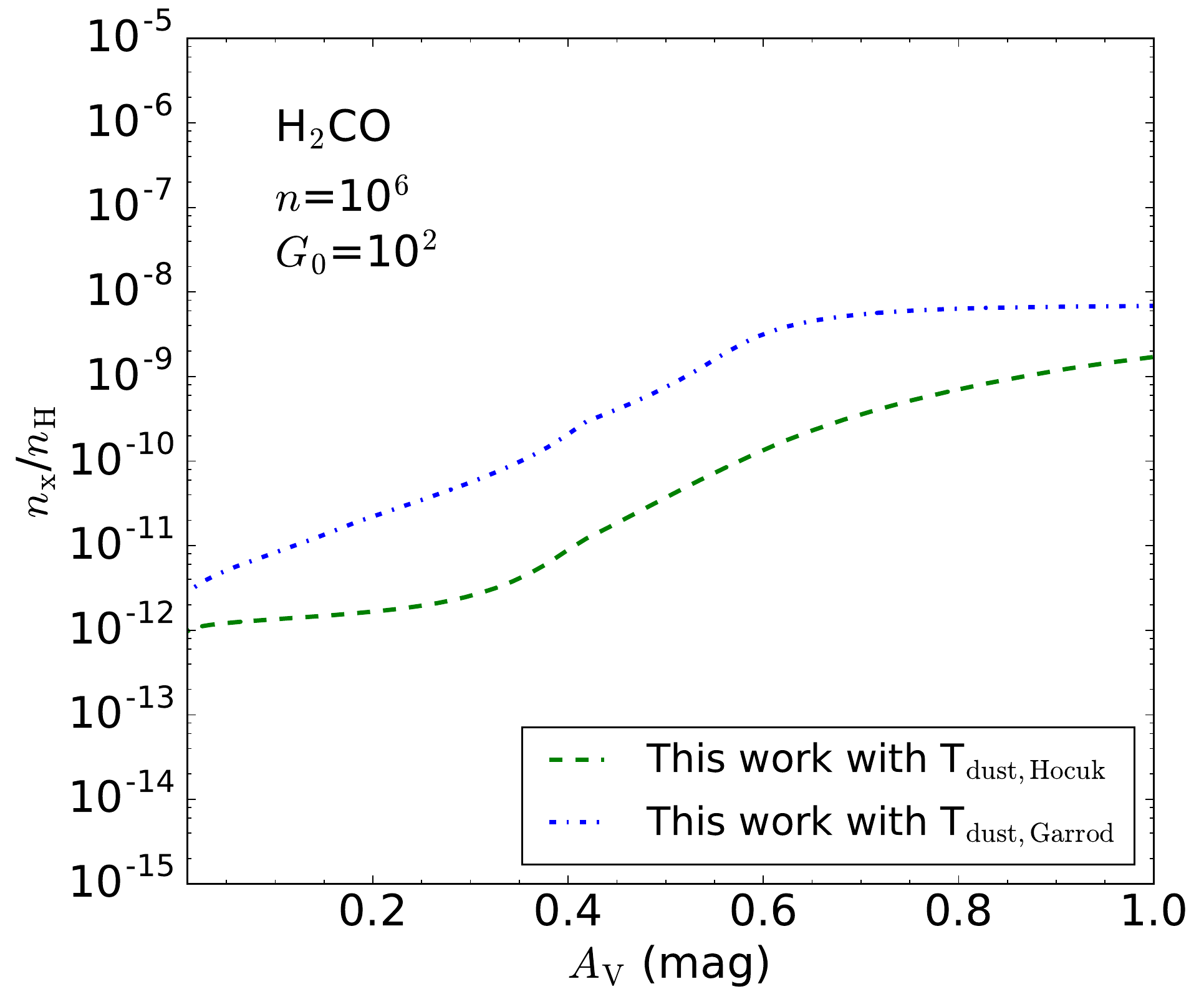} 
\includegraphics[scale=0.333, angle=0]{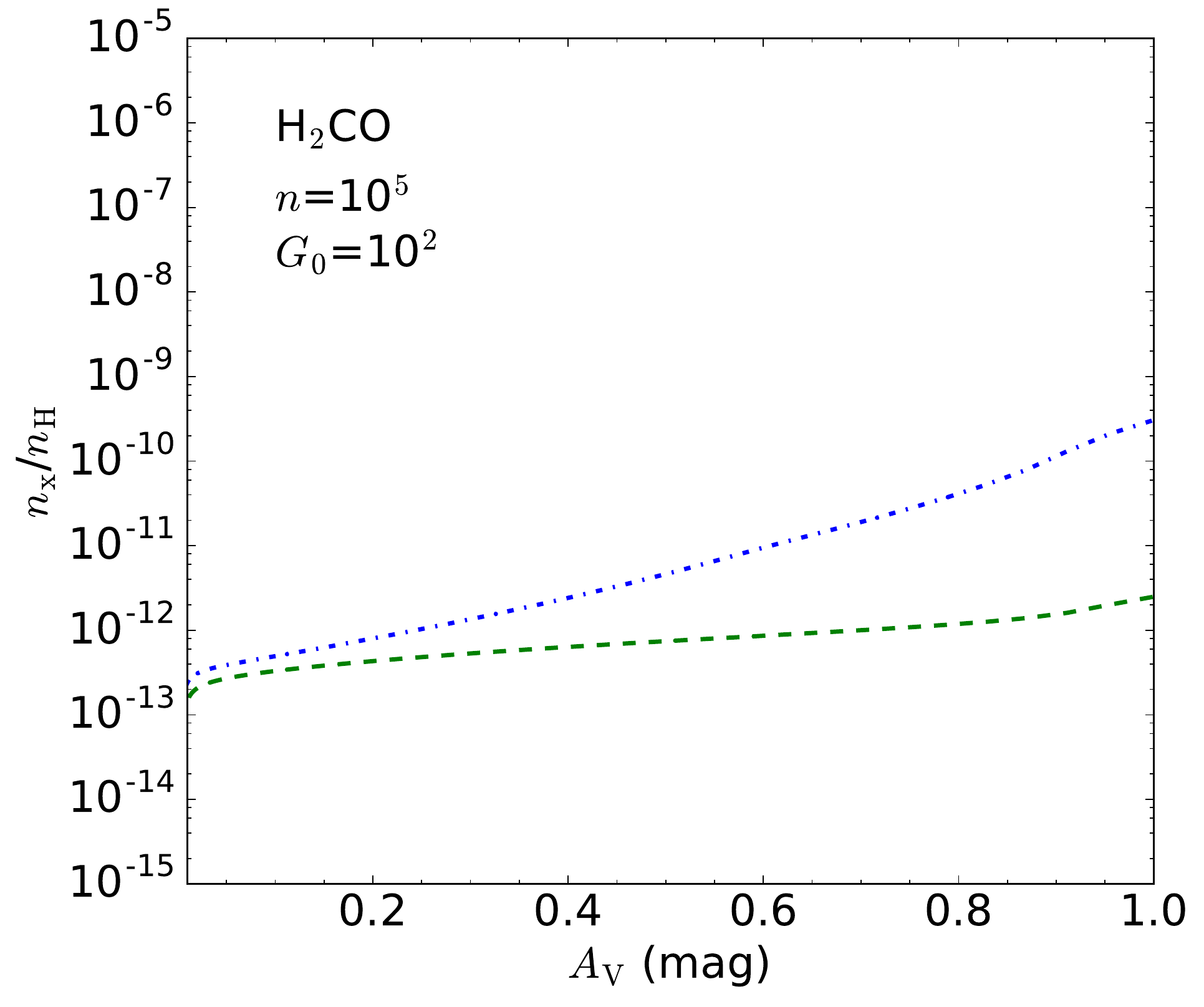}
\includegraphics[scale=0.33, angle=0]{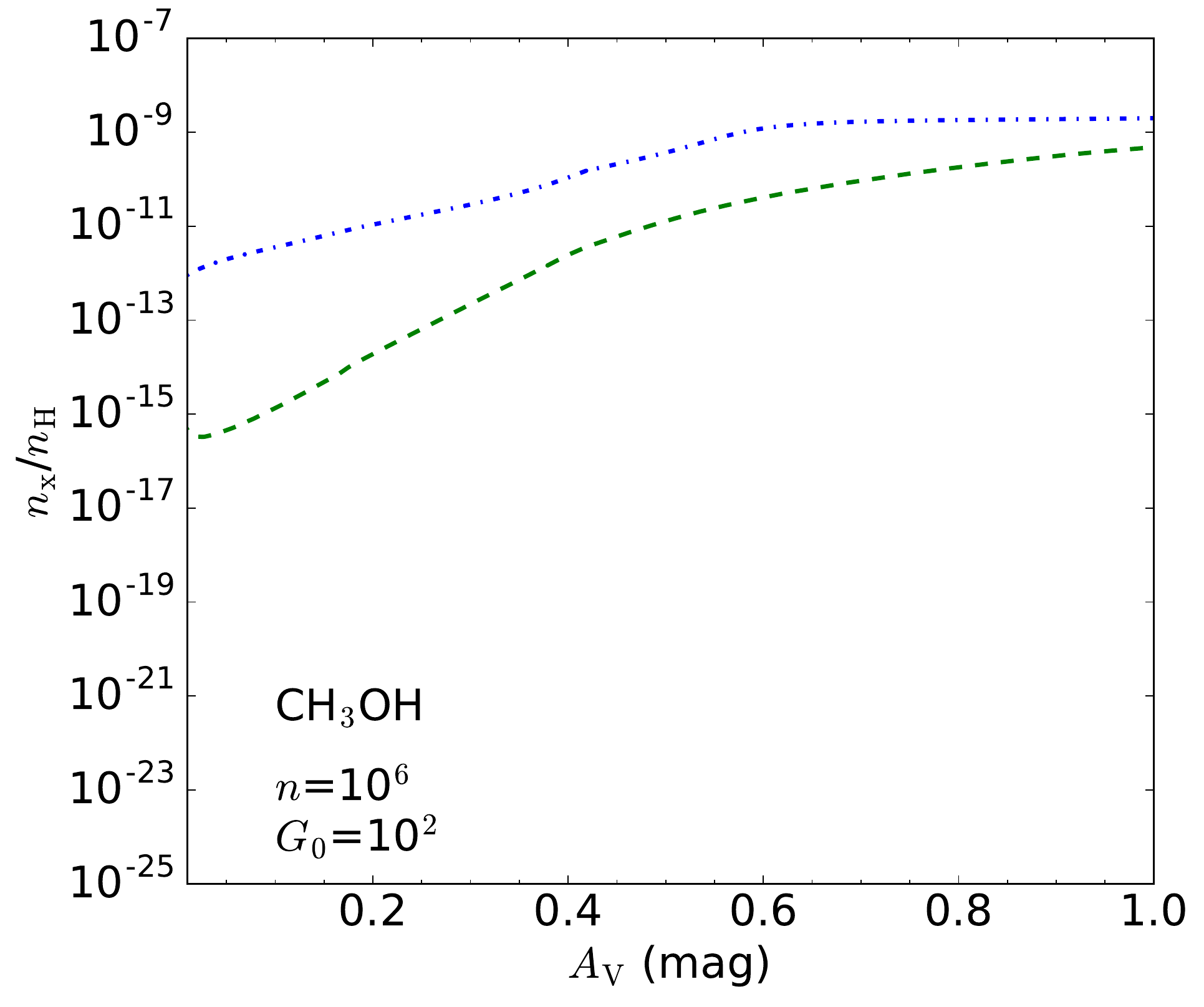} 
\includegraphics[scale=0.333, angle=0]{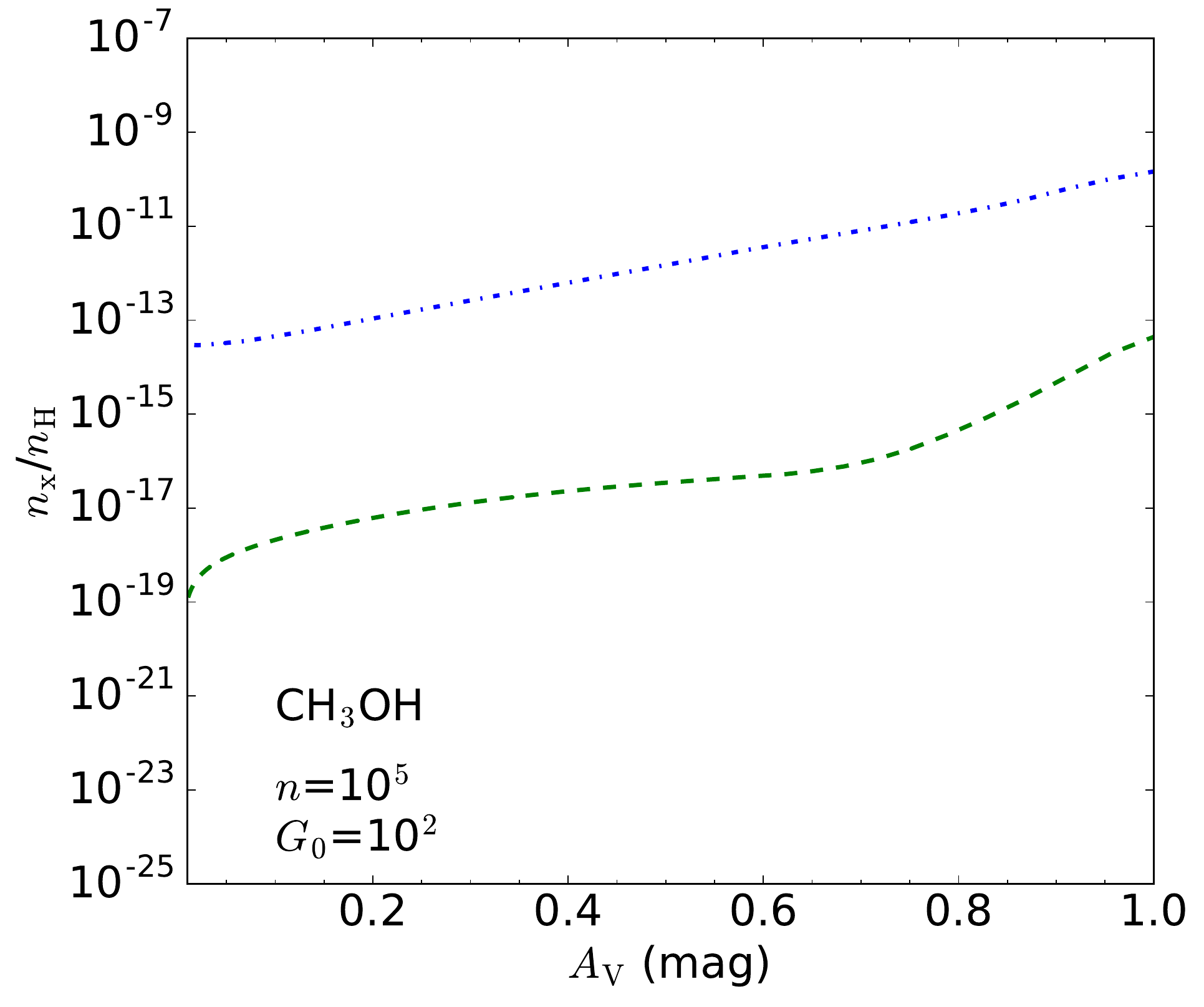}\\  
\caption{Abundances of H$_2$CO and CH$_3$OH obtained with the updated Meijerink PDR code using $T$$_{\mathrm{dust}}$ from Hocuk et al. (2017) (green dashed lines) and from Garrod $\&$ Pauly (2011) (blue dotted lines) for a model with $n$=10$^6$ cm$^{-3}$ and $G$$_{\mathrm{0}}$=10$^2$ (left column), and with $n$=10$^5$ cm$^{-3}$ and $G$$_{\mathrm{0}}$=10$^2$ (right column).}
\label{figure:density_effects_low_G0}
\end{figure}

\begin{figure}
\centering
\includegraphics[scale=0.42, angle=0]{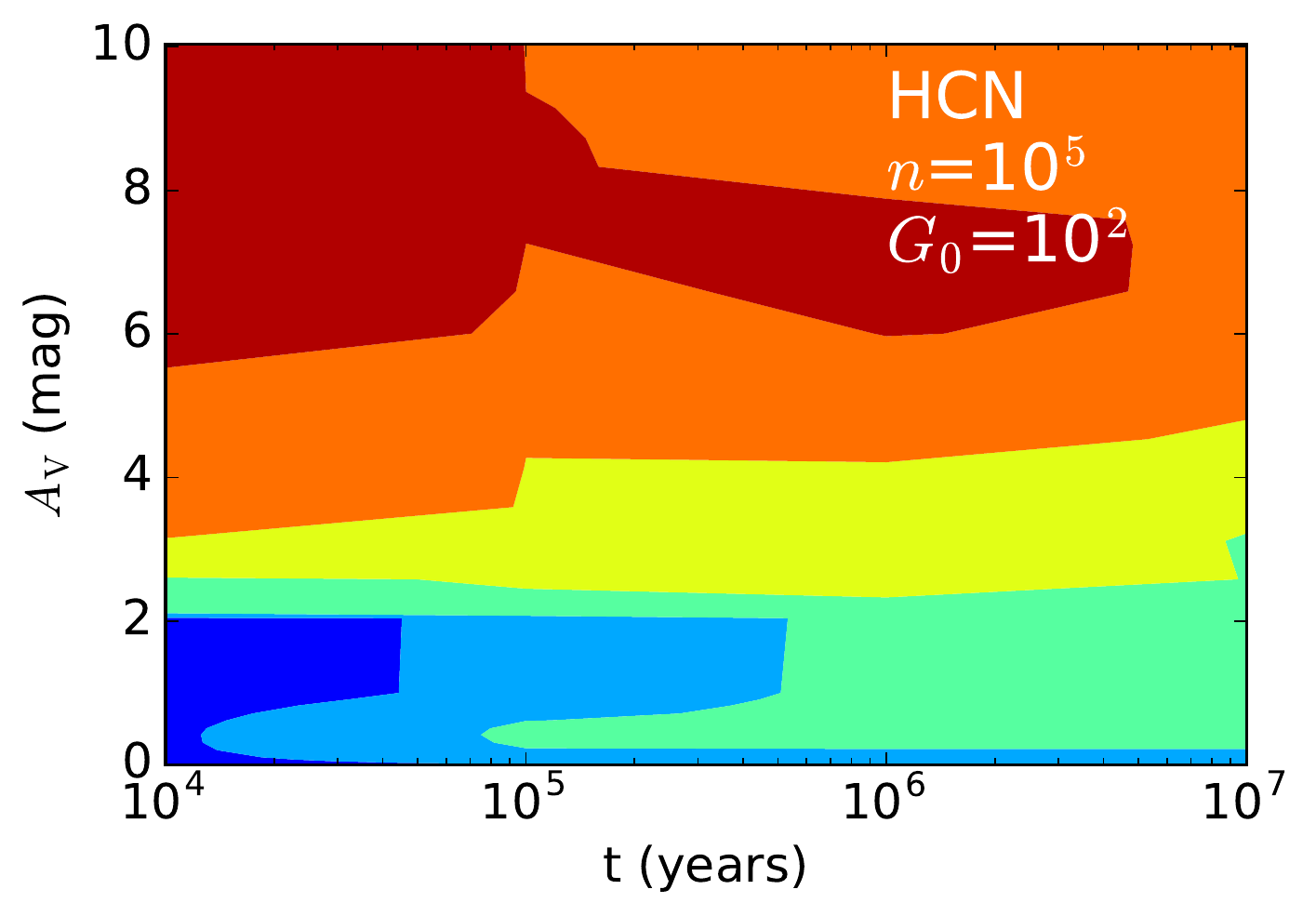}  \hspace{0.00cm}
\includegraphics[scale=0.42, angle=0]{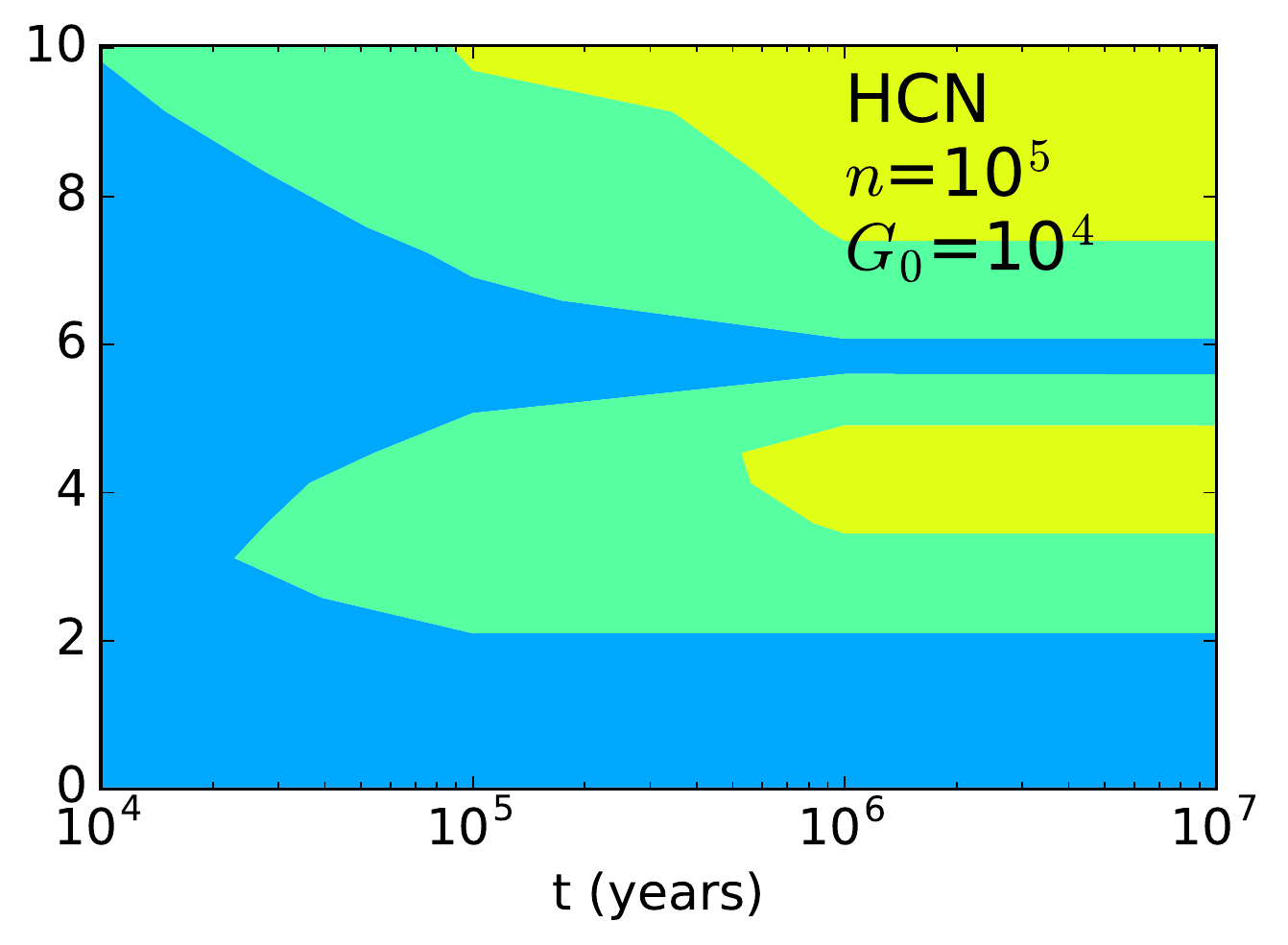}  \hspace{0.00cm}
\includegraphics[scale=0.42, angle=0]{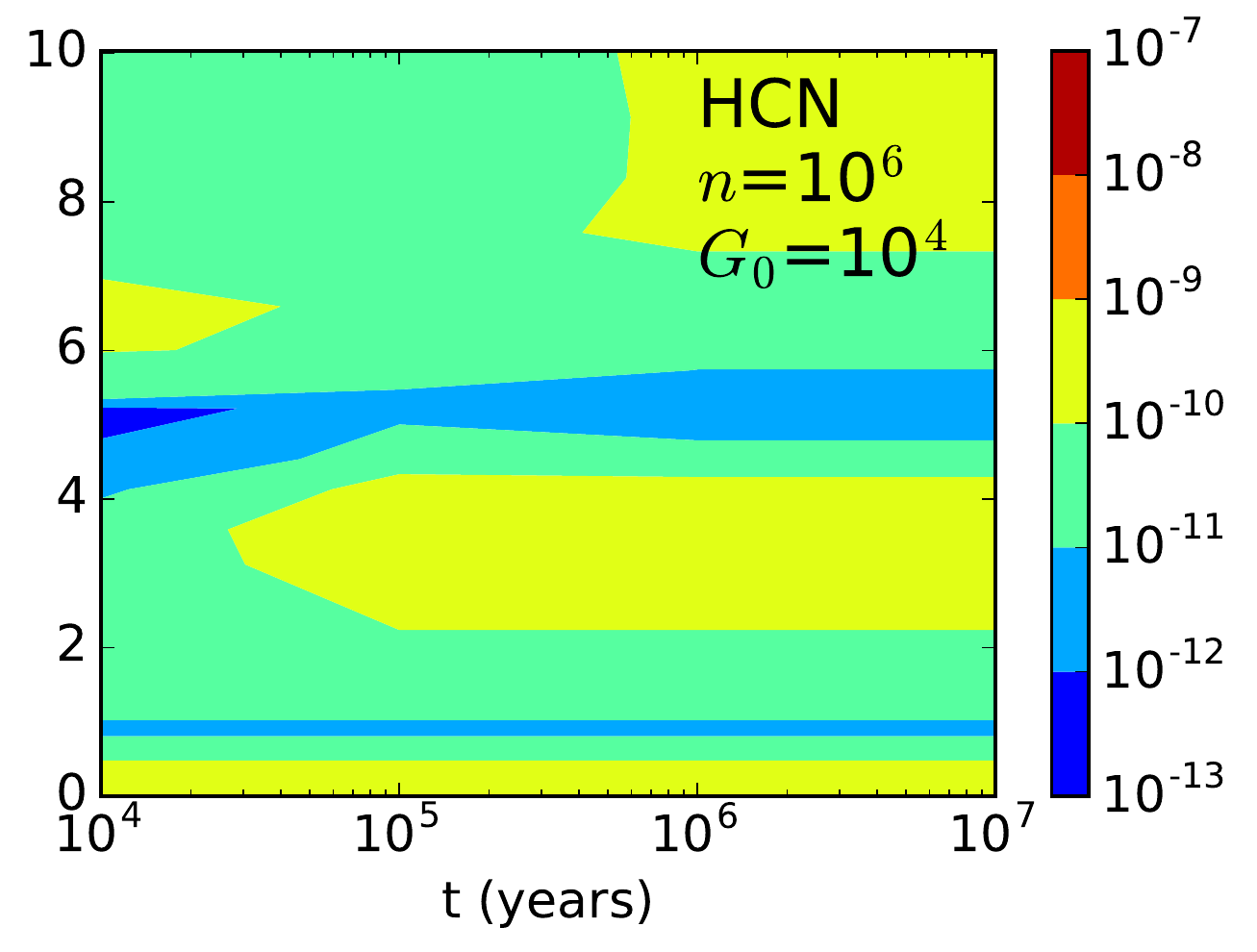}  \hspace{0.00cm}\\
\caption{Contour maps with the abundances of HCN for Models 1 (left panel), 2 (middle panel), and 3 (right panel) as a function of time (x-axis) and visual extinction (y-axis).}
\label{figure:contours5}
\end{figure}

\begin{figure}
\centering
\includegraphics[scale=0.285, angle=0]{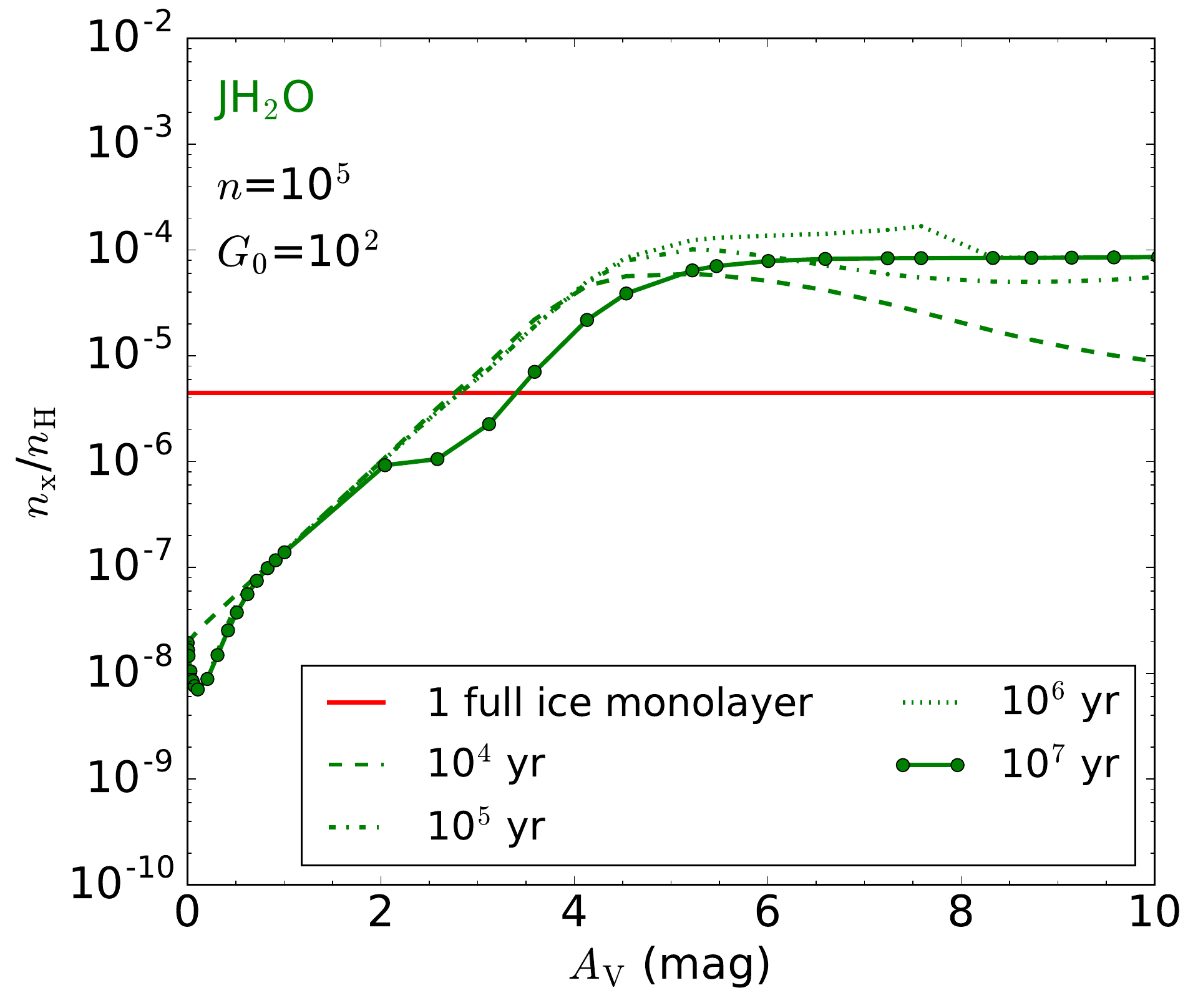}  \hspace{0.00cm}
\includegraphics[scale=0.285, angle=0]{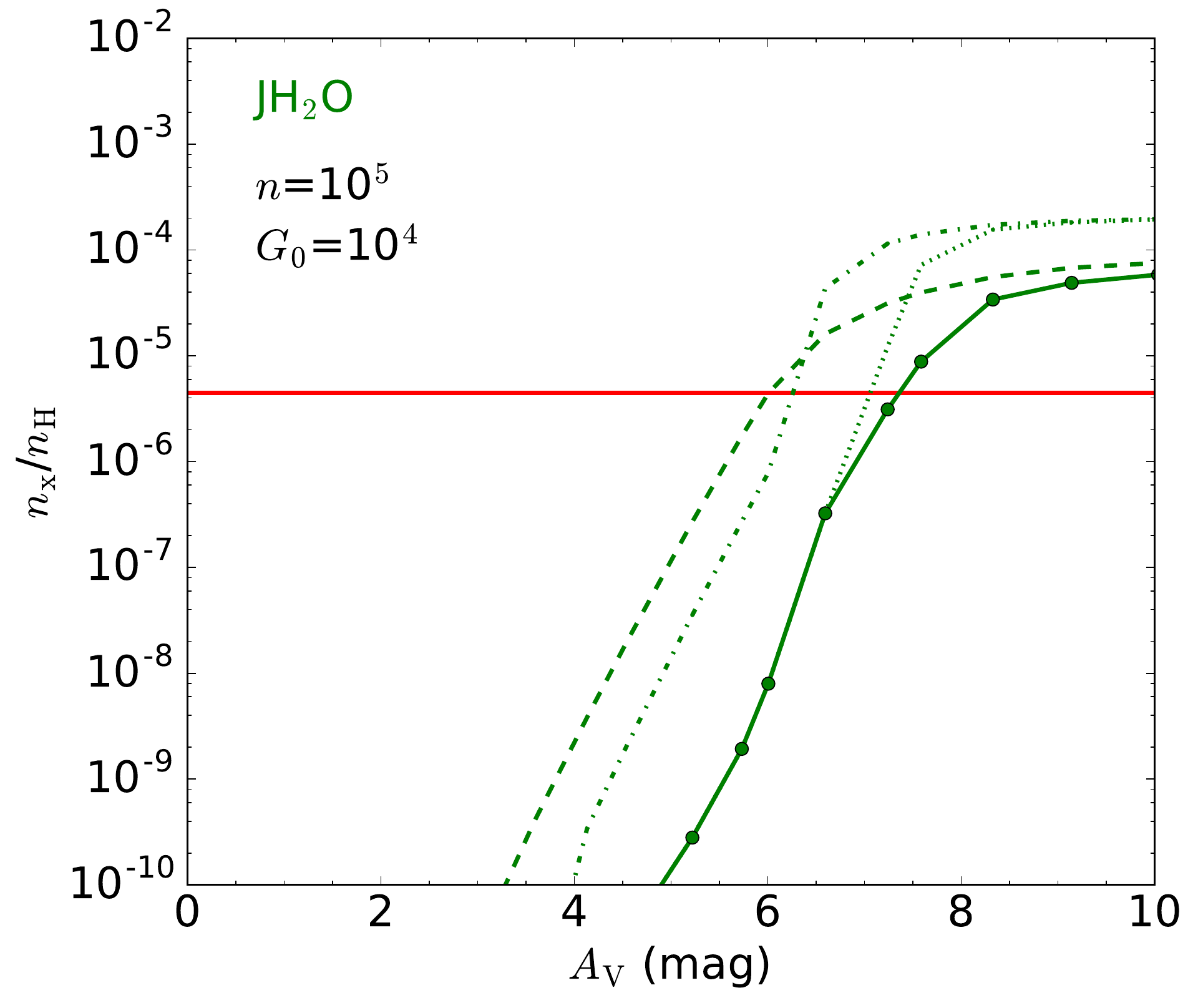}  \hspace{0.00cm}
\includegraphics[scale=0.285, angle=0]{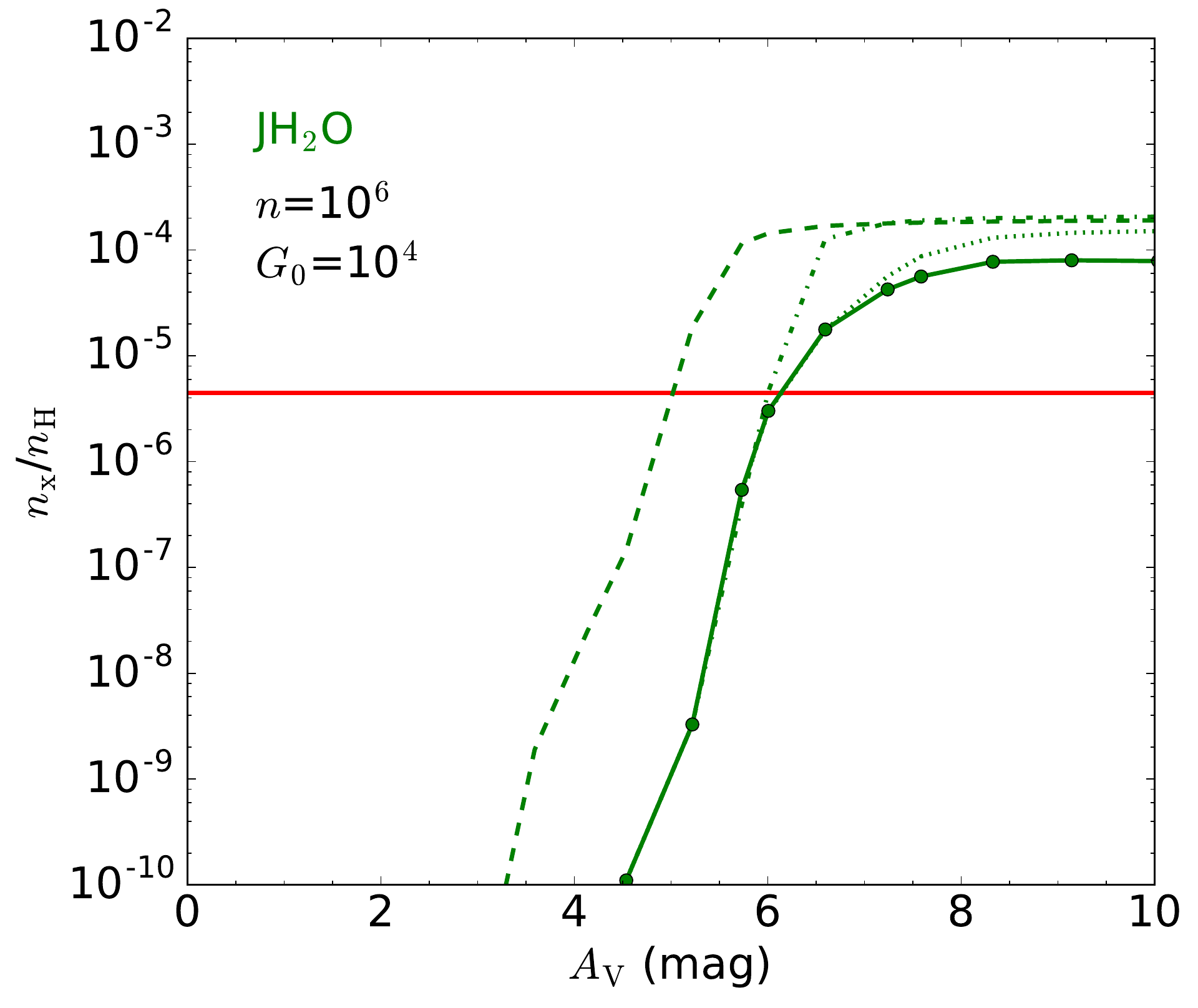}  \hspace{0.00cm}
\\
\caption{Fractional abundances of solid H$_2$O for Models 1 (left panel), 2 (middle panel), and 3 (right panel) as a function of the visual extinction at different timescales
: 10$^4$ yr (dashed line), 10$^5$ yr (dash-dotted line), 10$^6$ yr (dotted line), and 10$^7$ yr (solid-dotted line). 
J$i$ means solid $i$. The red line represents the number of possible adsorption sites on grain surfaces per cm$^{2}$.}
\label{figure:solid_H2O}
\end{figure}

\begin{figure}
\centering
\includegraphics[scale=0.285, angle=0]{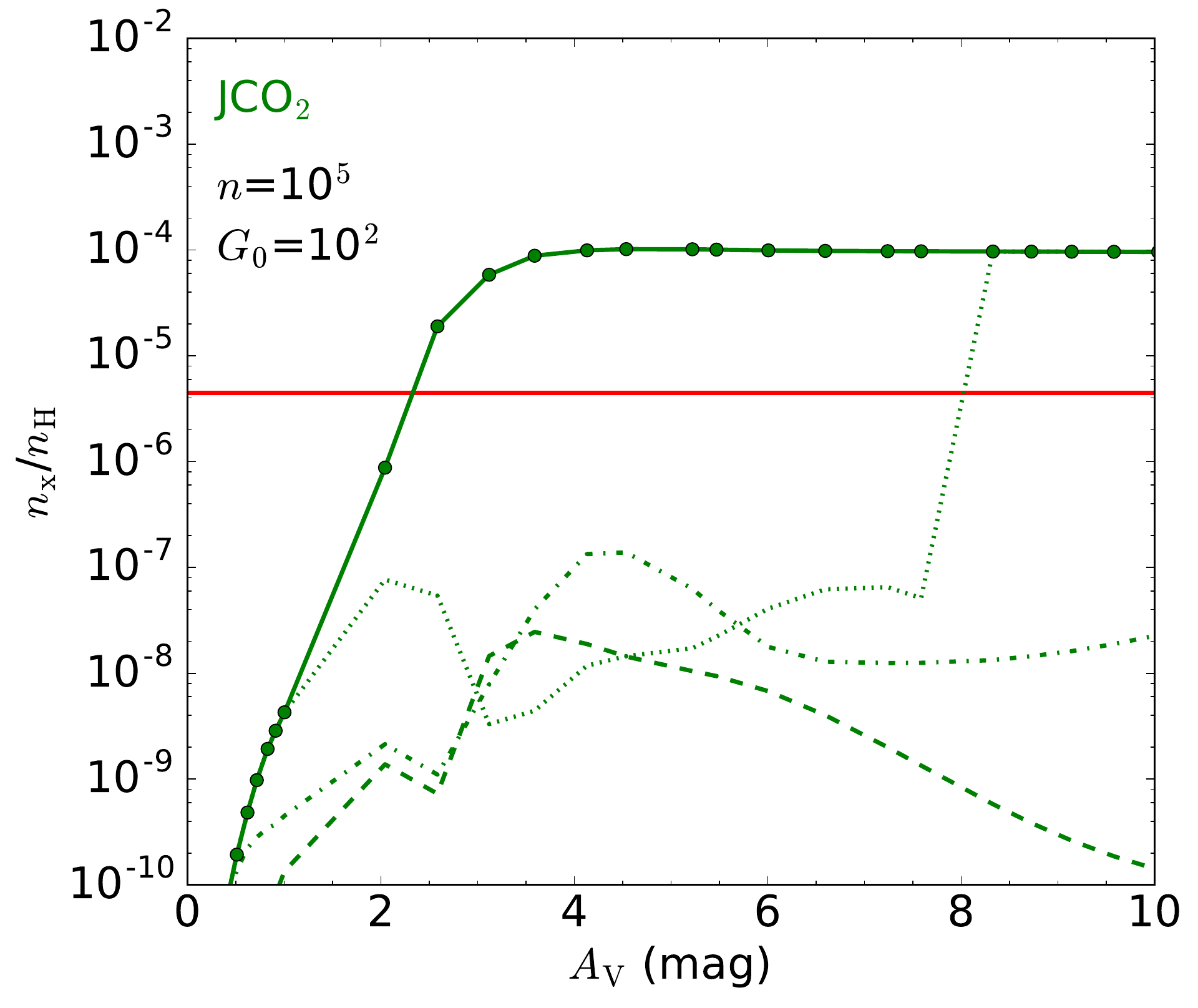}  \hspace{0.00cm}
\includegraphics[scale=0.285, angle=0]{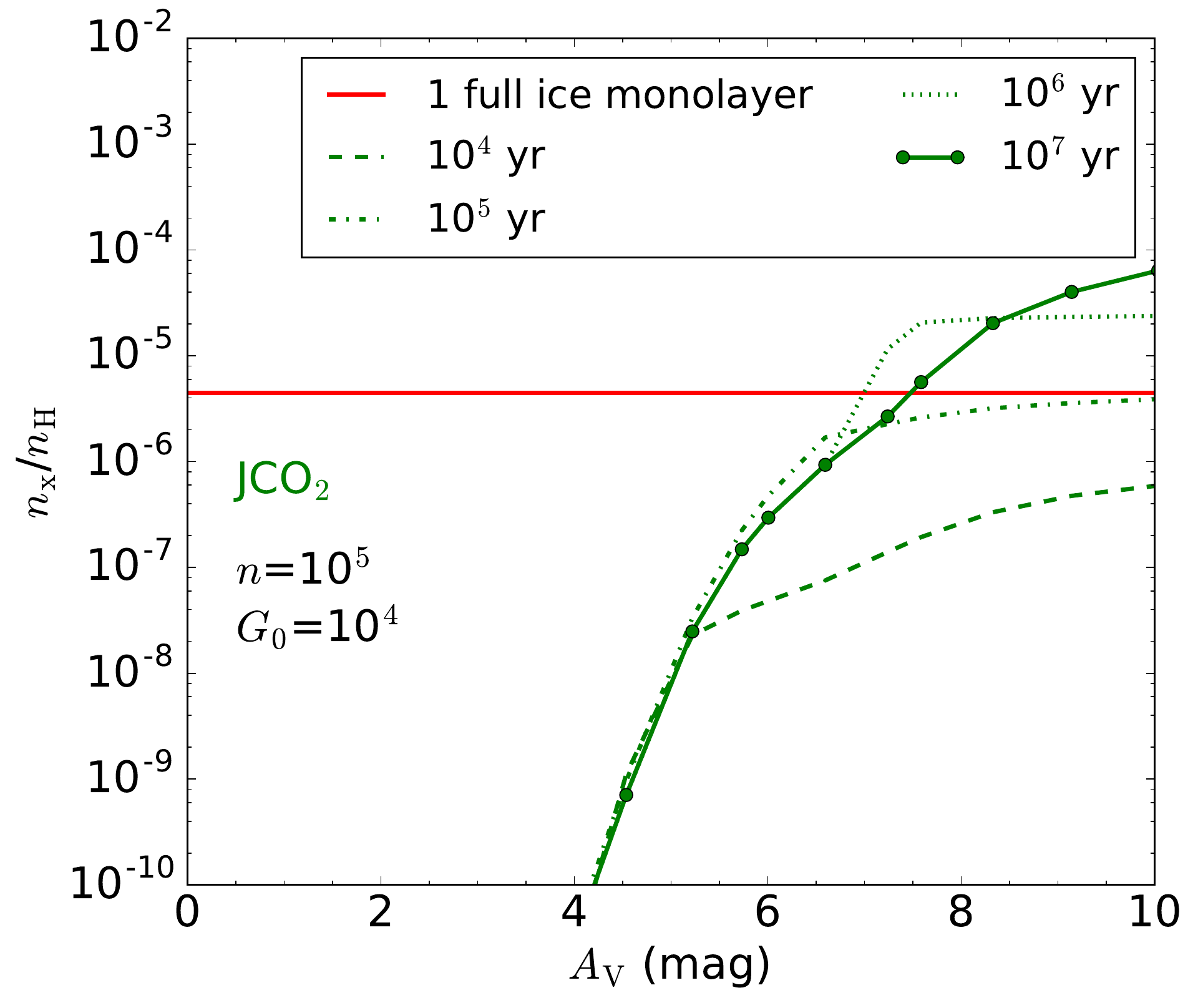}  \hspace{0.00cm}
\includegraphics[scale=0.285, angle=0]{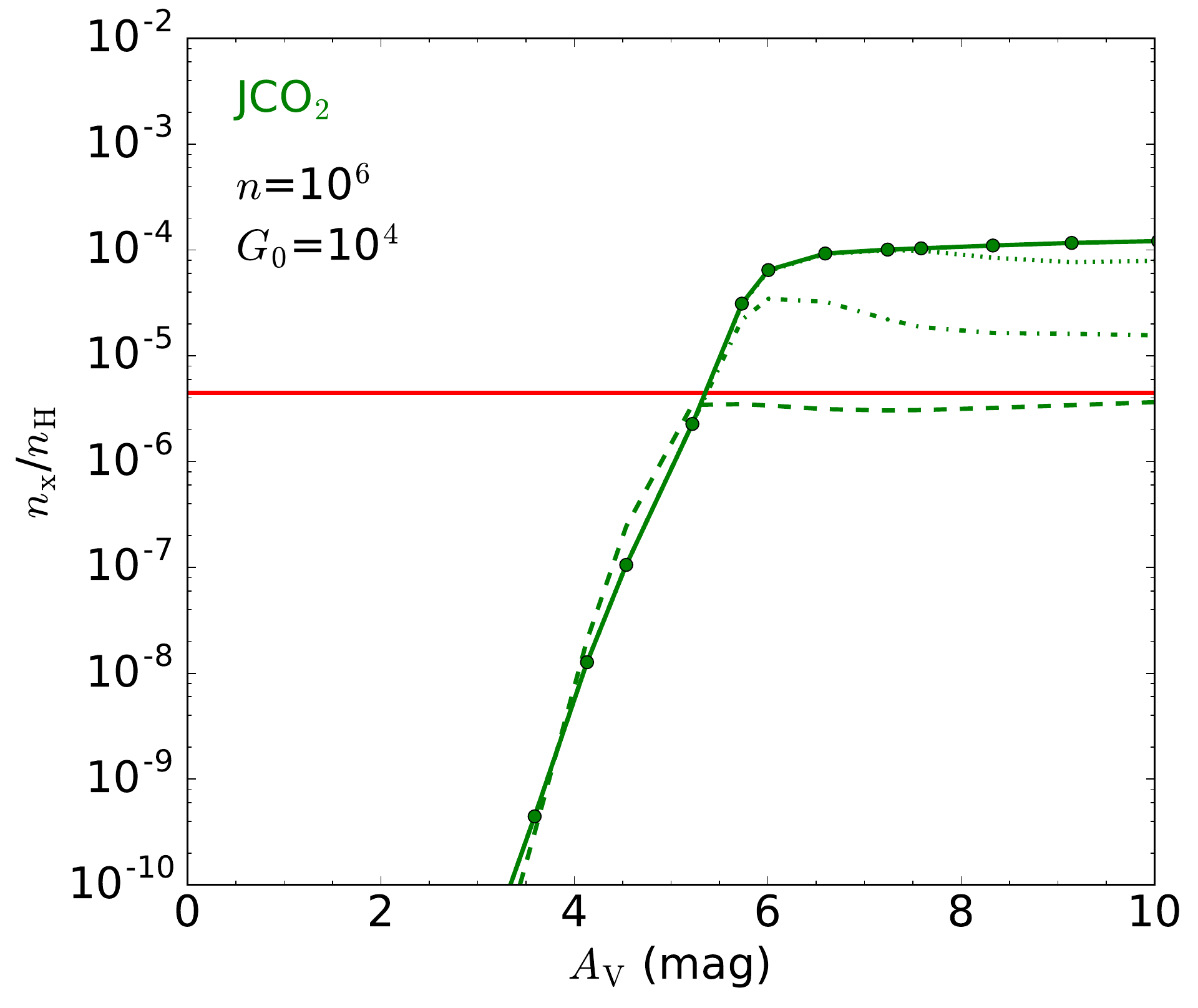}  \hspace{0.00cm}
\\
\caption{Fractional abundances of solid CO$_2$ for Models 1 (left panel), 2 (middle panel), and 3 (right panel) as a function of the visual extinction at different timescales: 10$^4$ yr (dashed line), 10$^5$ yr (dash-dotted line), 10$^6$ yr (dotted line), and 10$^7$ yr (solid-dotted line). 
J$i$ means solid $i$. The red line represents the number of possible adsorption sites on grain surfaces per cm$^{2}$.}
\label{figure:solid_CO2}
\end{figure}

\end{appendix}

\label{lastpage}

\end{document}